%% file: consensus_game.tex
\documentclass[bibtex]{fun_article}
\input{preamble}
\input{authors}

\title{\papertitle}
\ifdefined\anonymous\author{}\else
\author[1,*]{Casper van Elteren\orcid{0000-0001-9862-8936}}
\author[1]{Maria Kleshnina\orcid{0000-0002-5518-8317}}
\affil[1]{\affiliation}
\affil[*]{Correspondence to \href{mailto:casper.vanelteren@qut.edu.au}{casper.vanelteren@qut.edu.au}}
\fi
\date{}
\Keywords{collective decision-making, consensus, veto, emergence, institutions, Antarctic Treaty System}
\begin{document}
\input{sections/abstract}
\input{sections/main}
\clearpage
\bibliography{consensus_game}
\end{document}

\maketitle
\thispagestyle{firstpage}
\tableofcontents
\clearpage
\input{sections/supplement}
\clearpage
\bibliography{consensus_game}

%% file: preamble.tex
\usepackage{fontspec}
\newif\ifFunArticle
\makeatletter
\@ifclassloaded{fun_article}{\FunArticletrue}{\FunArticlefalse}
\makeatother
\usepackage{geometry}
\ifFunArticle
\makeatletter
\renewcommand\normalsize{\@setfontsize\normalsize{9}{11}}
\makeatother
\normalsize
\usepackage{fontawesome5}
\allsectionsfont{\sffamily\bfseries\color{DarkRed}}

\else
\fi
\usepackage{amsmath,amssymb,amsthm}
\usepackage{unicode-math}
\usepackage{graphicx}
\usepackage{booktabs,longtable,array,calc}
\usepackage{float}
\usepackage{caption}
\usepackage{microtype}
\usepackage[super,sort&compress,numbers]{natbib}
\usepackage{xurl}
\usepackage{hyperref}
\hypersetup{unicode=true,hidelinks}
\usepackage{bookmark}
\usepackage{etoolbox}
\usepackage{fvextra}
\DefineVerbatimEnvironment{Highlighting}{Verbatim}{breaklines,commandchars=\\\{\}}

\AtBeginEnvironment{longtable}{\small}

\newcommand{\unpublishedwork}{\ifdefined\anonymous author details withheld, in preparation\else C.v.E., in preparation\fi}
\newcommand{\companioncite}{\ifdefined\anonymous (author details withheld)\else\citep{vanElterenDecisionNetwork2026}\fi}

%% file: authors.tex
\newcommand{\papertitle}{How to Agree Without Agreeing}

\newcommand{\affiliation}{Securing Antarctica's Environmental Future (SAEF), School of Mathematics, Queensland University of Technology, Brisbane, Australia}

%% file: sections/abstract.tex
% At most 150 words for npj.
\begin{abstract}
Consensus lets institutions decide without a central authority, but a record of agreement reveals neither its basis nor its credibility. We model support, standing aside and veto as responses to a proposal. A proposal can pass without active support, and the member least willing to allow passage limits what can be adopted, however many members oppose the proposal. Letting proposers moderate their demands can ease agreement while undermining the prescribed punishment. Proposal restrictions or deviation costs can preserve that threat. Across the initialisations tested, learning agents agree frequently even though many reconstructed rules allow profitable departures. In our sample of Antarctic Treaty working papers, 80\% of proposals revising existing arrangements were adopted at submission, against 38\% of new proposals. Consensus can keep producing agreement on what already exists while new protection waits.
\end{abstract}

%% file: sections/main.tex
\section{Introduction}

In 2025, the Antarctic Treaty Consultative Meeting adopted a revised management plan for the emperor penguin colony at Amanda Bay, while Parties continued to disagree over special protection for the species \citep{ATCM2025FinalReport}. Understanding conservation under consensus requires knowing why members allow some proposals to pass while others remain unresolved.

Consensus is a rule of non-opposition, under which a decision passes when nobody raises a blocking objection \citep{Urfalino2014}. Many international organisations rely on this rule to reconcile competing national interests where no higher authority can impose an agreement. Its appeal lies partly in protecting members against decisions imposed against their will. Buchanan and Tullock argued that unanimity provides such protection at the cost of making collective decisions more difficult to reach \citep{BuchananTullock1962}. Maggi and Morelli subsequently showed that, without external enforcement, an international organisation's voting rule must itself be self-enforcing, potentially making unanimity the only rule that members will respect \citep{MaggiMorelli2006}. Consensus practices shape decisions in the Council of the European Union \citep{Heisenberg2005}, the World Trade Organization (WTO) \citep{Steinberg2002}, climate negotiations \citep{Rietig2023}, and the commission managing fisheries in the Southern Ocean \citep{Goldsworthy2022}. For more than sixty years, states that cannot even agree on who, if anyone, \emph{owns} Antarctica have governed the continent through consensus \citep{AntarcticTreatySecretariatATCM,AntarcticTreatyArticleIV}. These institutions continue to reach agreements despite sharply divergent interests. How much cooperation do these agreements actually reveal?

Under consensus, a member may voice support or stand aside, letting a proposal pass without endorsement. Standing aside can reflect mild support, reluctant acceptance or continuing opposition, yet each response can leave the same record of passage. A record of agreement can therefore conceal differences in what members wanted, how they influenced the decision, and what accepting it cost each member. Without recorded votes, analysts lose an important means of inferring individual preferences \citep{HageHug2016}. Granovetter showed that collective outcomes can provide a poor guide to individual dispositions \citep{Granovetter1978}, so an agreement need not reveal shared preferences. At the General Agreement on Tariffs and Trade (GATT), Steinberg showed how consensus concealed the greater bargaining power of the largest members behind formally equal rights to object \citep{Steinberg2002}. Novak found that European ministers sometimes avoid recording opposition to escape political blame at home \citep{Novak2013}, leaving dissatisfaction unrecorded.

To understand how such agreement arises, we begin with a simple comparison, in which each member evaluates a proposal against what would happen if the group failed to agree. If failure merely preserves the status quo, members willing to exercise their veto can prevent changes they consider unfavourable \citep{Tsebelis2002}. But disagreement need not be costless, nor must it leave existing arrangements unchanged. A proposer may moderate its demands until potential objectors prefer the revised proposal to the alternative, as in models of agenda setting \citep{RomerRosenthal1978}. A member may also let a proposal pass despite preferring a different outcome, because blocking it would impose immediate costs or put future cooperation at risk \citep{Urfalino2014,FudenbergMaskin1986}. These mechanisms can operate together, with moderation changing what is adopted and restraint allowing contested demands to stand. Under a rule that gives every member a veto, the formal power to block is shared, yet the incentives to exercise it can differ considerably. Who is willing to object can therefore matter more than how many members oppose a proposal.

The mechanisms producing agreement also matter for how securely it can be sustained. Members may tolerate unfavourable decisions when the prospect of future cooperation outweighs the immediate benefits of deviation \citep{FudenbergMaskin1986}. Linking cooperation across issues can further strengthen these incentives \citep{BernheimWhinston1990}. Enforcement, however, depends on what members would actually do following a disagreement. Punishment threats can lose credibility when the group would prefer to renegotiate rather than carry them out \citep{FarrellMaskin1989}. Agreement may consequently persist along the course of ordinary interaction while remaining vulnerable to departures from that course. Consensus also raises a question of institutional adaptability. Participation in modifying collective rules is a principle of enduring common-pool resource institutions \citep{Ostrom1990}, and institutional change often proceeds by layering new arrangements onto existing ones or adapting them to new purposes \citep{StreeckThelen2005}. Where every member can block a proposal, revising established commitments may be easier than agreeing on entirely new ones. An institution can therefore continue producing decisions without necessarily demonstrating either strategically stable cooperation or the capacity to respond to emerging problems.

Here, we use game theory to examine how members' choices to support a proposal, stand aside or veto it shape what can pass and what a record of agreement reveals. We bring established mechanisms of proposal adjustment and non-opposition into a common model and derive conditions under which these pathways produce the same record of agreement. A proposal can pass without active support from any member, and the member least willing to allow passage limits what can be adopted, whatever the number of opposed members. Allowing proposers to moderate their demands can ease agreement while giving them a profitable alternative to carrying out a prescribed punishment. We identify when restricting proposals or making departures costly preserves that threat. The analytical credibility window also shows how a valuable future can favour giving way during punishment. Across the initialisations tested, learners often agree even when a member could gain by departing from its reconstructed rule.

We examine these mechanisms in the records of the Antarctic Treaty System, where each Consultative Party present can block the adoption of substantive instruments at the Antarctic Treaty Consultative Meeting (ATCM) \citep{AntarcticTreatySecretariatRules2016} \citep[Decision 2, Rules 21, 24--25]{ATCM2025FinalReport}. Six decades of proposals, revisions, and adoptions provide a history of how agreement is negotiated \companioncite{}. The contrast between revisions and new protection extends beyond Amanda Bay. Of 1,595 working papers submitted to the ATCM since 1991, 366 entered our comparison of proposals to amend existing arrangements or establish new ones. Amendments led to an adopted instrument at the submission meeting considerably more often than proposals for new arrangements. The latter were more frequently deferred or agreed without an instrument than recorded as not agreed. Proposal histories and meeting records allow us to examine how agreement was negotiated and where it failed to materialise. Understanding consensus therefore requires looking beyond what institutions adopt to what they revise, defer, and leave unresolved.

\section{Results}

Consensus records whether a proposal passes, but passage alone does not reveal the preferences and incentives behind a decision. We first ask which responses let a proposal pass, distinguishing support, standing aside and veto (Figure~\ref{fig:routes}). We then ask whether the threat that sustains restraint survives when proposers can adjust their demands (Figure~\ref{fig:flexibility}). Learning experiments ask whether frequent agreement rests on rules that remain in each member's interest to follow at every decision point (Figures~\ref{fig:learning} and~\ref{fig:robustness}). Negotiation histories in the Antarctic Treaty System then show which contributions to agreement can be distinguished from the records (Figure~\ref{fig:record}).

\subsection{Standing aside can hide support or opposition}

\begin{figure*}[t]
\centering
\includegraphics[width=1\linewidth,height=0.65\textheight,keepaspectratio]{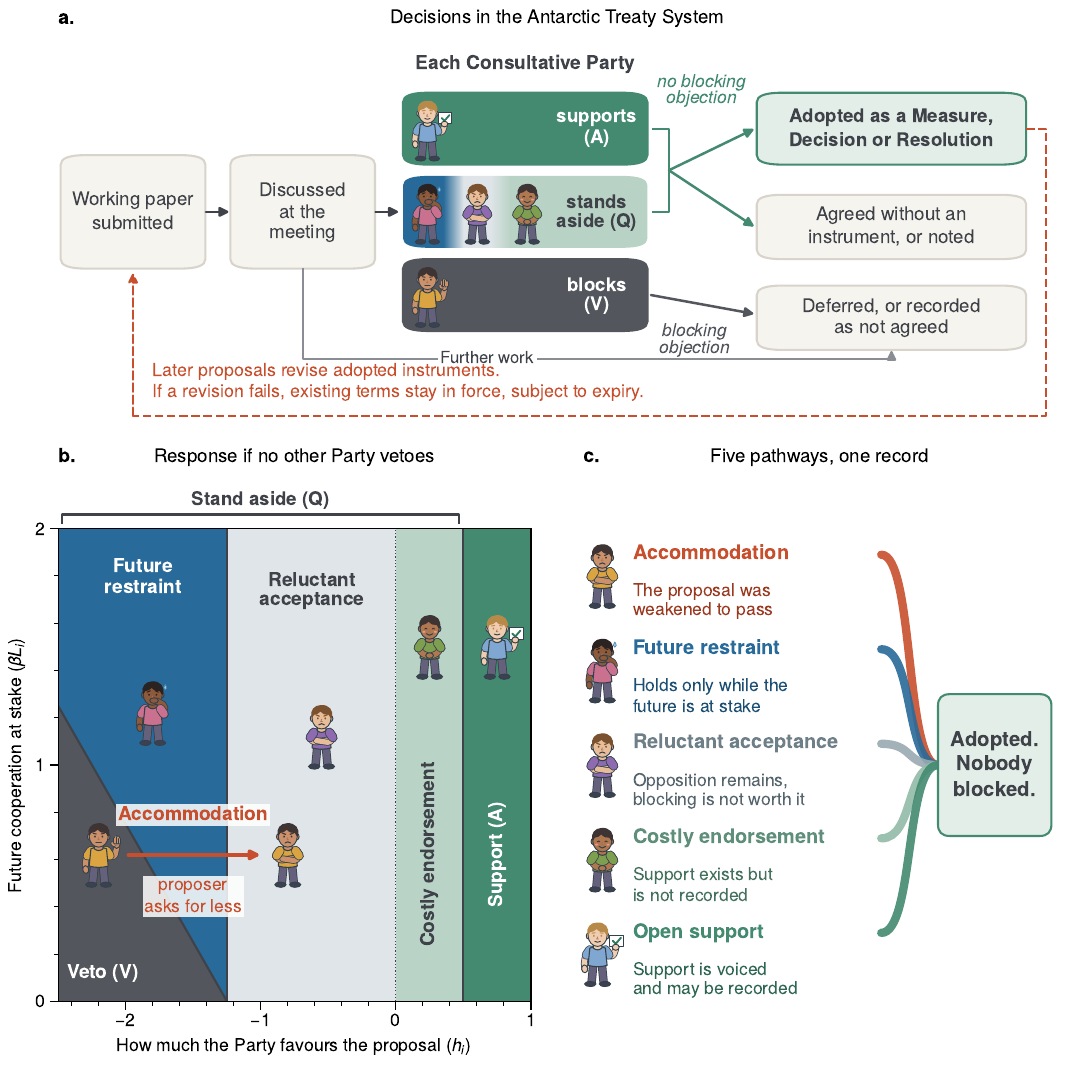}
\caption{Five pathways can contribute to the same record of passage. (a) Simplified decision process at the Antarctic Treaty Consultative Meeting (ATCM). Consultative Parties discuss a working paper and can support its proposal ($A$), stand aside ($Q$) or block it ($V$). Without a blocking objection, the proposal may be adopted as an instrument, agreed without one, noted or deferred for further work. Later proposals can revise adopted instruments. If a revision fails, its predecessor can remain in force, subject to expiry or separate renewal. (b) When no other Party vetoes, a Party's response depends on how much it favours the proposal ($h_{i}$, horizontal axis) and how much future cooperation a veto puts at stake ($\beta L_{i}$, vertical axis). The bracket spans three reasons for standing aside, including opposition held in check by the value of future cooperation. The orange arrow shows accommodation into reluctant acceptance, in which the proposer lowers its request until an objecting Party lets it pass. (c) Accommodation changes the proposal, whereas future restraint, reluctant acceptance, costly endorsement and open support describe responders' reasons for allowing passage. These pathways can coexist across Parties and after accommodation. The panels are schematic, with public-stance cost $\Delta_{i} = 0.5$, veto cost $c_{V,i} = 0.25$ and decision value $b_{i} = 0.5$ in (b).}
\label{fig:routes}
\end{figure*}

We write support, standing aside and veto as $A$, $Q$ and $V$, respectively. A proposal passes if nobody vetoes. In each round of the repeated game one Party proposes, the others respond after observing the proposal, and another round follows with probability $\chi$. Parties discount future payoffs by $\delta$, so the weight on the next round is $\beta=\delta\chi$ (Supplementary Section S2 gives the timing). In the two-Party variants below, proposer and responder roles alternate between rounds. We first hold the proposal fixed and consider the responders.

Conditional on no other Party vetoing, each responder weighs what it gets if the proposal passes against what follows if agreement fails. Relative to standing aside, support pays $h_{i} - \Delta_{i}$, where $h_{i}$ measures how much the Party favours the proposal and $\Delta_{i}$ is the cost of taking a public stance, whether supporting or vetoing. A Party favours the proposal when $h_{i}>0$ and opposes the proposal when $h_{i}<0$. A veto pays $G_{i} - \beta L_{i}$, where $G_{i}$ is the immediate gain from blocking relative to standing aside. Here $L_{i}$ is the difference between future payoffs after passage and after a veto (Methods and Supplementary Section S3). We take $L_{i}$ as given here and determine the continuation loss from what follows a veto in the later subsections. We specify these payoffs in the model and do not estimate them from the records.

These scores satisfy $G_{i} = - h_{i} - \Delta_{i} - c_{V,i} - b_{i}$, where passage value $b_{i} > 0$ is what the Party gains when a decision passes, stance cost $\Delta_{i} > 0$, additional veto cost $c_{V,i} \geq 0$ and $\beta L_{i} \geq 0$. Support and standing aside have the same continuation. Consequently, $h_{i} > \Delta_{i}$ makes support profitable and veto unprofitable relative to standing aside, so support beats both alternatives. When $h_{i} \leq \Delta_{i}$, standing aside is optimal exactly when $G_{i} \leq \beta L_{i}$, with ties at the boundaries. Supplementary Section S3 gives the full three-action comparisons.

The comparison separates five response regions in the fixed-proposal model (Figure~\ref{fig:routes} b). The inequalities define each region. The verbal descriptions name those regions, and the names in parentheses are the labels used in the figure. A Party supports a proposal when its alignment outweighs the cost of saying so ($h_{i} > \Delta_{i}$). A Party mildly favours the proposal when its alignment is positive but does not exceed that cost ($0 < h_{i} \leq \Delta_{i}$). It stands aside because endorsement yields no positive net gain (costly endorsement). A Party is mildly opposed when it is against the proposal but blocking offers no positive current gain ($h_{i} < 0$, $G_{i} \leq 0$), and it also stands aside (reluctant acceptance). A more strongly opposed Party gains from blocking now and stands aside only because a block would cost it future cooperation (future restraint),

\begin{equation}
0 < G_{i} \leq \beta L_{i},
\label{eq:future-restraint}
\end{equation}

and vetoes once the gain from blocking exceeds the discounted future loss ($G_{i}>\beta L_{i}$). At $G_{i}=\beta L_{i}$, veto and standing aside pay equally, and at $h_{i}=\Delta_{i}$, support and standing aside pay equally. For display, we assign these action ties to standing aside. Neutral alignment ($h_{i} = 0$) establishes neither support nor opposition, and such a Party stands aside. Standing aside therefore covers mild support as well as opposition. Increasing the future at stake can widen restraint without changing a Party's alignment.

The proposer can also change a proposal's bite $x\in[0,1]$, which measures how much it demands of the Parties it affects. Each Party has a largest bite it will let pass, ${\overline{x}}_{i}$, which rises with the future at stake. With complete information, the collective limit is

\begin{equation}
x^{\ast} = \min\left\{ 1,\min\limits_{i}{\overline{x}}_{i} \right\}.
\label{eq:bite-limit}
\end{equation}

The Party least willing to allow passage sets what can pass, whatever the number of opposed Parties. A proposer facing a veto can ask for less, moving the binding Party out of the veto region (Figure~\ref{fig:routes} b). We call this accommodation. It changes the proposal, while standing aside concerns responses to the proposal actually considered. Supplementary Section S6 gives the threshold derivation, and the Antarctic tourism stress test in S25 illustrates its dependence on who objects.

\subsection{A record of agreement does not reveal how it was reached}

The five response regions in Figure~\ref{fig:routes} b include veto. The five agreement pathways in Figure~\ref{fig:routes} c instead combine the four non-veto response regions with accommodation. Accommodation can precede any combination of non-veto responses, including unanimous support for the revised proposal. On the proposal actually considered, reluctant acceptance means standing aside despite opposition because blocking offers no positive current gain, whereas future restraint means standing aside because future losses offset a positive current gain from vetoing. Under costly endorsement a Party favours the proposal but does not voice support, whereas under open support it favours the proposal and says so. Different Parties can respond for different reasons, and the record of passage does not identify their contributions.

We call passage with at least one Party standing aside from the adopted proposal \emph{apparent consensus}. This definition does not require concealed opposition. A Party may stand aside despite favouring the proposal. Accommodation produces apparent consensus only if at least one Party stands aside from the adopted proposal.

Recorded responses can bound its share. In an audited sample of $M$ passed decisions, let $D$ have a documented stand-aside from the adopted proposal and $E$ have documented active support from every responder. The share of apparent consensus lies between $D/M$ and $1 - E/M$ (Supplementary Section S12.1). Passage alone, with neither response documented, leaves the share anywhere between zero and one. An objection to an earlier draft does not by itself establish a stand-aside from the adopted text.

\subsection{Flexible proposals can weaken the threat behind restraint}
\label{sec:flexibility}

\begin{figure*}[t]
\centering
\includegraphics[width=0.95\linewidth,height=0.45\textheight,keepaspectratio]{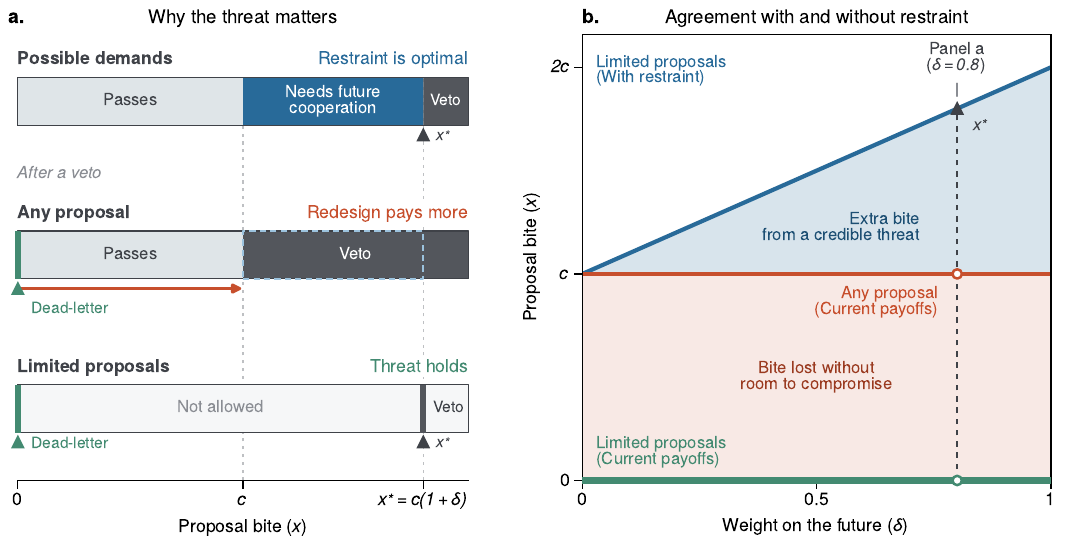}
\caption{Flexibility can weaken the threat supporting demanding agreements. (a) Each candidate strategy profile prescribes a target bite accepted in ordinary play and permanent dead-letter passage after a deviation. After any other proposal, the responder allows bite up to $c$ and vetoes a larger bite, since either response leads to punishment. The top strip compares the responder's incentives across possible target bites. Current payoffs sustain restraint up to veto cost $c$, and future cooperation extends the bound to $c(1+\delta)$. Dark triangles mark the largest target supported by this trigger, $x^{\ast}=c(1+\delta)=0.9$. At this boundary the responder is indifferent, and ties favour passage. After a veto, prescribed dead-letter passage gives both members zero. If any proposal is allowed, the arrow marks a profitable redesign. The dashed range would be vetoed during punishment. Restricting the menu to the target and the dead-letter removes this profitable departure. (b) Blue traces the largest target this trigger supports as the substantive option varies across restricted menus. For one fixed target, passage stays at that target wherever the trigger is credible. Orange and green show history-independent equilibria. Unrestricted proposals pass at $c$ each round, whereas a menu containing only the dead-letter and a target above $c$ permits only dead-letter passage. Shading compares bite under these specified equilibria without predicting which equilibrium play selects. The dashed vertical line at $\delta=0.8$ in (b) corresponds to (a), with circles marking zero bite and $c$ and a triangle marking $x^{\ast}$ on the blue boundary. Here $c=0.5$ and $\delta=0.8$ in (a), and the blue curve's endpoints at $\delta=0$ and $\delta=1$ are limiting bounds. Supplementary Section S8 gives the conditions.}
\label{fig:flexibility}
\end{figure*}

Making agreement easier can weaken the threat that sustains restraint. Here two members take turns as proposer in the repeated game, and another round always follows, so $\beta=\delta$. A passed proposal of bite $x$ gives the proposer $x$ and costs the responder $x$. The responder pays a total cost $0<c<1$ to veto. We consider a trigger strategy profile. In ordinary play the proposer offers a target bite $x$ and the responder stands aside. A veto or any other proposal ends substantive cooperation and starts permanent passage of the dead-letter, an instrument at zero bite with no substantive effect and zero payoff for both members. The responder gives up $x$ now but proposes in the next round, so standing aside gives the responder $-x/(1+\delta)$, compared with $-c$ for a veto. Avoiding the punishment can therefore sustain restraint at $c<x\leq c(1+\delta)$ within the feasible bite range (Figure~\ref{fig:flexibility} a, top, and b).

That threat must also be worth carrying out. If any proposal is allowed during punishment, the responder accepts a bite up to the veto cost because either response leaves the same future. The proposer can then replace the dead-letter with a modest substantive proposal and earn a positive payoff (Figure~\ref{fig:flexibility} a, middle). The prescribed punishment is no longer optimal for the proposer.

Restricting the menu to the original demand and the dead-letter removes this profitable departure and preserves the specified rule (Figure~\ref{fig:flexibility} a, bottom). For $0<\delta<1$, the largest target this trigger supports is $x^{\ast}=\min\{1,c(1+\delta)\}$, shown in Figure~\ref{fig:flexibility}. A minimum substantive bite or a charge for changing the prescribed proposal can also preserve the threat under the conditions in Supplementary Section S8. With unrestricted proposals, another equilibrium sustains passage every round at the current-payoff acceptance limit $c$ (Figure~\ref{fig:flexibility} b). Flexibility can therefore preserve agreement at lower bite even when the original threat fails.

\subsection{Frequent agreement need not imply credible rules}

\begin{figure*}[t]
\centering
\includegraphics[width=1\linewidth,height=0.65\textheight,keepaspectratio]{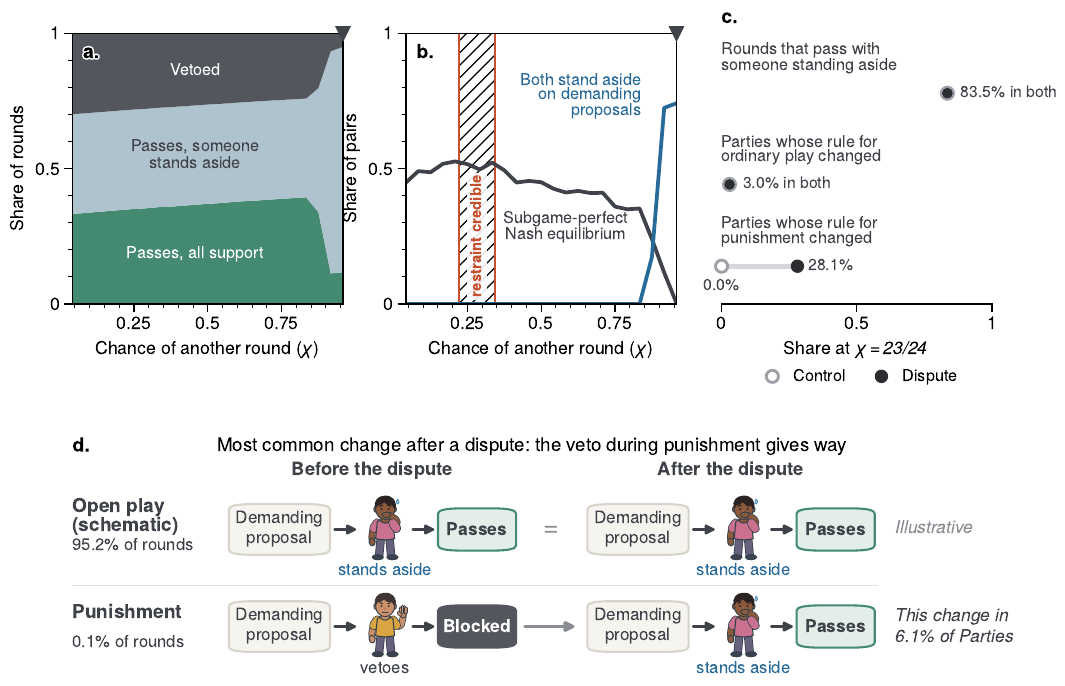}
\caption{Frequent agreement can rest on rules that learners rarely test. All panels use a gain of 0.15 from vetoing a demanding proposal and 1,000 pairs trained from zero initial action values for 2,000,000 rounds at each continuation probability $\chi = k/24$, $k = 1,\ldots,23$. (a) Outcome shares over the final 4,000 training rounds. Vetoed rounds include vetoes that lead to revision and vetoes that trigger punishment. (b) Share of pairs whose reconstructed rules form a subgame-perfect Nash equilibrium (pass the full incentive audit), and share in which both Parties stand aside on demanding proposals in open play. The hatched band marks where restraint in ordinary play backed by a punishment veto is credible. Triangles mark $\chi = 23/24$, the setting of (c) and (d). (c) Each trained pair resumes learning for 2,000 rounds in punishment (dispute) or in open play (control), with 20 replicates per pair on matched random streams. The top row gives the share of the last 1,000 resumed rounds that pass with someone standing aside. The lower rows give the share of Parties whose rule for ordinary play (open play and revision) or for punishment changed over the resumed rounds. (d) The most common rule change after a dispute. A Party that vetoed demanding proposals during punishment stands aside on them (6.1\% of Party-replicate records). The open-play row is schematic, and the 6.1\% does not condition on the open-play prescriptions shown. Percentages on the left give the shares of the final training rounds spent in open play and in punishment. Parameters are $\delta = 0.9$, $\rho = 0.65$ and $\tau = 0.25$.}
\label{fig:learning}
\end{figure*}

In the learning experiments, two Parties take turns proposing from a menu of two bites. Payoffs are fixed and the same for both Parties. Alignment falls with bite, $h(x)=h_{0}-\lambda x$, so the responder favours the modest proposal ($x=0.25$) and opposes the demanding one ($x=1$). The public mode records whether play is open, in revision or in punishment. We call open play and revision together ordinary play. A rule is a deterministic strategy specifying a proposal in each mode and a response to every available proposal. At the reference values, vetoing the demanding proposal gains 0.15 relative to standing aside. A veto of a demanding proposal in open play leads to revision, where only the modest proposal is available. Other vetoes trigger punishment, where passage restores open play with probability $\rho=0.65$. The discount factor is $\delta=0.9$. Methods and Supplementary Section S10 give the payoffs and the remaining parameter values.

Each learner updates its values for the available proposals and responses from the payoffs received during play, choosing actions with higher learned values more often (softmax choice with temperature $\tau=0.25$ and SARSA updates, Methods). Within each main-figure run, the payoff parameters, discount factor, recovery probability and temperature stay fixed, while the learning rate decreases with visits. Across experiments we vary the continuation probability $\chi$, the veto gain through the intercept $h_{0}$, the initial values and the training length (Methods and Supplementary Section S10.5).

\phantomsection\label{sec:learning-credibility-window}In this two-bite game, a valuable future can also make punishment less attractive. The symmetric rule prescribes passage of the demanding proposal with a stand-aside in open play, backed by a veto of demanding proposals during punishment. This rule is credible, meaning subgame perfect, only for $2/9\leq\chi\leq0.343$ (hatched band in Figure~\ref{fig:learning} b). Above this window, standing aside during punishment pays more than the prescribed veto because passage offers a chance to restore the relationship (Supplementary Section S10.1).

We reconstruct each Party's deterministic strategy by retaining its most likely action at every proposal and response node. The pair forms a strategy profile. The path check asks whether either Party gains by changing its action at nodes reachable from open play under that pair of rules. The full incentive audit tests whether the pair of rules is a subgame-perfect Nash equilibrium, meaning that neither Party can gain by changing its strategy after any history. It checks every feasible proposal and response for each Party in open play, revision and punishment, including responses to proposals neither Party prescribes. At each node, we check for a profitable one-shot deviation, in which a Party changes one action and then returns to its prescribed strategy while the other Party follows its strategy. Under the one-shot-deviation principle, a pair with no such deviation is a stationary subgame-perfect Nash equilibrium, to the numerical tolerance stated in Methods and under the model's assumptions. This classification concerns frozen prescriptions. Continued learning, including exploratory actions, is a separate process (Methods). Audits that instead freeze the learners' action probabilities, retaining exploration, are reported in Supplementary Section S10.6. A failed audit identifies an incentive to depart at a decision point, rather than a probability that observed agreement will collapse.

\begin{figure*}[t]
\centering
\includegraphics[width=0.92\linewidth,height=0.65\textheight,keepaspectratio]{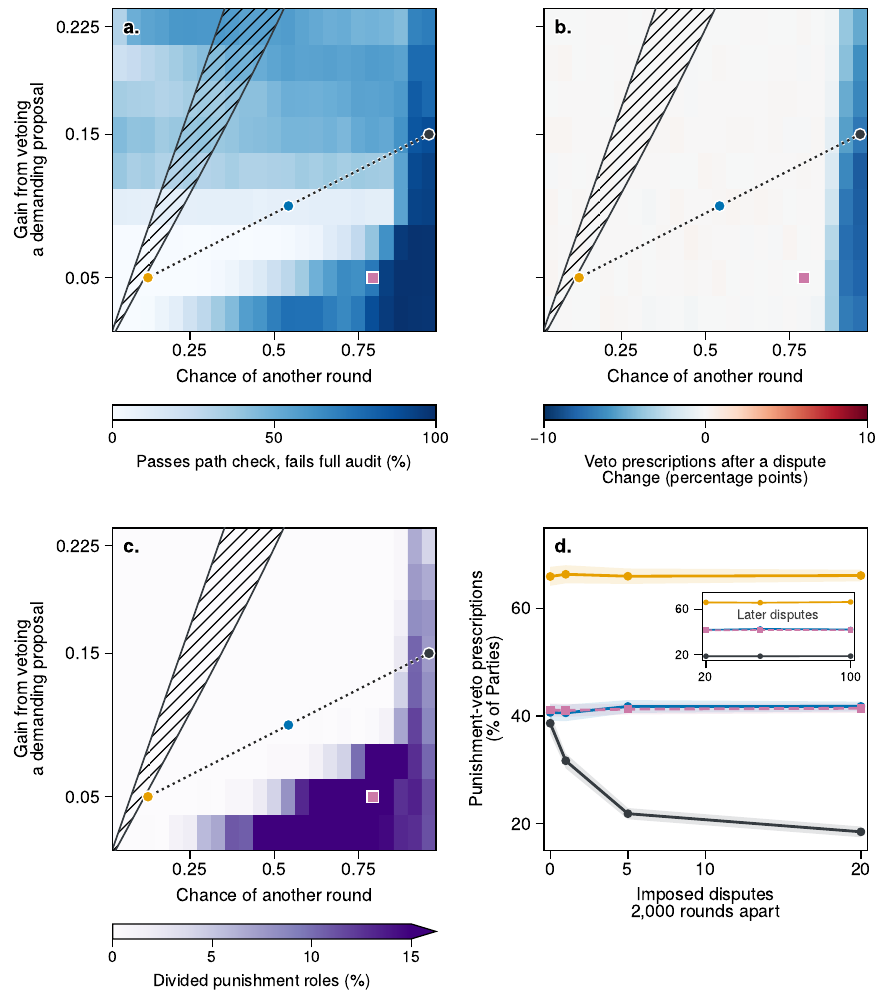}
\caption{With zero initial action values, frequent agreement can coexist with rules that fail the full incentive audit. Panels (a)--(c) show unsmoothed cell estimates with linear colour scales across 207 combinations of continuation probability $\chi$ and gain from vetoing a demanding proposal. Each setting has 1,000 pairs trained for 2,000,000 rounds. Hatching marks where the specified symmetric rule prescribing restraint backed by a punishment veto is credible. Coloured markers identify the settings compared in (d). Grey, blue and orange circles mark $(\chi,\mathrm{gain})=(23/24,0.15)$, $(13/24,0.10)$ and $(3/24,0.05)$, joined by the dotted line. The pink square marks $(19/24,0.05)$, where divided punishment roles are most common. (a) Share of pairs passing the path check but failing the full audit. (b) Change against matched controls in the share of Parties whose response rule vetoes the bite their own proposer rule prescribes during punishment, after one imposed dispute and 2,000 resumed rounds, in percentage points. The available bites remain fixed at 0.25 and 1, but the proposer continues learning which bite to choose. We check the response to the bite that each Party currently prescribes as proposer, so the response being checked can change as proposals change. Each pair has 20 replicates per arm. (c) Share of pairs matching the specified division of punishment roles. Both Parties stand aside on demanding proposals in ordinary play. During punishment, one proposes the demanding bite and vetoes a demanding proposal from the other Party. The other proposes the modest bite and stands aside on the demanding one. Colours saturate above 15\%. (d) Absolute share of Parties with the punishment-veto prescription defined in (b), after 0, 1, 5, 20, 50 or 100 imposed disputes. Each marked setting has 1,000 trained pairs and five continuation replicates per dose. Disputes start play in punishment 2,000 rounds apart, leaving 2,000 rounds after the last dispute. Learning continues between disputes, and no action is forced. Matched controls receive the same learning time without imposed disputes (Supplementary Section S10.13). The inset shows 20--100 disputes. Shading gives pointwise 95\% intervals from 5,000 bootstrap resamples of trained pairs. Other parameters follow Figure~\ref{fig:learning}.}
\label{fig:robustness}
\end{figure*}

Learners agree frequently even when their reconstructed rules allow profitable departures, under both zero and positive initial values. We test 207 combinations of continuation probability and the gain from vetoing, training 1,000 pairs for 2,000,000 rounds at each setting. Many pairs pass the check along their prescribed path while failing the full audit (Figure~\ref{fig:robustness} a). Across settings with $\chi\geq0.75$, this share ranges from 10.3\% to 99.2\% with zero initial values. The comparison of initial values retains the reference veto gain of 0.15. At $\chi=23/24$, positive initial values of $0.5/(1-\beta)$ or $1.25/(1-\beta)$ yield agreement in 76.6\% of the final 4,000 rounds. About 86\% of reconstructed pairs pass the path check, while 38.0--39.3\% pass the full audit (Supplementary Section S10.12). These results apply to the tested learning procedure, whose parameters are not calibrated to an institution.

With zero initial values at the same setting, agreement reaches 95.1\%, while 88.9\% of reconstructed pairs pass the path check and only 0.3\% pass the full audit (Figure~\ref{fig:learning} a, b). Apparent consensus rises from 36.4\% under positive starts to 83.5\% under zero starts. Both reconstructed rules prescribe standing aside on demanding proposals in 74\% of zero-start pairs. Up to $\chi=5/6$, by comparison, apparent consensus accounts for about 37\% of zero-start rounds, between 35\% and 53\% of pairs pass the full audit, and no pair has both Parties standing aside on demanding proposals. Both positive and zero starts therefore yield frequent agreement despite failed audits, while the extreme high-continuation pattern occurs only under zero starts.

The gap lies mainly outside ordinary play. At $\chi = 23/24$, 94\% of the 2,000 Parties could gain from a departure somewhere, compared with 5.7\% along the prescribed path. Punishment accounts for only 0.1\% of the final training rounds there, against 3.4\% at $\chi = 5/6$, the shorter future we use for comparison. Most profitable departures concern punishment, revision or responses to proposals the pair does not usually make. The scarcity of experience in these situations is consistent with the audit failures, although the comparison does not isolate its effect from exploration or state-specific learning rates. Supplementary Section S10.11 gives the node-level diagnostics.

Learners also fail to reach and retain the symmetric restraint rule where it is credible. None of 600 runs trained for 8,000 rounds inside the window finds it, nor do 300 runs trained for 100,000 rounds. When 600 pairs instead start at the credible rule, all leave it within 300 rounds, usually after a responder begins to veto a demanding proposal. The credible rule was therefore not retained under the learning dynamics considered here. This is a result about adaptation with exploration and noisy value updates, rather than a failure of the rule's equilibrium conditions. The experiments separating these effects are in Supplementary Section S10.8.

Longer training changes the result in some settings. At $\chi = 5/6$, apparent consensus falls from 84\% after 8,000 rounds to about 37\% from 512,000 rounds onwards, while the share of pairs passing the full audit rises from 0.8\% to 35.3\%. At $\chi = 23/24$, by contrast, apparent consensus stays between 81\% and 84\% across 100 matched pairs trained for 8,000 to 2,000,000 rounds, and none of these 100 pairs passes the full audit at those training lengths. Extending the same 100 high-continuation seeds to 200 million rounds reduces apparent consensus to 79.5\% and raises the full-pass share to 26\%. Frequent agreement can therefore arise well before a rule passes the full check, although these finite runs do not establish eventual convergence (Supplementary Section S10.12). These comparisons do not isolate experience as the cause, because policies, partners and learning rates change together. Methods and Supplementary Sections S10.5--S10.8 report the shorter experiments, exploratory-policy audits and learning diagnostics.

Because punishment is rarely visited at $\chi = 23/24$, a dispute there puts learners in situations their training seldom reached. To examine behaviour after disagreement, we restart each trained pair in punishment and compare it with a control restarted in ordinary play, retaining learned values and visit counts and using matched random streams (Figure~\ref{fig:learning} c). No action is forced. Responders veto at least once in 55.0\% of first dispute episodes, compared with 23.6\% of control episodes. After 2,000 resumed rounds, punishment rules change for 28.1\% of Parties against fewer than 0.1\% in controls, while ordinary-play rules change for 3.0\% in each arm. Over the final 1,000 rounds, apparent consensus is 83.5\% in both arms. The most common punishment-response change is to stand aside on a demanding proposal that the Party previously vetoed (6.1\%, Figure~\ref{fig:learning} d). At $\chi = 5/6$, by comparison, the same intervention changes punishment prescriptions little, and late apparent consensus is about 36.5\% in both arms (Figure~\ref{fig:robustness} b).

Across the 207 settings, we ask whether a Party's response rule would veto the bite its own proposer rule prescribes during punishment. One imposed dispute changes the share with this veto prescription by between −8.3 and +0.3 percentage points against matched controls. All 16 settings with a decline exceeding 5 points have $\chi \geq 11/12$. Late apparent-consensus rates, however, differ by less than 0.01 percentage points between arms in every setting (Figure~\ref{fig:robustness} b). Thus the dispute can change prescriptions that the aggregate agreement rate leaves almost invisible.

Credible restraint is nevertheless possible above the symmetric window when Parties divide punishment roles. One Party continues to make demanding proposals during punishment and stands ready to veto the other's demanding proposals. The other makes modest proposals and stands aside on the demanding ones. Along this pair's prescribed path, punishment costs the Party holding the threat nothing relative to ordinary play, because its demanding proposals still pass and it faces only modest proposals. At the reference gain, this specified pair passes the full audit above approximately $\chi = 0.338$ (Supplementary Section S10.10). Up to 52.2\% of trained pairs match the prescriptions defining these divided roles across the grid, but these are partial matches. At $\chi = 23/24$, none of the 75 matching pairs passes the full audit (Figure~\ref{fig:robustness} c).

Repeated disputes shift punishment responses towards standing aside at the high-continuation reference setting with zero initial values (Figure~\ref{fig:robustness} d). After 100 disputes, punishment-veto prescriptions fall to 18.6\%, compared with 38.8\% in matched controls, while apparent consensus remains at 83.6\% in both arms. The other marked settings show little change, and the decline largely disappears with positive initial values. With the higher positive initial values, one dispute changes the veto-prescription share by 0.06 percentage points (95\% interval $-0.24$ to $0.36$). These shifts do not establish accurate learning of continuation costs or convergence to credible rules (Supplementary Section S10.13).

\subsection{The Antarctic record can still tell some pathways apart}

\begin{figure*}[t]
\centering
\includegraphics[width=1\linewidth,height=0.65\textheight,keepaspectratio]{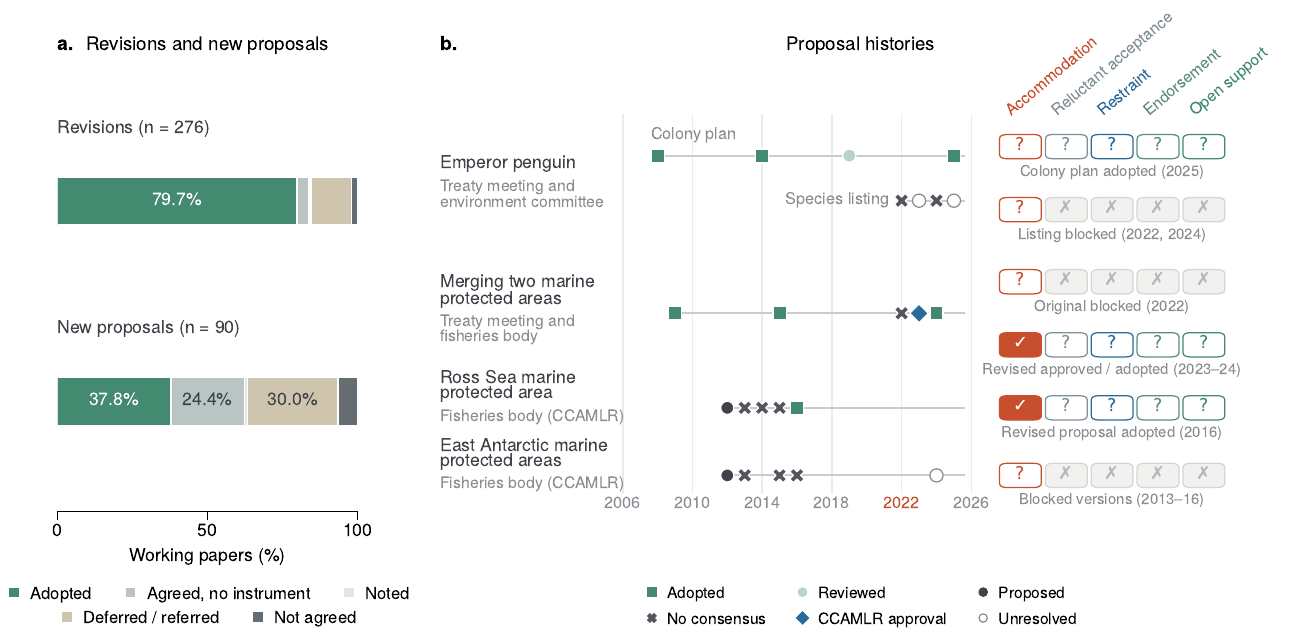}
\caption{The Antarctic record can still tell some pathways apart. (a) Revisions of existing instruments are more often adopted at the submission meeting than proposals for new arrangements. Bars show raw outcome shares for working papers at ATCM 16--47. After standardising over agenda route, era and site subject, the adoption gap is 28.0 percentage points (95\% interval 17.1--41.2, bootstrapping meetings). (b) Selected proposal histories connect objections, revisions and adoptions with the five pathways. Each matrix row gives its proposal object and decision stage. Ticks support an accommodation reading of revisions followed by approval or adoption, crosses exclude pathways for objectors on the blocked versions, and question marks leave pathways open. Earlier objections leave response pathways at later adoption open. The colony plan and species listing are separate proposals whose juxtaposition does not establish accommodation. The cases illustrate the model. Supplementary Sections S22.1 and S22.3 give the outcome checks and case evidence.}
\label{fig:record}
\end{figure*}

The Antarctic records preserve objections, changes to proposals and outcomes, but rarely reveal every Party's reasons for allowing passage. We examine what these traces support or rule out. The working-paper comparison uses submissions through ATCM 47 (2025). The ATCM 48/CEP 28 report (2026) supplies a later update to the emperor penguin history below, outside the frozen working-paper sample. We identified these patterns with the records in view, so the comparisons illustrate possible mechanisms rather than independently test the model (Methods).

\subsubsection{Revisions more often yield immediate adoption}

Of 1,595 working papers submitted to ATCM 16--47, 366 enter the classified comparison after outcome checks. Proposals to revise existing arrangements end in an adopted instrument at the submission meeting 80\% of the time (220/276), against 38\% for proposals to establish new arrangements (Figure~\ref{fig:record} a). The raw difference of 42 percentage points falls to 28 after standardising over agenda route, era and site subject (95\% interval 17 to 41). The source checks and coding sensitivities are reported in Methods and Supplementary Section S22.1.

The later outcomes qualify this difference. In the follow-up comparison restricted to instrument-seeking proposals with sufficient, resolved follow-up, 97\% of revisions and 83\% of new proposals yield an instrument within two meetings. The standardised difference is 10 points (95\% interval 1 to 20) in this restricted follow-up sample. At two meetings, high-confidence matching gives 7.8 points (−1.4 to 18.5) and alternative model-assisted coding gives 4.8 points (−5.1 to 15.0). Both intervals include zero. The immediate-adoption difference is therefore more secure than the difference within two meetings. The records do not establish an ultimate difference in adoption. Among new proposals, deferral or referral (30\%) and agreement without an instrument (24\%) are each more common than recorded non-agreement (7\%, compared with 2\% for revisions).

The model offers one explanation. A revision may demand only an incremental change to an arrangement already accepted. If the existing instrument has not expired, a failed revision can leave it in force, while a failed new proposal introduces no new instrument. A smaller additional demand may stay below every Party's blocking threshold. Other explanations remain possible. Many management-plan revisions follow the scheduled reviews required by Annex V \citep{ATCM1991AnnexV}. New proposals may be broader or establish precedents, and only arrangements that previously passed can be revised. The aggregate comparison does not separate these effects.

Emperor penguin protection illustrates the difference in proposal scope. In 2022 all CEP Members but one supported designating the species for special protection, so the Committee could not advise designation \citep{ATCM2022FinalReport}. In 2025 China co-proposed the revised management plan for the Amanda Bay colony, which passed, while continuing to oppose species listing \citep{ATCM2025FinalReport}. The plan revises an instrument adopted in 2008, whereas listing would establish a new protection for the species. The same objecting Party therefore accepts narrower site protection while opposing broader species protection. Amanda Bay is a separate proposal, rather than a demonstrated accommodation of the listing text.

\subsubsection{Proposal histories distinguish accommodation from
persistent blocks}

The pathways can coexist within a proposal history. A blocking objection rules out standing aside by that objector on that version at that decision stage. It does not classify other members or their later responses. Revision followed by adoption supports accommodation, although the adopted proposal can still combine open support with different reasons for standing aside. Four histories show what these sequences add (Figure~\ref{fig:record} b, Supplementary Section S22.3).

The emperor penguin listing remained unagreed from 2022 to 2024, and the 2025 statements record continuing disagreement without a new listing proposal \citep{ATCM2022FinalReport} \citep{ATCM2023FinalReport} \citep{ATCM2024FinalReport} \citep{ATCM2025FinalReport}. At ATCM 48 in 2026, China again considered designation premature and Russia supported that intervention. The Meeting agreed to continue considering appropriate protection measures at its next meeting \citep[paras. 93, 97; pp. 28--29]{ATCM2026FinalReport}. This later record updates the history but is outside the frozen adoption and statement comparisons.

The merger of ASPA 152 and 153 shows a different sequence. China and Russia objected in 2022 to a proposed expansion of the two marine protected areas, and CCAMLR did not approve the expansion \citep[p. 31]{CCAMLR2022Report}. The United States then retained the original areas and placed their upper boundary 20 m below the sea surface. CCAMLR approved the revised plan in 2023, and ATCM adopted it as ASPA 182 in 2024 \citep[p. 43]{CCAMLR2023Report} \citep[Measure 17, pp. 207--208]{ATCM2024FinalReport}. The sequence supports accommodation, although individual Members' responses at approval and Parties' responses at adoption remain unresolved.

For the Ross Sea marine protected area, proposals begun in 2012 underwent successive revisions before adoption in 2016. Changes met China's concerns in 2015 and Russia's remaining concerns in 2016 \citep[pp. 57--58]{CCAMLR2015Report} \citep{CCAMLR2016Report}. This supports accommodation, without establishing that every Member actively supported the final proposal. The East Antarctic proposal was also reduced, from seven areas in 2012 to three in 2015, but objections persisted. It remained a draft in 2025, with Russia seeking three separate proposals and China questioning its baseline data \citep{CCAMLR2012Report} \citep{CCAMLR2015Report} \citep[pp. 38--39]{CCAMLR2025Report}. Proposal adjustment can therefore precede either adoption or a continuing block.

A separate descriptive comparison finds increased recorded disagreement in Russia's sampled statements after 2022 relative to other Parties (Supplementary Section S22.2). These statements include reservations that did not block adoption. The model does not represent geopolitical change, and the comparison does not identify a change in the value of future cooperation.

The Antarctic histories provide complementary evidence of how objections and revisions shape what reaches adoption, while leaving many reasons for standing aside unresolved. The model and learning experiments show how consensus can sustain frequent agreement while concealing differences in strategic stability, and negotiation histories help reveal the constraints shaping collective decisions.

\section{Discussion}

How does one govern without a sovereign power? We studied how structured human groups decide by consensus when no member has formal authority to impose an agreement. Our case is the Antarctic Treaty System, where each Consultative Party present can block substantive ATCM instruments governing one of the largest global commons \citep[Decision 2, Rule 24]{ATCM2025FinalReport}.

We show that accommodation, reluctant acceptance, future restraint, costly endorsement and open support can each produce consensus, and that all five leave the same record of passage (Figure~\ref{fig:routes}). Making these pathways explicit shows what a record of agreement can and cannot reveal. The flexibility that lets a proposer weaken a proposal can also remove the threat behind restraint (Figure~\ref{fig:flexibility}).

Our model builds directly on work on governance and consensus decision-making. Accommodation comes from the study of agenda setting, where a proposer chooses a proposal subject to the votes needed for passage \citep{RomerRosenthal1978}. In consensus negotiations over an issue considered on its own, this can pull agreements towards what the least ambitious member will accept \citep{Underdal1980}. Future restraint comes from work on repeated cooperation. The promise of future cooperation can sustain behaviour that would not pay in one round \citep{FudenbergMaskin1986}, and linking interactions across issues can strengthen these incentives \citep{BernheimWhinston1990}. Punishment can become vulnerable to renegotiation if the members would all prefer a different continuation \citep{FarrellMaskin1989}. The same logic explains why members of an international organisation go along with decisions they dislike when cooperation must be self-enforcing \citep{MaggiMorelli2006}. In our model, the flexibility that allows accommodation also weakens the threat behind restraint, because offering an intermediate proposal beats carrying out the prescribed punishment. Restricting the menu, setting a minimum bite or making departures from the prescribed proposal costly can preserve that threat (Supplementary Sections S8--S9). Studies of consensus in practice show what silence can hide, which is where reluctant acceptance and costly endorsement sit. Under consensus, silence counts as consent \citep{Urfalino2014}, and it has concealed both the bargaining power of the largest members \citep{Steinberg2002} and the opposition of ministers who would rather not be blamed \citep{Novak2013}. What negotiators are willing to say on the record also depends on whether they bargain in public \citep{Stasavage2004}.

Others have already looked behind the record. Unanimous court panels can hide disagreement among judges \citep{Fischman2011}. In the Council of the European Union, consensus practice can mask contested bargaining \citep{Heisenberg2005}. In international organisations more broadly, consensus removes the votes from which analysts infer preferences \citep{HageHug2016}. In animal groups, a consensus can impose costs on the members whose preferred option lost \citep{ConradtRoper2005}. In Antarctic governance, gridlock over marine protected areas has been modelled as an institutional anti-commons \citep{Huang2026}. The gap we find between revisions and new proposals fits that account, since agreement on existing arrangements continues while new ones stall. Our contribution is to compare the incentives behind these pathways within the same model, show how proposal flexibility can undermine punishment credibility, and examine whether frequent agreement among learners rests on credible rules. These results also extend an earlier draft, which showed that agreement and costly objection can leave the same record under the rule of non-opposition (\unpublishedwork{}). The inherited distinctions between assent, silence and costly dissent (Figure~\ref{fig:routes} c) are detailed in Supplementary Section S31.

This changes how agreement should be read -- recorded agreement need not imply that an institution is stable. A proposal can pass without active support from any Party. The members remain in the treaty, so decisions can still be made while disagreement persists. Reaching consensus is much richer than the binary of veto or accept, and this richness can also hide fragility. Work on repeated cooperation shows that restraint can be sustained, and work on learning shows how groups can come to cooperate \citep{Kleshnina2023}. Our learners often reached rules that allowed few profitable departures along their prescribed paths but many at rarely visited nodes, especially punishment. Learners can hold mistaken expectations about situations they rarely reach, since play there seldom corrects those expectations \citep{FudenbergLevine1993}. With zero initial values, a dispute changed punishment rules while late agreement rates remained nearly identical to controls (Figure~\ref{fig:learning} c, d).

How far agreement and credibility separate depends on initial values and training duration. Positive initial values increase full-audit passes at high continuation while agreement remains frequent, and longer zero-start training also improves some rules (Supplementary Section S10.12). The \hyperref[sec:learning-credibility-window]{analytical credibility window} shows why a valuable future can weaken the prescribed punishment threat, since restoring cooperation can pay more than continuing to veto. The repeated-dispute response under zero starts is consistent with this incentive (Figure~\ref{fig:robustness} d). Where the future at stake is shorter, between a third and a half of pairs at the reference gain reach rules that pass the full audit, all with modest proposals, and a dispute barely changes punishment prescriptions (Figure~\ref{fig:robustness} b). The framework offers a basis for examining how an institution withstands disagreement.

The model also lets us look behind the curtain of decision-making, because it distinguishes which pathways are compatible with the recorded responses. The Antarctic final reports keep adoptions, deferrals and objections, and these are already enough to tell some pathways apart. Revisions more often yield an instrument at the submission meeting (Figure~\ref{fig:record} a), and the histories of single proposals separate accommodation from blocks that persist (Figure~\ref{fig:record} b). This approach can inform the study of other bodies that decide by consensus, from fisheries commissions \citep{Goldsworthy2022} to climate negotiations, where the Glasgow conference softened its language on coal before adoption, despite several delegations' continuing dissatisfaction \citep{Allan2021Glasgow}. Records that also kept stand-asides, maintained objections and who blocked would narrow the pathways consistent with agreement and provide evidence for testing why each member allowed passage or blocked a proposal.

A systematic synthesis of treaty impact evaluations found that, except in trade and finance, treaties had mostly failed to produce their intended effects \citep{Hoffman2022}, which has earned them a reputation as tigers without teeth. Our results suggest a complementary view. The strength of an agreement need not lie only in what its decisions achieve. It also lies in which decisions can be reached, and under what conditions. An agreement offers sovereign nations a means to talk to one another and to govern together where no government exists \citep{AbbottSnidal2000}.

Beyond human systems, the same question arises in animal groups, which face many of the same problems when they decide together \citep{ConradtList2009}. Honeybee scouts that favour one nest site headbutt scouts dancing for a rival site, and these stop signals let the swarm settle on one site before it moves \citep{Seeley2012}. Among African wild dogs, fewer sneezes are needed for departure when a dominant individual initiates the rally \citep{Walker2017}. These studies show how collective decisions can emerge from inhibitory interactions and unequal contributions. These cases also differ in a way the formalism here does not include. A swarm settles on a site without a leader, while among wild dogs the initiator's rank changes the departure quorum. Our model gives every Party the same veto, reflecting the formal blocking right of Consultative Parties present for substantive ATCM instruments. In human institutions, too, formal equality can coexist with unequal weight \citep{Steinberg2002}. How such hierarchies shape the pathway a group takes is a natural next step. Whether animal groups also make decisions through accommodation, reluctant acceptance or restraint remains open, and the model specifies what would need to be observed to find out.

These results have limits that further work could address. The main model stops at adoption and does not follow whether a Party that stood aside later implements the decision. In the Antarctic Treaty System, Measures intended to establish binding obligations on tourism insurance (Measure 4, 2004), environmental liability (Measure 1, 2005) and passenger landings (Measure 15, 2009) were adopted by consensus but have still not entered into force \citep{ATCM2005FinalReport} \citep[paras. 212--214]{ATCM2025FinalReport}. The first two have awaited completion of approval for more than twenty years. Adoption, entry into force and implementation are distinct outcomes. The discrete-instrument extension shows how this distinction can persist after a dispute. Formal passage can resume through an instrument that receives no active follow-through, leaving effective bite at zero even as recorded agreement recovers (Supplementary Sections S4--S5). Maggi and Morelli showed how the need for self-enforcing cooperation constrains the choice of decision rule \citep{MaggiMorelli2006}. Where acquiescence depends on future cooperation, a decline in its value could weaken the incentives for later approval or implementation even if the formal decision rule is unchanged. Whether this mechanism contributes to the observed delays remains to be tested. Modelling subsequent approval and implementation decisions would show when a Party that stood aside carries out the adopted decision.

When both Parties follow the same rule, the theory allows credible restraint only within a narrow window, alongside an equilibrium in which proposals stay modest. A division of roles keeps the threat credible above the window. After two million rounds, up to 52.2\% of trained pairs match the specified division of punishment roles across settings, as defined in the Results. Lowering the temperature makes stand-asides after demanding proposals more frequent at lower continuation values, but proposers often keep their demands modest and complete credible rules remain rare (Supplementary Section S10.5.1). The within-window diagnostics show why reducing exploration alone can be insufficient, since sampling noise can still move learners away from the credible rule (Supplementary Section S10.8). Whether larger groups, or groups that learn in other ways, reach restraint that holds after a dispute remains open. Exploratory membership experiments give limited support for protection through gradual entry, while informed initial values sustain frequent passage under the tested conditions (Supplementary Figure S13). These results do not establish that entrants learn durable cooperation from incumbents. Accommodation can also preserve passage as membership grows by reducing demands to remain within each Party's acceptance threshold (Supplementary Section S6).

The Antarctic patterns can be tested on new records. They were identified after the records were assembled, then measured in the selected documentary samples and checked against alternatives. The checks reported in the Results support these patterns within the stated samples and coding assumptions, although the gap after longer follow-up is less secure. An exploratory threshold model closely approximates aggregate adoption and non-agreement shares when whole meetings are held out. Estimates are less accurate for new proposals when predicting ATCM 44–47 from earlier meetings (Supplementary Section S22.4). Because the patterns were found with the records in view, predictions stated in advance would show whether they hold in other records, at later Treaty meetings and in other bodies that decide by consensus. The 2026 final report again records disagreement over listing, and the ATCM agreed to continue consideration at its next meeting \citep[paras. 93, 97; pp. 28--29]{ATCM2026FinalReport}.

Emperor penguins need sea ice that holds until their chicks can swim. Each year that ice is less certain, and each year the Treaty meeting adopts measures and revises management plans by consensus, while the species still waits for special protection. Our model shows how both can be true. If the old plan has not expired, a failed revision leaves it in force. When its incremental demands are small enough relative to passage benefits, blocking costs and future cooperation at stake, a Party can stand aside. The higher adoption rate of revisions is compatible with this explanation, alongside scheduled review, narrower scope and prior selection into agreement. A new protection that fails leaves only what was already in force, so a single Party content with that can block it however many Parties are in favour. Consensus can then keep producing agreement on what already exists, while new protection waits. The record looks like a system at work. How securely agreement is sustained depends on why Parties allowed proposals to pass and what they would do after disagreement.

\input{sections/submission_methods}

\section{Data availability}

The primary meeting reports, working papers and adopted instruments are public documents of the Antarctic Treaty Secretariat and the Commission for the Conservation of Antarctic Marine Living Resources. The accompanying reviewer package contains frozen copies of the derived inputs used here, including working-paper records and outcome judgments, instrument matches and source evidence, the sampled statements and coding outputs, proposal histories, topic records and simulation summaries. Its manifest records file hashes, source versions and remaining provenance gaps. These analysis inputs are distinct from the source-report corpus and from the companion decision-network study \companioncite{}.

\section{Code availability}

The accompanying reviewer package contains the analysis and figure scripts, tests, environment specification and build instructions. It uses frozen inputs by default, and rebuilding the working-paper frame from a separate source checkout requires an explicit option. The package README identifies the scripts and cached inputs needed to reproduce the reported analyses.

\section{Ethics}

The analysis uses public institutional documents and does not involve human participants or private personal data.

\ifdefined\anonymous \else
\section{Acknowledgments}

This work was supported by the Australian Research Council Special Research Initiative Securing Antarctica's Environmental Future (SR200100005), and by an Australian Research Council DECRA Fellowship awarded to Maria Kleshnina (DE250101223).

\section{Author contributions}

C.v.E.: Conceptualization, Methodology, Software, Formal analysis, Investigation, Data curation, Validation, Visualization, Project administration, Writing -- original draft, Writing -- review \& editing. M.K.: Conceptualization (supporting), Methodology, Funding acquisition, Writing -- review \& editing.

\section{Competing interests}

The authors declare no competing interests.

\fi

%% file: sections/submission_methods.tex
\section{Methods}

\protect\phantomsection\label{sec:methods}{}

\subsection{Model, response definitions and proposal choice}

\protect\phantomsection\label{sec:methods-model}{}

A consensus record establishes that no party maintained a veto. We use $A$, $Q$ and $V$ to denote active support, standing aside (acquiescence), and veto, respectively. A proposal passes when every responding party chooses either $A$ or $Q$. Passage with at least one $Q$ is apparent consensus. Proposal-side accommodation occurs when the proposer lowers bite or changes instrument class because a stronger alternative would provoke a veto or fail in implementation or ratification. Accommodation can produce unanimous support for the weaker proposal. It counts as apparent consensus only when at least one party stands aside from the actual proposal.

The proposal-response game first draws an issue or menu according to the specified public-state process. A proposer then chooses bite or instrument form, the responders observe the proposal and choose their actions simultaneously with one another, the institution records passage, and the public state updates from the proposal, response profile and continuation event. Removing proposal choice gives the procedural random-menu model. Fixing the proposal gives the response-only game. In the strategic-proposer models, $N$ counts responders after a separate proposer. In the response-only models and the Antarctic topic application, $N$ counts all participating responders. Thus $q^{N - 1}$ always concerns the focal responder's counterparts. Formal passage, later ratification and implementation are separate outcomes in institutions that distinguish these stages.

We distinguish the discount factor $\delta$ from the probability $\chi$ that the relationship continues. Their product $\beta = \delta\chi$ is the effective continuation weight. Models written with $\delta$ alone do not separately parameterise termination. Other parameters are defined within each variant. Shared symbols do not imply shared payoff assumptions.

For a fixed public proposal, party $i$ receives passage value $b_{i} > 0$, alignment payoff $h_{i}$ from support, pays stance cost $\Delta_{i} > 0$ for a visible position, and pays an additional $c_{V,i} \geq 0$ for veto. The stage reward is $h_{i}\sigma_{i} - \Delta_{i}\sigma_{i}^{2} - c_{V,i}\mathbb{1}_{\sigma_{i} = V} + b_{i}Y$, with $\sigma_{i} \in \left\{ 1,0, - 1 \right\}$ and passage indicator $Y$. Active support and standing aside have the same continuation. Writing $L_{i} = W_{i}^{C} - W_{i}^{P} \geq 0$, net support payoff $a_{i} = h_{i} - \Delta_{i}$ and net veto payoff $s_{i} = - h_{i} - \Delta_{i} - c_{V,i}$, the relative scores against standing aside when all other parties refrain are $a_{i}$, zero and $G_{i} - \beta L_{i}$, where $G_{i} = s_{i} - b_{i}$. When another party vetoes, the passage term disappears. The equivalent position cost is $k_{i} = 2\Delta_{i} + c_{V,i}$, so $s_{i} = - a_{i} - k_{i}$. Recovering alignment still requires the stance-cost decomposition. These costs represent expression and blocking, rather than estimates from the documentary counts \citep{Rietig2023} \citep{Panke2017} \citep{Gardiner2025} \citep{BartonRodet2015} \citep{Gaisbauer2020}.

The five-pathway illustration separates opposed current-payoff acquiescence ($h_{i} < 0$, $G_{i} \leq 0$), future restraint ($0 < G_{i} \leq \beta L_{i}$), costly endorsement ($0 < h_{i} < \Delta_{i}$) and open support ($h_{i} > \Delta_{i}$). Proposal accommodation can precede any non-veto response. It fixes $\Delta_{i} = 0.5$, $c_{V,i} = 0.25$ and $b_{i} = 0.5$, giving $G_{i} = - h_{i} - 1.25$. Action-indifference boundaries are assigned to standing aside for display. Neutral alignment $h_{i} = 0$ establishes neither opposition nor support. These illustrated regions do not supply response frequencies.

In the discrete-instrument variant, the proposer selects a binding instrument $B$ of nominal bite one or a hortatory instrument $H$ of bite $\alpha \in (0,1)$. Effective bite is one after $(B,A)$, zero after $(B,Q)$, $\alpha$ after either non-veto response to $H$, and zero after veto. A responder of stance $\theta \in \left\{ - 1, + 1 \right\}$ and unsigned salience $s$ receives $\theta sx - e\mathbb{1}_{A} - c\mathbb{1}_{V}$, with $e,c > 0$. The proposer receives $s_{P}x$, with $s_{P} > 0$. The displayed three-region map selects $H$ when the proposer is indifferent between dead-letter binding passage and rejected hortatory passage. In its alternating-agenda construction, both parties favour their own proposal direction. Cooperation prescribes $(H,Q)$, every deviation activates permanent $(B,Q)$ punishment, an off-path $B$ receives $Q$, and an off-path $H$ in punishment receives $V$. Commitment to $B$ in punishment or a positive rejection cost makes the otherwise weak punishment-proposer comparison strict.

The continuous-bite variant holds current-payoff parameters and continuation gaps fixed as $x \in \lbrack 0,1\rbrack$ and $\beta$ vary. An opposed responder suffers implementation loss $g_{i}x$ only upon passage, receives benefit $b_{i} \geq 0$, and pays $c_{V,i} > 0$ for veto. Active support costs an additional $e_{i} > 0$ and has the same continuation as standing aside. Under complete information and deterministic passage-favouring response selection, acceptance requires $x \leq \frac{b_{i} + c_{V,i} + \beta L_{i}}{g_{i}}$. A proposer whose return increases with accepted bite, and who prefers some passing offer to every rejected offer, selects the largest jointly acceptable bite. If its unconstrained optimum is lower, that optimum also limits the choice. The derivations and the consequences of changing threshold tie-breaks are in the Supplementary Information.

For the general strategic-proposer audit, a finite public agenda phase identifies the proposer and responder set, and an action-independent transition kernel selects the next phase. A pure strategy specifies a proposal in each public memory state $C,P$ and a complete response profile after every feasible proposal. Only the prescribed unanimous-restraint pair preserves $C$. Every deviation activates $P$. Compliance with the punishment prescription either preserves $P$ or returns to $C$, as specified, while punishment deviations remain in $P$. We solve the prescribed Bellman values and check unilateral alternatives for the cooperative proposer and responders, the punishment proposer and responders, and the responders after off-prescription proposals in both states. The finite implementation checks every available proposal-grid point and simultaneous responder action, holding other responders' actions fixed. A reduced $\left\{ Q,V \right\}$ audit applies to the three-action game only when $A$ is also checked or its dominance by $Q$ in current-plus-continuation value has been established. Bounded discounted payoffs, complete information, perfect public monitoring and perfect recall are maintained for the one-shot-deviation argument.

The four-institution comparison uses two alternating parties. Here $c>0$ is the responder\textquotesingle s total veto cost, distinct from the additional charge $c_{V,i}$ in the fixed-proposal game. The alternating game discounts future payoffs by $\delta$. The learning model also includes the probability $\chi$ of another round, giving effective continuation weight $\beta=\delta\chi$. Dead-letter passage has zero payoff. A passed substantive bite $x$ gives $(x, - x)$ to proposer and responder, and veto gives $( - k, - c)$, with $k \geq 0$ and $c > 0$. The candidate prescribes $(x,Q)$ in cooperation and dead-letter passage in permanent punishment. We compare unrestricted continuous redesign, the binding/hortatory menu, a minimum substantive bite, and a primitive lump-sum charge $K$ on a proposal other than the state-designated proposal. These are alternative institutions within the same specified trigger, not a search over every repeated-game equilibrium. Main Figure 2b compares the trigger with history-independent equilibria. With unrestricted proposals, offering $\min\{c,1\}$ each round and accepting proposals up to $c$ is subgame perfect when ties favour passage. With only the dead-letter and bites above $c$ available, no substantive proposal passes in such an equilibrium (Supplementary Section S8). Supplementary Figure S3 fixes $\frac{x}{c} = 1.5$ for its penalty comparison.

\subsection{Payoff learning and exact incentive audits}

\protect\phantomsection\label{sec:methods-learning}{}

Two parties alternate proposer and responder roles whenever their relationship continues. The proposal bite comes from a finite menu. In open mode $O$, veto is licensed when $x \geq x_{V}$ and moves play to revision $R$. Revision restricts the next proposal to $x \leq x_{R}$. Passage returns to $O$, while a further veto activates punishment $P$. An unlicensed veto also activates $P$. Passage in $P$ repairs the relationship with probability $\rho$ and otherwise leaves it in $P$. A non-veto response outside punishment returns to $O$. With probability $\chi$ another round follows and roles alternate. Otherwise the next episode starts in $O$ with a randomly chosen proposer. Learned values and visit counts persist.

Alignment is $h(x) = h_{0} - \lambda x$. The proposer receives $b + x$ after passage and $- k$ after veto. The responder receives $b + h(x) - \Delta$, $b$, and $- h(x) - \Delta - c_{V}$ after $A,Q,V$, respectively. The exact two-bite specification uses $x \in \left\{ 0.25,1 \right\}$, $x_{R} = 0.25$, $x_{V} = 1$, $h_{0} = 0.45$, $\lambda = 1$, $b = 0.25$, $\Delta = 0.10$, $c_{V} = 0.05$, $k = 0.05$, $\delta = 0.9$ and $\rho = 0.65$.

Each party stores separate action values $Q_{i}(s,a)$ for its proposer and responder contexts. This value notation is distinct from the stand-aside action $Q$. Available actions follow softmax probabilities proportional to $\exp(\frac{Q_{i}(s,a)}{\tau})$. Lower values of the temperature $\tau$ favour actions with larger learned values more strongly, while positive temperature gives alternatives a chance to be tried. It changes how actions are selected, not the payoffs or the exact credibility window. Realised state-action-reward-state-action learning (SARSA) updates the previous action value toward $r_{i} + \delta Q_{i}(s',a')$ \citep{SuttonBarto2018}. The continuation term is zero at episode termination. The implementation also treats the final pending update of a finite run as terminal because no subsequent action is sampled. This affects the final stored values and possibly their modal projection, but not the already sampled trajectory. The realised update uses $\delta$ because continuation has already been sampled. Expected Bellman equations instead use $\beta = \delta\chi$. The reported reference runs use $\alpha_{n} = \frac{\alpha_{0}}{\sqrt{1 + \frac{n}{4000}}}$, with visit count $n$ incremented before the update, base rate $\alpha_{0} = 0.05$ and temperature $\tau = 0.25$ for both parties and roles. Neutral starts set values and visit counts to zero. The illustrative veto-prone path alone uses a 2,000-visit decay scale.

Main Figure 3a starts every action value at zero and uses seeds 0--999 at each of 23 continuation probabilities $\chi = \frac{k}{24}$, $k = 1,\ldots,23$, with a gain of 0.15 from vetoing a demanding proposal. Each run lasts 2,000,000 rounds. Outcome shares and mean bite are calculated over rounds 1,996,001--2,000,000, then averaged with equal weight across pairs. Seed numbers are reused across continuation values. The compiled replay reproduces the stored symmetric modal rule, apparent-consensus share and per-Party full-audit classification for all 23,000 endpoints. The intervention in Figure 3c uses $\chi = \frac{23}{24}$, selected after inspecting the grid for a large dispute effect on punishment prescriptions. It is a descriptive follow-up to that grid, not an independently selected confirmatory test. The $\chi = \frac{5}{6}$ comparison uses the same horizon and learning settings. All draws use NumPy's \nolinkurl{default_rng(seed)}. The initial proposer is random. The independently seeded unit is the trained pair, not a round.

A round is classified as full support after $A$, apparent consensus after $Q$, a filtering veto after veto of a licensed demanding open-mode proposal, and breakdown after any other veto. Breakdown routes play to $P$ and does not mean permanent termination. Mean bite includes rejected proposals. The archived flow diagram from 8,000-round training pools all 1,000 pairs at $\chi = \frac{5}{6}$ and retains 3,999 post-burn-in rounds per run, omitting the last because its next observed mode is unavailable. The complement of punishment visitation therefore combines open and revision play. It is not an open-mode share or a statistic restricted to the most frequent endpoint rule.

For main Figures 3 and 4, each Party's deterministic modal rule retains that Party's most likely action at every node. We audit the resulting pair against one another, allowing the two rules to differ. For comparisons with the symmetric-rule enumeration, we instead form a symmetric modal pure rule by averaging the two Parties' action probabilities at each node and retaining the most likely action. Action values themselves are not averaged. Exact ties favour weak over demanding proposals and $A$ over $Q$ over $V$ for responses. There are two proposal choices in $O$ and $P$, one in $R$, and three response choices at five available mode--proposal combinations, giving $2^{2}3^{5} = 972$ symmetric pure stationary rules. The strong-restraint candidate demands bite one and receives $Q$ in $O$, and offers the weak bite with $A$ in $R$ and $P$. It also prescribes $A$ after an off-path weak open proposal and $V$ after an off-path demanding punishment proposal. Asymmetric, mixed, nonstationary and history-dependent candidate profiles are outside this symmetric enumeration. The per-Party learned-rule audit in Figures 3 and 4 permits asymmetric deterministic pairs. It does not audit the learners' exploratory policies or establish convergence of learning.

The candidate phase map audits that specified rule on 81-point continuation and recovery grids, with continuation spacing $0.0125$. Continuous credibility boundaries use 80 bisections in the brackets $\lbrack 0,0.30\rbrack$ and $\lbrack 0.30,1\rbrack$. This is distinct from enumerating all 972 rules at each of 21 equally spaced continuation values over $\lbrack 0,1\rbrack$ and recoveries $\rho \in \left\{ 0,0.25,0.65,1 \right\}$ in the selected-equilibria table. The later window diagnostic enumerates all 972 rules on the 101-point grid $\chi = 0,0.01,\ldots,1$ at $\rho = 0.65$. Exact candidate auditing, exhaustive finite-class enumeration and learned endpoint frequencies are separate calculations.

For the symmetric projection, the modal-path margin is the negative of the largest one-shot deviation gain at proposal nodes reachable from either open starting role and response nodes following the modal rule's prescribed proposals. The full margin uses every proposer role and public mode and responses to every available proposal, including proposals the rule never makes. Each departure changes one action and then returns to the rule. Continuation values solve the Bellman equations, with $\beta = 0.75$ in the reference sample. The full check therefore exceeds an audit of paths reached by starting in punishment. The implementation accepts a full rule when its largest gain is at most $10^{- 9}$ and uses $10^{- 12}$ for modal-path classification. For rules that pass the modal path but fail the full check, the deviation breakdown selects the single largest positive gain per rule and weights it by the number of runs producing that rule. A failed check establishes an incentive at a specified node, not a probability of institutional collapse or a failure of either original stochastic policy. For Figure 3, the pair audit solves continuation values under the two own rules, checks every Party's alternative proposal and response at all available nodes, and accepts the pair when neither Party can gain more than $10^{- 9}$. For the reported pair-level path checks, we traverse the states reachable with positive transition probability from either open-play starting role under the pair's prescribed actions. We check proposal alternatives at those states and response alternatives after the prescribed proposal, using the same $10^{- 9}$ tolerance. Passing the full check identifies a stationary subgame-perfect Nash equilibrium in the reconstructed deterministic rules, to the stated numerical tolerance. A path-only pass does not establish subgame perfection or verify the agents' beliefs.

\subsection{Learning sensitivity, horizon and interventions}

The sensitivity to initial action values repeats the main learning procedure at $\chi \in \{5/6,11/12,23/24\}$ with seeds 0--999 and two million rounds per condition. Additional runs extend seeds 0--99 at $\chi=23/24$ to twenty million and two hundred million rounds. Supplementary Section S10.12 specifies the initial values, temperature scaling, stationary-policy benchmark and matched comparisons. These checks were selected after examining the reference learners' action values and visit counts.

\protect\phantomsection\label{sec:methods-learning-checks}{}

The one-at-a-time sensitivity varies eight parameters at $\chi = \frac{5}{6}$, retaining the other reference settings. Lower and upper values are $\Delta \in \left\{ 0.05,0.20 \right\}$, $c_{V} \in \left\{ 0,0.10 \right\}$, $b \in \left\{ 0.15,0.35 \right\}$, $k \in \left\{ 0,0.10 \right\}$, $\rho \in \left\{ 0.40,0.90 \right\}$, $\delta \in \left\{ 0.80,0.95 \right\}$, $\alpha_{0} \in \left\{ 0.02,0.10 \right\}$ and $\tau \in \left\{ 0.10,0.50 \right\}$. Running the common reference once gives 17 conditions, with 200 seeds, 0--199, reused in each. Every run has neutral values, 8,000 rounds and a 4,000-round burn-in. Learning-rate changes retain the 4,000-visit scale, and temperature changes apply to both parties and roles. Each endpoint is projected and audited in its own game. The complete 972-rule search is repeated for each condition. Wilson intervals for full-pass proportions use trained runs as the unit.

A separate continuation--temperature sensitivity uses $\tau \in \left\{ 0.25,0.10,0.05 \right\}$ for both parties and roles at $\chi \in \left\{ 0,0.15,0.20,\frac{2}{9},0.25,0.30,\frac{1}{3},0.35,0.50,\frac{2}{3},0.75,\frac{5}{6},1 \right\}$. Each cell has 100 neutral-start runs, using seeds 0--99, 8,000 rounds, a 4,000-round burn-in and the reference learning rate and payoffs. Seed numbers are reused across cells. Conditional stand-aside frequencies divide the total number of stand-asides after demanding open proposals by the total number of those proposals. Demanding-proposal frequencies divide demanding open proposals by all open rounds. Their 95\% percentile intervals use 2,000 bootstrap samples of whole runs within each cell, with generator seed 20261103. Replicates with no qualifying proposals have undefined conditional frequencies and are omitted from that interval. Endpoint rules are projected and audited as above. This finite-grid sensitivity separates conditional responses, proposal exposure and complete-rule credibility. It does not estimate a sharp learning threshold or a vanishing-exploration limit.

A separate experiment follows 30 matched seeds, 0--29, for 100,000 rounds at $\chi = \frac{5}{6}$. One exploration schedule keeps $\tau = 0.25$. The other uses $\tau_{t} = 0.25\max(0.1,\min(1,\frac{8000}{\max(1,t)}))$ for zero-based round $t$, reaching a floor of $0.025$. Both start from neutral values with the reference visit-dependent learning rate and coincide through 8,000 rounds. Values are retained at 8,000, 20,000, 50,000 and 100,000 rounds. Apparent consensus at a checkpoint is the fraction of $Q$ responses in the preceding 4,000 rounds. Means and medians across runs are distinct summaries. The primary audit freezes each party's own softmax policy at the checkpoint temperature and checks all one-shot departures using the original payoffs, with no reward for exploration. A separate audit uses the symmetric modal projection. Intermediate snapshots follow the preceding update but precede the current reward. The final snapshot includes the terminal update. An intermediate 8,000-round snapshot need not equal a separately truncated 8,000-round endpoint.

The matched training-length comparison uses neutral-start runs with the same seeds at 8,000, 32,000, 128,000, 512,000, 1,000,000 and 2,000,000 rounds. It includes 1,000 pairs at $\chi = \frac{5}{6}$ and 100 pairs at $\chi = \frac{23}{24}$, with a veto gain of 0.15 and the main learning settings. Each horizon retains its own finite-run terminal update. Outcome shares use the preceding 4,000 rounds, and each Party's own endpoint rule is projected and audited as above. Training horizons sharing a seed are matched observations, not independent pairs. For the action-value accuracy diagnostic at $\chi = \frac{5}{6}$, we freeze both Parties' endpoint softmax policies and solve their expected Bellman equations. We compare each legal proposer and responder action value with the learned value, weight these entries equally within each pair and take the median pair's mean absolute error. Punishment errors restrict the entries to that mode. This diagnostic uses all 1,000 pairs at each of the five horizons through 1,000,000 rounds. It measures agreement with the current frozen exploratory policies, not with an equilibrium, and does not isolate why values changed during training.

The main Figure 3 resumed-learning intervention copies both parties' action values and state--action visit counts from each 2,000,000-round endpoint at $\chi = \frac{23}{24}$ into two arms, starting one in $P$ and the control in $O$. Counts are retained directly by the compiled training kernel so that learning resumes at the trained rate. No action is forced. Each arm runs for 2,000 further rounds without burn-in using the original payoffs, temperatures and learning settings. For trained seed $s = 0,\ldots,999$, each of 20 replicates $r = 0,\ldots,19$ gives both arms the fresh seed $10,000,000 + 20s + r$. They share the initial proposer draw and random streams. Realised actions can differ as their modes and values diverge. Values and counts are recopied for every arm and replicate. The independent training units remain the 1,000 original pairs. Supplementary Section S10.11 repeats this comparison at $\chi = 5/6$ and $\chi = 11/12$ with a gain of 0.15, using the saved endpoints and the same seeds. Both settings were chosen before these comparisons were run. For the separate supplementary time-resolved comparison from 8,000-round endpoints, we replay the same intervention seeds and retain every round outcome, checking each replicate against its stored outcome shares, first-episode veto count and length, and final modal rule. The grey baseline uses the last 100 training rounds once per pair. Round zero is the first resumed round in both arms, with a freshly sampled proposer after the terminal training update, so the baseline and resumed trajectories are separated at zero. For the longer trajectory retained with the data, we calculate the shares of all four outcomes in non-overlapping bins of ten rounds before restart, single rounds from zero through nine, five rounds from ten through 49, 25 rounds from 50 through 199, and 100 rounds from 200 through 1,999. No bin crosses the restart. At each bin we average within each pair across its 20 replicates, then across pairs. The retained differences subtract the matched control share from the dispute share within each bin, in percentage points. The four differences therefore sum to zero. For the apparent-consensus series retained with the data, pointwise 95\% percentile intervals use 2,000 bootstrap resamples of the 1,000 trained pairs, retaining both arms and all replicates together, with generator seed 20261103. These intervals describe each time bin rather than simultaneous coverage of the trajectory.

The first episode ends at the first sampled failure of continuation, including that last round. Its veto count counts $V$ responses rather than the fraction of rounds with a veto. Longer-run outcome shares average all 2,000 resumed rounds. The final modal rule is projected after the terminal update and counted as changed if any of its seven free prescriptions differs from the training endpoint, including at unvisited nodes. This endpoint comparison does not count all changes during learning. Equal replication makes averaging all replicate records equivalent to averaging within each trained pair and then across pairs.

Figure 3b reports the share of pairs passing the per-Party full audit and the share in which both rules prescribe $Q$ after a demanding open proposal. Figure 3c compares matched dispute and control replicates at $\chi = \frac{23}{24}$. Its outcome statistic is the apparent-consensus share over the final 1,000 resumed rounds. Each Party's ordinary-play rule comprises its open proposal, both open responses and its revision response. Its punishment rule comprises its punishment proposal and both punishment responses. A Party is counted as changed in a group if at least one prescription in that group differs between the training endpoint and the end of recovery. Shares weight both Parties equally within each replicate, then replicates equally within each pair and pairs equally across the 1,000 seeds. Figure 3d illustrates the most frequent individual prescription change after a dispute, from veto to stand-aside in response to a demanding punishment proposal. The 6.1\% share for that change and the 28.1\% share changing any punishment prescription have different numerators. Mode shares pool the final 4,000 training rounds across all 1,000 pairs. The fixed training budget is not a convergence criterion. The reported best-response rate for stand-aside prescriptions after recovery conditions on the Party actually prescribing that response to a demanding open proposal. Each Party in each replicate contributes at most one such prescription. There are 29,687 qualifying prescriptions in controls and 29,686 after dispute, of which seven and six, respectively, admit a profitable alternative under the partner's reconstructed rule. Pairs with no stand-aside prescription are excluded from this conditional rate.

In the standard intervention, a new episode resets the public mode to $O$. In the separately reported 8,000-round carry variant, it starts in the mode sampled as the previous episode's next mode. Carry applies to both arms and all episode boundaries, including revision as well as punishment. Episode endings still have zero continuation in the SARSA target and a newly sampled proposer. The standard and carry variants of that supplementary comparison both use the same 8,000-round training endpoints and replicate seeds. Their final audits use the original episodic game, with $\beta = \delta\chi$ and zero value at termination. They do not audit an infinite-horizon game assigning continuation value to carried modes. Resumed learning measures response to entering $P$. The separate one-shot audit measures incentives to depart from a frozen rule.

The settings sweep in main Figure 4 crosses $\chi = k/24$, $k = 1,\ldots,23$, with nine alignment intercepts $h_{0} = 0.375,0.400,\ldots,0.575$. The gain from vetoing the demanding proposal rather than standing aside is $0.60 - h_{0}$. It runs from 0.025 to 0.225 and equals 0.15 at the reference. Each of the 207 settings trains seeds 0--999 for 2,000,000 rounds with a 1,996,000-round burn-in and the reference learning settings. This sweep uses the same training horizon as Figure 3. The separate short-training supplementary analyses retain their original horizons. Values and accumulated visit counts are retained at intervention, so longer training also changes the subsequent visit-dependent learning rate. Each trained pair is then resumed in $P$ and in $O$ with 20 replicates per arm, using the seeds, value copies and visit counts of the reference intervention. All panels project each Party's rule from its own action values, as in Figure 3. Panel (a) counts pairs whose endpoint rules admit no profitable one-shot departure along the path reached from $O$ but fail the full audit. Panel (b) records, after 2,000 resumed rounds, whether each Party's rule prescribes $V$ after its own proposal in $P$, averaged over the two Parties. Breakdown shares refer to the final 4,000 training rounds. Apparent consensus after resumption refers to the final 1,000 resumed rounds. In panel (c), a pair is counted when both rules prescribe $Q$ after the demanding open proposal, one rule proposes the demanding bite in $P$ and prescribes $V$ after it, and the other proposes the weak bite in $P$ and prescribes $Q$ after the demanding bite. Other prescriptions are unrestricted, so a counted pair need not pass the full audit. The audited pair departs from the strong-restraint candidate only in $P$. In $P$, one rule proposes the demanding bite and the other prescribes $Q$ after the demanding bite. Each party's largest one-shot gain against the other's rule is computed on 199 equally spaced values of $\chi$ from 0.005 to 0.995. The lower edge of the passing range uses 40 bisections.

The repeated-dispute experiment in main Figure 4d retains the saved two-million-round values and visit counts at the four marked settings. Each setting uses seeds 0--999, five continuation replicates per arm and doses $d\in\{0,1,5,20,50,100\}$. For $d>0$, we impose punishment at rounds $0,2000,\ldots,2000(d-1)$ and run for $2000d$ rounds, leaving 2,000 rounds after the last imposed start. Dose zero runs for 2,000 rounds without an imposition. Each control runs for the same duration without imposed mode changes. Seed $s$ and replicate $r$ use stream $260,000,000+5s+r$ in both arms and across doses. A scheduled start changes only the public mode before action choice, preserving proposer identity and any pending SARSA transition. No proposal or response is forced. Natural episode endings retain reward-only updates and resets to ordinary play, and the final horizon retains its terminal update. Shares are averaged across replicates within pairs and then across pairs. Pointwise 95\% intervals use 5,000 bootstrap resamples of whole trained pairs. Panel (d) shows absolute prescription shares in the dispute arm. Matched-control differences are reported in Supplementary Section S10.13. The primary measure reads each Party\textquotesingle s response to its own prescribed punishment proposal, as in main Figure 4b. A separate diagnostic fixes the demanding punishment proposal, and realised outcome shares use the final 1,000 rounds. Supplementary Section S10.13 reports sensitivity checks and exact value comparisons. Earlier supplementary repeated-dispute curves retain the segmented protocol with 100 exposure rounds and 2,000 recovery rounds, whose segment length and terminal-update frequency change with dispute count.

Inside the credibility window, we compare neutral starts with values initialised to the strong-restraint candidate's exact action values from its Bellman equations. At each of $\chi = 0.25$, $0.30$ and $\frac{1}{3}$, 200 seeded pairs run for 8,000 rounds, with reference learning and exploration and zero initial visits. Modal rules are checked every 50 rounds in the candidate-initialised experiment. A neutral-start success requires demanding open play with $Q$ and a full equilibrium pass. Zero successes among 200 runs gives the one-sided 95\% binomial bound $1 - 0.05^{\frac{1}{200}}$ separately at each continuation value. The long-window experiment uses seeds 0--49 for each continuation value, each exploration schedule and each initialisation, giving 600 runs in total, 300 per initialisation. It runs for 100,000 rounds, checks modal rules every 500 rounds and reports ten non-overlapping 10,000-round outcome blocks. Early/late comparisons average the blocks across seeds within a condition. Its demanding-proposal share counts demanding proposals in open play divided by all rounds. The probability bound does not pool different continuation values into a common-probability sample.

The accompanying diagnostics separate exploration from sampling noise. A constant-rate Bernoulli continuation estimator starts at the true $\chi$, uses rate $0.05$, and takes 400,000 draws for each of $\chi = 0.25$, $0.30$, $\frac{1}{3}$ and $\frac{5}{6}$ using one generator seeded one. After discarding the first 1,000 estimates, it reports standard deviation, 5th/95th percentiles and the fraction below $\frac{2}{9}$. This approximation holds other action values exact. A mixed-response comparison replaces only the demanding open response with probabilities $(A,Q,V) = (0,q,1 - q)$, freezes the other candidate prescriptions, evaluates exact behavioural values, and finds the first nonnegative margin on $q = 0,0.001,\ldots,1$.

Our damped iteration of logit responses \citep{McKelveyPalfrey1995} uses exact action values instead of learned values, damping $0.20$, change tolerance $10^{- 10}$ and at most 50,000 iterations. Separate candidate-start calculations use temperatures $0.25$, $0.10$, $0.05$, $0.02$ and $0.01$. The ascending branch starts at $0.002$, increases in steps of $0.0005$ toward $0.05$ (excluded), warm-starts from the preceding solution, and stops at the first modal change or nonconvergence. Its reported thresholds are grid limits for the modal prescription of a stochastic policy, not exact continuous cutoffs or convergence to a pure equilibrium.

Additional candidate-initialised learning at $\chi = 0.30$ varies temperature over $0.10$, $0.05$, $0.02$ and $0.01$ at $\alpha_{0} = 0.05$, and base rate over $0.02$, $0.01$ and $0.005$ at temperatures $0.25$ and $0.01$. Together with the three core conditions, these give 13 conditions of 200 runs each. Exit and return times are detected only at the sampled modal checkpoints. Median exit times exclude non-exits and are conditional on an observed exit. These finite diagnostics do not establish a vanishing-noise selection limit.

The additional five-bite version uses $x \in \left\{ 0.2,0.4,0.6,0.8,1 \right\}$, revision bites no greater than $0.6$ and licensed-veto threshold $0.8$, with the remaining reference payoff and recovery parameters. Its figure crosses 13 continuation values with temperatures $0.15$, $0.20$, $0.25$, $0.30$, $0.40$ and $0.55$, using seeds 0--9 and the last 4,000 rounds of each 8,000-round run. The supplementary table selects $\tau = 0.25$. The smaller two-bite illustration uses seeds 0--9 at $\chi = \frac{5}{6}$. These samples are reported separately. Reused seeds overlap the larger reference caches and are not independent replications.

The deliberately selected veto-prone five-bite path uses $\chi = 0.85$, seed three, 5,000 rounds, temperature $0.25$, base rate $0.05$ and the 2,000-visit decay scale. It starts proposer values at zero and adds $0.35$ to every responder veto value. Although its configuration has a 2,500-round burn-in, the figure displays all 5,000 rounds in blocks of 100. It is not a randomly selected representative trajectory.

\subsection{Additional response, agenda and private-type variants}

\protect\phantomsection\label{sec:methods-variants}{}

The response-only variants retain the full three-action stage payoff. The unchanged-proposal construction repeats the same payoffs indefinitely. The alternating construction instead switches which of two parties faces alignment $+ H$ or $- H$, with cooperative profiles $(Q,A)$ and $(A,Q)$ and a punishment path prescribing one consequential veto. It defines $s = H - \Delta - c_{V}$ and $a = H - \Delta$, requiring the punishment checks $s \geq b$ and $a \geq 0$. The $N$-party extension uses a balanced cycle disadvantaging one party per proposal and each party once per cycle. Its membership comparison changes future agenda opportunities and differs from adding possible vetoers to a fixed proposal. The fixed-proposal membership comparison assumes conditionally independent counterpart restraint with common probability $q$.

The separate binary accept/veto illustration sets the acceptance cost on an adverse issue to $c = 0.35$, the benefit on a favourable issue to $v = 0.70$, and payoff after veto to zero. It contrasts an unchanged issue with alternating issues under grim punishment. The three-round tree uses $\delta = 0.60$ and distinguishes a known terminal date, with strict terminal veto incentives and no terminal cooperation value, from a window into an indefinite relationship with continuation beyond the display. These are illustrative feasibility comparisons rather than estimates of Antarctic patience.

The procedural adaptive-menu model has no strategic proposer. Nature draws uniformly from an active menu containing permanent aligned proposals, permanent cheap contested proposals and removable expensive contested proposals. Both parties accept the first two classes. Only the disadvantaged party vetoes an expensive proposal. Any veto gives both parties zero, and removes an expensive proposal in either memory state. Any unprescribed action activates permanent punishment in which both parties veto every proposal. Continuation values are computed recursively from smaller menus, starting with the menu containing only permanent proposals. The complete audit covers acceptance, prescribed veto and punishment at every reachable menu. The six-proposal example has two aligned proposals paying $(0.4,0.4)$, two cheap proposals paying $(0.2, - 0.2)$ and its reverse, and two expensive proposals paying $(0.6, - 0.6)$ and its reverse. The larger threshold illustration uses 250 proposals in each class, with contested directions equally split. Effective weight $\beta = \delta\chi$ controls incentives, while $\chi$ separately controls exposure to later draws and the probability of reaching a pruned menu. This binary-action model does not distinguish active support from standing aside.

The direct-cost feedback variant draws private veto payoffs independently from a continuously differentiable distribution $F$ and imposes a type-independent expected continuation advantage $L(q)$. Choosing restraint costs $gx$ even if another party vetoes, giving the symmetric response map $T(q;x,\beta) = F\left( bq^{N - 1} + \beta L(q) - gx \right)$. This differs from an implementation loss incurred only on passage, whose current payoff index is $(b - gx)q^{N - 1}$. The private-type model sets $x = 0$, where the two maps coincide. We iterate adaptive expectations $q' = T(q)$ and classify regular fixed points using the analytic derivative. This selection rule is an assumption. An imposed continuation function, including $L(q) = 1$, does not by itself implement a sequential equilibrium of a repeated private-information game. Proposal-choice calculations condition on a selected differentiable stable branch and a positive increasing return $R(x)$ on passage, with zero on failure.

The normal private-type illustrations use $N = 5$, $b = 1$ and $s_{i}|P \sim \mathcal{N}(\mu_{P},\sigma_{P}^{2})$. They compare constant continuation, $0.4 + 1.2q$, $0.4 + 2.4q^{2}$, and $0.4 + 1.2S_{12}(q)$, where $S_{12}(q) = \left( 1 + \exp( - 12(q - 0.5)) \right)^{- 1}$. All equal one at $q = 0.5$. The phase maps vary mean and standard deviation on their displayed axes, comparing $\delta = 0$ and $0.6$ for constant continuation. The shape comparison fixes $\delta = 0.6$.

Stable fixed points receive the mass of common initial beliefs in their basins under the stated iteration, with $q_{0} \sim \text{ Beta}\left( m\kappa,(1 - m)\kappa \right)$. An unstable fixed point bounds adjacent basins, a unique stable point receives weight one, and fold equalities remain boundaries. This law represents analyst uncertainty about a common starting belief across decision episodes, not private-belief heterogeneity or Bayesian learning. At selected root $q_{k}$, current-payoff restraint probability is $r_{k} = F_{P}\left( bq_{k}^{N - 1} \right)$, and active-support probability is $p_{A} = F_{P}\left( - 2\Delta - c_{V} \right)$ for common costs. Conditional independence gives unanimous support $p_{A}^{N}$, passage $\sum_{k}w_{k}q_{k}^{N}$, and continuation-supported passage $\sum_{k}w_{k}\left( q_{k}^{N} - r_{k}^{N} \right)$. The current-payoff comparison holds the selected fixed point constant rather than resolving the system at zero continuation. We report unconditional continuation-supported passage and conditional shares only with positive passage denominators. The illustrative probability comparison fixes $\mu_{P} = 0.7$, $\sigma_{P} = 0.25$, $m = 0.5$, $\kappa = 8$ and 251 equally spaced discount factors from zero to $0.999$.

The outcome-opposition extension instead holds non-opposed parties silent, gives opposed party $i$ loss $g_{i} > 0$ only if everyone remains silent and passage occurs, zero if someone else vetoes, and cost $c_{i} > 0$ whenever it vetoes. Silence then gives $- \pi_{i}g_{i}$, where $\pi_{i}$ is the probability all others remain silent. Its complete-information result assumes at least one opponent and $g_{i} > c_{i}$ for every opponent. This is a separate payoff from expressive veto, which can reward a veto position even after another party has blocked passage.

For the documented-response bound, $M$ counts passed decisions in an audited sample, $D$ those with documentation of at least one stand-aside from the adopted proposal, and $E$ those with documentation of active support from every party. The categories must be disjoint. Reservations about a clause or earlier draft are insufficient unless they establish non-endorsement of the adopted proposal. Assigning the undocumented decisions either unanimous support or at least one stand-aside yields the attainable bounds $\frac{D}{M}$ and $1 - \frac{E}{M}$. This is a sample-composition bound, not a confidence interval. Selected histories do not provide a representative denominator.

\subsection{Documentary sources and reconstruction}

\protect\phantomsection\label{sec:methods-sources}{}

Main Figure 5 compares working-paper outcomes and selected proposal histories. We identified these patterns after assembling the records. They do not estimate private preferences, proposal bite, veto costs or the value of future cooperation, and do not constitute prospective tests of the model. The separate statement comparison is reported in Supplementary Section S22.2. The unsigned topic-profile stress test is a separate exercise in assigning unobserved directions to documentary magnitudes.

We used a machine-assisted reconstruction of Antarctic Treaty meeting records developed for a companion study of institutional decision networks \companioncite{}. The primary sources are the Secretariat's final reports and inventories of papers and adopted instruments \citep{AntarcticTreatySecretariatATCM}. Reports were converted to OCR-derived text and divided into agenda or topic units. Model-assisted readings recorded papers, instruments, relationships and supporting passages. Separate disposition readings recorded meeting treatment of papers. Paper identifiers retain meeting, type and number. Modern instrument identifiers retain kind, number and adoption year, because numbering resets annually. Early Recommendations retain meeting and sequence. Paper inventory records reconcile editions using meeting, type and number rather than treating every language or revision edition as a separate paper.

The working-paper frame was assembled from the reconstruction's 13 October 2026 rerun and 15 October figure-refresh snapshot, with the separately frozen 20 October 2026 operative input used for the predecessor comparison. The retained inputs include paper inventories, paper--instrument links and dispositions, as well as the packaged working-paper frame. The statement sampler instead uses the repaired final-report segment file from the apparent-consensus project, and the prominence calculation uses its separately packaged actor--paper--topic records. These are distinct derived inputs. A newly repaired report corpus or replacement extractor does not retrospectively identify the source versions used by the historical extractions. Exact source-tier and hash correspondence for every historical reading is not completely recoverable from the inspected records.

Identifier checks resolve paper references against the inventory and instrument names against the dated instrument list. An untyped paper can receive a type when only one inventory type has its meeting-specific number. An undated modern instrument can receive a year from an explicitly identified meeting. Quote checks and targeted model-assisted keep, fix or drop readings accompany unresolved or flagged links. Quotation checks provide textual anchoring rather than validation of the relationship. Strong contribution labels and weak shared-subject associations remain different evidence. Database identity, shared subject matter and agreement with another extracted graph do not validate proposal adoption. A subsequent documentary review examined source support for the 124 cases in the bounded object/stage review described below (Supplementary Section S22.5).

The 15 October snapshot assembled dispositions from full-item model readings and targeted source reviews, recording submission-meeting categories separately from longitudinal outcomes and retaining mixed aggregate outcomes as unclear. Source-reviewed categories took precedence during repair integration. Other replacements required consistent returned categories and agreement with the retained classification or explicit acceptance, followed by documented Codex-assisted source adjudications.

Paper--instrument associations and institutional dispositions are separate observations. In the reconstruction, the relation label ``adoption'' identifies a same-meeting association. A match on a protected site's title can connect records without establishing adoption of the paper's requested plan or revision. CEP advice and subsequent ATCM adoption can also concern different stages. Report selection and missed paper references can limit coverage before relationship coding begins.

The completed 13--15 October 2026 agenda-unit rerun records a local model served as \nolinkurl{qwen3.8-27b}, with FP8 quantisation, vLLM 0.31.0, codebook 1.1.0, temperature 1.0, top-p 0.95, top-k 20 and a 98,304-token context limit, without a fixed request-level sampling seed. Later codebook-1.2 repairs used a 131,072-token context limit and constrained final JSON responses. The assembled records also incorporate retained instrument readings and targeted Codex-assisted source adjudication. An immutable model checkpoint revision and exact model settings for those Codex readings were not recoverable. The served identifier is reported as recorded. The directory label and architecture name do not establish an immutable weight version. These execution records do not establish extraction accuracy.

\subsection{Working-paper titles, outcomes and follow-up}

\protect\phantomsection\label{sec:methods-working-papers}{}

The frame contains 1,595 working papers submitted to ATCM 16--47 (1991--2025). The unit is a working paper at its submission meeting, not a unique instrument, a Party response or a complete negotiation lineage. Information papers and other document types are outside this frame. Repeated submissions remain separate papers, and several papers can contribute to one instrument. We retain this frozen frame so that source corrections and outcome sensitivities concern the same papers. The statement frame separately ends at ATCM 46 (2024). The available ATCM 48/CEP 28 report (2026) is used for a later emperor penguin case update, not added to the adoption or statement comparisons \citep{ATCM2026FinalReport}.

We classified titles with ordered, case-insensitive rules. Report titles are removed first, including annual, progress, chair and convener reports and summaries. Next come procedural papers on prior assessment, comprehensive or initial environmental evaluation, and lists of working papers. Informational expressions such as an update on a subject, new records or a new centre are assigned to the residual category. Among remaining titles, revision takes precedence over new proposal. Revision cues include revision, amendment, modification, renewal (excluding renewable), renumbering, merger, replacement of a named instrument or rule, updated management plans or guidelines, review of a management plan or annex, and extension of expiry or designation. New-proposal cues include new (excluding New York, New Zealand and New College), designation, establishment, creation, a proposal for an arrangement, addition or inclusion, and a draft Measure, Decision, Resolution or Recommendation. Unmatched titles remain other. The exact expressions and their precedence are retained in \nolinkurl{lineage_passage.py}. These labels describe title language and need not establish legal novelty.

The predecessor consistency check asks whether an associated instrument acts on an earlier instrument in the frozen operative layer. It excludes bulk and spent-instrument actions. The original implementation retains the first unflagged same-meeting instrument link per paper before making that comparison. Agreement between title classes and predecessor links is a comparison of documentary classifications, not independent validation of title or outcome accuracy.

Submission-meeting outcomes began with machine-assisted dispositions in the frozen reconstruction. The original frame overrode a disposition whenever it found an unflagged same-meeting adoption association. Endorsement or agreement without an instrument, noting, deferral or referral, and non-agreement remained distinct. Withdrawal and unclear or absent outcomes were excluded. This originally retained 276 revisions and 91 new proposals. We preserve that cohort definition, subject to the unresolved-object exclusion below, rather than adding previously excluded papers selectively when an instrument can be found.

The original instrument audit records whether a paper sought an instrument, a missed adoption at submission, and an eventual instrument concerning the same object, with lag and confidence judgments. It added 13 positive working-paper outcomes, six revisions and seven new proposals, all retained in the present analysis. These are paper-level corrections, not distinct instruments. Papers originally classified as adopted are treated as instrument-seeking, including revisions whose adoption is subsequently found to have been deferred. Other sought judgments distinguish an instrument request from discussion, information, an intersessional process, internal advice or a review requiring no change. The automatic sensitivity searches paper text for adoption, draft-instrument, designation, listing or attached-plan language and rejects explicit no-change language.

We then assessed the union of 98 disagreements between saved dispositions and link-based adoption assignments, 15 weak-link-only adoptions and the 13 positive corrections, totalling 124 papers after overlap. For each, the source ledger distinguishes requested object and version, CEP advice or referral, and final ATCM action. Adoption requires an instrument adopting at least one identified primary requested instrument object. Reported amendments are allowed, but a shared site or subject, an extension of an existing plan, or an earlier-stage recommendation is insufficient. This does not establish that every original clause survived or that the instrument entered into force. For bundled proposals the ledger identifies which objects passed. Additional checks of nine originally unclear papers remain supplementary and do not enlarge the primary denominator. The initial assessment and subsequent panel review were model-assisted and not blind to existing outcome assignments. Exact model versions and generation settings are not recoverable for every earlier judgment. The field name \nolinkurl{sought_hand} does not establish human coding. Outcomes outside this bounded review retain their original classifications.

The reviewed comparison reverses four same-meeting revision adoptions and excludes ATCM 16 WP 41 because the record does not resolve its individual contribution to a subgroup tourism instrument. The primary denominator is therefore 276 revisions and 90 new proposals. For ATCM 28 WP 37, the primary object is the revised Ardley Island management plan. Its separately requested extension of the old plan is counted only in an alternative-outcome sensitivity. Further sensitivities assign the unresolved tourism paper either outcome, or exclude the entire union of 111 originally disputed or weak-link-only assignments while preserving the 13 positive corrections. These are coding or sample-restriction checks, not independent validation or causal comparisons. The frozen frame is unchanged. An explicit, source-backed overlay supplies the revised categories and lags. Missing frozen inputs or a missing overlay raise an error rather than triggering a reconstruction from another checkout.

Follow-up ends at ATCM 47. Direct report chains move the Byers Peninsula and Haswell Island revisions to adoption one meeting later and the Ardley Island revision to four meetings later. The later disposition of ATCM 21 WP 24 remains unresolved. We retain it as non-adoption at submission but excluded from positive-window and Kaplan--Meier denominators, rather than treated as never adopted. The unresolved ATCM 16 WP 41 is also excluded from those denominators. Lag zero denotes adoption at submission. Positive lags are coded meeting steps to a later instrument, not elapsed years or observed bargaining rounds. The audit explicitly counts the 2000 special meeting as the next step for two entries. For a window of $w$ meetings, the implemented denominator retains only papers with $47 - m \geq w$, where $m$ is the submission meeting number, even if a more recent paper was adopted earlier. These fixed-window estimates compare cohorts with complete follow-up. Windows are zero, one, two, three and five meetings. The stored Kaplan--Meier calculation instead censors unadopted papers at the end of available follow-up. It is not the source of the main fixed-window percentages. A later instrument may have changed scope, so a match does not establish adoption of the original text unchanged.

\subsection{Working-paper statistical comparison}

\protect\phantomsection\label{sec:methods-paper-statistics}{}

For each outcome, we estimate the revision-minus-new risk difference within strata defined by agenda route, era and site subject. Route is CEP when the recorded agenda field begins with CEP, and ATCM otherwise. Eras are before 1995 (meetings below 19), 1995--2004 (19--27), and 2005 onwards (28--47). Site subjects are detected from title references to protected or managed areas, management plans or arrangements, historic sites or monuments, visitor sites or site guidelines, and their recorded abbreviations. These controls describe the record and do not make the groups exchangeable.

Writing stratum sizes as $n_{R,s}$ and $n_{N,s}$ and outcome proportions as $p_{R,s}$ and $p_{N,s}$, we use weights $w_{s} = n_{R,s}\frac{n_{N,s}}{n_{R,s} + n_{N,s}}$ and the standardised difference

\begin{equation}
D = \frac{\sum_{s}w_{s}\left( p_{R,s} - p_{N,s} \right)}{\sum_{s}w_{s}}.
\end{equation}Only strata containing both types contribute. We use 2,000 bootstrap
draws of submission meetings with replacement, retain all papers in each drawn meeting and recalculate the weights. Intervals are the 2.5th and 97.5th percentiles of finite estimates, with random seed zero. The stored two-sided permutation comparison separately shuffles revision/new labels within strata 2,000 times and uses an add-one correction, allowing $10^{- 12}$ for numerical equality. It is a descriptive association check, not random assignment of proposal type.

Sensitivity comparisons retain site and non-site subjects separately, exclude the online ATCM 43 meeting, restrict to 2005 onwards, distinguish single from multiple sponsors, and vary instrument-seeking and outcome inclusion rules. Corrected comparisons include the curated instrument-seeking subset, the automatic text rule, inclusion of uncertain sought judgments and inclusion of unclear outcomes. A strict match sensitivity retains only high-confidence instrument corrections and follow-up matches. The archived alternative model-assisted sought/match coding substitutes those judgments where available and weights sampled originally adopted revisions by the ratio of all originally adopted revisions to those coded, retaining unit weight elsewhere. The same direct object/stage corrections and exclusions are then applied to both match variants, preserving the original sampling weights. Unreviewed alternative judgments remain unchanged. Its sample sizes count observed rows rather than weighted totals, and its intervals also resample meetings. The two-meeting gaps are standardised differences. Sensitivity intervals that include zero must remain visible with the results. Scheduled review, prior selection into adoption, scope and agenda composition remain alternative explanations.

We also fit an exploratory threshold model to the 276 revisions and 90 new proposals in the primary denominator, asking whether the accommodation threshold can approximate their recorded outcomes. Each paper $j$ carries an unobserved residual demand $R_{j} = \mu_{j} + \epsilon_{j}$, with standard logistic $\epsilon_{j}$. Agreement is recorded when $R_{j} \leq 0$, deferral or referral when $0 < R_{j} \leq a$, and non-agreement when $R_{j} > a$, with $a > 0$. A separate logistic regression distinguishes instrument adoption from agreement without an instrument, and a constant probability covers noting. Both regressions use proposal type, agenda route, site subject and era. Slopes carry a quadratic penalty of one, and intercepts are unpenalised. Half-count smoothing is applied to the noting probability and to the comparison models built from outcome frequencies. We score predictions by the mean negative log predictive probability. One scheme holds out each whole meeting in turn, and the other trains on ATCM 16--43 and predicts ATCM 44--47. Both are repeated after removing training papers whose titles match a held-out paper once case and spacing are normalised. The comparison models are outcome frequencies by proposal type and an unrestricted multinomial model with the same covariates (Supplementary Section S22.4).

\subsection{Additional checks of documentary coding}
\label{sec:methods-documentary-panel}
We completed a further documentary review using separately tasked AI agents and a human coder. Two source reviewers divided the 124 working papers in the bounded object/stage review and examined the cited report passages, distinguishing the requested object, CEP advice and final ATCM action. They retained source locators and quotations for each judgment. This review was not blind to existing outcomes and used raw report text or archived report extracts, with the source layer recorded for each case.

Statement recoding and extraction-coverage procedures are reported in Supplementary Section S22.5.

\subsection{Selection and interpretation of proposal histories}

\protect\phantomsection\label{sec:methods-cases}{}

The four cases in main Figure 5b were selected to illustrate contrasts, not sampled to estimate pathway frequencies. We retain forum, proposal object, version, recorded response and outcome. CEP advice, CCAMLR approval and ATCM adoption are different stages. A recorded objection preventing agreement excludes a stand-aside explanation for that objecting Party on that version at that decision point. It does not classify other Parties, establish private motives or exclude reluctant acceptance or future restraint when a revised proposal is later adopted. The case panel distinguishes adoption, CCAMLR approval, reported non-agreement and unresolved discussion. Its five-pathway matrix is an interpretive comparison. Ticks indicate a supported reading, crosses exclusions for objectors at documented blocking episodes, and question marks unresolved pathways. The penguin rows separate adoption of the Amanda Bay colony plan in 2025, for which all five pathways remain unresolved, from the species-listing blocks in 2022 and 2024. On those blocked listing versions, accommodation remains unresolved and the other four pathways are excluded only for the objectors. The merger rows separate the blocked 2022 plan from approval of the revised plan in 2023 and its adoption in 2024. The revision sequence supports accommodation, while the Parties' final responses remain unidentified. Crosses apply only to the objectors on the blocked 2022 version. No full Party response profile or exclusive causal mechanism is inferred from the matrix.

The case evidence follows emperor-penguin species protection alongside the separate Amanda Bay site plan. It also follows the combination of ASPAs 152 and 153 into ASPA 182 and the Ross Sea and East Antarctic marine-protected-area proposals. The reports and stage-specific source references are retained with the supplementary histories \citep{ATCM2022FinalReport} \citep{ATCM2023FinalReport} \citep{ATCM2024FinalReport} \citep{ATCM2025FinalReport} \citep{CCAMLR2012Report} \citep{CCAMLR2013Report} \citep{CCAMLR2014Report} \citep{CCAMLR2015Report} \citep{CCAMLR2016Report} \citep{CCAMLR2022Report} \citep{CCAMLR2023Report} \citep{CCAMLR2025Report}. The Amanda Bay plan is a narrower, separate legal object, not a demonstrated revision of the species-listing draft. No new species-designation proposal was tabled in 2025, so continued disagreement that year is not another observed veto on a submitted listing. Changes in fishing access, boundaries, research arrangements, duration and scope supply no measured scalar bite or uniquely identified limiting responder. Differences between Ross Sea and East Antarctica prevent treating their histories as a controlled comparison.

\subsection{Documentary prominence and sign scenarios}

\protect\phantomsection\label{sec:methods-prominence}{}

The topic-profile exercise uses working papers from 2015 through 2024. The 2020 meeting did not occur. It reads the packaged actor--paper--topic table and a 29-Party Consultative membership list. The table derives from the documentary database's document summary, retains year, paper, party, topic and paper type, and deduplicates those records. Rebuilding it restricts to working papers, expands the recorded submitting Parties, normalises the aliases Korea (ROK) and Russian Federation, and retains Consultative Parties. Information papers are outside this action-paper frame. The records link papers to broad agenda topics, not to a particular adopted proposal, amendment path or instrument. They supply heterogeneous stress-test inputs rather than reconstructed proposal-level stakes or positions.

Let $n_{iP}$ be party $i$`s working-paper authorships on topic $P$, $n_{i}$ its authorships across the corpus, $n_{P}$ all parties' authorships on the topic, and $n$ all authorships. Relative topic specialisation is $\text{RTS}_{iP} = \frac{\frac{n_{iP}}{n_{i}}}{\frac{n_{P}}{n}}$. Prominence is

\begin{equation}
a_{iP} = \frac{\log(1 + \text{RTS}_{iP})}{c_{0.95}},
\end{equation}where $c_{0.95}$ is the 95th percentile of $\log(1 + \text{RTS})$
among strictly positive actor--topic cells in the full corpus before topic filtering. Values above one remain uncapped. Denominators and scale are calculated before removing topics with fewer than 30 actor--paper observations. The Secretariat umbrella label \nolinkurl{ALL} remains in the denominators but not among the eight substantive topic profiles. The all-zero party receives prominence zero on every topic and remains in the 29-party response system. The ratio measures relative documentary specialisation. A party with few submissions can have high prominence from one topic paper. It does not identify preference direction, influence, expertise, policy stakes, bargaining power or a position on adopted text. A zero cell leaves political direction open. Heat-map clustering makes similar documentary profiles adjacent, not political blocs.

The analyst assigns direction $z_{iP} \in \left\{ - 1, + 1 \right\}$ and transfer $\rho_{P} \in \lbrack 0,1\rbrack$ in the location model $s_{iP} = \mu_{0}(P) - \eta\rho_{P}a_{iP}z_{iP} + \varepsilon_{iP}$, with conditionally independent residuals $\varepsilon_{iP} \sim \mathcal{N}(0,\tau^{2})$. Positive signs lower veto-payoff locations and move parties toward restraint. The baseline fixes $N = 29$, $b = 1$, $\mu_{0}(P) = 0.6$, $\eta = 0.4$, $\rho_{P} = 1$, $\tau = 0.15$, $m = 0.5$, $\kappa = 8$ and $L(q) = 1$ across topics. These are illustrative assumptions, not estimates of preferences, topic relevance or continuation payoffs. Lower transfer compresses sign sensitivity toward the baseline location.

Starting from the same scalar initial belief for every party, the heterogeneous map iterates $q_{i}' = F_{i}\left( b\prod_{j \neq i}q_{j} + \delta L\left( \overline{q} \right) \right)$, where $\overline{q} = \sum_{i}\frac{q_{i}}{N}$ and $F_{i}$ is the normal distribution at the assigned location. The focal party is excluded from its own current-passage product. At a selected vector limit, passage is $\prod_{i}q_{i}$. Removing continuation while holding that limit fixed gives current-payoff passage $\prod_{i}F_{i}\left( b\prod_{j \neq i}q_{j} \right)$. Their difference is continuation-supported passage $P_{C}$. Resolving at $\delta = 0$ is a different counterfactual. The threshold $\delta_{50}$ separates selected passage below one half from passage at least one half, under this maintained selection protocol. It is not an estimate of institutional patience.

The tourism profile has positive prominence for 17 parties. The other 12 remain in the unanimity products but their signs do not alter their locations. Opposition count $k$ therefore counts negative signs only among positive-prominence parties. For each $k = 0,\ldots,17$, the placement design evaluates opposition among the $k$ most prominent and $k$ least prominent parties and up to 96 randomly drawn subsets, sampling without replacement within each subset and discarding duplicate subsets at that count. It also evaluates 256 independent fair-sign profiles (seed 20260729). The resulting 1,309 random placements and 1,601 design rows include ordered and fair-sign profiles. These rows need not all be distinct sign vectors. Random bands describe a constructed assignment distribution, not confidence intervals. Ordered endpoints are sharp for the transition conditional on common-scalar starts, common response parameters apart from location, permutation-invariant selection and convergence. Ordered maximum-$P_{C}$ paths provide a sensitivity range, not sharp bounds.

The four named scenarios are broad alignment, one negative sign on the most prominent party, alternating signs down the positive-prominence ranking starting with a negative sign, and negative signs on the most prominent third, rounded up. The balanced topic-by-scenario decomposition allocates sums of squared deviations to topic, sign scenario and their interaction. Eight alternative menus replace the named scenarios with four independent fair-sign masks applied by prominence rank (seed 20260729). These comparisons diagnose the chosen scenario menu rather than the information supplied by the documents.

The named scenarios use 128 equal-probability Beta mid-quantile draws of the common starting belief. The fair-sign integration uses 48 belief draws per sign profile and 512 sign profiles arranged as eight independently scrambled Sobol batches of 64, with seeds 1407--1414. Across-batch uncertainty uses a $t$ multiplier of 2.365 for seven degrees of freedom. Comparing the first 32 profiles per batch with all 64 checks integration resolution. This numerical integration uncertainty is separate from the random-placement bands and from statistical uncertainty about historical behaviour.

Co-sponsorship components are transitive closures of direct joint-working-paper links among parties with positive prominence. The component-equality restriction assumes a common direction within each component. The graph itself does not reveal that direction, and agreement on a paper need not extend to the adopted proposal. Under the restriction, observing one party's proposal-level direction fixes its component's sign. We solve the remaining component assignments and compare the worst remaining transition range for either observed sign with the unrestricted component-equality range. Zero-prominence parties remain in the full response system but outside this graph.

Prominence sensitivity replaces relative specialisation with each party's share of a topic's authorships, again transformed by $\log(1 + x)$ and scaled by the 95th percentile of positive full-corpus cells. Profile ablations preserve the positive-cell pattern while either permuting positive values within topics, assigning their topic mean, or drawing lognormal values matched to the pooled positive log-moments. There is one levelled profile and five shuffled and synthetic replicates (seed 20260729). Unresolved transition crossings remain missing rather than being assigned a boundary. Transition averages use only resolved draws. One-at-a-time parameter grids are $\mu_{0} \in \left\{ 0.5,0.55,0.6,0.65,0.7 \right\}$, $\eta \in \left\{ 0.2,0.3,0.4,0.5,0.6 \right\}$, $\tau \in \left\{ 0.10,0.125,0.15,0.20,0.25 \right\}$, $m \in \left\{ 0.3,0.4,0.5,0.6,0.7 \right\}$ and $\kappa \in \left\{ 2,4,8,16,32 \right\}$, holding the remaining baseline parameters fixed.

For each topic, the mean share averages the ordered placement gap over nontrivial positive-prominence opposition counts as a fraction of the full transition bound. The widest share is the largest count-specific fraction. For maximum $P_{C}$, the reference is the range spanned by the ordered placements, not a sharp bound. Within-count placement variance is averaged with weights proportional to the number of retained random rows at each count and divided by the variance of all those rows. Count, mean opposed prominence and its coefficient of variation are separated by averaging sequential explained-variance increments over every predictor ordering in ordinary least squares, with standardised predictors and an intercept, for stratified random draws with $2 \leq k \leq 16$. Within-count slopes describe how the transition changes with mean opposed prominence. Concentration uses weights proportional to prominence in the opposed set and population-divisor variance for its coefficient of variation. The subset-mean skew calculation is for simple random subsets without replacement. Its sign change belongs to the sampling design, not a political effect.

\subsection{Numerical solution and checks}

\protect\phantomsection\label{sec:methods-numerics}{}

The scalar solver brackets roots of $T(q) - q$ using a mesh supplemented by extrema located from the analytic derivative $T'(q) - 1$. The extrema split intervals that could hide a nearby unstable/stable root pair. Each returned root is refined to bracket width $10^{- 11}$ in $q$ and classified using the analytic derivative. This derivative-aware search improves fold resolution but does not guarantee completeness for arbitrary smooth continuation functions with unresolved derivative oscillations.

The general party-specific solver checks every initial-belief draw with strict residual tolerance $10^{- 10}$ and a 2,000-iteration budget. A draw exhausting that budget survives only if its residual is at most $2 \times 10^{- 7}$. Otherwise the solver raises an error. Diagnostics retain fallback counts and maximum residuals. The batched sign-placement calculations instead request tolerance $10^{- 9}$ and up to 10,000 iterations, retaining the same fold-residual ceiling. Monotonicity alone does not guarantee convergence of simultaneous vector iteration, so selected-limit claims require the recorded convergence checks.

The scalar-versus-party-specific transition comparison uses direct bisection to tolerance $10^{- 5}$, with 512 initial-belief draws for the heterogeneous calculation. Passage thresholds 0.1, 0.5 and 0.9 are compared as a numerical sensitivity. Batched placement calculations start from 61 equally spaced discount factors over $\lbrack 0,0.999\rbrack$ and subdivide each transition bracket into 17 steps, reporting its midpoint with resolution approximately $9.8 \times 10^{- 4}$. The scenario-menu sensitivity uses 41 grid points and eight subdivisions. Maximum-$P_{C}$ resolution is checked by recalculating the 36 ordered tourism paths on 61, 121 and 241 points.

Checks require nonnegative Beta basin weights summing to one, $0 \leq P_{C} \leq P_{\text{pass}} \leq 1$, and $P_{C} = 0$ when continuation is absent. Monte Carlo draws of initial beliefs and private types provide a comparison with analytic basin weights and outcome probabilities. The ordered-placement check exhausts all $2^{17} = 131,072$ tourism sign vectors against location-order bounds. It does not numerically solve every vector's transition. The general continuation solver is compared with the constant and linear wrappers. These are model and numerical checks, not validation of the documentary preference mapping.

%% file: sections/supplement.tex
\textbf{Contents and relation to the main text}

This supplement gives the mathematical derivations and supporting numerical and documentary results behind \emph{How to Agree Without Agreeing}. Main Methods describes the core model, simulation and documentary procedures. Additional statement-comparison procedures are in Section S22.2, with documentary checks in Section S22.5. The \hyperref[sec:si-response]{fixed-proposal game}, \hyperref[sec:si-five-pathways]{five pathways}, \hyperref[sec:si-bite]{proposal threshold} and \hyperref[sec:si-observation-bound]{documented-response bound} support main-text Figure 1. The \hyperref[sec:si-strategic]{proposal-redesign conditions} support Figure 2. The \hyperref[sec:si-learning]{learning game} and \hyperref[sec:si-learning-current]{main-text simulations and interventions} support Figures 3 and 4. The \hyperref[sec:si-record]{Antarctic record analyses} support Figure 5. The \hyperref[sec:si-unsigned]{unsigned documentary profiles} supply the separate tourism stress test. Additional model variants and numerical checks retain their stated assumptions, and their results are not all properties of the learning game.

\section{Accommodation and restraint}

A consensus record establishes that no party maintained a veto. Unanimous active support remains unknown. We use $A$, $Q$, and $V$ to denote active support, standing aside (acquiescence), and veto, respectively. In the full response game, party $i$ chooses $\sigma_{i} \in \left\{ A,Q,V \right\}$. A proposal passes when every party chooses either $A$ or $Q$.

\textbf{Response-side restraint} occurs when at least one party chooses $Q$ on the proposal that reaches the table. Passage with at least one $Q$ is \textbf{apparent consensus}. The definition concerns behaviour toward the actual proposal.

\textbf{Proposal-side accommodation} occurs when the proposer lowers the proposal's bite or changes its instrument class because a stronger alternative would provoke a veto or fail in implementation or ratification. Accommodation can occur before submission or after a blocked draft is revised. It can produce unanimous support for the weaker proposal. Such an outcome conceals counterfactual opposition. It counts as apparent consensus only when at least one party chooses $Q$.

\begin{longtable}[]{@{}
  >{\raggedright\arraybackslash}p{(\linewidth - 4\tabcolsep) * \real{0.2581}}
  >{\raggedright\arraybackslash}p{(\linewidth - 4\tabcolsep) * \real{0.2581}}
  >{\raggedright\arraybackslash}p{(\linewidth - 4\tabcolsep) * \real{0.4839}}@{}}
\caption{Proposal adjustment and response composition under
non-opposition.}\label{tab:si-mechanisms}\tabularnewline
\toprule\noalign{}
\endfirsthead
\endhead
\bottomrule\noalign{}
\endlastfoot
\textbf{Proposal choice} & \textbf{Response profile} &
\textbf{Interpretation} \\
No accommodation & No restraint & Unanimous active support for the
proposer's preferred proposal \\
Accommodation & No restraint & Accommodated agreement in which all parties
support the weaker actual proposal \\
No accommodation & Restraint & Apparent consensus through restraint on
the preferred proposal \\
Accommodation & Restraint & Accommodated apparent consensus in which the
proposer weakens the proposal and at least one party still acquiesces \\
\end{longtable}

These mechanisms answer different questions. Accommodation explains why a harder proposal disappears before formal consideration. Restraint explains why a party refrains without actively supporting the proposal under consideration. This can conceal an objection, but $Q$ can also be optimal for a positively aligned party that avoids the cost of active support. A binary passage record can pool all four cells.

\section{Timing keeps the two mechanisms distinct}

\protect\phantomsection\label{sec:si-timing}{}

The proposal-response game has five stages.

\begin{longtable}[]{@{}
  >{\centering\arraybackslash}p{(\linewidth - 4\tabcolsep) * \real{0.0804}}
  >{\raggedright\arraybackslash}p{(\linewidth - 4\tabcolsep) * \real{0.2590}}
  >{\raggedright\arraybackslash}p{(\linewidth - 4\tabcolsep) * \real{0.6610}}@{}}
\caption{Timing of proposal choice, response and
continuation.}\label{tab:si-timing}\tabularnewline
\toprule\noalign{}
\endfirsthead
\endhead
\bottomrule\noalign{}
\endlastfoot
\textbf{Stage} & \textbf{Actor} & \textbf{Move} \\
0 & Nature or procedure & An issue or menu $z_{t}$ arrives according
to $\mu(z_{t}|s_{t})$. \\
1 & Proposer $p$ & After observing the available issue, choose bite
$x_{t} \in \mathcal{X}(z_{t})$ or instrument $g_{t}$. \\
2 & Responders $i = 1,\ldots,N$ & Choose $A$, $Q$, or $V$
simultaneously. \\
3 & Institution & Record passage $Y_{t} = 1$ iff no responder chose
$V$. Record ratification and implementation separately when the
institution distinguishes ratification from implementation. \\
4 & Public state & Update $s_{t + 1}$ from the proposal, response
profile, public outcome, and any exogenous continuation event. \\
\end{longtable}

See main Methods (``Model, response definitions and proposal choice'') for the model and procedures.

\begin{longtable}[]{@{}
  >{\raggedright\arraybackslash}p{(\linewidth - 4\tabcolsep) * \real{0.2150}}
  >{\raggedright\arraybackslash}p{(\linewidth - 4\tabcolsep) * \real{0.3178}}
  >{\raggedright\arraybackslash}p{(\linewidth - 4\tabcolsep) * \real{0.4674}}@{}}
\caption{Payoff and continuation assumptions of the model
variants.}\label{tab:si-model-assumptions}\tabularnewline
\toprule\noalign{}
\begin{minipage}[b]{\linewidth}\raggedright
\textbf{Model}
\end{minipage} & \begin{minipage}[b]{\linewidth}\raggedright
\textbf{Current payoff assumption}
\end{minipage} & \begin{minipage}[b]{\linewidth}\raggedright
\textbf{Choice and continuation assumption}
\end{minipage} \\
\midrule\noalign{}
\endfirsthead
\toprule\noalign{}
\begin{minipage}[b]{\linewidth}\raggedright
\textbf{Model}
\end{minipage} & \begin{minipage}[b]{\linewidth}\raggedright
\textbf{Current payoff assumption}
\end{minipage} & \begin{minipage}[b]{\linewidth}\raggedright
\textbf{Choice and continuation assumption}
\end{minipage} \\
\midrule\noalign{}
\endhead
\bottomrule\noalign{}
\endlastfoot
Fixed proposal & Passage adds $b_{i}$, and stance adds alignment and
action costs & The proposal is given, and $A$ and $Q$ share
continuation \\
Continuous implementation & Policy loss $g_{i}x$ is incurred upon
passage & Under complete information, a passage-seeking proposer chooses
bite \\
Direct-cost feedback & Choosing restraint incurs $gx$, even after
another veto & Independent private types with a stated adaptive-expectations
selection rule \\
Discrete instrument & Effective implementation depends on both
instrument and response & A proposer selects $B$ or $H$, and the
punishment proposer may be indifferent \\
Public-state game & Payoffs and transitions are specified at every node
& Strategic proposal and response deviations receive a complete audit \\
Private types & The response payoff has zero proposal bite & Independent
private types and common-belief basin selection \\
Learning & Public modes determine menus and continuation transitions &
Realised SARSA updates select behaviour, followed by a separate
equilibrium audit \\
\end{longtable}

\begin{figure}[htbp]
\centering
\includegraphics[width=1\linewidth,height=0.65\textheight,keepaspectratio]{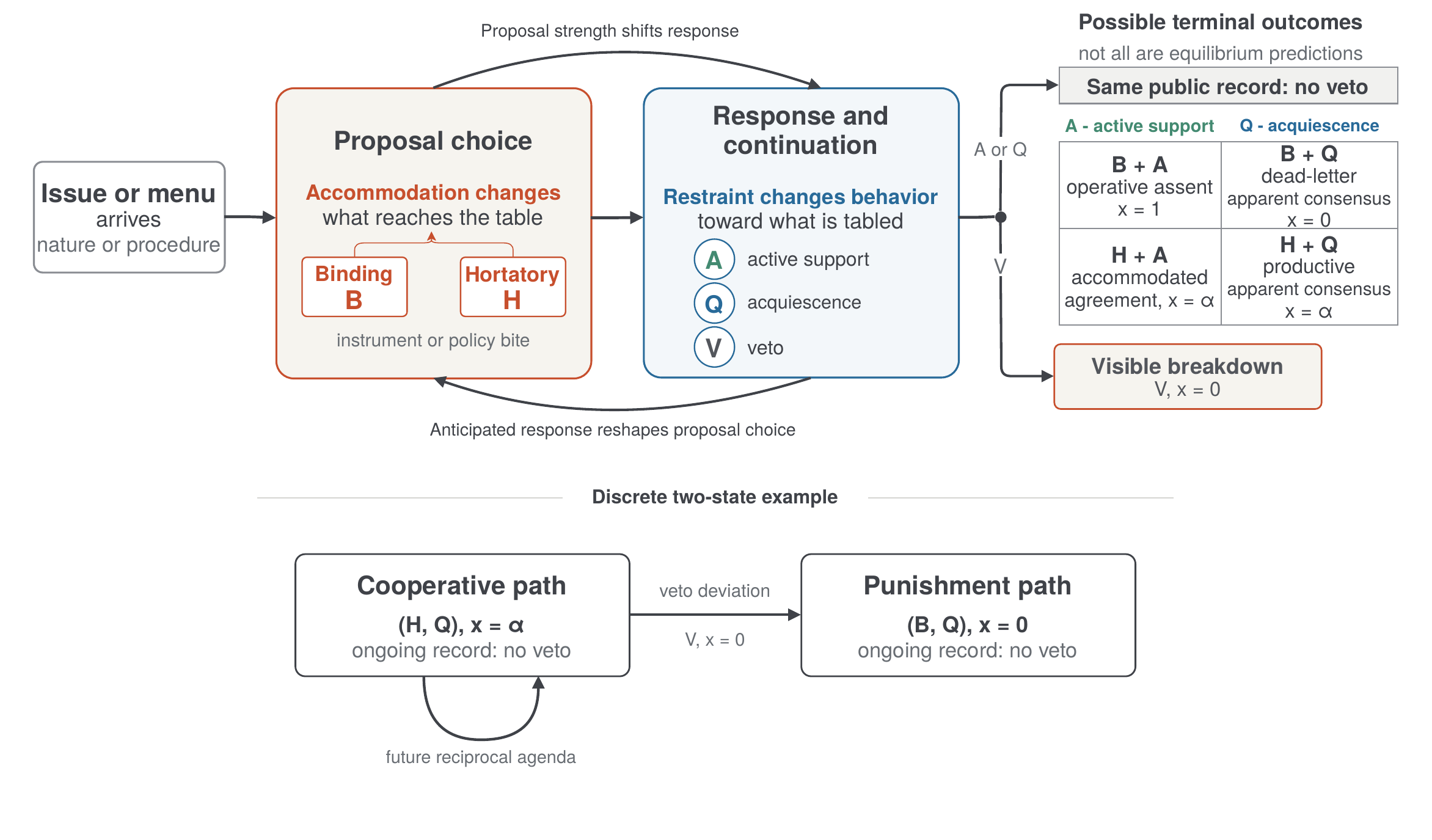}
\caption{The agenda-response process and its terminal outcomes. Proposal choice and response follow the arrival of an issue or menu. The proposal stage permits accommodation, and the response and continuation stage permits restraint. The matrix of instruments $B,H$ and responses $A,Q$ lists possible terminal outcomes sharing a no-veto record and supplies no equilibrium prediction. $V$ produces visible breakdown. Accommodation counts as apparent consensus only when at least one party acquiesces to the actual proposal. The lower loop previews the repeated-game comparison. Its $(B,Q)$ punishment rests on a weak proposer tie unless commitment or a rejection cost makes the comparison strict.}
\label{fig:model-ownership}
\end{figure}

\section{Fixed-proposal game: restraint by responders}

\protect\phantomsection\label{sec:si-response}{}

Active support means advancing a position, which is more than allowing adoption \citep{Rietig2023}. Preparing positions and obtaining instructions can require resources even when a delegation supports a proposal \citep{Panke2017}, while Antarctic practitioners report time and resource constraints on science-policy engagement \citep{Gardiner2025}. Experiments also show that charging for political comments reduces expression \citep{BartonRodet2015}. We represent the additional effort of actively advancing a position relative to standing aside by a stance cost, following an expression-cost formulation in opinion models \citep{Gaisbauer2020}. Blocking is distinguished by the higher standard for objection described in consensus practice \citep{Rietig2023}. These sources motivate the distinction between expression and blocking costs, not numerical estimates of either.

Fix a public proposal and suppose every other party refrains. Party $i$ receives passage value $b_{i} > 0$. Active support carries alignment payoff $h_{i}$ and stance cost $\Delta_{i} > 0$. Veto carries additional cost $c_{V,i} \geq 0$. Let $W_{i}^{C}$ denote continuation after restraint and $W_{i}^{P}$ continuation after veto, and define

\begin{equation}
L_{i} = W_{i}^{C} - W_{i}^{P} \geq 0.
\end{equation}The three payoffs, conditional on every other party refraining, are

\begin{equation}
u_{i}(A) = h_{i} - \Delta_{i} + b_{i} + \beta W_{i}^{C},\quad u_{i}(Q) = b_{i} + \beta W_{i}^{C},\quad u_{i}(V) = - h_{i} - \Delta_{i} - c_{V,i} + \beta W_{i}^{P}.
\end{equation}Write the net stance payoffs as

\begin{equation}
a_{i} = h_{i} - \Delta_{i},\quad s_{i} = - h_{i} - \Delta_{i} - c_{V,i},
\end{equation}so $R_{i}(A) = a_{i}$, $R_{i}(Q) = 0$, and $R_{i}(V) = s_{i}$. In
general, net passage value is $B_{i}(x) = b_{i} + v_{i}(x)$. The fixed-proposal game here takes $v_{i} = 0$, so $B_{i} = b_{i}$. The current reward $R_{i}\left( \sigma_{i} \right) + B_{i}(x)Y$ is an exact rewriting for every response profile, including profiles with another party's veto.

An equivalent two-parameter form uses the combined position cost $k_{i} = 2\Delta_{i} + c_{V,i} > 0$, giving $s_{i} = - a_{i} - k_{i}$. This indexed $k_{i}$ is local to the response-payoff mapping, while the proposer models use $k$ for rejection cost. Neither $\left( a_{i},s_{i} \right)$ nor $\left( a_{i},k_{i} \right)$ alone identifies alignment $h_{i}$, because recovering $h_{i} = a_{i} + \Delta_{i}$ requires the stance-cost decomposition. We retain $h_{i}$, $\Delta_{i}$, and $c_{V,i}$ whenever the analysis separates alignment from the costs of expressing that alignment. Common-cost variants below set $\Delta_{i} = \Delta$.

Relative to acquiescence, active support scores $a_{i}$, acquiescence scores zero, and veto scores

\begin{equation}
G_{i} - \beta L_{i},\quad G_{i} = s_{i} - b_{i} = - h_{i} - \Delta_{i} - c_{V,i} - b_{i}.
\end{equation}This local menu assumes that every other party refrains and that $A$
and $Q$ share continuation. In the general implementation model, $G_{i} = s_{i} - B_{i}(x)$ under the same conditions. Dynamic solutions retain the full current rewards and the continuation specified at each state.

Without restricting the relationship between the three scores, the complete best-response conditions are

\begin{equation}
\begin{aligned}
A \in \text{ BR}_{i} & \Longleftrightarrow h_{i} - \Delta_{i} \geq \max(0,G_{i} - \beta L_{i}), \\
Q \in \text{ BR}_{i} & \Longleftrightarrow 0 \geq \max(h_{i} - \Delta_{i},G_{i} - \beta L_{i}), \\
V \in \text{ BR}_{i} & \Longleftrightarrow G_{i} - \beta L_{i} \geq \max(0,h_{i} - \Delta_{i}).
\end{aligned}
\end{equation}The fixed-proposal model links $G_{i}$ and $h_{i}$ through
$G_{i} = - h_{i} - \Delta_{i} - c_{V,i} - b_{i}$. Together with $b_{i} > 0$, $\Delta_{i} > 0$, $c_{V,i} \geq 0$ and $\beta L_{i} \geq 0$, this makes the support--veto comparison redundant once support beats acquiescence. The following proposition uses these restrictions.

\begin{proposition}[Fixed-proposal best response]

If $h_{i} > \Delta_{i}$, active support is party $i$`s unique best response. If $h_{i} = \Delta_{i}$, support and acquiescence tie and both strictly beat veto. If $h_{i} < \Delta_{i}$, acquiescence is a best response exactly when $G_{i} \leq \beta L_{i}$, and is uniquely optimal when the inequality is strict. Veto is uniquely optimal exactly when $G_{i} > \beta L_{i}$. Continuation makes acquiescence optimal against a currently profitable veto exactly when $0 < G_{i} \leq \beta L_{i}$, and is uniquely optimal when $0 < G_{i} < \beta L_{i}$.

\end{proposition}

\begin{proof}

Subtracting $u_{i}(Q)$ from $u_{i}(A)$ gives $h_{i} - \Delta_{i}$. Subtracting $u_{i}(Q)$ from $u_{i}(V)$ gives $G_{i} - \beta L_{i}$. For $h_{i} \geq \Delta_{i}$, support also strictly beats veto because $u_{i}(A) - u_{i}(V) = 2h_{i} + b_{i} + c_{V,i} + \beta L_{i} > 0$. For $h_{i} < \Delta_{i}$, acquiescence strictly beats support, leaving the comparison with veto. Moreover, $G_{i} > \beta L_{i}$ implies $h_{i} < - \left( \Delta_{i} + c_{V,i} + b_{i} + \beta L_{i} \right) < 0$, so support cannot beat a profitable veto. The stated comparisons follow. At $G_{i} = \beta L_{i}$, acquiescence and veto tie. No parameter values satisfying the restrictions produce a three-action tie.

\end{proof}

For $h_{i} < \Delta_{i}$, current-payoff acquiescence occupies $G_{i} \leq 0$, while continuation-supported restraint occupies $0 < G_{i} \leq \beta L_{i}$. Both actions print as $Q$. A no-veto outcome is apparent consensus when every party refrains and at least one party occupies one of these $Q$ regions.

\subsection{Five pathways in the main-text figure}

\protect\phantomsection\label{sec:si-five-pathways}{}

See main Methods (``Model, response definitions and proposal choice'') for the model and procedures.

\begin{longtable}[]{@{}
  >{\raggedright\arraybackslash}p{(\linewidth - 4\tabcolsep) * \real{0.2212}}
  >{\raggedright\arraybackslash}p{(\linewidth - 4\tabcolsep) * \real{0.3366}}
  >{\raggedright\arraybackslash}p{(\linewidth - 4\tabcolsep) * \real{0.4424}}@{}}
\caption{Five pathways represented in main-text Figure
1.}\label{tab:si-pathways}\tabularnewline
\toprule\noalign{}
\begin{minipage}[b]{\linewidth}\raggedright
\textbf{Pathway}
\end{minipage} & \begin{minipage}[b]{\linewidth}\raggedright
\textbf{Condition or change}
\end{minipage} & \begin{minipage}[b]{\linewidth}\raggedright
\textbf{What passage leaves unknown}
\end{minipage} \\
\midrule\noalign{}
\endfirsthead
\toprule\noalign{}
\begin{minipage}[b]{\linewidth}\raggedright
\textbf{Pathway}
\end{minipage} & \begin{minipage}[b]{\linewidth}\raggedright
\textbf{Condition or change}
\end{minipage} & \begin{minipage}[b]{\linewidth}\raggedright
\textbf{What passage leaves unknown}
\end{minipage} \\
\midrule\noalign{}
\endhead
\bottomrule\noalign{}
\endlastfoot
Accommodation & The proposer weakens a demand that would otherwise be
blocked. & Whether the actual proposal receives $A$ or $Q$, and how
much stronger the forgone proposal was. \\
Reluctant acceptance & $h_{i} < 0$ and $G_{i} \leq 0$. The party
chooses $Q$. & Opposition remains, but a veto has no positive current
gain. \\
Future restraint & $0 < G_{i} \leq \beta L_{i}$. The party chooses
$Q$. & A positive current veto gain is offset by the future at
stake. \\
Costly endorsement & $0 < h_{i} < \Delta_{i}$. The party chooses
$Q$. & Positive alignment does not produce active support. \\
Open support & $h_{i} > \Delta_{i}$. The party chooses $A$. &
Passage alone does not record that support was voiced. \\
\end{longtable}

The five response regions in main-text Figure 1b include veto, whereas the five agreement pathways in Figure 1c instead combine the four non-veto response regions with accommodation. The response conditions assume all other parties refrain, $b_{i} > 0$, $\Delta_{i} > 0$, $c_{V,i} \geq 0$, $\beta L_{i} \geq 0$, and identical continuation after $A$ and $Q$, as in the fixed-proposal game. At $h_{i} = \Delta_{i}$, $A$ and $Q$ tie, whereas at $G_{i} = \beta L_{i}$, $Q$ and $V$ tie. The figure assigns the boundaries to standing aside for display. A party with $h_{i} = 0$ is neutral in alignment, so neither positive support nor opposition follows from its silence.

\section{Discrete-instrument game: accommodation by proposers}

\protect\phantomsection\label{sec:si-discrete}{}

The binding-hortatory game lets the proposer choose an instrument whose implementation depends on the response.

The proposer chooses a binding instrument $B$ with nominal bite one or a hortatory instrument $H$ with bite $\alpha \in (0,1)$. The responder then chooses $A$, $Q$, or $V$. Binding force requires active follow-through, so effective bite is

\begin{equation}
x(g,\sigma) = \begin{cases}
1 & \text{ if  }g = B\text{  and  }\sigma = A \\
0 & \text{ if  }g = B\text{  and  }\sigma = Q \\
\alpha & \text{ if  }g = H\text{  and  }\sigma \in \left\{ A,Q \right\} \\
0 & \text{ if  }\sigma = V.
\end{cases}.
\end{equation}The responder has stance $\theta \in \left\{ + 1, - 1 \right\}$ and
salience $s \geq 0$. Here $s$ is an unsigned magnitude, distinct from the net veto payoff $s_{i}$ in the fixed-proposal game. Its payoff is

\begin{equation}
u_{R}(g,\sigma) = \theta sx(g,\sigma) - e\mathbb{1}_{\sigma = A} - c\mathbb{1}_{\sigma = V},
\end{equation}where $e,c > 0$. The proposer favours positive bite and receives
$u_{P} = s_{P}x(g,\sigma)$ for $s_{P} > 0$.

\begin{proposition}[Discrete best responses]

For an aligned responder, $\theta = + 1$, the selected outcome is $(B,A)$ with bite one when $s > e$, and $(H,Q)$ with bite $\alpha$ when $s < e$. For an opposed responder, $\theta = - 1$, the outcome is $(H,Q)$ when $\alpha s < c$. When $\alpha s > c$, both $(B,Q)$ and $(H,V)$ give the proposer zero effective bite. Selecting open veto in that region requires an explicit proposer tie-break toward $H$ or a payoff that favours public rejection over an unratified binding instrument.

\end{proposition}

\begin{proof}

Against $H$, an aligned responder obtains $\alpha s - e$, $\alpha s$, and $- c$ from $A,Q,V$. So $Q$ strictly dominates. Against $B$, the payoffs are $s - e,0, - c$, so $A$ wins for $s > e$ and $Q$ wins for $s < e$. The proposer then chooses $B$ in the first region and $H$ in the second. Against $B$, an opposed responder obtains $- s - e,0, - c$, so $Q$ strictly dominates. Against $H$, the payoffs are $- \alpha s - e, - \alpha s, - c$. So $Q$ wins for $\alpha s < c$ and $V$ wins for $\alpha s > c$. In the last region, the proposer obtains zero after either $B$ or $H$, which creates the stated tie.

\end{proof}

Writing signed salience as $h = \theta s$, and imposing the tie-break toward $H$, yields the familiar three-region map:

\begin{longtable}[]{@{}
  >{\raggedleft\arraybackslash}p{(\linewidth - 4\tabcolsep) * \real{0.2791}}
  >{\centering\arraybackslash}p{(\linewidth - 4\tabcolsep) * \real{0.2559}}
  >{\raggedright\arraybackslash}p{(\linewidth - 4\tabcolsep) * \real{0.4652}}@{}}
\caption{Selected outcomes in the discrete-instrument game, conditional
on the stated proposer tie-break.}\label{tab:si-discrete-regions}\tabularnewline
\toprule\noalign{}
\endfirsthead
\endhead
\bottomrule\noalign{}
\endlastfoot
$h > e$ & $(B,A)$ & operative agreement, bite one \\
$- \frac{c}{\alpha} < h < e$ & $(H,Q)$ & apparent consensus, bite
$\alpha$ \\
$h < - \frac{c}{\alpha}$ & $(H,V)$ & open breakdown, zero bite \\
\end{longtable}

The middle region combines accommodation and restraint. Its binding counterfactual either needs costly active follow-through or dies in ratification. Its hortatory actual proposal passes through acquiescence. Unlike the continuous-bite game below, this game separates formal passage from implementation, because $Q$ permits passage under either instrument but implements positive bite only under $H$.

\section{A changing agenda sustains restraint in the discrete game}

\protect\phantomsection\label{sec:si-discrete-repeated}{}

The discrete game also admits a changing-agenda construction with discount factor $\delta \in \lbrack 0,1)$. Two symmetric parties alternate proposer and responder roles forever. Each proposer favours its own direction with salience $s$, while the current responder opposes that direction. Write $a = \alpha s$ for the magnitude of the hortatory policy payoff and assume $a > c$.

In cooperative state $C$, the proposer chooses $H$ and the responder chooses $Q$. The role payoffs are $+ a$ and $- a$. After an off-path $B$ in $C$, the responder chooses $Q$. That proposal-response pair and every other deviation move play to punishment state $P$. In $P$, each current proposer chooses $B$ and each responder chooses $Q$ forever. This binding instrument passes but never enters into force, so both parties receive zero policy payoff. An off-path $H$ in state $P$ meets $V$, the responder's strict best response when $a > c$. Every action deviation in $P$ leaves play in $P$.

\begin{proposition}[Alternating instruments can sustain restraint]

The stated strategy forms a subgame-perfect equilibrium (SPE) when
\begin{equation}
a \leq c(1 + \delta),\text{  or equivalently  }\delta \geq \frac{a}{c} - 1.
\end{equation}For a fixed $\delta$, continuation changes veto into acquiescence on
$c < a \leq c(1 + \delta)$. A continuation-supported interval exists for some $\delta < 1$ exactly when $c < a < 2c$. The punishment proposal is only a weak best response because $(B,Q)$ and an off-path $(H,V)$ both give the proposer zero effective bite.

\end{proposition}

\begin{proof}

Let $V_{R}$ and $V_{P}$ denote cooperative values at the start of responder and proposer periods. They solve
\begin{equation}
V_{R} = - a + \delta V_{P},\quad V_{P} = a + \delta V_{R},
\end{equation}so
$V_{R} = - \frac{a}{1 + \delta}$ and $V_{P} = \frac{a}{1 + \delta}$. A responder who vetoes obtains $- c$ and then zero forever. The restraint condition $V_{R} \geq - c$ gives the stated threshold. Active support costs an additional $e$ and cannot improve on $Q$. On path, the proposer receives $V_{P} > 0$ and rejects a deviation to the zero-bite punishment.

After an off-path $B$ in state $C$, $Q$ strictly beats $A$ and $V$ because the common punishment continuation cancels and $0 > \max\left\{ - s - e, - c \right\}$. The same comparison holds after $B$ in state $P$. After $H$ in $P$, veto strictly beats $A$ and $Q$ because $a > c$. The proposer receives zero after either $B$ followed by $Q$ or $H$ followed by $V$, so $B$ is a weak best response. These comparisons cover both roles, both public states, and every one-shot action deviation.

\end{proof}

\begin{longtable}[]{@{}
  >{\raggedright\arraybackslash}p{(\linewidth - 6\tabcolsep) * \real{0.2083}}
  >{\raggedright\arraybackslash}p{(\linewidth - 6\tabcolsep) * \real{0.1750}}
  >{\raggedright\arraybackslash}p{(\linewidth - 6\tabcolsep) * \real{0.2083}}
  >{\raggedright\arraybackslash}p{(\linewidth - 6\tabcolsep) * \real{0.4083}}@{}}
\caption{Sequential incentive comparisons for alternating opposed
proposals.}\label{tab:si-discrete-audit}\tabularnewline
\toprule\noalign{}
\begin{minipage}[b]{\linewidth}\raggedright
\textbf{History}
\end{minipage} & \begin{minipage}[b]{\linewidth}\raggedright
\textbf{Actor}
\end{minipage} & \begin{minipage}[b]{\linewidth}\raggedright
\textbf{Deviation}
\end{minipage} & \begin{minipage}[b]{\linewidth}\raggedright
\textbf{Sequential comparison}
\end{minipage} \\
\midrule\noalign{}
\endfirsthead
\toprule\noalign{}
\begin{minipage}[b]{\linewidth}\raggedright
\textbf{History}
\end{minipage} & \begin{minipage}[b]{\linewidth}\raggedright
\textbf{Actor}
\end{minipage} & \begin{minipage}[b]{\linewidth}\raggedright
\textbf{Deviation}
\end{minipage} & \begin{minipage}[b]{\linewidth}\raggedright
\textbf{Sequential comparison}
\end{minipage} \\
\midrule\noalign{}
\endhead
\bottomrule\noalign{}
\endlastfoot
Cooperation & Proposer & $H \rightarrow B$ &
$V_{P} = \frac{a}{1 + \delta} > 0$ versus zero punishment value \\
Cooperation & Responder & $Q \rightarrow V$ &
$- \frac{a}{1 + \delta} \geq - c$ exactly at the stated threshold \\
Cooperation after $B$ & Responder &
$Q \rightarrow \left\{ A,V \right\}$ &
$0 > \max\left\{ - s - e, - c \right\}$ after the common punishment
continuation cancels \\
Punishment after $B$ & Responder &
$Q \rightarrow \left\{ A,V \right\}$ &
$0 > \max\left\{ - s - e, - c \right\}$ \\
Punishment proposal & Proposer & $B \rightarrow H$ & zero after
$(B,Q)$ and zero after $(H,V)$ \\
Punishment after $H$ & Responder &
$V \rightarrow \left\{ A,Q \right\}$ &
$- c > \max\left\{ - a - e, - a \right\}$ because $a > c$ \\
\end{longtable}

The weak proposer tie becomes strict under either of two conditions. A punishment-stage commitment can restrict the feasible menu to $B$. Alternatively, charge the proposer any rejection cost $k > 0$ after $(H,V)$ while leaving $(B,Q)$ at zero. The proposer then strictly prefers $B$ in state $P$, while $a > 0$ preserves its strict on-path preference for $(H,Q)$.

Fixed roles remove this mechanism. If the same party always responds to the same opposed hortatory proposal, cooperation gives $- \frac{a}{1 - \delta}$ while immediate veto gives $- c$ followed by zero. The inequality $\frac{a}{1 - \delta} \leq c$ has no solution when $a > c$. Alternating proposal direction supplies the future gain, while identical repetition supplies none.

\begin{figure}[htbp]
\centering
\includegraphics[width=1\linewidth,height=0.65\textheight,keepaspectratio]{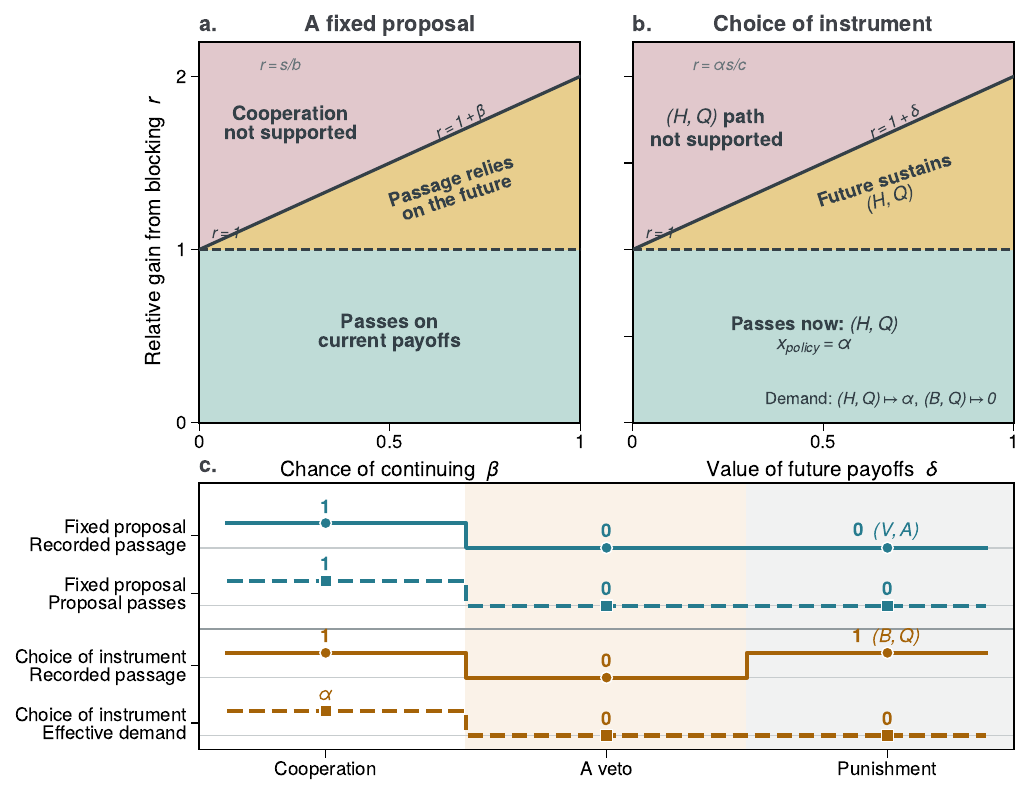}
\caption{Model constructions. Panels (a) and (b) show parameter regions supporting each specified path, conditional on the stated punishment checks, including $H - \Delta \geq 0$ in panel (a), and omit equilibrium selection. In panel (a), $H$ is alignment magnitude; in the instrument game, $H$ denotes the hortatory instrument. Panel (c) separates the public record $Y$ from effective bite after a deviation. Fixed-proposal punishment $(V,A)$ makes both zero. The strategic-instrument path $(H,Q) \rightarrow (H,V) \rightarrow (B,Q)$ restores $Y$ from zero to one while bite remains zero. $(B,Q)$ is a weak proposer best response unless commitment or a rejection cost makes it strict.}
\label{fig:repeated-game-outcomes}
\end{figure}

\section{Continuous-bite game: accommodation and restraint share a
threshold}

\protect\phantomsection\label{sec:si-bite}{}

The continuous-bite game isolates the algebra behind accommodation. A passage-seeking proposer chooses bite $x \in \lbrack 0,1\rbrack$. The current-payoff parameters and continuation gaps $L_{i} \geq 0$ are held fixed as bite and $\beta$ vary. Under complete information, we fix deterministic response strategies with passage-favouring selection whenever unanimous restraint is an equilibrium. An opposed responder $i$ suffers policy loss $g_{i}x$ with $g_{i} > 0$, receives passage benefit $b_{i} \geq 0$, and pays $c_{V,i} > 0$ for veto. Active support also costs $e_{i} > 0$ and is strictly dominated by acquiescence for this responder. Conditional on every other responder refraining, the payoffs are

\begin{equation}
u_{i}\left( Q|x \right) = b_{i} - g_{i}x + \beta W_{i}^{C},\quad u_{i}\left( V|x \right) = - c_{V,i} + \beta W_{i}^{P},\quad u_{i}\left( A|x \right) = u_{i}\left( Q|x \right) - e_{i}.
\end{equation}\begin{proposition}[Continuous accommodation threshold]

Responder $i$ acquiesces weakly exactly when
\begin{equation}
x \leq {\overline{x}}_{i} = \frac{b_{i} + c_{V,i} + \beta L_{i}}{g_{i}}.
\end{equation}Suppose the proposer's payoff increases strictly with accepted bite and
some accepted proposal is preferred to every rejected proposal. Under the maintained deterministic response selection and passage-favouring tie-breaks at threshold equalities, the optimal bite with one responder is $x^{\ast} = \min\left\{ 1,{\overline{x}}_{i} \right\}$. With several opposed responders, the weakest threshold binds:
\begin{equation}
x^{\ast} = \min\left\{ 1,\min\limits_{i}{\overline{x}}_{i} \right\}.
\end{equation}\end{proposition}

\begin{proof}

The difference $u_{i}\left( Q|x \right) - u_{i}\left( V|x \right)$ equals $b_{i} + c_{V,i} + \beta L_{i} - g_{i}x$. Every responder acquiesces when $x$ lies below every threshold. Since acceptance is deterministic and some accepted proposal dominates every rejected proposal, the proposer chooses the largest bite in this intersection. If a responder vetoes at equality and the binding threshold lies inside $(0,1)$, the proposer has a supremum but no maximizer on the continuous action set. A finite instrument grid restores a largest strictly acceptable choice.

\end{proof}

\begin{corollary}[An off-path veto can change policy]

Suppose $\min\limits_{i}{\overline{x}}_{i} < 1$ and the proposer chooses the strongest feasible bite. No veto occurs on path, yet the veto threat lowers the adopted bite from one to $\min\limits_{i}{\overline{x}}_{i}$. So the no-veto record conceals opposition to stronger counterfactual proposals.

\end{corollary}

\begin{proof}

Any bite above the minimum threshold triggers at least one veto. The proposer avoids that response by choosing the threshold bite or its closest feasible value below the threshold. The threat changes $x^{\ast}$ although the equilibrium response profile contains no veto.

\end{proof}

In net-payoff notation, this model sets $s_{i} = - c_{V,i} < 0$, giving the cap $\frac{b_{i} - s_{i} + \beta L_{i}}{g_{i}}$. The general fixed-proposal payoff $s_{i}$ need not be negative. The cap is a feasibility boundary before it is an optimal proposal. More generally, if the proposer's payoff among passing proposals increases up to $x^{u}$ and decreases thereafter, its preferred feasible bite is $\min\left\{ x^{u},x^{\ast} \right\}$, provided inducing rejection is no better. Accommodation means that the veto constraint lowers this choice below $x^{u}$. Feasibility alone does not establish that adjustment.

\textbf{Threshold model versus feedback model}

The continuous-bite game composes thresholds under complete information and fixed pure response selection. With current-payoff parameters and continuation gaps fixed, its strongest-feasible-bite rule is piecewise linear in $\beta$, and a finite instrument grid introduces steps. The feedback model below adds a different mechanism. Proposal bite shifts an aggregate restraint map, aggregate restraint changes passage probability, and that probability feeds back into proposal choice. Multiplicity and folds then matter for agenda setting.

\section{Proposal bite and expected restraint feed back}

\protect\phantomsection\label{sec:si-feedback}{}

See main Methods (``Additional response, agenda and private-type variants'') for the model and procedures.

Nature draws each responder's private veto payoff $s_{i}$ independently from a continuously differentiable distribution $F$ with density $f > 0$ on the relevant interval. Let $q$ denote the common probability assigned to any one responder refraining. Conditional independence makes expected current passage value $bq^{N - 1}$. Let $L(q)$ denote an imposed, type-independent expected continuation advantage from restraint over veto, already averaging over the other parties' responses. This section uses an alternative \textbf{direct cost of restraint}: choosing restraint incurs $gx$, where $g > 0$, even if another party vetoes. A symmetric response equilibrium conditional on that continuation specification solves

\begin{equation}
q = T(q;x,\beta) = F\left( bq^{N - 1} + \beta L(q) - gx \right).
\end{equation}This differs from the preceding implementation model, where policy loss
is incurred only upon passage. That model would instead give $T_{\text{pass }}(q;x,\beta) = F\left( (b - gx)q^{N - 1} + \beta L(q) \right)$. The results below concern the displayed direct-cost map $T$. Both maps coincide at $x = 0$, as used in the private-type model below.

\textbf{Continuation is a maintained input}

The function $L(q)$ specifies an expected continuation advantage. For example, suppose only unanimous restraint preserves cooperation and any veto activates the same punishment. If $\tilde{L} = W^{C} - W^{P}$ is the fixed continuation difference between those two states, independent counterpart responses give
\begin{equation}
L(q) = q^{N - 1}\tilde{L},\quad q = F\left( \left( b + \beta\tilde{L} \right)q^{N - 1} - gx \right).
\end{equation}A constant $L(q) = 1$ is a separate continuation specification. The
fixed-point analysis characterizes current responses conditional on the imposed $L$. It does not establish a sequential equilibrium of a repeated private-information game. That implementation requires a complete dynamic strategy and its deviation checks.

\begin{lemma}[Bite lowers restraint on a stable branch]

At any regular stable fixed point $q_{s}(x)$,
\begin{equation}
\frac{dq_{s}}{dx} = \frac{- \left( gf\left( z_{s} \right) \right)}{1 - T_{q}\left( q_{s};x,\beta \right)} < 0,\quad z_{s} = bq_{s}^{N - 1} + \beta L\left( q_{s} \right) - gx.
\end{equation}So a stronger proposal lowers equilibrium restraint along every regular
stable branch.

\end{lemma}

\begin{proof}

Define $G(q,x) = T(q;x,\beta) - q$. The response map satisfies $T_{x} = - gf(z) < 0$. The implicit-function theorem gives $d\frac{q_{s}}{dx} = - \frac{G_{x}}{G_{q}} = - g\frac{f\left( z_{s} \right)}{1 - T_{q}}$. Stability makes the denominator positive.

\end{proof}

Let $\Phi$ and $\varphi$ denote the standard-normal cumulative distribution function and density. For a normal location family with mean veto payoff $\mu$ and standard deviation $\sigma$,

\begin{equation}
T(q;x,\beta) = \Phi(\frac{bq^{N - 1} + \beta L(q) - gx - \mu}{\sigma}).
\end{equation}Only the index $\mu + gx$ contains the mean type and proposal bite.
Increasing bite by $\Delta x$ shifts the response map by the same amount as increasing mean veto payoff by $g\Delta x$. This equivalence locates the phase boundary while leaving equilibrium selection open.

Equivalently, parameterize the interior equilibrium curve by $q$:

\begin{equation}
x(q) = \frac{bq^{N - 1} + \beta L(q) - F^{- 1}(q)}{g},\quad x'(q) = \frac{T_{q} - 1}{gf\left( F^{- 1}(q) \right)}.
\end{equation}Stable branches have $x'(q) < 0$. A regular fold occurs where this
slope vanishes and its derivative is nonzero. This representation exposes the same geometry without solving separately for each bite.

\begin{proposition}[A stronger proposal can remove high restraint at a fold]

Assume $F$ and $L$ are twice continuously differentiable near $\left( q_{f},x_{f} \right)$. Let $G\left( q_{f},x_{f} \right) = 0$, $G_{q}\left( q_{f},x_{f} \right) = 0$, and $G_{qq}\left( q_{f},x_{f} \right) < 0$. Since $G_{x} = - gf < 0$, a stable high-restraint root and its adjacent unstable root exist locally for $x < x_{f}$ and disappear for $x > x_{f}$. A gradual rise in bite can produce a discrete loss of the high-restraint equilibrium.

\end{proposition}

\begin{proof}

Writing $z_{f} = bq_{f}^{N - 1} + \beta L\left( q_{f} \right) - gx_{f}$, a second-order expansion around the fold gives
\begin{equation}
x - x_{f} = \frac{G_{qq}\left( q_{f},x_{f} \right)}{2gf\left( z_{f} \right)}\left( q - q_{f} \right)^{2} + o\left( \left( q - q_{f} \right)^{2} \right).
\end{equation}The coefficient is negative. So two nearby roots exist on the
$x < x_{f}$ side and none exists on the $x > x_{f}$ side. Their slopes lie on opposite sides of one, so one root is stable and the other unstable under the stated adaptive process.

\end{proof}

The fold result concerns equilibrium availability. A path claim needs an adjustment rule. Under the specified iteration, a system on the high branch loses that fixed point at $x_{f}$. If the low-restraint root is the only remaining attractor reached from the post-fold state, beliefs then move toward that root. The model supports the term fold or tipping. Metastability requires noisy transitions and vanishing-noise residence-time results.

Let the proposer receive increasing policy return $R(x) > 0$ when the proposal passes and zero when it fails. Conditional on selecting a differentiable stable branch $q_{s}(x)$, its expected payoff is

\begin{equation}
\Pi(x) = R(x)q_{s}(x)^{N}.
\end{equation}\begin{proposition}[The proposer trades bite against passage]

On a regular stable branch, an interior proposer optimum satisfies
\begin{equation}
\frac{R'(x)}{R(x)} = \frac{Ngf\left( z_{s} \right)}{q_{s}\left\lbrack 1 - T_{q}\left( q_{s};x,\beta \right) \right\rbrack}.
\end{equation}The left side is the proportional return to bite. The right side is the
proportional passage loss induced by stronger opposition. Passage risk becomes steep as the branch approaches a fold from the stable side.

\end{proposition}

\begin{proof}

Differentiate $\log\Pi = \log R + N\log q_{s}$ and substitute the derivative from the preceding lemma. At an interior optimum the derivative equals zero. As $T_{q}$ approaches one from below, the denominator on the right approaches zero.

\end{proof}

\begin{corollary}[Continuation slack can finance greater bite]

Holding a target fixed point $q$ constant, a marginal increase in
continuation strength permits \begin{equation}
dx = \frac{L(q)}{g}d\beta
\end{equation}without
changing the responder's payoff index. If $L(q) = \ell$ is constant, the response map depends on $(\beta\ell - gx)$, so the compensation holds for every $q$. The proposer can then spend extra continuation slack on greater bite while preserving the entire response map and its basin structure.

\end{corollary}

\begin{proof}

At fixed $q$, the only changing terms in the response index are $\beta L(q) - gx$. Setting their total differential to zero yields the expression. Constant $L$ makes the equality hold pointwise over the complete response map.

\end{proof}

This corollary establishes a feasible compensation. A global monotonicity theorem for the proposer's optimum requires further restrictions on $R$, $F$, $L$, and the equilibrium-selection rule.

\section{Strategic proposal choice changes punishment credibility}

\protect\phantomsection\label{sec:si-strategic}{}

The strongest interaction appears off the cooperative path. A fixed-proposal trigger can prescribe a consequential veto after a deviation. Once a proposer can redesign that punishment proposal, sequential rationality requires an additional proposal-stage test. The following result supplies the complete test for any finite responder set within a binary trigger memory.

Let $\omega \in \Omega$ denote a finite public agenda phase. It identifies the current proposer $p(\omega)$ and responder set $\mathcal{R}(\omega)$. Public memory is $m \in \left\{ C,P \right\}$. At $(\omega,m)$, the proposer chooses $x \in \mathcal{X}_{m}(\omega)$. Every responder observes $x$, and the responders simultaneously choose $\sigma_{j} \in \left\{ Q,V \right\}$. Write $\sigma = \left( \sigma_{j} \right)_{j \in \mathcal{R}(\omega)}$ for their response profile and $\sigma^{Q} = (Q)_{j \in \mathcal{R}(\omega)}$ for unanimous restraint. Stage payoff $u_{i}(\omega,m,x,\sigma)$ may depend on public memory and the complete response profile. The theorem is stated for this reduced action set. Embedding it in a game with actions $\left\{ A,Q,V \right\}$ requires either adding $A$ to the deviation checks or establishing that $Q$ dominates $A$ in total current-plus-continuation value after every history. Current-payoff dominance suffices when the two actions induce the same continuation. The next agenda phase follows the action-independent kernel $M\left( \omega'|\omega \right)$.

A pure public strategy specifies $x_{m}(\omega) \in \mathcal{X}_{m}(\omega)$ and a complete response profile $r_{m}(\omega,x)$ after every feasible proposal. Impose $r_{C}\left( \omega,x_{C}(\omega) \right) = \sigma^{Q}$ and define $\sigma_{P}(\omega) = r_{P}\left( \omega,x_{P}(\omega) \right)$. In $C$, only $\left( x_{C},\sigma^{Q} \right)$ preserves $C$, and every other proposal-response pair moves play to $P$. In $P$, only $\left( x_{P},\sigma_{P} \right)$ leads to $\eta(\omega) \in \left\{ C,P \right\}$, and every other pair remains in $P$. So $\eta = P$ is permanent punishment and $\eta = C$ is one-period punishment followed by recovery. Write

\begin{equation}
{\overline{V}}_{i}^{m}(\omega) = \sum_{\omega'}M\left( \omega'|\omega \right)V_{i}^{m}(\omega').
\end{equation}The prescribed values solve

\begin{equation}
V_{i}^{C}(\omega) = u_{i}\left( \omega,C,x_{C},\sigma^{Q} \right) + \delta{\overline{V}}_{i}^{C}(\omega),\quad V_{i}^{P}(\omega) = u_{i}\left( \omega,P,x_{P},\sigma_{P} \right) + \delta{\overline{V}}_{i}^{\eta}(\omega).
\end{equation}\begin{theorem}[Strategic-proposer trigger test with $N$ responders]

Assume bounded payoffs, complete information, perfect public monitoring, perfect recall, $\delta \in \lbrack 0,1)$, and finite menus, or compact menus for which the displayed suprema exist. The specified pure public strategy is subgame perfect if and only if all six deviation families below hold in every agenda phase. The responder families apply to every $j \in \mathcal{R}(\omega)$.

\begin{equation}
u_{j}\left( x_{C},\sigma^{Q} \right) + \delta{\overline{V}}_{j}^{C} \geq u_{j}\left( x_{C},\left( \sigma_{j},\sigma_{- j}^{Q} \right) \right) + \delta{\overline{V}}_{j}^{P}\quad\text{ for every  }\sigma_{j} \neq Q,\quad\text{ (R-C) }
\end{equation}\begin{equation}
r_{Cj}(\omega,x) \in \arg\max\limits_{\sigma_{j} \in \left\{ Q,V \right\}}u_{j}\left( \omega,C,x,\left( \sigma_{j},r_{C, - j}(\omega,x) \right) \right)\quad\text{ for every  }x \neq x_{C},\quad\text{ (R-CO) }
\end{equation}\begin{equation}
u_{j}\left( x_{P},\sigma_{P} \right) + \delta{\overline{V}}_{j}^{\eta} \geq u_{j}\left( x_{P},\left( \sigma_{j},\sigma_{P, - j} \right) \right) + \delta{\overline{V}}_{j}^{P}\quad\text{ for every  }\sigma_{j} \neq \sigma_{Pj},\quad\text{ (R-P) }
\end{equation}\begin{equation}
r_{Pj}(\omega,x) \in \arg\max\limits_{\sigma_{j} \in \left\{ Q,V \right\}}u_{j}\left( \omega,P,x,\left( \sigma_{j},r_{P, - j}(\omega,x) \right) \right)\quad\text{ for every  }x \neq x_{P},\quad\text{ (R-PO) }
\end{equation}\begin{equation}
u_{p}\left( x_{C},\sigma^{Q} \right) + \delta{\overline{V}}_{p}^{C} \geq \sup\limits_{x \neq x_{C}}\left\{ u_{p}\left( x,r_{C}(\omega,x) \right) + \delta{\overline{V}}_{p}^{P} \right\},\quad\text{ (P-C) }
\end{equation}\begin{equation}
u_{p}\left( x_{P},\sigma_{P} \right) + \delta{\overline{V}}_{p}^{\eta} \geq \sup\limits_{x \neq x_{P}}\left\{ u_{p}\left( x,r_{P}(\omega,x) \right) + \delta{\overline{V}}_{p}^{P} \right\}.\quad\text{ (P-P) }
\end{equation}We suppress arguments $(\omega,m)$ inside the inequalities. At
off-path proposal nodes, every response profile leads to $P$, so continuation cancels in R-CO and R-PO. Each of those families requires the prescribed response profile to form a Nash equilibrium after the observed proposal. The result audits this automaton and leaves response-equilibrium selection and the wider repeated-game equilibrium set open.

\end{theorem}

\begin{proof}

Each inequality is necessary because it excludes a one-shot deviation at a distinct proposal or response information set. The responder families hold the other responders' actions fixed, as Nash optimality requires. Together the six families cover the proposer, every responder, both memory states, every on-path proposal, and every off-path proposal. Bounded discounted payoffs give continuity at infinity. The one-shot-deviation principle makes the six families sufficient.

\end{proof}

The theorem makes the interaction transparent. Define the proposer's enforcement gap and best redesign gain by

\begin{equation}
\Gamma_{p} = \delta({\overline{V}}_{p}^{\eta} - {\overline{V}}_{p}^{P}),\quad G_{p} = \sup\limits_{x \neq x_{P}}\left\lbrack u_{p}\left( x,r_{P}(\omega,x) \right) - u_{p}\left( x_{P},\sigma_{P} \right) \right\rbrack.
\end{equation}For responder $j$, define its largest punishment-stage action gain

\begin{equation}
D_{j} = \sup\limits_{\sigma_{j} \neq \sigma_{Pj}}\left\lbrack u_{j}\left( x_{P},\left( \sigma_{j},\sigma_{P, - j} \right) \right) - u_{j}\left( x_{P},\sigma_{P} \right) \right\rbrack,\quad\Gamma_{j} = \delta({\overline{V}}_{j}^{\eta} - {\overline{V}}_{j}^{P}).
\end{equation}\begin{corollary}[Proposal-redesign credibility condition]

The punishment proposer test holds exactly when
\begin{equation}
G_{p} \leq \Gamma_{p}.
\end{equation}The on-prescription punishment responder test
R-P holds exactly when $D_{j} \leq \Gamma_{j}$ for every responder. The off-path proposal-response tests R-PO remain separate static best-response requirements. Proposal redesign destroys the specified veto trigger exactly when the best redesign gives the proposer more than it loses by evading the prescribed punishment. This deviation is accommodation when a weaker accepted proposal supplies the largest gain.

\end{corollary}

\begin{proof}

Subtract the prescribed punishment value from each side of P-P and every R-P inequality. Under permanent punishment, $\eta = P$, so every enforcement gap equals zero. Any strictly profitable acceptable redesign then invalidates the candidate trigger.

\end{proof}

\begin{corollary}[Unanimity creates two regime-specific thresholds]

Suppose $N$ responders face scalar bite $x$. Responder $i$ pays $x$ after unanimous restraint, receives zero through restraint after another veto, and pays veto cost $c_{i} > 0$ after choosing $V$. Define its cooperative continuation loss as $L_{i} = {\overline{V}}_{i}^{C} - {\overline{V}}_{i}^{P} \geq 0$. Then unanimous restraint on the cooperative path requires
\begin{equation}
x_{C} \leq {\overline{x}}_{C} = \min\limits_{i}\left\{ c_{i} + \delta L_{i} \right\}.
\end{equation}After a proposal deviation has already activated punishment,
continuation cancels. Under a passage-favouring tie-break, the largest bite that every responder accepts is
\begin{equation}
{\overline{x}}_{P} = \min\limits_{i}c_{i}.
\end{equation}Assume passed bite gives
the proposer $x$, while a failed proposal and the dead-letter instrument give zero. An unrestricted continuous menu defeats the permanent dead-letter punishment-proposer test whenever ${\overline{x}}_{P} > 0$. A minimum substantive bite restores that test weakly when
\begin{equation}
{\overline{x}}_{P} < m \leq x_{C} \leq {\overline{x}}_{C}.
\end{equation}With
maximum bite $x_{\max}$, a lump-sum proposal-deviation penalty restores that test weakly when
\begin{equation}
K \geq \min\left\{ {\overline{x}}_{P},x_{\max} \right\}
\end{equation}and
${\overline{x}}_{P} < x_{C} \leq \min\left\{ {\overline{x}}_{C},x_{\max} \right\}$.

These bounds close the cooperative-responder and punishment-proposer comparisons. The other deviation families in the theorem must also hold for subgame perfection.

\end{corollary}

\begin{proof}

Against unanimous restraint, responder $i$ compares $- x_{C} + \delta{\overline{V}}_{i}^{C}$ with $- c_{i} + \delta{\overline{V}}_{i}^{P}$. The comparison gives $x_{C} \leq c_{i} + \delta L_{i}$ for every responder. After a proposal deviation, all response actions lead to $P$, so the response game is static. The all-$Q$ profile is a Nash equilibrium exactly when $x \leq \min\limits_{i}c_{i}$. For $x > \min\limits_{i}c_{i}$, a profile with exactly one vetoer $k$ is a Nash equilibrium whenever $c_{k} \leq x$, because allowing passage gives that vetoer no gain and every restraining responder avoids a redundant veto cost. No profile with multiple vetoers is a Nash equilibrium because any one vetoer can switch to $Q$ while another veto still blocks passage. The passage-favouring selection uses all-$Q$ through ${\overline{x}}_{P}$ and one veto above ${\overline{x}}_{P}$. The proposer's largest accepted redesign is $\min\left\{ {\overline{x}}_{P},x_{\max} \right\}$. A bite floor removes that redesign set, while $K$ offsets its supremum.

\end{proof}

Different responders can attain the two minima. Holding the existing $c_{i}$ and $L_{i}$ fixed, adding a responder weakly lowers each cap. The cooperative cap tightens, while the largest punishment-stage redesign that can pass also falls. Membership can change continuation values, so this is a parameter-fixed comparative static. Unanimity makes the minimum responder threshold in each regime an obstacle to current passage or a source of commitment against accommodation.

The permanent-punishment case also corrects a tempting shortcut. Continuation inside the same absorbing state cannot generate a positive $\Gamma_{p}$ because it appears on both sides of the proposal comparison. Recovery after compliance helps only when ${\overline{V}}_{p}^{\eta} > {\overline{V}}_{p}^{P}$. In the normalised alternating game, recovery makes today's proposer tomorrow's responder, so ${\overline{V}}_{p}^{C} = - \frac{x}{1 + \delta}$. A punishment round that also recovers gives ${\overline{V}}_{p}^{P} = \delta\frac{x}{1 + \delta}$, so $\Gamma_{p} = - \delta x$, which is negative for $\delta > 0$. The fourth case below uses a primitive lump-sum penalty on proposal deviation. A terminal exclusion mechanism would add a public state and require a separate audit.

\subsection{Four institutional cases in one alternating game}

Set $N = 1$ to compare four proposal institutions in closed form. Normalize dead-letter passage $(D,Q)$ to zero. Let $c > 0$ and $k \geq 0$. A passed substantive proposal of bite $x$ gives the current proposer $x$ and the current responder $- x$, while a veto gives them $- k$ and $- c$, respectively. Roles alternate each round. The candidate path prescribes $(x,Q)$ in $C$ and $(D,Q)$ in $P$. Its cooperative role values are

\begin{equation}
V_{p}^{C} = \frac{x}{1 + \delta},\quad V_{j}^{C} = - \frac{x}{1 + \delta},\quad V_{i}^{P} = 0.
\end{equation}The cooperative responder restrains exactly when
$x \leq c(1 + \delta)$. Current payoffs already sustain passage at $x \leq c$, and continuation changes veto into restraint on $c < x \leq c(1 + \delta)$.

\begin{corollary}[Proposal institutions determine whether the trigger survives]

In the normalised alternating game, the four proposal institutions have the following implications.

\begin{enumerate}
\item With unrestricted continuous redesign, $\mathcal{X} = \lbrack 0,1\rbrack$, $c > 0$, and no additional sanction, the dead-letter punishment fails, because the punishment proposer chooses an accepted $y \in (0,\min\left\{ c,1 \right\}\rbrack$ instead. \item The menu $\left\{ B,H \right\}$ is the discrete case with dead-letter passage $(D,Q) = (B,Q)$ and substantive bite $x = \alpha s$ under $(H,Q)$. If $\alpha s > c$, an off-path $H$ meets $V$. The candidate is a weak SPE when $c < \alpha s \leq c(1 + \delta)$. A rejection cost $k > 0$ makes $(B,Q)$ strictly better than $(H,V)$ in punishment. \item A minimum substantive bite $m > c$ removes every acceptable redesign. The candidate is a weak SPE whenever $m \leq x \leq c(1 + \delta)$, and $k > 0$ makes rejected proposal deviations strict. If $m \leq c$, an accepted redesign remains available. \item Under unrestricted redesign with maximum bite one and $c \leq 1$, suppose an institutional rule charges the proposer a lump-sum penalty $K$ for selecting a proposal other than the state-designated $x$ in $C$ or $D$ in $P$. The penalty restores the candidate when $K \geq c$ and $c < x \leq \min\left\{ c(1 + \delta),1 \right\}$. For a general maximum bite, the conditions are $K \geq \min\left\{ c,x_{\max} \right\}$ and $c < x \leq \min\left\{ c(1 + \delta),x_{\max} \right\}$.
\end{enumerate}

\end{corollary}

\begin{proof}

The responder comparison gives $x \leq c(1 + \delta)$. After a proposal deviation into permanent punishment, the responder accepts $y$ exactly when $y \leq c$. The punishment proposer gains the largest accepted bite available in its menu. Free redesign gives a positive gain. The discrete and minimum-bite menus remove that accepted option, and $K$ offsets its supremum. In the case with instruments $B$ and $H$, $(B,Q)$ gives zero while $(H,V)$ gives $- k$. Cases 2 and 3 also satisfy P-C because $\frac{x}{1 + \delta} > 0$ exceeds the dead-letter payoff zero and every rejected-deviation payoff $- k$. In Case 4, any proposal deviation gives at most $c - K \leq 0 < \frac{x}{1 + \delta}$, which establishes P-C.

\end{proof}

\begin{corollary}[Agreement without the dead-letter trigger]
Proposals range over $[0,1]$, with $c>0$, $k\geq0$ and $0\leq\delta<1$. The history-independent strategy offers $q=\min\{c,1\}$ each round and accepts proposals $y\leq c$. This strategy is a subgame-perfect equilibrium when ties favour passage. Every proposal on its path passes at bite $q$.
\end{corollary}

\begin{proof}
Actions leave continuation unchanged under this strategy. The responder therefore compares $-y$ with $-c$ and accepts exactly when $y\leq c$, resolving equality in favour of passage. The proposer obtains $q>0$, every other accepted proposal gives at most $q$, and rejection gives $-k\leq0$. These comparisons hold after every history.
\end{proof}

This equilibrium establishes that agreement can continue without the dead-letter threat. It does not establish which equilibrium play selects after the candidate trigger fails.

\begin{corollary}[History-independent play passes at most $c$]
Under history-independent strategies, no subgame-perfect equilibrium passes a proposal with bite above $c$. If the menu contains only the dead-letter and substantive bites above $c$, no substantive proposal passes on the path of such an equilibrium. Offering the dead-letter each round and accepting proposals $y\leq c$ is one such equilibrium.
\end{corollary}

\begin{proof}
Actions leave continuation unchanged under a history-independent profile. The responder therefore compares $-y$ with $-c$ and strictly prefers veto when $y>c$. When every substantive bite exceeds $c$, each substantive proposal is vetoed and returns $-k\leq0$ to the proposer, while the dead-letter passes and returns zero. Offering the dead-letter is therefore optimal after every history.
\end{proof}

\begin{figure}[htbp]
\centering
\includegraphics[width=1\linewidth,height=0.65\textheight,keepaspectratio]{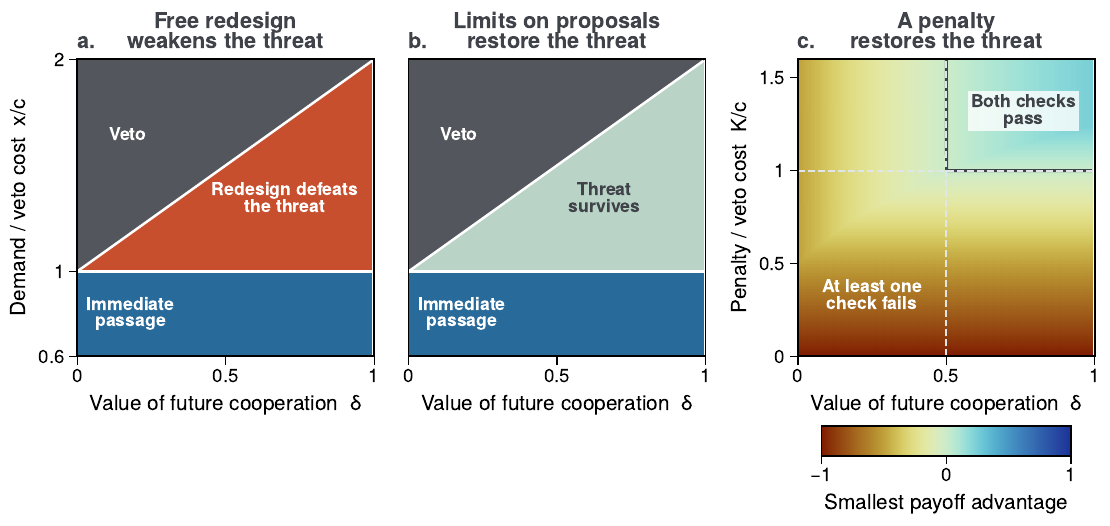}
\caption{Credibility in the normalised alternating game. Panels (a) and (b) show node-level incentives. Blue denotes immediate responder acceptance and grey denotes veto. The middle band shows whether free redesign defeats the trigger or proposal restrictions sustain it. Blue alone establishes neither target optimality nor punishment credibility. Panel (c) fixes $\frac{x}{c} = 1.5$ and plots $\min\left\{ 1 - \frac{x/c}{1 + \delta},\frac{K}{c} - 1 \right\}$, and the dark contour marks zero. $K$ is a lump-sum charge for selecting a non-designated proposal. The figure reports incentive conditions and omits equilibrium-selection probabilities.}
\label{fig:flexibility-credibility}
\end{figure}

\textbf{The bounded claim}

Menu expansion can raise or lower cooperation. A weak option can create immediate agreement when a rigid menu would fail. Flexibility defeats this particular trigger only when the best acceptable redesign is worth more than the enforcement gap attached to evading punishment. A bite floor can repair punishment credibility while making the cooperative proposal too strong to sustain. The result compares institutions within one specified two-state pure public strategy.

\section{Complete flexibility can crowd out on-path veto}

The same proposal stage changes the adaptive-menu interpretation. Under complete information and the deterministic response rule above, a passage-seeking proposer can select the strongest jointly acceptable bite. When that cap lies below its unconstrained preferred bite, the outcome combines accommodation with no veto. The veto remains consequential because it caps bite. On-path menu pruning remains absent.

\begin{proposition}[Mechanism crowding under complete information]

Assume complete information and a continuous action set $\left\lbrack x_{\min},1 \right\rbrack$, and fix a pure response strategy after every feasible proposal. Suppose at least one bite $x$ is strictly accepted by every responder. If some jointly accepted proposal gives the proposer more than every proposal inducing rejection, every optimal proposal passes. Prescribed on-path veto and endogenous accommodation cannot both arise under this deterministic response selection.

\end{proposition}

\begin{proof}

The fixed pure response strategies give every proposal a deterministic passage outcome. The proposer can select a jointly accepted bite and obtain a payoff strictly above every rejection payoff. Any proposal that prescribes at least one veto yields a lower payoff and cannot be optimal.

\end{proof}

On-path veto can return if a minimum viable bite exceeds a responder's threshold or a discrete menu omits jointly acceptable proposals. Private information and mixed response strategies can instead make passage uncertain. With either source of uncertainty, the proposer may prefer the expected return from a stronger proposal to certain weak passage. Mixed responses can permit on-path veto even under complete information.

The binding-hortatory game supplies the discrete-instrument friction. The feedback model supplies private-type uncertainty. The procedural random-menu model considered later instead assigns proposals without a strategic proposer and audits a public response strategy.

\section{Apparent consensus rises as accommodation falls}

\protect\phantomsection\label{sec:si-learning}{}

The preceding results characterize credible strategies. The behavioural layer asks which strategy parties learn when they begin with uncertain expectations and adapt to realised payoffs. It selects between proposal-side accommodation, response-side restraint, and veto along the histories parties actually experience. Both mechanisms remain unchanged.

See main Methods (``Payoff learning and exact incentive audits'') for the model and procedures.

The learning game retains the response payoffs used throughout this note. Let $h(x) = h_{0} - \lambda x$ denote alignment. The proposer receives $b + x$ after passage and $- k$ after veto. The responder receives

\begin{equation}
u_{R}\left( A|x \right) = b + h(x) - \Delta,\quad u_{R}\left( Q|x \right) = b,\quad u_{R}\left( V|x \right) = - h(x) - \Delta - c_{V}.
\end{equation}Small bites can invite active support, intermediate bites restraint, and
large bites veto. The public transition rule then determines whether a veto filters the agenda or damages the relationship.

\subsection{Forgiveness bounds credible restraint}

\protect\phantomsection\label{sec:si-forgiveness}{}

The exact two-bite specification uses bites $0.25$ and $1$, limits revision to $0.25$, and licenses an open-mode veto only at bite $1$. It fixes $h_{0} = 0.45$, $\lambda = 1$, $b = 0.25$, $\Delta = 0.10$, $c_{V} = 0.05$, and $k = 0.05$. Consider the symmetric stationary candidate

\begin{longtable}[]{@{}
  >{\raggedright\arraybackslash}p{(\linewidth - 6\tabcolsep) * \real{0.1923}}
  >{\centering\arraybackslash}p{(\linewidth - 6\tabcolsep) * \real{0.1731}}
  >{\centering\arraybackslash}p{(\linewidth - 6\tabcolsep) * \real{0.2115}}
  >{\raggedright\arraybackslash}p{(\linewidth - 6\tabcolsep) * \real{0.4231}}@{}}
\caption{The symmetric two-bite candidate rule in the open, revision and
punishment modes.}\label{tab:si-candidate-rule}\tabularnewline
\toprule\noalign{}
\endfirsthead
\endhead
\bottomrule\noalign{}
\endlastfoot
\textbf{Mode} & \textbf{Bite} & \textbf{Response} & \textbf{Next mode,
conditional on continuation} \\
Open $O$ & $1$ & $Q$ & $O$ \\
Revision $R$ & $0.25$ & $A$ & $O$ \\
Punishment $P$ & $0.25$ & $A$ & $O$ with probability $\rho$,
otherwise $P$ \\
\end{longtable}

An off-path weak bite in the open mode receives $A$, and an off-path strong bite in punishment receives $V$. These prescriptions complete every proposal-response node in the two-bite game. Write $W_{m}^{p}$ and $W_{m}^{r}$ for the values of entering mode $m$ as proposer and responder. The incentive comparisons need only the continuation differences

\begin{equation}
d_{p} = W_{O}^{p} - W_{P}^{p},\quad d_{r} = W_{O}^{r} - W_{P}^{r},\quad t = \beta(1 - \rho).
\end{equation}Open and punishment proposer payoffs differ by $1.25 - 0.50 = 0.75$,
and their responder payoffs differ by $0.25 - 0.35 = - 0.10$. Subtracting their Bellman equations gives

\begin{equation}
d_{p} = 0.75 + td_{r},\quad d_{r} = - 0.10 + td_{p},\quad d_{p} = \frac{0.75 - 0.10t}{1 - t^{2}}.
\end{equation}Open and revision proposer values also satisfy
$W_{O}^{p} - W_{R}^{p} = 0.75$, since both prescribe passage followed by open play.

\begin{proposition}[Forgiving recovery makes restraint non-monotone]

Fix $\delta = 0.9$ and $\rho = 0.65$. The strong-restraint candidate passes every one-shot deviation check exactly when
\begin{equation}
\frac{2}{9} \leq \chi \leq \chi_{U},\quad\chi_{U} = \frac{480}{0.9\left( 780 + \sqrt{601680} \right)} \approx 0.3428296401.
\end{equation}The binding conditions, expressed in effective continuation
$\beta = \delta\chi$, are
\begin{equation}
\beta \geq 0.2,\quad 0.15 \geq \beta\frac{\rho(0.75 - 0.10\beta(1 - \rho))}{1 - \beta^{2}(1 - \rho)^{2}}.
\end{equation}Continuation first makes restraint credible, but sufficiently strong
continuation makes the prescribed off-path veto incredible because passage can repair punishment.

\end{proposition}

\begin{proof}

After the strong open proposal, restraint gives $0.25 + \beta W_{O}^{p}$ and a licensed veto gives $0.40 + \beta W_{R}^{p}$. Their comparison is $0.75\beta \geq 0.15$. After the off-path strong punishment proposal, veto gives $0.40 + \beta W_{P}^{p}$. Restraint gives $0.25 + \beta\left\lbrack \rho W_{O}^{p} + (1 - \rho)W_{P}^{p} \right\rbrack$. Veto requires $0.15 \geq \beta\rho d_{p}$. Substitution of $d_{p}$ at $\rho = 0.65$ reduces this inequality to

\begin{equation}
7\beta^{2} - 780\beta + 240 \geq 0.
\end{equation}Its smaller root is $\frac{480}{780 + \sqrt{601680}}$, and the larger
root lies outside $\beta \in \lbrack 0,0.9\rbrack$. This gives the displayed upper bound on $\chi$.

The remaining comparisons do not tighten this interval. The open proposer's margin over the weak bite is $0.75$. At a weak bite, active support beats acquiescence by $0.10$ and beats veto by $0.70 + \beta d_{p}$ in open or revision mode, and by $0.70 + \beta\rho d_{p}$ in punishment. At a strong bite, acquiescence beats active support by $0.65$ under the common passage continuation. The punishment proposer's margin is $0.55 + \beta\rho d_{r}$. Here $d_{p} > 0$ and $d_{r} = - 0.10 + td_{p} \geq - 0.10$, so this last margin is at least $0.55 - 0.10\beta\rho > 0$. These comparisons exhaust the other nodes and action alternatives.

\end{proof}

\textbf{What the exact search establishes}

The proposition audits one specified policy. Its complete one-shot audit gives the analytic credibility interval above and confirms it on the $0.0125$ continuation grid, from $\chi = 0.225$ through $0.3375$. That grid audits the specified candidate. The separate exhaustive search evaluates all 972 symmetric pure stationary profiles on the selected parameter grids described in main Methods. Passing the complete audit excludes profitable arbitrary unilateral deviations by the one-shot-deviation principle. Asymmetric, mixed, nonstationary and history-dependent candidate profiles remain outside the enumeration.

\subsection{On-path checks can hide threats}

\protect\phantomsection\label{sec:si-path-audit}{}

See main Methods (``Payoff learning and exact incentive audits'') for the model and procedures.

The historical ten-seed two-bite illustration at $\chi = \frac{5}{6}$ selects the strongest bite followed by restraint. In its evaluation window, apparent consensus averages $99.98\%$ and mean bite averages $0.99994$. The projected symmetric modal rule has a positive modal-path margin of $0.4125$ in every seed, yet its full margin is negative in every seed and averages $- 0.687$. The projected rule passes every on-path one-shot comparison and fails the complete off-path audit.

The larger \hyperref[sec:si-learning-audit]{1,000-run reference audit} is reported below. The ten-run illustration and its figure remain a separate, smaller sample.

The distinction is substantive. A subgame-perfect equilibrium requires parties to be willing to carry out a punishment after any history. Payoff learning may rarely visit the punishment nodes needed to expose an incredible threat.

\subsection{Earlier two-bite learning samples}

\protect\phantomsection\label{sec:si-learning-current}{}

See main Methods (``Payoff learning and exact incentive audits'') for the model and procedures.

Selected rows of the primary sweep follow. Outcome columns report mean percentages across runs, and bite includes accepted and rejected proposals.

\begin{longtable}[]{@{}
  >{\centering\arraybackslash}p{(\linewidth - 10\tabcolsep) * \real{0.1274}}
  >{\centering\arraybackslash}p{(\linewidth - 10\tabcolsep) * \real{0.1820}}
  >{\centering\arraybackslash}p{(\linewidth - 10\tabcolsep) * \real{0.1820}}
  >{\centering\arraybackslash}p{(\linewidth - 10\tabcolsep) * \real{0.1820}}
  >{\centering\arraybackslash}p{(\linewidth - 10\tabcolsep) * \real{0.1820}}
  >{\centering\arraybackslash}p{(\linewidth - 10\tabcolsep) * \real{0.1456}}@{}}
\caption{Outcome shares and proposal bite in the current two-bite
learning samples.}\label{tab:si-learning-grid}\tabularnewline
\toprule\noalign{}
\endfirsthead
\endhead
\bottomrule\noalign{}
\endlastfoot
\textbf{$\chi$} & \textbf{Full support (\%)} & \textbf{Apparent
consensus (\%)} & \textbf{Filtering veto (\%)} & \textbf{Breakdown (\%)}
& \textbf{Mean bite} \\
$0$ & 33.74 & 36.73 & 27.49 & 2.04 & 0.578 \\
$\frac{1}{3}$ & 36.77 & 36.85 & 24.30 & 2.09 & 0.542 \\
$\frac{2}{3}$ & 38.53 & 39.18 & 21.23 & 1.06 & 0.528 \\
$\frac{3}{4}$ & 19.34 & 70.04 & 10.39 & 0.24 & 0.767 \\
$\frac{5}{6}$ & 10.79 & 85.91 & 3.26 & 0.04 & 0.869 \\
$\frac{11}{12}$ & 16.10 & 82.63 & 1.27 & 0.003 & 0.845 \\
$1$ & 15.34 & 84.66 & 0.0003 & 0 & 0.831 \\
\end{longtable}

The 1,000-run reference sample at $\chi = \frac{5}{6}$ has mean apparent-consensus share $84.0967\%$ and mean bite $0.855439$. The 756 runs whose final projected rule prescribes a demanding proposal followed by $Q$ average $99.6490\%$ apparent consensus in the final 4,000 rounds. The sample combines different learned behaviours. These samples are reported separately from the historical ten-run illustration and five-bite results. Reused two-bite seeds overlap the larger reference sample.

\subsection{Which learned rules fail the audit}

\protect\phantomsection\label{sec:si-learning-audit}{}

This subsection retains the earlier reference sample and its symmetric rule projection, rather than the current main Figure 3 audit. Different runs can produce the same projected rule. The 1,000 endpoints contain 60 distinct rules. The cached audit is \nolinkurl{data/emergent_reference_audit_n1000.csv}, with projected-rule identities in \nolinkurl{data/emergent_policy_landscape_seed_summary_n1000.csv}.

See main Methods (``Payoff learning and exact incentive audits'') for the model and procedures.

\begin{longtable}[]{@{}
  >{\raggedright\arraybackslash}p{(\linewidth - 8\tabcolsep) * \real{0.3390}}
  >{\centering\arraybackslash}p{(\linewidth - 8\tabcolsep) * \real{0.1186}}
  >{\centering\arraybackslash}p{(\linewidth - 8\tabcolsep) * \real{0.1695}}
  >{\centering\arraybackslash}p{(\linewidth - 8\tabcolsep) * \real{0.2034}}
  >{\centering\arraybackslash}p{(\linewidth - 8\tabcolsep) * \real{0.1695}}@{}}
\caption{Modal-path and full-node audit outcomes for the 1,000-run
reference sample.}\label{tab:si-learning-audit}\tabularnewline
\toprule\noalign{}
\endfirsthead
\endhead
\bottomrule\noalign{}
\endlastfoot
\textbf{Prescribed open play} & \textbf{Runs} & \textbf{Fails modal
path} & \textbf{Passes path only} & \textbf{Passes full audit} \\
Weak bite, $A$ & 203 & 2 & 197 & 4 \\
Weak bite, $Q$ & 31 & 31 & 0 & 0 \\
Demanding bite, $Q$ & 756 & 0 & 756 & 0 \\
Demanding bite, $V$ & 10 & 10 & 0 & 0 \\
Total & 1,000 & 43 & 953 & 4 \\
\end{longtable}

Among the 957 runs that pass the modal-path check, 756 prescribe a stand-aside and 201 prescribe full support. All 756 stand-aside rules and 197 of the full-support rules fail the full audit. The selected departure is a veto in 511 runs, a weaker proposal in 384, withholding a prescribed veto in 43, a stronger proposal in 12, and active support in three. These categories describe the largest gain, not every profitable departure. The other 43 runs already fail in usual play and are outside this breakdown.

The most frequent endpoint, code 757 in the cache, occurs in 107 runs. It proposes the demanding bite in both $O$ and $P$, the weak bite in $R$, and prescribes $A$ after every weak bite and $Q$ after either demanding bite. Its modal-path margin is $0.4125$. After a demanding proposal in $P$, changing $Q$ to $V$ gains $0.15$, giving full margin $- 0.15$. This gain includes continuation under the prescribed rule. The four fully credible endpoints instead share code 56, which proposes the weak bite in every mode, supports weak proposals, and vetoes demanding proposals. It is the only fully credible candidate among the 972 symmetric pure stationary rules at these parameters.

In this earlier reference sample, visitation is pooled across all 1,000 runs. These counts do not describe the current main Figure 3c. Of 3,999,000 recorded rounds, 1,527 occur in $P$ ($0.03818\%$), 124,529 in $R$ ($3.11400\%$), and 3,872,944 in $O$ ($96.84781\%$). The complement of the dispute share, $99.96182\%$, combines $O$ and $R$. It is not the open-mode share or a visitation estimate restricted to the 107 runs with code 757. The corresponding 4,000-round summaries give a dispute share of $0.038175\%$ across all runs.

The audit concerns projected symmetric pure stationary rules under the stated game. It does not identify either learner's original stochastic policy with that projection, establish that these are the only possible equilibrium forms, or measure a probability of agreement collapsing. A failed check establishes an incentive to depart at a specified node.

\subsection{Sensitivity to payoff and learning parameters}

\protect\phantomsection\label{sec:si-parameters}{}

See main Methods (``Learning sensitivity, horizon and interventions'') for the model and procedures.

\begin{longtable}[]{@{}
  >{\raggedright\arraybackslash}p{(\linewidth - 8\tabcolsep) * \real{0.3103}}
  >{\centering\arraybackslash}p{(\linewidth - 8\tabcolsep) * \real{0.1207}}
  >{\centering\arraybackslash}p{(\linewidth - 8\tabcolsep) * \real{0.2069}}
  >{\centering\arraybackslash}p{(\linewidth - 8\tabcolsep) * \real{0.1724}}
  >{\centering\arraybackslash}p{(\linewidth - 8\tabcolsep) * \real{0.1897}}@{}}
\caption{Learning outcomes and modal-rule audits under parameter
changes.}\label{tab:si-parameter-sensitivity}\tabularnewline
\toprule\noalign{}
\endfirsthead
\endhead
\bottomrule\noalign{}
\endlastfoot
\textbf{Parameter} & \textbf{Value} & \textbf{Apparent consensus (\%)} &
\textbf{Full pass (/200)} & \textbf{Path fail (/200)} \\
Reference & -- & 83.53 & 2 & 9 \\
Stance cost $\Delta$ & 0.05 & 70.72 & 2 & 6 \\
Stance cost $\Delta$ & 0.20 & 99.38 & 0 & 1 \\
Veto cost $c_{V}$ & 0 & 71.09 & 2 & 13 \\
Veto cost $c_{V}$ & 0.10 & 94.54 & 0 & 4 \\
Passage value $b$ & 0.15 & 49.67 & 11 & 10 \\
Passage value $b$ & 0.35 & 95.44 & 1 & 3 \\
Rejection cost $k$ & 0 & 85.76 & 0 & 9 \\
Rejection cost $k$ & 0.10 & 83.75 & 1 & 6 \\
Recovery $\rho$ & 0.40 & 85.75 & 0 & 8 \\
Recovery $\rho$ & 0.90 & 83.20 & 0 & 5 \\
Discount $\delta$ & 0.80 & 66.77 & 3 & 23 \\
Discount $\delta$ & 0.95 & 86.85 & 2 & 8 \\
Base learning rate $\alpha_{0}$ & 0.02 & 89.59 & 0 & 1 \\
Base learning rate $\alpha_{0}$ & 0.10 & 76.56 & 7 & 14 \\
Temperature $\tau$ & 0.10 & 60.41 & 7 & 11 \\
Temperature $\tau$ & 0.50 & 42.35 & 0 & 142 \\
\end{longtable}

The unlisted remainder passes the modal-path check only. The largest full-pass share is $\frac{11}{200} = 5.5\%$ at $b = 0.15$ (95\% Wilson interval $3.10\%$--$9.58\%$). Apparent consensus ranges from $42.353\%$ to $99.3825\%$. At $\tau = 0.50$, 142 of 200 rules fail the modal-path check. Frequent agreement is sensitive to the parameters, while fully credible projections remain uncommon in these conditions. No condition has a symmetric pure equilibrium prescribing a demanding open proposal followed by $Q$. The condition $\Delta = 0.20$ has eight fully credible candidates, supported by the weak-bite $A$--$Q$ payoff tie, and each other condition has one. This finite sweep does not cover joint parameter changes or other learning rules.

\subsubsection{Choice noise across continuation values}

\protect\phantomsection\label{sec:si-temperature-continuation}{}

\begin{figure}[htbp]
\centering
\includegraphics[width=1\linewidth,height=0.65\textheight,keepaspectratio]{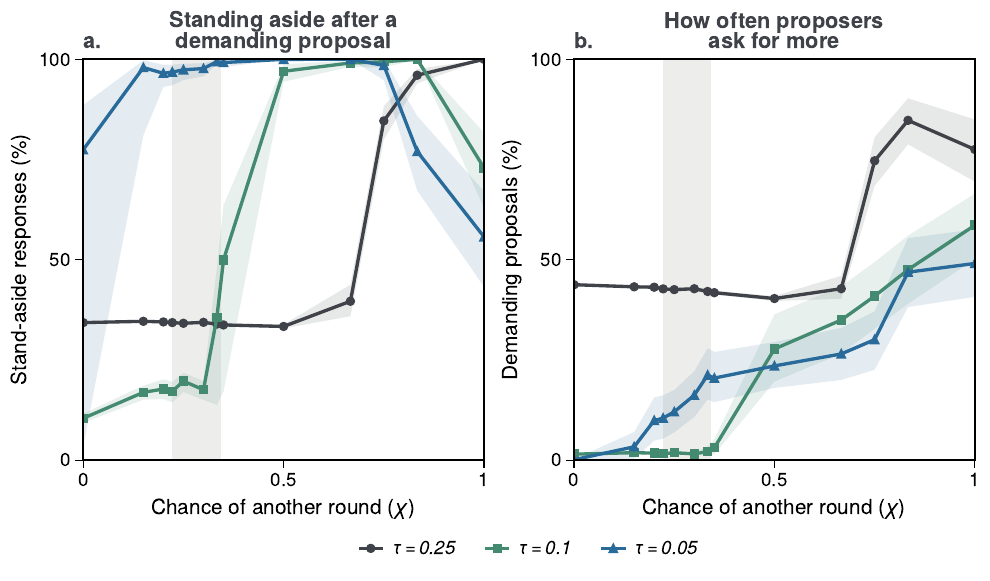}
\caption{Lower exploration changes responses and proposals together. \textbf{a,} Stand-asides as a percentage of responses to demanding open proposals. \textbf{b,} Demanding proposals as a percentage of all open rounds. Each point pools visits over the final 4,000 rounds of 100 independently seeded, 8,000-round runs. Colours and symbols identify the shared proposer and responder temperature. Bands give 95\% percentile intervals from 2,000 bootstrap samples of whole runs. Grey shading marks the exact credibility window of the specified strong-restraint candidate at $\rho = 0.65$ and $\delta = 0.9$. Lines connect the sampled continuation values. Conditional responses alone do not establish credibility of the complete learned rule.}
\label{fig:si-temperature-continuation}
\end{figure}

Reducing exploration changes both what responders accept and what proposers ask for. Across 100 neutral-start runs per condition, lowering $\tau$ from $0.25$ to $0.05$ raises the pooled stand-aside frequency after demanding open proposals at $\chi = 0.30$ from 34.4\% to 97.7\% (Figure~\ref{fig:si-temperature-continuation} a). Demanding proposals, however, fall from 42.8\% to 16.2\% of open play (Figure~\ref{fig:si-temperature-continuation} b). Only 34 of the 100 cooler runs make any demanding open proposal during the measurement period. The high conditional response frequency describes the proposals that are made, with runs weighted by their number of such proposals.

The distinction is especially important when demanding proposals are rare. At $\tau = 0.05$ and $\chi = 0$, they account for only 399 of 400,000 open rounds, spread across 42 runs. Of those 399 responses, 77.4\% are stand-asides, with a wide 95\% run-bootstrap interval of 3.5\%--88.6\%. No final modal rule at that condition prescribes demanding open play with restraint. Sparse exposure and persistent learned values can accompany a high conditional response frequency even without any prospect of future cooperation.

The complete-rule audit gives a more demanding test of whether learners recover credible restraint. At $\tau = 0.05$, 12, 18 and 31 of 100 final modal rules prescribe a demanding open proposal with a stand-aside at $\chi = 0.25$, $0.30$ and $\frac{1}{3}$, respectively. Of these, only one passes the full audit, at $\chi = \frac{1}{3}$. At $\tau = 0.10$ and $0.25$, none at these three continuation values prescribes demanding open play with restraint and passes the full audit. Cooler choice brings the conditional response closer to the restraint candidate without generally recovering a complete credible rule. The diagnostics in Section~\ref{sec:si-learners-leave} separate exploration from noise in learned values. This finite continuation grid establishes neither a sharp learning threshold nor a vanishing-noise selection result.

See main Methods (``Learning sensitivity, horizon and interventions'') for the protocol. \nolinkurl{temperature_continuation_check.py} writes the per-run counts, summaries and manifest to \nolinkurl{data/temperature_continuation_check/}. \nolinkurl{fig_temperature_continuation.py} plots the cached summaries.

\subsection{Longer training at high continuation}

\protect\phantomsection\label{sec:si-horizon}{}

See main Methods (``Learning sensitivity, horizon and interventions'') for the model and procedures.

\begin{longtable}[]{@{}
  >{\raggedright\arraybackslash}p{(\linewidth - 10\tabcolsep) * \real{0.1876}}
  >{\centering\arraybackslash}p{(\linewidth - 10\tabcolsep) * \real{0.1563}}
  >{\centering\arraybackslash}p{(\linewidth - 10\tabcolsep) * \real{0.1563}}
  >{\centering\arraybackslash}p{(\linewidth - 10\tabcolsep) * \real{0.1876}}
  >{\centering\arraybackslash}p{(\linewidth - 10\tabcolsep) * \real{0.1563}}
  >{\centering\arraybackslash}p{(\linewidth - 10\tabcolsep) * \real{0.1563}}@{}}
\caption{Long-training outcome summaries and separate stochastic-policy
and modal-rule audits.}\label{tab:si-learning-horizon}\tabularnewline
\toprule\noalign{}
\endfirsthead
\endhead
\bottomrule\noalign{}
\endlastfoot
\textbf{Exploration} & \textbf{Mean apparent (\%)} & \textbf{Median
apparent (\%)} & \textbf{Median largest one-shot gain} &
\textbf{Stochastic full pass (/30)} & \textbf{Modal full pass (/30)} \\
Constant & 75.12 & 98.26 & 0.34464 & 0 & 3 \\
Decreasing & 73.33 & 100 & 0.28563 & 0 & 0 \\
\end{longtable}

All entries describe the 100,000-round endpoint. The smallest largest one-shot gain across the 30 stochastic policies is $0.10142$ with constant exploration and $0.09941$ with decreasing exploration. So every stochastic policy fails the full check. Applying the symmetric modal projection instead gives three full passes with constant exploration (seeds 13, 24 and 28), all the weak-bite rule, and none with decreasing exploration. The apparent-consensus medians of $98.26\%$ and $100\%$, alongside means of $75.12\%$ and $73.33\%$, show the uneven distribution across runs. The mean at 8,000 rounds is $83.335\%$ for these same seeds. Longer training does not establish convergence or a general increase in agreement.

Positive-temperature softmax assigns positive probability to every feasible action when values are finite. After a demanding proposal, support pays $0.65$ less than standing aside and has the same continuation. After a weak proposal, standing aside pays $0.10$ less than support, again with the same continuation. Keeping positive probability on these inferior responses makes a stochastic policy exploitable even if its continuation estimates are accurate. Exact equilibrium failure has an exploration component. Numerical acceptance still depends on whether the gain exceeds the audit tolerance of $10^{- 9}$. The gains reported above exceed that tolerance substantially, but their presence alone does not isolate errors in learned expectations. The modal audit separately checks the most likely prescriptions after removing exploratory mixing.

\subsection{Resuming learning after an imposed dispute}

\protect\phantomsection\label{sec:si-intervention}{}

See main Methods (``Learning sensitivity, horizon and interventions'') for the model and procedures.

\begin{longtable}[]{@{}
  >{\raggedright\arraybackslash}p{(\linewidth - 10\tabcolsep) * \real{0.1587}}
  >{\raggedright\arraybackslash}p{(\linewidth - 10\tabcolsep) * \real{0.1429}}
  >{\centering\arraybackslash}p{(\linewidth - 10\tabcolsep) * \real{0.1587}}
  >{\centering\arraybackslash}p{(\linewidth - 10\tabcolsep) * \real{0.1905}}
  >{\centering\arraybackslash}p{(\linewidth - 10\tabcolsep) * \real{0.1905}}
  >{\centering\arraybackslash}p{(\linewidth - 10\tabcolsep) * \real{0.1587}}@{}}
\caption{Learning resumed from open play or an imposed dispute, under
two episode-boundary conventions.}\label{tab:si-dispute-intervention}\tabularnewline
\toprule\noalign{}
\endfirsthead
\endhead
\bottomrule\noalign{}
\endlastfoot
\textbf{Episode boundary} & \textbf{Start} & \textbf{First-episode
vetoes} & \textbf{Changed rule (/20,000)} & \textbf{Apparent consensus
(\%)} & \textbf{Full pass (/20,000)} \\
Reset to $O$ & Control $O$ & 0.24835 & 2,514 (12.57\%) & 83.9775 &
108 (0.54\%) \\
Reset to $O$ & Dispute $P$ & 0.80125 & 7,320 (36.60\%) & 83.9410 &
112 (0.56\%) \\
Carry mode & Control $O$ & 0.24835 & 2,776 (13.88\%) & 83.9287 & 120
(0.60\%) \\
Carry mode & Dispute $P$ & 0.80125 & 8,205 (41.025\%) & 83.8840 & 123
(0.615\%) \\
\end{longtable}

The paired first-episode difference is $0.55290$ vetoes, with a mean episode length of $6.0317$ rounds in both arms. The mean within-episode veto fractions, a different statistic, are $4.2547\%$ in control and $19.4874\%$ after a dispute. These replicate endpoints share 1,000 trained pairs. They are not additional independent training samples.

Both arms retain high apparent-consensus shares and very low full-pass shares. Before resumption, the four fully credible training endpoints contribute 80 of 20,000 replicate records per arm. After resumption the full-pass share is $0.54\%$--$0.615\%$. The small differences in these descriptive counts do not establish equal credibility or an improvement caused by the intervention. The original-game audit interpretation and the distinction from resumed learning are specified in main Methods.

\subsection{Learners leave the credible rule}

\protect\phantomsection\label{sec:si-learners-leave}{} Inside the credibility window the strong-restraint candidate is one of two symmetric pure stationary equilibria. In the other, every mode offers the weak bite, which receives active support, and a strong bite would be vetoed. This weak-bite rule is the only symmetric pure stationary equilibrium of the one-shot game. On the enumerated continuation grid it remains an equilibrium throughout $\lbrack 0,1\rbrack$, always with full margin $0.10$, the gain from active support over acquiescence at the weak bite. The candidate's full margin never exceeds $0.032$, reached at $\chi = 0.27$. It is $0.019$ at $\chi = 0.25$ and $0.30$ and $0.004$ at $\chi = \frac{1}{3}$.

Learners from neutral starts do not find the candidate. None of the 600 short-window runs or the 300 long-window runs has a matching final projection, and the long-window runs never match it at the sampled checkpoints. Under constant exploration, the mean of no outcome share changes by more than $0.011$ between the first and last 10,000-round blocks. Under decreasing exploration, demanding proposals in open play account for less than 0.05\% of all rounds in every run after round 50,000, and about 99\% of the final 10,000 rounds pass with active support.

At each of the three tested continuation values, zero successes in 200 independently seeded runs gives a one-sided 95\% upper bound of approximately $1.5\%$ on a final candidate match under this training procedure. The bound applies separately at each value, not to a pooled common-probability sample of all 600 runs.

See main Methods (``Learning sensitivity, horizon and interventions'') for the model and procedures.

All 600 seeded pairs leave the candidate at a sampled checkpoint, 70\% within 100 rounds and all within 300. In 587 the first change is a veto of the strong open bite, where the candidate prescribes restraint. By round 8,000, 454 pairs project to the weak-bite equilibrium. Of the other 146, 109 still offer the strong bite, which the responder vetoes. Over the final 4,000 rounds, every outcome share of the seeded pairs lies within $0.003$ of the neutral-start runs at the same $\chi$. Over 100,000 rounds, one of 300 seeded pairs is observed back at the candidate, at one checkpoint.

Restraint in usual play is reciprocal. A responder that lets a strong bite pass gives up $0.15$ now. In return it keeps the open menu for its own next proposal, which is worth $0.75$ more than the weak bite that revision would allow. Discounted by $\beta$, the trade pays only for $\beta \geq 0.2$, that is $\chi \geq \frac{2}{9}$, the lower edge of the window. At the three tested continuation values, relationships last $\frac{1}{1 - \chi}$ rounds on average, between 1.33 and 1.50. Two thirds to three quarters end after their first round. A learner judges the future from the relationships it has seen. As a rough guide, with its other values exact, restraint beats the veto in a learner's estimates only while its running estimate of the share of relationships that continue exceeds $\frac{2}{9}$. The constant-rate diagnostic described in main Methods gives that estimate a standard deviation of about $0.07$. At $\chi = 0.30$ its 5--95\% range, $0.18$ to $0.42$, is wider than the window. The estimate lies below $\frac{2}{9}$ for 36\%, 15\% and 7\% of the time at $\chi = 0.25$, $0.30$ and $\frac{1}{3}$. At $\chi = \frac{5}{6}$ its 5--95\% range is $0.73$ to $0.92$, far above $\frac{2}{9}$.

Restraint also depends on the other party. In the mixed-response diagnostic, restraint pays only for $q \geq 0.935$, $0.848$ and $0.804$ at $\chi = 0.25$, $0.30$ and $\frac{1}{3}$. At the candidate's exact values, softmax choice at $\tau = 0.25$ gives $q$ between $0.50$ and $0.55$. Proposers keep offering the strong bite as long as $q \geq 0.422$. Responders start to veto before proposers give up the strong bite, which matches the first change in 587 of the 600 seeded pairs.

Exploration alone moves learners off the candidate at the reported temperature. Along the ascending logit branch, the last sampled temperatures retaining the candidate modal prescription are $0.003$, $0.012$ and $0.019$ at $\chi = 0.25$, $0.30$ and $\frac{1}{3}$. These limits lie at least 13 times below the reported $\tau = 0.25$. At the next sampled temperatures, the iteration selects a stochastic policy whose modal prescription is the weak-bite rule. Sampling noise in the learned values moves learners off the candidate even below these limits. At $\chi = 0.30$ and $\tau = 0.01$, all 200 seeded pairs have left by the 450-round checkpoint, as they do at $\tau = 0.02$, $0.05$ and $0.10$. A responder that exercises restraint and sees the relationship end moves its value of restraint toward the passage value $0.25$. Holding the update rate at $\alpha = 0.05$, three such endings in a row erase the margin of $0.0525$ by which restraint beats the veto at $\chi = 0.30$. The value of the veto barely moves, because at $\tau = 0.01$ the veto is rarely tried.

Slower learning separates exploration from noise. Lowering the base rate $\alpha_{0}$ tenfold, from $0.05$ to $0.005$, raises the median observed exit checkpoint at $\tau = 0.25$ from 100 to 800, roughly in proportion to $\frac{1}{\alpha_{0}}$, the pattern of a steady drift away from the candidate. At $\tau = 0.01$ the same change raises the conditional median among observed exits from 100 to 1,750. With the slowest learning at $\tau = 0.01$, 11 of 200 pairs retain the candidate at every sampled checkpoint through 8,000 rounds, and the latter median uses the 189 observed exits. Exit times that grow faster than $\frac{1}{\alpha_{0}}$ indicate noise acting on a rule that noise-free logit response would hold.

Within the window, the credible restraint rule exists but rarely persists under this learning process, and then only with almost no exploration and slow learning. In some models with several equilibria, noisy adaptation selects equilibria according to their resistance to mistakes \citep{KandoriMailathRob1993} \citep{Young1993}. The persistence of the weak-bite rule in our runs is consistent with this pattern. They do not establish the vanishing-noise limit that a formal selection result requires. \nolinkurl{seeded_credible_check.py} and \nolinkurl{window_horizon_check.py} write these results to \nolinkurl{data/seeded_credible_check/} and \nolinkurl{data/window_horizon_check/}.

\subsection{Apparent consensus rises with bite}

See main Methods (``Learning sensitivity, horizon and interventions'') for the model and procedures.

\begin{longtable}[]{@{}
  >{\centering\arraybackslash}p{(\linewidth - 8\tabcolsep) * \real{0.1875}}
  >{\centering\arraybackslash}p{(\linewidth - 8\tabcolsep) * \real{0.2411}}
  >{\centering\arraybackslash}p{(\linewidth - 8\tabcolsep) * \real{0.1875}}
  >{\centering\arraybackslash}p{(\linewidth - 8\tabcolsep) * \real{0.2054}}
  >{\centering\arraybackslash}p{(\linewidth - 8\tabcolsep) * \real{0.1786}}@{}}
\caption{Selected continuation values in the five-bite learning
sweep.}\label{tab:si-five-bite}\tabularnewline
\toprule\noalign{}
\endfirsthead
\endhead
\bottomrule\noalign{}
\endlastfoot
\textbf{Continuation $\chi$} & \textbf{Apparent consensus} &
\textbf{Mean bite} & \textbf{Filtering veto} & \textbf{Breakdown} \\
$0.000$ & $0.488$ & $0.599$ & $0.170$ & $0.087$ \\
$0.667$ & $0.602$ & $0.634$ & $0.120$ & $0.039$ \\
$0.750$ & $0.930$ & $0.852$ & $0.023$ & $0.003$ \\
$0.833$ & $0.9996$ & $0.951$ & $0.0002$ & $0.00005$ \\
\end{longtable}

Across these continuation settings, no-veto outcomes and proposal bite rise together. At low continuation, veto and revision still shape the realised path. At high continuation under this learning process, responders select restraint and proposers select harder bites. Continuation can increase apparent consensus while reducing proposal-side accommodation.

\begin{figure}[htbp]
\centering
\includegraphics[width=1\linewidth,height=0.65\textheight,keepaspectratio]{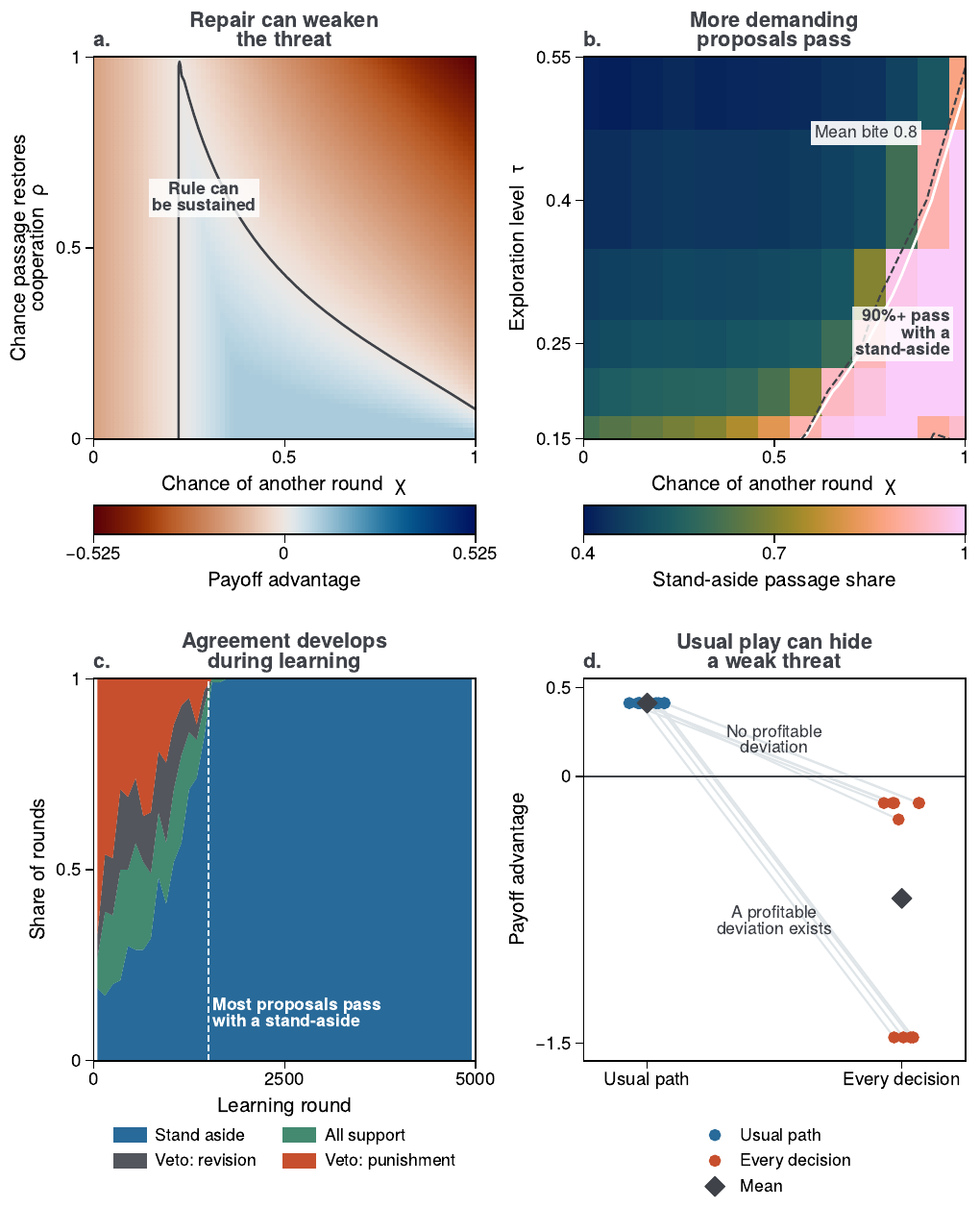}
\caption{Exact credibility and learned selection in the three-mode game. Panel (a) shows the full-audit margin of the two-bite strong-restraint candidate, and the zero contour bounds credibility. Panel (b) reports apparent-consensus shares under the five-bite learning process, and the contours mark a $0.9$ share and mean bite $0.8$. Panel (c) follows one veto-prone path at $\chi = 0.85$. Panel (d) compares the modal-path and full-node audits in the exact two-bite game at $\chi = \frac{5}{6}$. Panels (b) and (d) average ten seeds, while panel (c) is one deliberately selected transition path.}
\label{fig:emergent-consensus}
\end{figure}

Veto can occur on the route to the learned convention. In the illustrative veto-prone path in panel (c), veto occupies $51.6\%$ of the first 1,000 rounds and breakdown $35.8\%$. Restraint reaches $67.8\%$ between rounds 1,000 and 1,500, then becomes effectively universal through round 5,000 while mean bite remains near $0.77$ to $0.79$. Licensed vetoes filter demanding proposals into revision, while unlicensed or repeated vetoes damage the relationship. Cooperation can follow either transition, but through different public states.

\textbf{The bounded behavioural claim}

The simulations establish conditional selection under the stated finite menus, payoff parameters, SARSA rule, temperature, initialisation, and horizon. Historical estimation and characterisation of other learning processes remain outside the exercise. The finite path supports a persistent learned convention. A metastability result would require dwell-time and escape-time limits as experimentation vanishes.

The executable specification in \nolinkurl{emergent_consensus.py} produces the exact audit and learning paths. \nolinkurl{run_emergent_consensus_analysis.py} writes the source tables in \nolinkurl{data/emergent_*.csv}, and \nolinkurl{fig_emergent_consensus_learning.py} reads those tables to produce Figure~\ref{fig:emergent-consensus}.

\subsection{Disputes across settings}

\protect\phantomsection\label{sec:si-robustness}{} The figures in this subsection retain earlier diagnostic sweeps and projection conventions. They do not reproduce the current two-million-round, per-Party analyses in main Figure 4. Each heat map has 1,000 trained pairs per cell. Main Methods specifies the current design, and Supplementary Section S10.11 distinguishes the newer results. Hatched regions mark where the symmetric strong-restraint candidate passes the full audit, and white points mark the reference setting. Differences between two percentages are given in percentage points (pp). Shares of rules count projected prescriptions. They are not observed frequencies of vetoes.

\begin{figure}[htbp]
\centering
\includegraphics[width=1\linewidth,height=0.65\textheight,keepaspectratio]{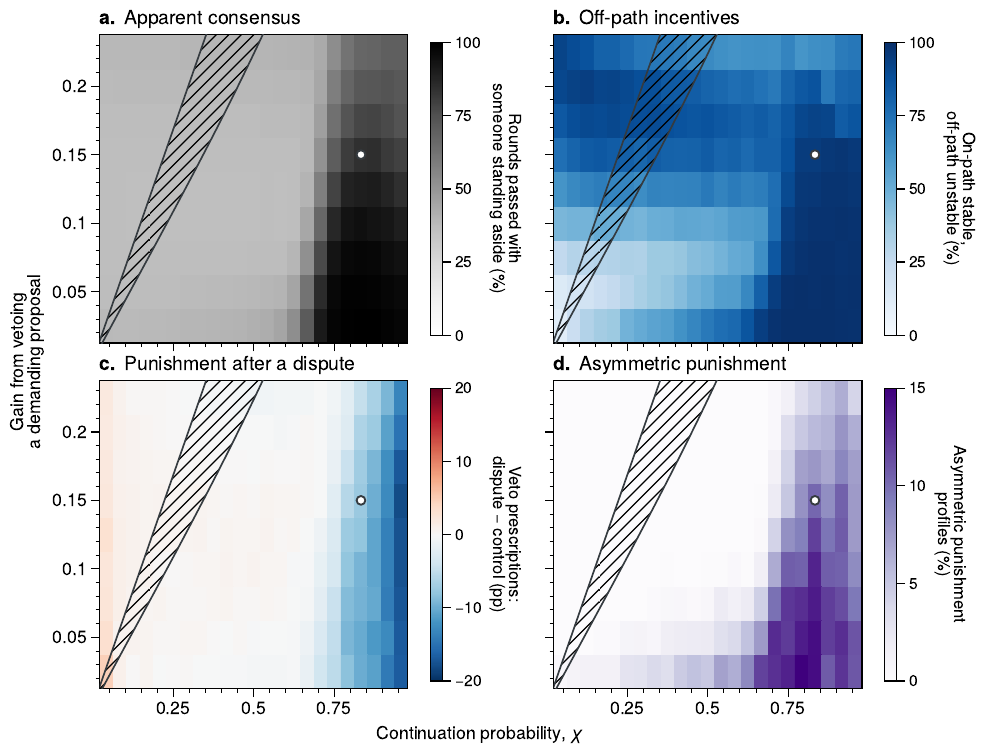}
\caption{Learning and one imposed dispute across the chance of another round and the gain from vetoing a demanding proposal. (a) Rounds that pass with someone standing aside, over the final 4,000 training rounds. (b) Symmetric modal rules that pass the modal-path check but fail the full audit. (c) Rules that veto during punishment after one imposed dispute and 2,000 resumed rounds, minus matched controls, in percentage points. (d) Pairs whose separately projected rules divide the punishment roles. Each cell has 1,000 trained pairs and 20 replicates per arm. These panels belong to the earlier diagnostic sweeps, rather than the current main Figure 4a--c.}
\label{fig:si-robustness-alignment}
\end{figure}

Apparent consensus rises between $\chi = 2/3$ and $0.79$ at every gain (Figure~\ref{fig:si-robustness-alignment} a). Above $\chi = 0.7$, a larger gain from vetoing lowers apparent consensus. Rules that pass the modal-path check but fail the full audit make up 22\% to 100\% of pairs across cells (Figure~\ref{fig:si-robustness-alignment} b). They exceed 85\% wherever $\chi \geq 0.75$ and the gain is at most 0.15. From $\chi = 0.79$, one dispute lowers rules that veto during punishment by 2 to 18 pp (Figure~\ref{fig:si-robustness-alignment} c). Below $\chi = 0.7$, it changes them by at most 1.5 pp. Divided punishment roles appear in up to 15\% of pairs, mostly where $\chi \geq 0.7$ (Figure~\ref{fig:si-robustness-alignment} d).

The divided roles rest on an exact property of one pair of rules. The pair departs from the symmetric strong-restraint candidate only in $P$. There the Party holding the threat proposes the demanding bite, and the other Party stands aside on the demanding bite. Both keep the candidate's other prescriptions. The Party holding the threat keeps the candidate's veto of a demanding proposal in $P$. The other Party never makes that proposal there. At the reference payoffs, punishment costs the Party holding the threat nothing. As proposer it receives $b + 1 = 1.25$ in both $O$ and $P$, because its demanding proposal passes in both. As responder it receives $b = 0.25$ after standing aside on a demanding proposal in $O$ and $b + h(0.25) - \Delta = 0.35$ after supporting a weak proposal in $P$. A veto that prolongs punishment does not lower its continuation value. Punishment is costly for the other Party, whose proposals in $P$ pass only at the weak bite. Its incentives set the lower edge of the passing range. At the reference gain, the pair passes the full audit at all 132 grid values from $\chi = 0.340$ to $0.995$, and bisection puts the lower edge at $\chi = 0.338$. The symmetric window ends at $\chi = 0.343$, just above that edge. Learned pairs counted in Figure~\ref{fig:si-robustness-alignment} d match this pair in the prescriptions that define the roles. Their other prescriptions can differ, so a counted pair need not pass the full audit.

\begin{figure}[htbp]
\centering
\includegraphics[width=1\linewidth,height=0.65\textheight,keepaspectratio]{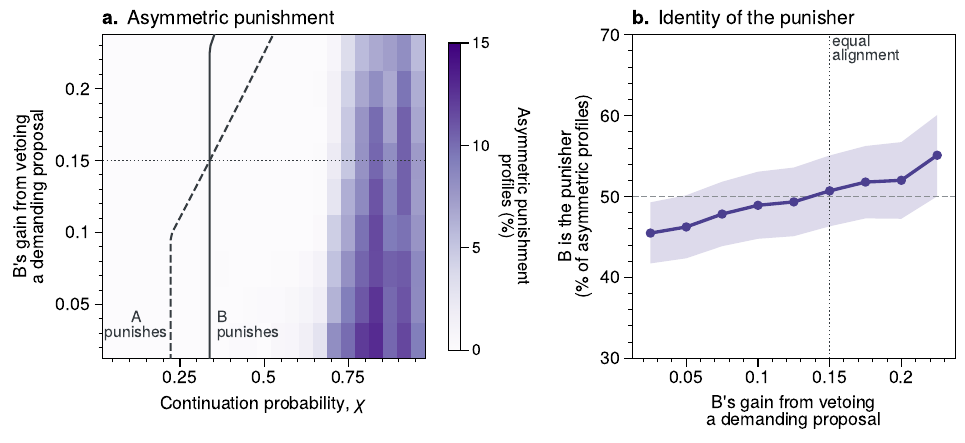}
\caption{Divided punishment roles when the two Parties gain differently from a veto. Party A keeps the reference gain of 0.15 and Party B's gain varies. (a) Pairs whose separately projected rules divide the punishment roles. The solid line marks the lower edge of the passing range for the audited pair with B holding the threat, and the dashed line the edge with A holding it. These lines concern that pair only. The dotted line marks equal gains. (b) Share of divided pairs in which B holds the threat, pooled over $\chi \geq 0.75$, with 95\% Wilson intervals. Each cell has 1,000 trained pairs.}
\label{fig:si-robustness-asymmetric}
\end{figure}

Unequal gains change who holds the threat only modestly (Figure~\ref{fig:si-robustness-asymmetric}). Divided roles are more common when B gains less from a veto. Where $\chi \geq 0.75$, they occur in 11.1\% of pairs at B's smallest gain and in 6.2\% at its largest. Among these pairs, B holds the threat in 45.5\% at its smallest gain and in 55.1\% at its largest. The exact passing range of the audited pair depends on which Party gives way. With A giving way, its lower edge stays at $\chi = 0.338$. With B giving way, it rises from $2/9$ to $0.50$ as B's gain grows from 0.025 to 0.225.

\begin{figure}[htbp]
\centering
\includegraphics[width=1\linewidth,height=0.65\textheight,keepaspectratio]{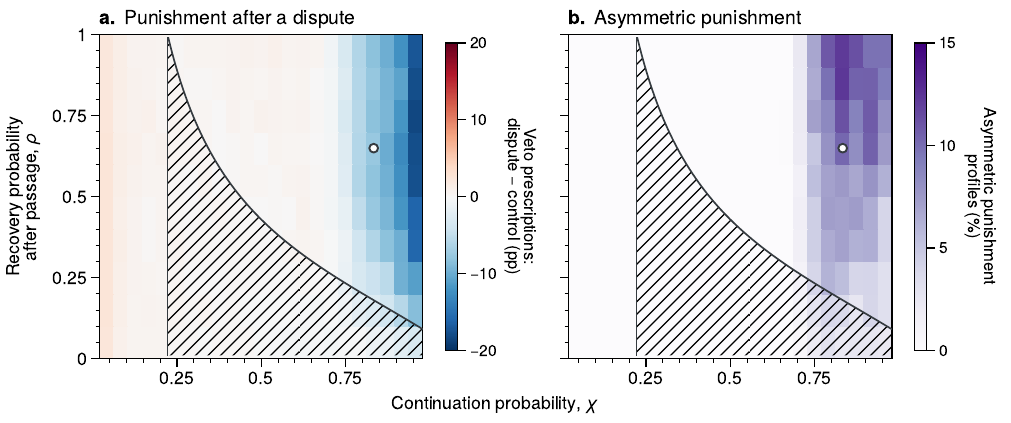}
\caption{One imposed dispute across continuation and recovery. Recovery $\rho$ is the chance that passage during punishment restores ordinary play. (a) Rules that veto during punishment after one imposed dispute and 2,000 resumed rounds, minus matched controls, in percentage points. (b) Pairs whose separately projected rules divide the punishment roles before the dispute. Each cell has 1,000 trained pairs and 10 replicates per arm, at the reference gain of 0.15.}
\label{fig:si-robustness-recovery}
\end{figure}

Recovery sets how far one dispute weakens the threat (Figure~\ref{fig:si-robustness-recovery} a). At $\chi = 23/24$, one dispute lowers rules that veto during punishment by 3 pp at $\rho = 0.05$ and by 16 to 18 pp for $\rho \geq 0.35$. With little chance of recovery, standing aside during punishment gains little. The theory then keeps the punishment veto credible up to longer futures, as the hatched region shows. Divided roles become more common as $\rho$ rises, from at most 3\% of pairs at $\rho = 0.05$ to 12\% for $\rho \geq 0.85$ (Figure~\ref{fig:si-robustness-recovery} b). Apparent consensus after the dispute stays within 0.05 pp of the control in every cell.

\begin{figure}[htbp]
\centering
\includegraphics[width=1\linewidth,height=0.65\textheight,keepaspectratio]{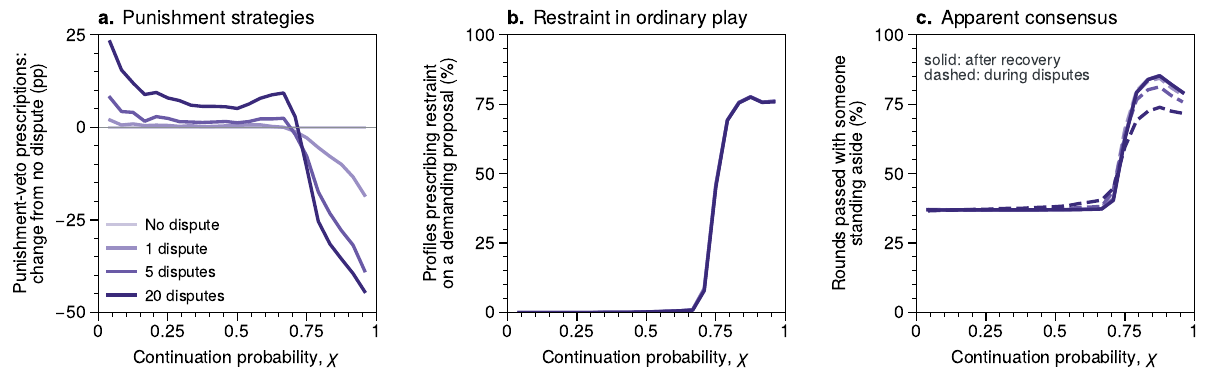}
\caption{Repeated disputes across continuation values. Each trained pair resumes for 100 rounds split into 1, 5 or 20 segments that start in punishment, or one segment in ordinary play without a dispute. It then learns for 2,000 rounds from ordinary play. (a) Change in rules that veto during punishment after recovery, relative to no dispute, in percentage points. (b) Rules that stand aside on a demanding proposal in ordinary play after recovery. (c) Rounds that pass with someone standing aside. Solid lines cover the final 1,000 recovery rounds and dashed lines the 100 rounds with imposed disputes. Shares of rules and shares of rounds have different denominators. Each continuation value has 1,000 trained pairs and 10 replicates per number of disputes.}
\label{fig:si-robustness-repeated}
\end{figure}

Repeated disputes move the threat in opposite directions at short and long futures (Figure~\ref{fig:si-robustness-repeated} a). For $\chi \geq 0.75$, more disputes lower the share of rules that veto during punishment further. At $\chi = 23/24$, 1, 5 and 20 disputes lower it from 47\% to 28\%, 8\% and 2\%. For $\chi \leq 2/3$, 20 disputes raise it by 5 to 23 pp. With little future at stake, the veto's gain within the round outweighs the chance of restoring usual play. Restraint in ordinary play changes by at most 0.9 pp in any cell (Figure~\ref{fig:si-robustness-repeated} b). Apparent consensus dips while disputes are imposed (Figure~\ref{fig:si-robustness-repeated} c). At $\chi = 5/6$, it falls from 84\% to 72\% of rounds during 20 disputes. After recovery it is within 0.7 pp of the level without disputes in every cell. Divided roles rise slightly with more disputes, from 10.2\% to 13.1\% of pairs at $\chi = 5/6$.

\begin{figure}[htbp]
\centering
\includegraphics[width=1\linewidth,height=0.65\textheight,keepaspectratio]{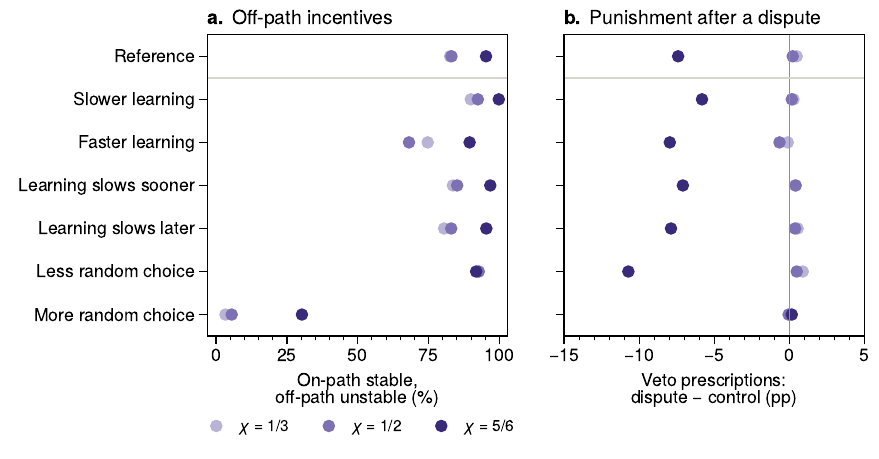}
\caption{Sensitivity to learning settings at three continuation values. (a) Symmetric modal rules that pass the modal-path check but fail the full audit. (b) Change in rules that veto during punishment after one imposed dispute, against matched controls, in percentage points. Slower and faster learning use base rates $\alpha_{0} = 0.02$ and $0.10$. Learning slows sooner or later with decay scales of 1,000 and 16,000 visits. Less and more random choice use $\tau = 0.10$ and $0.50$. The reference uses $\alpha_{0} = 0.05$, a decay scale of 4,000 visits and $\tau = 0.25$, at the reference gain of 0.15. Each point has 1,000 trained pairs and 20 replicates per arm.}
\label{fig:si-robustness-settings}
\end{figure}

The dispute effect at $\chi = 5/6$ changes little with the learning rate or the speed at which learning slows (Figure~\ref{fig:si-robustness-settings} b). There one dispute lowers rules that veto during punishment by 5.8 to 8.0 pp across these settings, against 7.4 pp at the reference. Less random choice ($\tau = 0.10$) strengthens the effect to 10.7 pp. More random choice ($\tau = 0.50$) removes it, with a change of 0.2 pp. A veto then sent play to punishment in 3.3\% of the final 4,000 training rounds, against 0.04\% at the reference. Learners have then met punishment often before the dispute, and only 3\% to 30\% of rules pass the modal-path check while failing the full audit (Figure~\ref{fig:si-robustness-settings} a). At $\chi = 1/3$ and $1/2$, the dispute changes these rules by less than 1 pp in every setting. Apparent consensus after the dispute stays within 0.02 pp of the control in all 21 cells.

\nolinkurl{robustness_experiment.py} runs the sweeps and writes the pair tables. \nolinkurl{band_edges.py} and \nolinkurl{split_edges.py} compute the exact edges. \nolinkurl{fig_robustness.py} draws these figures, and \nolinkurl{scripts/plot_robustness_veto.py} draws main Figure 4.

\subsection{Two-million-round per-Party audits and diagnostics}

\protect\phantomsection\label{sec:si-current-pair-audits}{}

This subsection distinguishes the current main Figures 3 and 4 from the earlier learning samples and projections retained above. Main Methods gives the payoff, learning, projection and audit procedures. Current endpoints reconstruct each Party's own deterministic rule. The independently seeded unit is the trained pair, and intervention replicates are nested within each pair. These frozen-rule audits do not classify the original exploratory policies or establish convergence of continued learning.

At $\chi = \frac{23}{24}$ and a gain of 0.15 from vetoing a demanding proposal, 95.0824\% of evaluation rounds pass, 11.5474\% with full support and 83.5350\% with a stand-aside. Each of 1,000 pairs contributes its final 4,000 of 2,000,000 training rounds. Three of the 1,000 reconstructed deterministic pairs pass the full audit, while 889 pass the prescribed-path audit, of which 886 fail elsewhere. Thus 88.9\% pass the path check and 88.6\% pass that check only. Those percentages count pairs, unlike the round shares.

Pooling the two learners' action probabilities into a single symmetric rule changes the object audited. At $\chi = \frac{5}{6}$, that construction gives 79.0\% full passes, compared with 35.3\% when each Party's own rule is audited. The average can hide a profitable departure by one Party. The older symmetric-projection results in S10.4 therefore cannot substitute for these per-Party classifications.

Of the 2,000 Parties at $\chi = \frac{23}{24}$, 94\% have a profitable one-shot departure somewhere, and 5.7\% do along the prescribed path. The selected largest profitable departure changes a proposal during punishment for 711 Parties, adds a punishment veto for 599 and drops one for 87. Other Parties have different largest departures or none. These are node-specific, frozen-rule comparisons. They do not count observed defections. Punishment accounts for 0.1\% of training rounds there, compared with 3.4\% at $\chi = \frac{5}{6}$.

The imposed-dispute comparison has 20 replicates in each arm for every trained pair. The first episode contains a veto in 11,004 of 20,000 dispute replicates (55.02\%) and 4,728 controls (23.64\%). Endpoint punishment prescriptions change in 11,233 of 40,000 Party--replicate records (28.0825\%) against 18 controls (0.045\%). Ordinary-play prescriptions change in 2.975\% and 2.9775\%, respectively. The demanding-punishment response changes from veto to stand-aside in 2,451 records (6.1275\%). This count does not condition on the open-play prescriptions illustrated in main Figure 3d. Apparent consensus over the final 1,000 of 2,000 resumed rounds is 83.52729\% after a dispute and 83.52773\% in controls.

\begin{figure}[htbp]
\centering
\includegraphics[width=\linewidth]{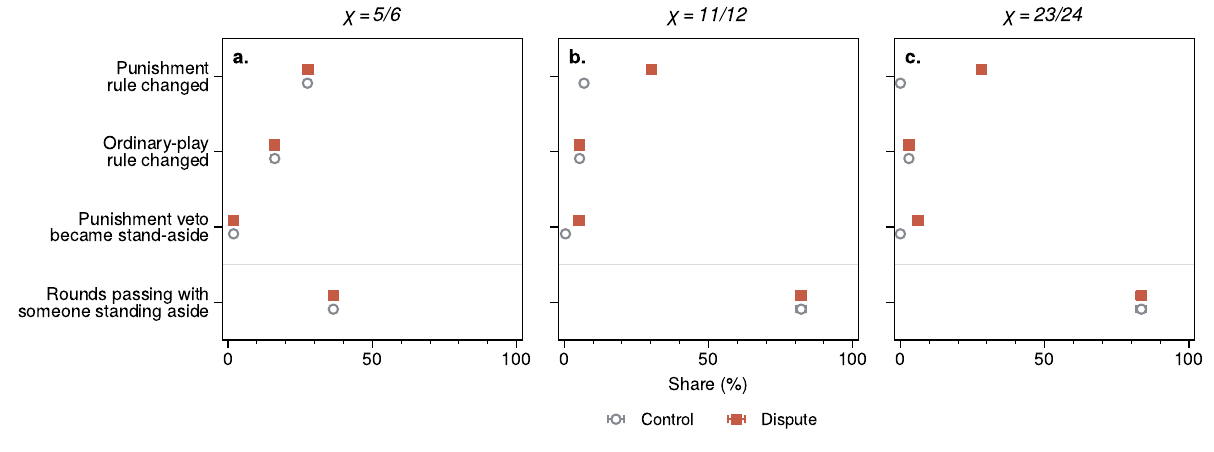}
\caption{The dispute comparison across continuation values, with all other Figure 3 settings unchanged. Open circles show controls resumed in ordinary play, and filled squares show pairs resumed in punishment. The rows count Parties whose punishment prescriptions changed, Parties whose ordinary-play prescriptions changed, Parties whose response to a demanding punishment proposal changed from veto to stand-aside, and rounds that pass with someone standing aside. Prescription changes compare each Party's deterministic rule before and after resumption. The final row averages the last 1,000 resumed rounds. Horizontal bars give pointwise 95\% percentile intervals from 5,000 bootstrap draws of the 1,000 trained pairs, retaining both Parties and all matched replicates within each pair. Some intervals are narrower than their markers.}
\label{fig:si-figure2-sensitivity}
\end{figure}

The Figure 3 comparison also uses $\chi=5/6$ and $\chi=11/12$, with the gain from vetoing a demanding proposal fixed at 0.15 (Figure~\ref{fig:si-figure2-sensitivity}). These continuation values were fixed before reconstructing the additional comparisons from the saved two-million-round agents. Each setting retains 1,000 trained pairs and 20 matched replicates per arm, with action values and visit counts carried into 2,000 resumed rounds. The saved seeds are reused, so these comparisons remain descriptive sensitivity checks.

At $\chi=11/12$, punishment prescriptions change for 30.27\% of Parties after a dispute and 6.88\% in controls, an excess of 23.39 percentage points with a paired 95\% interval of 22.25 to 24.53. At $\chi=5/6$, the corresponding shares are 27.65\% and 27.53\%, with an excess of 0.12 points and an interval of $-0.11$ to 0.35. The main-text setting, $\chi=23/24$, has an excess of 28.04 points with an interval of 26.85 to 29.20. Ordinary-play changes and late apparent-consensus shares remain close between the matched arms at each setting. The dispute-specific excess of punishment-rule changes therefore extends to a neighbouring high-continuation setting, with substantial parameter dependence across the comparisons.

A veto after a fixed demanding proposal in punishment and a veto after the Party's own prescribed punishment proposal are different measures. The former declines from 46.55\% before resumption to 39.8925\% in the dispute arm, against 46.53\% in controls. The latter is 38.65\% before resumption, 31.3925\% after a dispute and 38.6475\% in controls. The latter node can change when a Party changes its prescribed proposal, and it supplies the main Figure 4 comparison. Neither measure is an observed veto frequency.

Given the reconstructed rule pair, dropping the fixed demanding-proposal punishment veto pays more for 43.5\% of the Parties that hold this veto. Parties drop it in 22\% of cases where dropping pays and 17\% where it does not. After recovery, 99.98\% of prescriptions to stand aside on a demanding ordinary-play proposal are best replies to the partner's reconstructed rule in both arms. These comparisons use deterministic projections. Incentives against exploratory partner behaviour can differ, so the observed changes do not establish that learning corrects a particular audit failure. At $\chi = \frac{5}{6}$, the own-proposal veto contrast is +0.11 percentage points and late apparent consensus is about 36.5\% in both arms.

The separate repeated-dispute experiment has ten replicates per pair, 100 exposure rounds and 2,000 recovery rounds. The share vetoing after its own prescribed punishment proposal is 38.65\%, 31.455\%, 21.605\% and 17.93\% after zero, one, five and twenty disputes. Apparent consensus remains approximately 83.5\% over the final 1,000 recovery rounds. These values concern the current per-Party reconstruction, rather than the earlier 47\%--28\%--8\%--2\% sequence in S10.10.

The matched training-length experiment audits each Party's separately reconstructed rule at six horizons. The following counts were reproduced from the cached proposer and responder action values, without rerunning training. Each reference-horizon row contains the same 1,000 seeds at $\chi = \frac{5}{6}$, and each high-continuation row contains the same 100 seeds at $\chi = \frac{23}{24}$.

\begin{longtable}[]{@{}
  >{\raggedright\arraybackslash}p{(\linewidth - 6\tabcolsep) * \real{0.2174}}
  >{\raggedright\arraybackslash}p{(\linewidth - 6\tabcolsep) * \real{0.2609}}
  >{\raggedright\arraybackslash}p{(\linewidth - 6\tabcolsep) * \real{0.2609}}
  >{\raggedright\arraybackslash}p{(\linewidth - 6\tabcolsep) * \real{0.2609}}@{}}
\caption{Per-Party audits across matched training horizons. A path-only
pass satisfies checks at nodes reachable from open play under the
reconstructed pair but fails the full-node audit. Horizons within a seed
are matched, rather than independent training samples.}\label{tab:si-current-horizon-audits}\tabularnewline
\toprule\noalign{}
\begin{minipage}[b]{\linewidth}\raggedright
\textbf{Training rounds}
\end{minipage} & \begin{minipage}[b]{\linewidth}\raggedright
\textbf{Full passes, $\chi = \frac{5}{6}$ (/1,000)}
\end{minipage} & \begin{minipage}[b]{\linewidth}\raggedright
\textbf{Path-only passes, $\chi = \frac{5}{6}$ (/1,000)}
\end{minipage} & \begin{minipage}[b]{\linewidth}\raggedright
\textbf{Full passes, $\chi = \frac{23}{24}$ (/100)}
\end{minipage} \\
\midrule\noalign{}
\endfirsthead
\toprule\noalign{}
\begin{minipage}[b]{\linewidth}\raggedright
\textbf{Training rounds}
\end{minipage} & \begin{minipage}[b]{\linewidth}\raggedright
\textbf{Full passes, $\chi = \frac{5}{6}$ (/1,000)}
\end{minipage} & \begin{minipage}[b]{\linewidth}\raggedright
\textbf{Path-only passes, $\chi = \frac{5}{6}$ (/1,000)}
\end{minipage} & \begin{minipage}[b]{\linewidth}\raggedright
\textbf{Full passes, $\chi = \frac{23}{24}$ (/100)}
\end{minipage} \\
\midrule\noalign{}
\endhead
\bottomrule\noalign{}
\endlastfoot
8,000 & 8 & 827 & 0 \\
32,000 & 12 & 795 & 0 \\
128,000 & 37 & 689 & 0 \\
512,000 & 236 & 455 & 0 \\
1,000,000 & 299 & 492 & 0 \\
2,000,000 & 353 & 527 & 0 \\
\end{longtable}

At $\chi = \frac{5}{6}$, the share passing the full audit rises from 0.8\% to 35.3\%. The path-only share changes from 82.7\% to 52.7\%, with a lower intermediate value of 45.5\% at 512,000 rounds, so it does not fall monotonically. Apparent consensus is 84.0967\%, 81.46655\%, 66.114875\%, 36.9862\%, 36.501575\% and 36.513425\% at the respective horizons. At $\chi = \frac{23}{24}$, apparent consensus ranges from 80.88275\% to 84.2965\%, with zero full passes among 100 pairs at every endpoint. The three full passes in the larger high-continuation sample have seeds 278, 409 and 868, outside this matched 0--99 subset.

The cached-Q verification covers all 6,600 endpoints. Per-Party classifications match the prior reviewer records, pooled symmetric codes match their original horizon tables, and both full-audit implementations agree for every distinct pair encountered. The 1,100 endpoints overlapping the main two-million-round grid also match its per-Party audit seed by seed. Provenance records bind each endpoint to its source NPZ row and hash.

Against each pair's exact frozen-policy values, the median pair's mean absolute action-value error at $\chi = \frac{5}{6}$ falls from 1.9104 after 8,000 rounds to 0.0751 after 1,000,000 rounds. For punishment actions it falls from 2.1624 to 0.1064. These are estimates of learned-value accuracy, not equilibrium-pass rates. The source is \nolinkurl{data/panel_a_training_length_2026-11-05/reference_value_accuracy_medians.csv}. Policies, partners, visitation and visit-dependent learning rates change during training, so the comparison does not isolate broader experience as a causal explanation.

\subsection{Initial values, exploration and extended training}
\label{sec:si-exploration}

We tested whether the high-continuation pattern depends on initial action values and limited exploration. All conditions retain the main game's payoffs, $\delta=0.9$, $\rho=0.65$, alignment intercept $h_0=0.45$ and a gain of 0.15 from vetoing a demanding proposal. At each of $\chi \in \{5/6,11/12,23/24\}$, we reuse seeds 0--999 for two million rounds with the same SARSA kernel and visit-dependent learning rate. Each condition starts with zero visit counts, and outcome shares average the final 4,000 rounds with equal weight across pairs. The checks were chosen after inspecting the reference results and describe sensitivity within this learning procedure.

Alongside zero initial action values, we set every legal action value to $0.5/(1-\beta)$ or $1.25/(1-\beta)$, where $\beta=\delta\chi$. These positive starts give initially untried actions higher values, with the upper level corresponding to the largest stage reward received indefinitely. Another condition starts at the exact action values of a numerically obtained stationary logit equilibrium at $\tau=0.25$. These conditions retain the reference temperature. A separate zero-start condition scales temperature as $\tau=0.25(1-0.75)/(1-\beta)$, giving $\tau=0.3571$ at $\chi=11/12$ and $0.4545$ at $\chi=23/24$. At $\chi=5/6$ this equals the reference, so the condition is not repeated.

For the stationary benchmark, we solve the expected Bellman equations under both Parties' policies and update each node's action probabilities towards the softmax of its exact action values. Iteration starts from uniform probabilities over legal actions and uses a damping factor of 0.2 until the largest update is at most $10^{-10}$. This numerical procedure does not establish uniqueness or convergence of the learning dynamics. Its stationary round shares account for episode termination and restart in open play. At $\tau=0.25$, benchmark agreement is 75.6\%, 76.0\% and 76.2\% at the respective continuation values. We audit each learner's own modal rule using the original payoffs and a one-shot gain tolerance of $10^{-9}$. Proximity to a noisy equilibrium and credibility of its deterministic modal projection are different criteria.

\begin{table}[H]
\centering
\caption{Sensitivity to initial values and exploration after two million rounds. Each row contains 1,000 pairs. Agreement and apparent consensus are percentages of the final 4,000 rounds, full passes count reconstructed rule pairs, and low visits give the percentage of legal state--action entries tried fewer than twenty times over the complete run. Scaled-temperature conditions start from zero values.}\label{tab:si-exploration}
\begin{tabular}{@{}llrrrr@{}}
\toprule
$\chi$ & Initial values or change & Agreement & Apparent & Full passes & Low visits \\
\midrule

$5/6$ & Zero & 75.9 & 36.5 & 35.3 & 0.1 \\
$5/6$ & $0.5/(1-\beta)$ & 76.0 & 36.5 & 37.0 & 0.0 \\
$5/6$ & $1.25/(1-\beta)$ & 76.0 & 36.5 & 37.1 & 0.0 \\
$5/6$ & Equilibrium & 76.0 & 36.5 & 34.2 & 0.0 \\
$11/12$ & Zero & 93.4 & 82.1 & 11.6 & 60.8 \\
$11/12$ & $0.5/(1-\beta)$ & 76.4 & 36.4 & 32.7 & 0.0 \\
$11/12$ & $1.25/(1-\beta)$ & 76.4 & 36.4 & 32.1 & 0.0 \\
$11/12$ & Equilibrium & 76.4 & 36.4 & 30.9 & 0.0 \\
$11/12$ & Scaled temperature & 72.0 & 39.8 & 0.0 & 0.0 \\
$23/24$ & Zero & 95.1 & 83.5 & 0.3 & 70.9 \\
$23/24$ & $0.5/(1-\beta)$ & 76.6 & 36.4 & 38.0 & 0.0 \\
$23/24$ & $1.25/(1-\beta)$ & 76.6 & 36.4 & 39.3 & 0.0 \\
$23/24$ & Equilibrium & 76.6 & 36.4 & 39.4 & 0.0 \\
$23/24$ & Scaled temperature & 71.7 & 41.1 & 0.0 & 0.0 \\
\bottomrule
\end{tabular}
\end{table}

At $\chi=23/24$, positive starts at the reference temperature reduce agreement from 95.1\% to 76.6\% and apparent consensus from 83.5\% to 36.4\%, while full-audit passes rise from 0.3\% to 38.0--39.3\% (Table~\ref{tab:si-exploration}). Under the positive starts, 86.4\% and 86.3\% of pairs pass the path check, respectively, compared with 88.9\% under zero starts. The share of legal action entries tried fewer than twenty times falls from 70.9\% to zero. Starting at the logit equilibrium gives similar outcomes. At $\chi=11/12$, the same changes also reduce agreement and improve full-audit passes, whereas the outcome shares at $\chi=5/6$ change little. Initial values therefore materially affect the continuation comparison after two million rounds. Many reconstructed pairs still fail the audit under positive starts, so increased sampling does not by itself establish complete-rule credibility.

Scaling temperature changes the noisy-policy benchmark as well as exploration during learning. At $\chi=23/24$, learned agreement is 71.7\%, compared with 71.7\% at that temperature's stationary benchmark, but none of the reconstructed pairs passes the full audit. This condition cannot be interpreted as an isolated test of sampling at a fixed behavioural temperature.

The extended zero-start runs retain seeds 0--99 at $\chi=23/24$, with each horizon receiving its own final terminal update. At two million rounds these pairs have 95.1\% agreement, 84.3\% apparent consensus and no full-audit passes. At twenty million rounds the corresponding shares are 94.1\%, 83.9\% and 22\%, and at two hundred million rounds they are 92.3\%, 79.5\% and 26\%. The percentage of legal action entries tried fewer than twenty times falls from 71.0\% to 62.0\% and then 56.7\%. These matched observations show partial change with longer training, without establishing eventual convergence or a separate limiting outcome at high continuation.

The zero-start conditions reproduce all 3,000 cached reference endpoints exactly. Replaying seeds 0 and 999 in every two-million-round condition reproduces the stored action values, visits and outcome shares, and a separate Bellman calculation reproduces the full-audit classification for all 14,200 endpoints. The experiment is implemented in \nolinkurl{scripts/exploration_check.py}, with pair-level results, summaries and retained endpoints in \nolinkurl{data/exploration_check_2026-11-11/}.

\subsection{Continuous learning between repeated disputes}
\label{sec:si-continuous-disputes}

Main Figure 4d uses punishment starts 2,000 rounds apart, retaining continuous learning and natural episode endings as specified in Methods. The four marked settings each contain 1,000 trained pairs and five continuation replicates per arm. Additional checks use the same reference setting with optimistic initial values, lower discount factors of 0.1, 0.5 and 0.7, and slow punishment recovery of $\rho=0.05$ at $\chi=20/24$. The optimistic-start and slow-recovery checks each contain 1,000 pairs, and each lower-discount check contains 500. All conditions reuse existing two-million-round endpoints. These exploratory settings were selected after earlier results were inspected. Lower discount and changed initial values affect training as well as the subsequent intervention.

At the reference setting, 100 disputes reduce the punishment-veto prescription share from 38.84\% in matched controls to 18.58\%, a difference of $-20.26$ percentage points (pointwise 95\% interval $-21.99$ to $-18.51$). Fixed-demanding-proposal veto probability falls from 38.05\% to 18.81\%. Against that same proposal, stand-aside probability rises from 42.39\% without disputes to 63.87\% after 100 disputes, while support falls from 19.90\% to 17.32\%. Late veto events instead rise from 4.962\% to 5.012\% of rounds, a difference of 0.050 points (0.015 to 0.089). Apparent consensus is 83.58\% in both arms. All other tested settings have 100-dispute primary intervals including zero. These unresolved differences do not establish equivalence. The earlier slow-recovery increase under segmented exposure therefore does not establish an increase under this intervention.

We replay all five replicates of the 1,000 reference pairs in both arms at doses 0, 20 and 100 and solve the expected Bellman equations under each endpoint's frozen stochastic policies. The calculation uses the original episodic game after external impositions cease. These values do not assume equilibrium and do not evaluate an environment with continuing scheduled disputes. A separate node-value calculation agrees for every evaluated responder. We compare veto with standing aside at the same demanding punishment proposal. The immediate veto advantage is 0.15. After 100 disputes, the mean exact continuation contribution is $-0.0586$, leaving a mean exact veto advantage of 0.0914. The corresponding learned advantage is $-1.9687$. Mean absolute error in this value difference is 2.4785, against 0.4302 in matched controls. Exact values favour veto over standing aside for 75.62\% of Party--replicate records, against 24.30\% in controls. This comparison concerns those response alternatives rather than a full best response over every action.

\begin{figure}[htbp]
\centering
\includegraphics[width=1\linewidth,height=0.75\textheight,keepaspectratio]{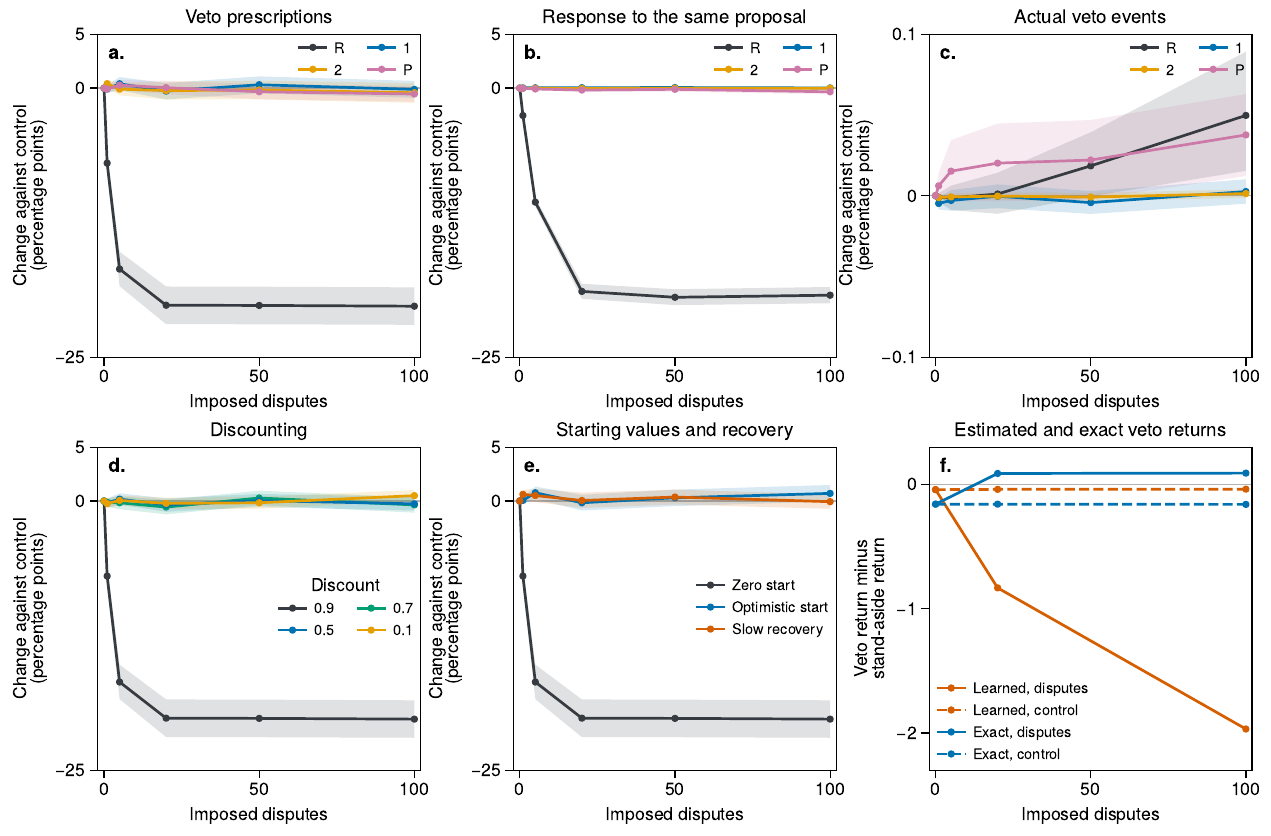}
\caption{Punishment responses change without accurate learning of veto returns. All differences subtract matched controls given the same learning time. (a) Change in own-proposal punishment-veto prescriptions at the marked Figure 4 settings. (b) Change in veto choice probability against the same demanding punishment proposal. (c) Change in actual veto events over the final 1,000 rounds. (d) Reference-setting prescription changes at different discount factors. (e) Reference-setting changes with zero or optimistic initial values, alongside slow recovery at $\chi=20/24$. (f) Learned and exact veto-minus-stand-aside values at the demanding punishment proposal, averaged across Parties and replicates. Solid lines show dispute arms and dashed lines show controls. Exact values use the final frozen stochastic policies after external impositions cease. Shading in (a)--(e) gives pointwise 95\% whole-pair bootstrap intervals. The vertical scale in (c) is expanded to show small changes in realised veto frequency.}
\label{fig:si-continuous-disputes}
\end{figure}

With initial values $1.25/(1-\beta)$, a single imposed dispute changes the own-proposal punishment-veto prescription share by 0.06 percentage points (pointwise 95\% interval $-0.24$ to $0.36$), from 22.15\% in controls to 22.21\% in the dispute arm. This sensitivity uses the continuous-learning protocol, with five replicates of each of the 1,000 pairs. For the zero-start reference pairs after 100 disputes, the deterministic modal-pair audit passes in 1.22\% of dispute replicates against 0.40\% in controls. This audit and the stochastic-policy value calculation concern different frozen policies. The decline in veto prescriptions therefore does not establish accurate learning of continuation costs or convergence to credible cooperation. Policies and partners change during learning, and the diagnostics do not isolate the contributions of sampling, estimation noise and changing policies to the value errors. Figure~\ref{fig:si-continuous-disputes} separates the changes in prescriptions, fixed-proposal responses and realised vetoes.

The scheduled implementation matches 24 Python simulations across natural continuation probabilities of zero, $23/24$ and one, with six exact empty-schedule replays of the original compiled kernel. Visit counts match the intervention duration and cached training arrays remain unchanged. The design, pair-level outcomes, summaries and verification records are in \nolinkurl{data/continuous_disputes_2026-11-11/}. The experiment uses \nolinkurl{scripts/check_continuous_disputes.py}, and the value diagnostic uses \nolinkurl{scripts/audit_continuous_dispute_values.py}.

\section{Scope of the analytical and learning results}

The response-only and strategic-proposer models make different parts of the decision process explicit. The tables summarise their results under the respective assumptions. The response-only derivations follow below.

\textbf{Response incentives and proposal choice}

\begin{longtable}[]{@{}
  >{\raggedright\arraybackslash}p{(\linewidth - 4\tabcolsep) * \real{0.2798}}
  >{\raggedright\arraybackslash}p{(\linewidth - 4\tabcolsep) * \real{0.2119}}
  >{\raggedright\arraybackslash}p{(\linewidth - 4\tabcolsep) * \real{0.5086}}@{}}
\caption{Scope of the analytical results across the model
variants.}\label{tab:si-analytical-scope}\tabularnewline
\toprule\noalign{}
\begin{minipage}[b]{\linewidth}\raggedright
\textbf{Object}
\end{minipage} & \begin{minipage}[b]{\linewidth}\raggedright
\textbf{Result type}
\end{minipage} & \begin{minipage}[b]{\linewidth}\raggedright
\textbf{Established result}
\end{minipage} \\
\midrule\noalign{}
\endfirsthead
\toprule\noalign{}
\begin{minipage}[b]{\linewidth}\raggedright
\textbf{Object}
\end{minipage} & \begin{minipage}[b]{\linewidth}\raggedright
\textbf{Result type}
\end{minipage} & \begin{minipage}[b]{\linewidth}\raggedright
\textbf{Established result}
\end{minipage} \\
\midrule\noalign{}
\endhead
\bottomrule\noalign{}
\endlastfoot
Fixed one-shot proposal & Analytical result & Support, current-payoff
acquiescence, and continuation-supported restraint are distinct actions
or mechanisms under the stated payoff. \\
Unchanged repetition & Analytical result & Repeating the same
unfavorable proposal leaves the veto floor intact. Patience alone adds
no restraint below that floor. \\
Alternating two-party agenda & Analytical result & With
$s = H - \Delta - c_{V}$, a credible one-veto punishment supports the
cooperative path when $\beta \geq \frac{s}{b} - 1$, under $s \geq b$
and $H - \Delta \geq 0$. \\
Balanced rotating $N$-party agenda & Analytical result & The threshold
is $s \leq b\sum_{k = 0}^{N - 1}\beta^{k}$. The theorem covers the
balanced rotation only. \\
Procedural adaptive menu & Analytical result & Prescribed vetoes remove
expensive proposals and raise later passage. Nature selects from the
menu, and the model omits a proposer stage. \\
Alternating discrete proposer roles & Weak SPE & Changing proposal
direction sustains $(H,Q)$ when $\delta \geq \alpha\frac{s}{c} - 1$,
and the punishment proposal rests on a payoff tie. \\
Continuous accommodation & Analytical result & The weakest responder
threshold determines the strongest feasible bite under complete
information. \\
Private-type bite feedback & Analytical, stated selection & Bite lowers
restraint on regular stable branches and can remove a high-restraint
root at a nondegenerate fold. \\
\end{longtable}

\textbf{Repeated strategic-proposal and selection results}

\begin{longtable}[]{@{}
  >{\raggedright\arraybackslash}p{(\linewidth - 4\tabcolsep) * \real{0.2798}}
  >{\raggedright\arraybackslash}p{(\linewidth - 4\tabcolsep) * \real{0.2119}}
  >{\raggedright\arraybackslash}p{(\linewidth - 4\tabcolsep) * \real{0.5086}}@{}}
\caption{Scope of the learning and documentary sensitivity
results.}\label{tab:si-numerical-scope}\tabularnewline
\toprule\noalign{}
\begin{minipage}[b]{\linewidth}\raggedright
\textbf{Object}
\end{minipage} & \begin{minipage}[b]{\linewidth}\raggedright
\textbf{Result type}
\end{minipage} & \begin{minipage}[b]{\linewidth}\raggedright
\textbf{Established result}
\end{minipage} \\
\midrule\noalign{}
\endfirsthead
\toprule\noalign{}
\begin{minipage}[b]{\linewidth}\raggedright
\textbf{Object}
\end{minipage} & \begin{minipage}[b]{\linewidth}\raggedright
\textbf{Result type}
\end{minipage} & \begin{minipage}[b]{\linewidth}\raggedright
\textbf{Established result}
\end{minipage} \\
\midrule\noalign{}
\endhead
\bottomrule\noalign{}
\endlastfoot
$N$-responder strategic proposer & Analytical result & Six families of
responder and proposer tests are necessary and sufficient for the
specified pure public trigger automaton. \\
Two regime-specific thresholds & Analytical result & The minimum
responder threshold in each regime caps cooperative bite or the best
punishment-stage redesign that can pass. Different responders may
bind. \\
Four proposal institutions & Analytical result & Free redesign defeats
the dead-letter trigger. A discrete menu, a bite floor, or a primitive
proposal-deviation penalty can restore it under the stated bounds. \\
Three-mode forgiving trigger & Analytical, specified policy & Recovery
creates an intermediate continuation window in which continuation first sustains
restraint and later undermines the veto threat sustaining restraint. \\
Two-bite stationary policy class & Exhaustively enumerated & Every one
of 972 symmetric pure stationary policies receives a complete one-shot
deviation audit. \\
Payoff-based adaptive selection & Simulation result & Under the stated
SARSA process, apparent consensus and policy bite rise together, and a
modal rule can pass every on-path comparison and fail the full-node
audit. \\
Richer repeated proposer games & Open & Private information, mixed or
asymmetric strategies, arbitrary history dependence,
response-equilibrium selection, imperfect public monitoring, and
endogenous agenda transitions require separate analyses. \\
\end{longtable}

See main Methods (``Model, response definitions and proposal choice'') for the model and procedures.

\section{What the public record identifies}

A no-veto record identifies the response condition

\begin{equation}
Y = \mathbb{1}\left\{ \sigma_{i} \neq V\text{  for every  }i \right\} = 1.
\end{equation}The record leaves each $A$ versus $Q$ response unidentified. Once
proposal choice is endogenous, it also leaves the proposer's unconstrained choice unidentified. The record can conceal two objects:

\begin{itemize}
\item The reason for choosing $Q$, which can be opposition to the actual proposal or avoidance of active-support costs despite positive alignment. \item Opposition to a stronger counterfactual proposal that accommodation removes before the vote.
\end{itemize}

Observed instrument text can reveal actual bite only if researchers measure its provisions. The counterfactual proposal remains unidentified without drafts, amendments, withdrawals, or bargaining records. Formal passage also leaves ratification and implementation unresolved when later action determines effective bite.

These distinctions change the interpretation of apparent consensus. Accommodated agreement with unanimous $A$ is active support for the actual proposal, despite hidden opposition to a stronger alternative. Apparent consensus requires at least one $Q$ toward the actual proposal. It does not imply that any party opposes it or needs continuation to refrain. Accommodated apparent consensus combines a weaker proposal with at least one acquiescent response.

Adaptive selection changes how often these observationally equivalent records arise while leaving their identification content unchanged. Even when learning produces an almost universal no-veto convention, the record alone leaves active support, restraint, and the harder and weaker proposals considered along the path unidentified.

\subsection{Documented responses bound apparent consensus}

\protect\phantomsection\label{sec:si-observation-bound}{}

Let an audited sample contain $M > 0$ passed decisions. Of these, let $D$ have documentation establishing that at least one party stood aside from the adopted proposal, and let $E$ have documentation establishing active support by every party. These categories must be disjoint, with $D + E \leq M$. A reservation about one clause or an earlier draft is not sufficient unless it establishes non-endorsement of the adopted proposal. Let $\theta_{\text{app}}$ be the share of passed decisions in this sample with at least one $Q$.

\begin{proposition}[Bounds from documented responses]

Without restrictions on the responses in the remaining $M - D - E$ decisions,
\begin{equation}
\frac{D}{M} \leq \theta_{\text{app }} \leq 1 - \frac{E}{M}.
\end{equation}Both
endpoints are attainable. With passage records alone, $D = E = 0$, so the identified interval is $\lbrack 0,1\rbrack$.

\end{proposition}

\begin{proof}

Every one of the $D$ documented non-endorsement decisions contributes to apparent consensus, and none of the $E$ unanimously supported decisions does. Assigning unanimous $A$ to every undocumented decision gives the lower endpoint. Assigning at least one $Q$ to each gives the upper endpoint. Both assignments preserve every observed passage record. With integer counts, the attainable shares between the endpoints occur in steps of $\frac{1}{M}$.

\end{proof}

This is a bound on the composition of the audited sample. It does not give a sampling confidence interval or an estimate of institution-wide prevalence. Selected proposal histories do not supply a representative denominator $M$. Accommodation followed by unanimous support belongs to the $E$ category if that support is documented. Weakening a proposal alone does not establish apparent consensus. \nolinkurl{observability_bound.py} implements the endpoints, with checks in \nolinkurl{test_observability_bound.py}.

\section{Additional response and agenda results}

The following sections give the response-only equilibria, procedural agenda variants and private-type selection model, followed by the Antarctic analyses and numerical checks.

\section{One proposal permits opposite outcomes}

We now derive the one-shot equilibria of the fixed-proposal game. For a public proposal, party $i$ chooses $\sigma_{i} \in \left\{ A,Q,V \right\} = \left\{ 1,0, - 1 \right\}$. The proposal passes when no party chooses $V$. Its stage payoff is

\begin{equation}
U_{i}\left( \sigma_{i},\sigma_{- i} \right) = h_{i}\sigma_{i} - \Delta_{i}\sigma_{i}^{2} - c_{V,i}\mathbb{1}_{\sigma_{i} = V} + b_{i}\mathbb{1}_{\text{pass}}.
\end{equation}Here $h_{i}$ measures signed alignment, $\Delta_{i} > 0$ is the cost
of taking a visible stance, $c_{V,i} \geq 0$ is the additional cost of veto, and $b_{i} > 0$ is the value of passage. Conditional on every other party refraining, the three payoffs are

\begin{equation}
h_{i} - \Delta_{i} + b_{i},\quad b_{i},\quad - h_{i} - \Delta_{i} - c_{V,i}.
\end{equation}After another party vetoes, the passage term disappears. The three
payoffs then become $h_{i} - \Delta_{i}$, $0$, and $- h_{i} - \Delta_{i} - c_{V,i}$.

\begin{proposition}[One proposal supports passage or breakdown]

For two symmetric parties with $h_{i} = h$, $\Delta_{i} = \Delta$, $b_{i} = b$, and $c_{V,i} = c_{V}$, the strict pure-strategy one-shot equilibria are

\begin{longtable}[]{@{}
  >{\raggedright\arraybackslash}p{(\linewidth - 4\tabcolsep) * \real{0.2898}}
  >{\centering\arraybackslash}p{(\linewidth - 4\tabcolsep) * \real{0.2150}}
  >{\raggedright\arraybackslash}p{(\linewidth - 4\tabcolsep) * \real{0.4954}}@{}}
\caption{Pure response equilibria for two identical parties in the
one-shot game.}\label{tab:si-response-equilibria}\tabularnewline
\toprule\noalign{}
\endfirsthead
\endhead
\bottomrule\noalign{}
\endlastfoot
$h > \Delta$ & $(A,A)$ & unanimous active support \\
$- \left( \Delta + c_{V} \right) < h < \Delta$ & $(Q,Q)$ &
current-payoff acquiescence \\
$- \left( b + \Delta + c_{V} \right) < h < - \left( \Delta + c_{V} \right)$
& $(Q,Q)$ and $(V,V)$ & passage and breakdown coexist \\
$h < - \left( b + \Delta + c_{V} \right)$ & $(V,V)$ & veto is
too attractive for passage \\
\end{longtable}

In the coexistence region, the same proposal and preferences support either public outcome. For two asymmetric objectionable parties with veto payoffs $0 < s_{i},s_{j} < b$, passage risk dominates breakdown exactly when $b > s_{i} + s_{j}$. The reduced game with actions $\left\{ Q,V \right\}$ also has a mixed equilibrium with restraint probabilities $q_{i} = \frac{s_{j}}{b}$ and $q_{j} = \frac{s_{i}}{b}$.

\end{proposition}

\begin{proof}

Support beats acquiescence exactly when $h_{i} > \Delta_{i}$. If every other party refrains, acquiescence beats veto exactly when $h_{i} > - \left( b_{i} + \Delta_{i} + c_{V,i} \right)$. After another veto, acquiescence beats veto exactly when $h_{i} > - \left( \Delta_{i} + c_{V,i} \right)$. Combining the inequalities gives the four regions. In the reduced coexistence game with actions $\left\{ Q,V \right\}$, choosing $V$ against $Q$ loses $b - s_{i}$, while choosing $Q$ against $V$ loses $s_{i}$. The risk-dominance comparison
\begin{equation}
\left( b - s_{i} \right)\left( b - s_{j} \right) > s_{i}s_{j}
\end{equation}reduces to $b > s_{i} + s_{j}$. Equality at an action-indifference
boundary creates weak best responses. Equality $b = s_{i} + s_{j}$ instead means neither pure equilibrium risk dominates. Both pure equilibria remain strict when $0 < s_{i},s_{j} < b$. Equating party $i$`s mixed-response payoffs, $bq_{j} = s_{i}$, and the same for $j$, gives the displayed mixed equilibrium.

\end{proof}

The coexistence result is a possibility statement. It does not select one outcome, infer private preferences from a public record, or claim that every consensus game has multiple equilibria.

\section{Membership raises the restraint threshold}

Let $N \geq 2$ denote the total number of responding parties. Let the symmetric private veto payoff be $s \in (0,b)$, and suppose each party believes independently that any one counterpart refrains with probability $q$. All other $N - 1$ parties refrain with probability $q^{N - 1}$.

\begin{corollary}[Unanimity creates an exponential burden]

Acquiescence is a best response exactly when \begin{equation}
bq^{N - 1} \geq s.
\end{equation}The
belief threshold is
\begin{equation}
q_{N}^{\ast} = \left( \frac{s}{b} \right)^{\frac{1}{N - 1}},
\end{equation}which
approaches one as $N$ grows. Holding $q_{N}^{\ast} = \bar{q}$ fixed
requires \begin{equation}
b_{N} = s\bar{q}^{- (N - 1)}.
\end{equation}At $\bar{q} = \frac{1}{2}$, this
becomes $b_{N} = s2^{N - 1}$.

\end{corollary}

\begin{proof}

The expected passage benefit from acquiescence is $bq^{N - 1}$, while veto secures $s$. Solving their equality gives the threshold. The limit and the expression for $b_{N}$ follow directly.

\end{proof}

At $N = 29$, the symmetric case requires $\frac{b}{s} = 2^{28} \approx 2.7 \times 10^{8}$ to place the boundary at $q = \frac{1}{2}$. This calculation isolates conditional independence and a common belief. Communication, correlated expectations, proposal choice, heterogeneous veto payoffs, and membership-dependent continuation values can change the comparison.

\begin{figure}[htbp]
\centering
\includegraphics[width=\linewidth]{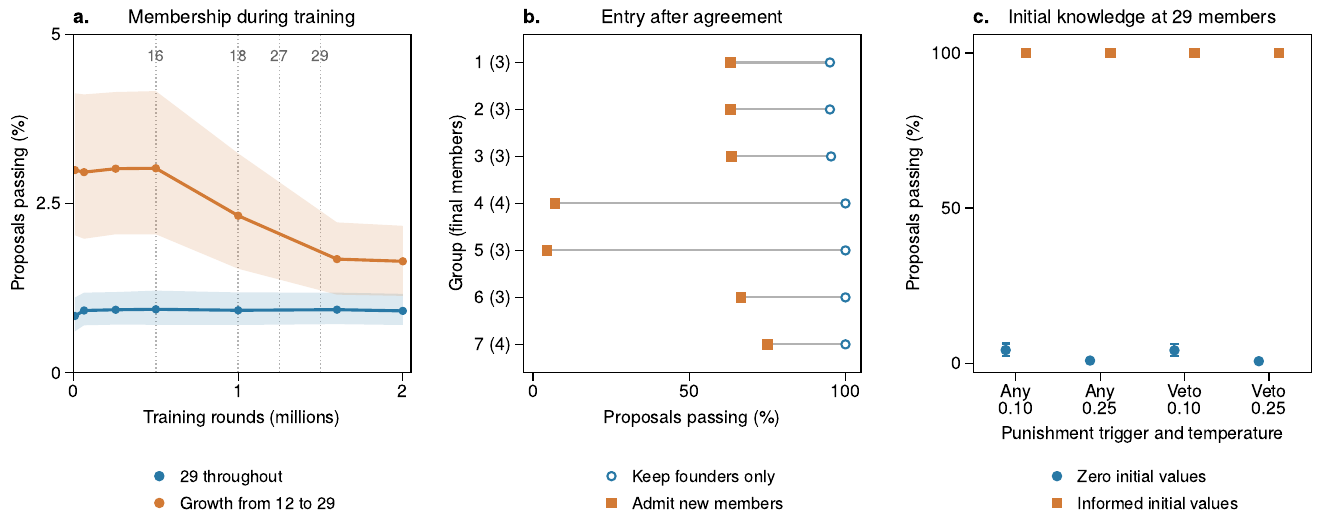}
\caption{Membership growth and prior knowledge in learned agreement. In this exploratory game, the disadvantaged role rotates among members. Payoffs are $h_i s_i-0.2s_i^2-0.1\mathbf{1}\{a_i=V\}+\mathbf{1}\{\text{no veto}\}$, with stance $s_i\in\{1,0,-1\}$ and alignment $h_i=-1.8$ for the disadvantaged member and $+1.8$ otherwise. Independent SARSA uses discount and relationship continuation 0.95, softmax choice, and learning rate $0.05/\sqrt{1+v/4000}$ after $v$ visits. Relationship resets restore ordinary public memory. (a) Mean proposal passage in 50 paired seed blocks for fixed membership at 29 and growth from 12 to 29 members. Dotted lines mark admissions, with the resulting membership above each line. Shading gives pointwise 95\% bootstrap intervals from resampling whole groups. Passage is measured over the final 4,000 rounds at each checkpoint, and connecting lines join the saved checkpoints. (b) Agreement qualifies after passage of at least 90\% in each of three consecutive 10,000-round windows. Passage after admission compared with continued learning by the same founders for each qualifying case. Seven of 72 two-member founder cases qualified across five seed blocks, and none of 72 twelve-member cases qualified. Passage averages the final three 10,000-round windows of the last attempted growth stage, with the founder control using the same future round draws and duration. Parentheses give final membership. These cases are selected by qualification and share seed blocks. (c) Passage after two million rounds in groups of 29, comparing zero initial action values with exact action values of an audited deterministic candidate strategy. Each condition has 12 paired seed blocks, and error bars give whole-group 95\% bootstrap intervals. Any denotes punishment following any departure from prescribed behaviour, while Veto denotes punishment following a veto. The numbers give choice temperature. All informed starts passed every proposal in the final 4,000-round observation window, which does not establish that the learned strategy is an equilibrium.}
\label{fig:si-membership-conventions}
\end{figure}

Exploratory learning in a separate rotating-agenda game gives limited support for protection through gradual entry (Figure~\ref{fig:si-membership-conventions}). Starting with 12 members and growing to 29 gives higher mean passage than starting with 29, but passage remains low and declines along the growth trajectory. Among the selected small groups that first established agreement, admitting new members reduces passage relative to continuing with the founders alone. Initial knowledge has a larger effect under the tested conditions, with groups initialised from an audited candidate strategy passing every proposal in the final observation window. These results distinguish preserving informed behaviour from learning agreement through membership growth, and the observed passage shares do not establish equilibrium credibility or historical effects of enlargement.

\section{An unchanged issue retains veto}

Suppose the same proposal and stage payoffs recur indefinitely. Acquiescence forever gives $\frac{b}{1 - \delta}$, while veto forever gives $\frac{s_{i}}{1 - \delta}$. A party can secure the veto path after every history under the maintained payoff.

\begin{proposition}[Identical repetition adds no restraint below the veto floor]

If $s_{i} > b$, no repeated-game strategy can sustain perpetual acquiescence by party $i$ on the unchanged proposal while respecting individual rationality. The comparison remains $b \geq s_{i}$ for every $\delta \in \lbrack 0,1)$.

\end{proposition}

\begin{proof}

Veto gives $s_{i}$ against every action profile of the other parties in every round. The discounted minmax floor is $\frac{s_{i}}{1 - \delta}$. Perpetual acquiescence gives $\frac{b}{1 - \delta}$. Discounting applies the same multiplier to both paths.

\end{proof}

This obstruction uses the payoff guarantee supplied by veto. It does not state that every known finite or infinite repeated game collapses to its one-shot outcome.

Changing proposal direction can create the missing continuation loss. In the full three-action alternating game, two parties take turns facing favourable and unfavourable proposals with alignment $+ H$ and $- H$, respectively, where $H > 0$. This alignment magnitude is distinct from the hortatory instrument $H$ in the discrete-instrument game. The cooperative profiles are $(Q,A)$ and $(A,Q)$. A deviation activates one consequential veto, $(V,A)$ or $(A,V)$. Define

\begin{equation}
s = H - \Delta - c_{V},\quad a = H - \Delta.
\end{equation}The punishment profile is a stage Nash equilibrium when $s \geq b$ and
$a \geq 0$. Let $\beta \in \lbrack 0,1)$ combine discounting and any exogenous probability that the relationship continues.

\begin{proposition}[A changing agenda can support restraint]

Under $s \geq b$ and $a \geq 0$, the alternating cooperative
strategy is subgame perfect exactly when \begin{equation}
b(1 + \beta) \geq s.
\end{equation}Current payoffs support restraint at $s \leq b$. Continuation changes
veto into restraint on $b < s < 2b$ for sufficiently large $\beta$. No $\beta < 1$ supports this candidate when $s \geq 2b$. The stationary one-veto breakdown strategy remains subgame perfect wherever the punishment profile is credible.

\end{proposition}

\begin{proof}

For a focal party, cooperative values at its unfavourable and favourable states are
\begin{equation}
C_{\text{bad}} = \frac{b + \beta(a + b)}{1 - \beta^{2}},\quad C_{\text{good}} = \frac{a + b + \beta b}{1 - \beta^{2}}.
\end{equation}Punishment values are
\begin{equation}
P_{\text{bad}} = \frac{s + \beta a}{1 - \beta^{2}},\quad P_{\text{good}} = \frac{a + \beta s}{1 - \beta^{2}}.
\end{equation}A cooperative-path veto gives $s + \beta P_{\text{good}}$. The
comparison $C_{\text{bad}} \geq s + \beta P_{\text{good}}$ reduces to $b(1 + \beta) \geq s$. The conditions $s \geq b$ and $a \geq 0$ make the punishment actions optimal. The remaining deviations do not tighten this condition.

\end{proof}

\begin{corollary}[Balanced rotation extends the continuation horizon]

In a symmetric $N$-party agenda that disadvantages one party per proposal and each party once per cycle, the same construction is subgame
perfect exactly when \begin{equation}
s \leq b\sum_{k = 0}^{N - 1}\beta^{k},
\end{equation}subject
to the same punishment checks. With $s > b$, some $\beta < 1$ supports restraint exactly when $s < Nb$.

\end{corollary}

\begin{proof}

Relative to punishment, cooperation adds $b - s$ on a party's unfavourable proposal and $b$ on each of its next $N - 1$ favourable proposals. Recursion around the cycle gives
\begin{equation}
D_{\text{bad}} = \frac{b\sum_{k = 0}^{N - 1}\beta^{k} - s}{1 - \beta^{N}}.
\end{equation}The deviation constraint is $D_{\text{bad}} \geq 0$.

\end{proof}

The rotating result uses a balanced agenda. Adding parties in this construction adds favourable future proposals, while adding parties to the fixed-proposal game adds possible vetoers. These comparative statics answer different questions.

\begin{figure}[htbp]
\centering
\includegraphics[width=1\linewidth,height=0.65\textheight,keepaspectratio]{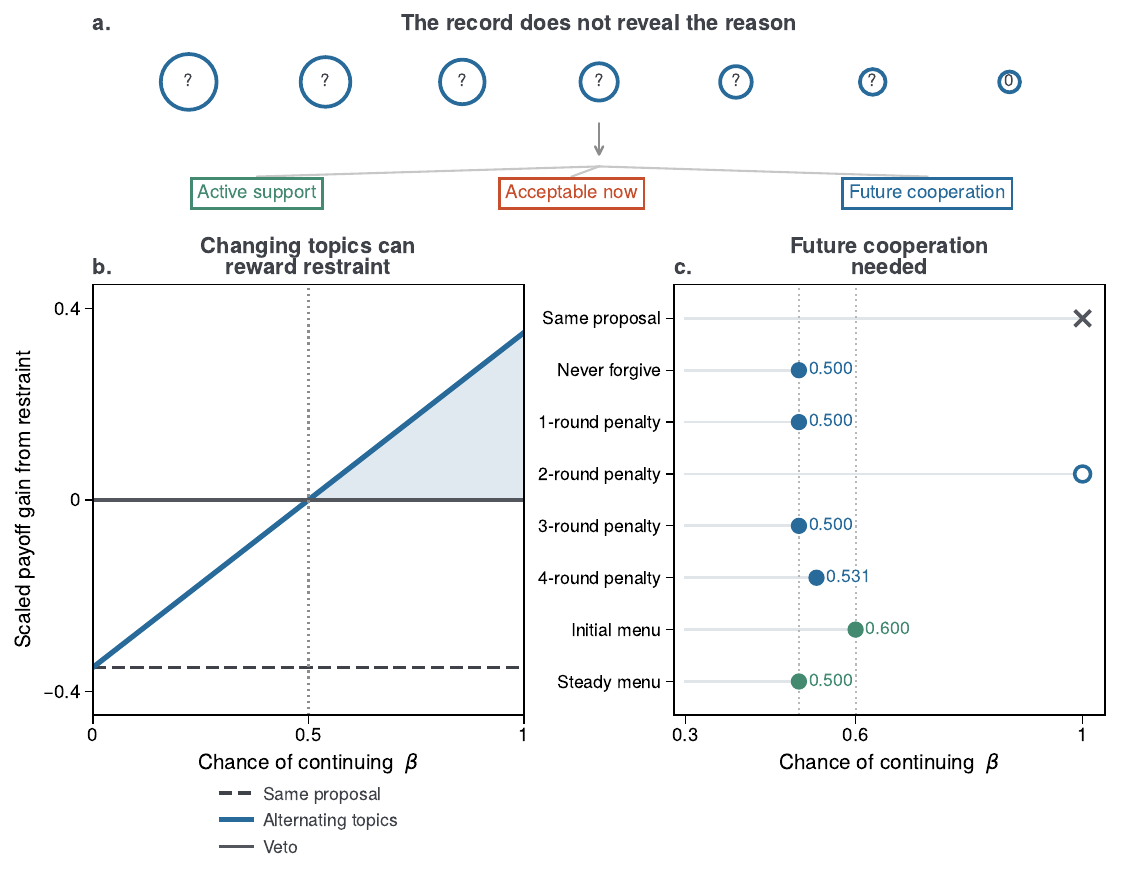}
\caption{Mechanism audit for repeated interaction. (a) Active support, current-payoff acquiescence, and continuation-supported restraint share a no-veto record, and circle area denotes unsigned attention. (b) A separate binary accept/veto game has acceptance cost $c = 0.35$ on an adverse issue, benefit $v = 0.70$ on a favourable issue, and zero payoff after veto. The plotted incentive numerators are $- c$ for an unchanged adverse proposal and $- c + \beta v$ for alternation under grim punishment. Alternating restraint is credible at $\beta \geq \frac{c}{v} = 0.50$. These are not full discounted values. (c) Complete-audit thresholds for the indicated rules. Blue uses that binary alternating game. Green uses 250 aligned, 250 cheap, and 250 expensive proposals, with contested directions equally split and acceptance payoffs of $0.4$ on aligned proposals and $\pm 0.2$ or $\pm 0.6$ on contested proposals. Filled points pass every check. An open point denotes a limiting threshold at $\beta \rightarrow 1$, and a cross denotes infeasibility for $\beta < 1$. These are illustrative feasibility calculations and do not estimate ATS patience.}
\label{fig:archive-repeated-game-signal}
\end{figure}

A displayed sequence of meetings can represent a known finite game or a window into an indefinite relationship. Under a known terminal date, strict terminal veto incentives and no terminal cooperation value yield backward unraveling. An indefinite agenda retains a continuation payoff after the displayed window.

\begin{figure}[htbp]
\centering
\includegraphics[width=1\linewidth,height=0.65\textheight,keepaspectratio]{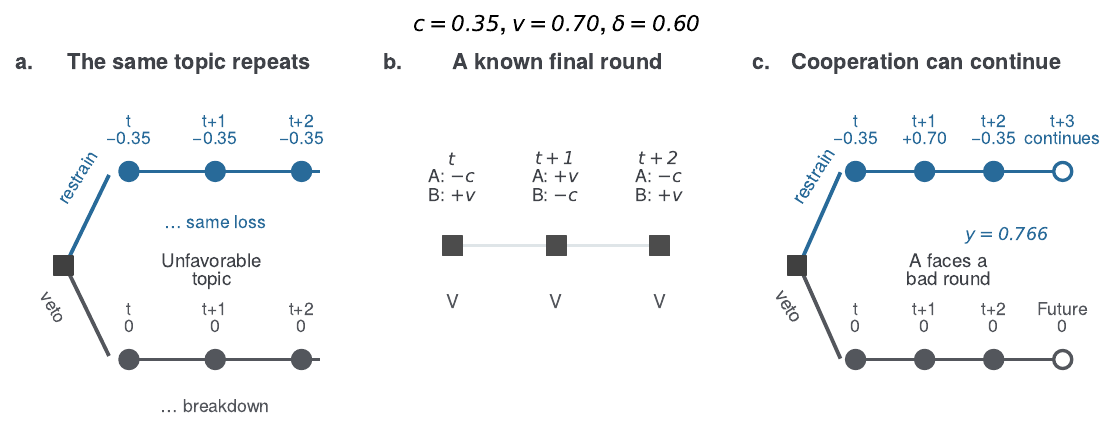}
\caption{Three repeated-decision cases with $c = 0.35$, $v = 0.70$, and $\delta = 0.60$. (a) Repeating the same unfavourable proposal gives restraint value $- \frac{c}{1 - \delta} = - 0.875$, whereas veto gives zero. (b) With a known three-round horizon, the terminal veto propagates backwards, and $V$ marks the induced action in each round. (c) In a three-round window into an indefinite alternating relationship, the continuation value is $y = \frac{v - \delta c}{1 - \delta^{2}} = 0.766$. Restraint is then worth $- c + \delta v - \delta^{2}c + \delta^{3}y = 0.109$, compared with zero from veto. The comparison depends on continuation beyond the displayed window and on the stated terminal incentives.}
\label{fig:archive-three-round-tree}
\end{figure}

\section{Prescribed failures prune the menu}

See main Methods (``Additional response, agenda and private-type variants'') for the model and procedures.

Let $P > 0$ count permanent proposals, and let $(a,b)$ count the remaining expensive proposals favouring each party. Each expensive proposal gives zero current payoff under its prescribed veto in this construction. If $R_{i}$ denotes party $i$`s total current payoff across permanent proposals, solving the Bellman equation for the current menu gives

\begin{equation}
W_{i}(a,b) = \frac{R_{i} + \beta aW_{i}(a - 1,b) + \beta bW_{i}(a,b - 1)}{P(1 - \beta) + a + b}.
\end{equation}Terms with a zero proposal count are omitted from the recursion. The
boundary value is $W_{i}(0,0) = \frac{R_{i}}{\left\lbrack P(1 - \beta) \right\rbrack}$, and the resulting values support the complete-strategy comparisons below.

\begin{proposition}[Pruning raises later passage without preference change]

Let $n_{S}$ count aligned proposals and let $n_{A,L} + n_{B,L}$ count cheap contested proposals. At menu state $(a,b)$, the next-draw shares are
\begin{equation}
p_{A} = \frac{n_{S}}{P + a + b},\quad p_{C} = \frac{n_{A,L} + n_{B,L}}{P + a + b},\quad p_{B} = \frac{a + b}{P + a + b}.
\end{equation}Every prescribed veto lowers $a + b$ by one. Conditional on the
audited strategy and continued interaction, proposal breakdown falls and passage rises as pruning proceeds. The cheap-contested share among passed proposals is constant when the two permanent passage classes keep fixed proportions.

\end{proposition}

\begin{proof}

The transition rule removes one expensive proposal after each prescribed veto and leaves every permanent proposal in the menu. The numerators of $p_{A}$ and $p_{C}$ remain fixed while their common denominator falls. The numerator of $p_{B}$ falls with the denominator. Fixed permanent-class proportions preserve $\frac{p_{C}}{p_{A} + p_{C}}$.

\end{proof}

\begin{longtable}[]{@{}
  >{\raggedright\arraybackslash}p{(\linewidth - 4\tabcolsep) * \real{0.2738}}
  >{\centering\arraybackslash}p{(\linewidth - 4\tabcolsep) * \real{0.1548}}
  >{\raggedright\arraybackslash}p{(\linewidth - 4\tabcolsep) * \real{0.5714}}@{}}
\caption{Proposal classes and payoffs in the finite-menu pruning
example.}\label{tab:si-pruning-payoffs}\tabularnewline
\toprule\noalign{}
\begin{minipage}[b]{\linewidth}\raggedright
\textbf{Proposal class}
\end{minipage} & \begin{minipage}[b]{\linewidth}\centering
\textbf{Count}
\end{minipage} & \begin{minipage}[b]{\linewidth}\raggedright
\textbf{Payoffs upon acceptance by both parties}
\end{minipage} \\
\midrule\noalign{}
\endfirsthead
\toprule\noalign{}
\begin{minipage}[b]{\linewidth}\raggedright
\textbf{Proposal class}
\end{minipage} & \begin{minipage}[b]{\linewidth}\centering
\textbf{Count}
\end{minipage} & \begin{minipage}[b]{\linewidth}\raggedright
\textbf{Payoffs upon acceptance by both parties}
\end{minipage} \\
\midrule\noalign{}
\endhead
\bottomrule\noalign{}
\endlastfoot
Aligned & 2 & $(0.4,0.4)$ for each proposal \\
Cheap contested & 2 & $(0.2, - 0.2)$ and $( - 0.2,0.2)$, one
proposal each \\
Expensive contested & 2 & $(0.6, - 0.6)$ and $( - 0.6,0.6)$, one
proposal each \\
\end{longtable}

\protect\phantomsection\label{tbl:si-pruning-spec}{}

The complete audit gives $\beta^{\ast} = 0.5805673$. At the initial menu, aligned passage, cheap-contested passage, and proposal breakdown each occur with probability one third. After both expensive proposals leave, the two passage classes each occur with probability one half. In a continuing sequence, removing both expensive proposals takes $\frac{6}{2} + \frac{5}{1} = 8$ draws in expectation.

Write $\beta = \delta\chi$, where $\chi$ is the probability of another decision after the current outcome. The product controls incentives. The raw probability $\chi$ separately controls whether the sequence lasts long enough to reach the pruned menu.

% Flush earlier floats before this figure so longtable output cannot push
% the figure and its caption below the page boundary.
\clearpage
\begin{figure}[htbp]
\centering
\includegraphics[width=1\linewidth,height=0.65\textheight,keepaspectratio]{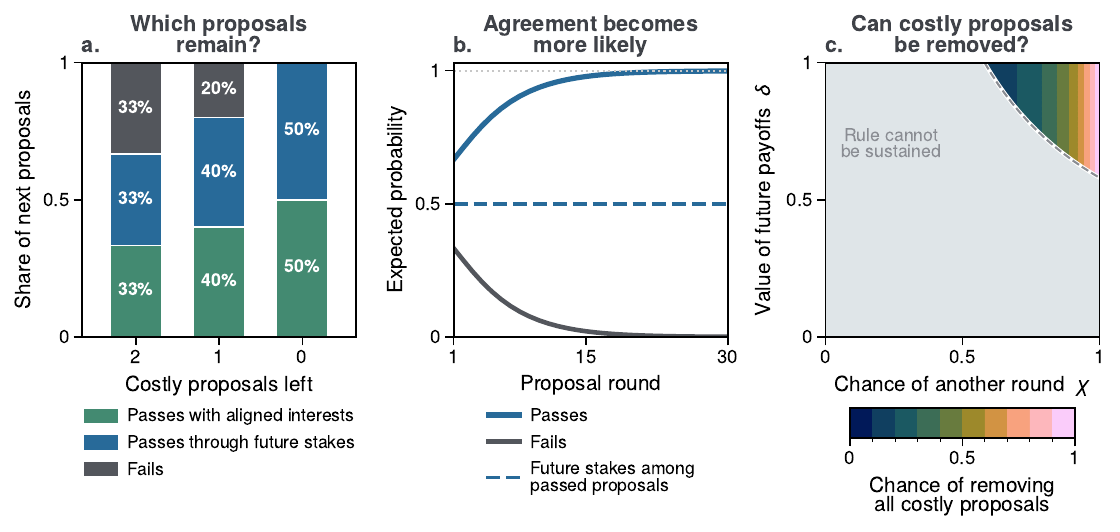}
\caption{Procedural proposal failures can raise later passage while preferences remain fixed. (a) Prescribed vetoes remove expensive proposals, shifting the next draw from equal shares of aligned passage, continuation-supported passage, and breakdown to equal shares of the two passage classes. (b) Conditional on the candidate strategy passing its complete audit, expected passage approaches one as breakdown proposals leave the menu, while continuation-supported passage remains half of all passage. (c) The strategy is credible above $\delta\chi \geq \beta^{\ast} = 0.5805673$, shown by the dashed boundary. Colour gives the probability that interaction lasts long enough to reach the pruned menu, and the grey region fails the complete audit. This game draws proposals procedurally and does not model strategic proposal choice.}
\label{fig:archive-proposal-pruning}
\end{figure}

This result establishes rising conditional passage. The adaptive-menu game contains accept and veto actions but does not identify active support versus acquiescence, so it does not by itself establish rising apparent consensus.

\section{Private types preserve aggregate multiplicity}

Set proposal bite to zero in the feedback model and let the private veto payoff satisfy $s_{i}|P \sim F_{P}$. Here $P$ indexes the proposal or topic under consideration, distinct from the punishment-state label. Retain the imposed expected continuation advantage $L(q)$. A symmetric response equilibrium conditional on this specification solves

\begin{equation}
q = T_{P}(q) = F_{P}\left( bq^{N - 1} + \delta L(q) \right).
\end{equation}For the normal location family
$s_{i}|P \sim \mathcal{N}(\mu_{P},\sigma_{P}^{2})$,

\begin{equation}
T_{P}(q) = \Phi(\frac{bq^{N - 1} + \delta L(q) - \mu_{P}}{\sigma_{P}}).
\end{equation}The derivative is

\begin{equation}
T'_{P}(q) = \varphi(\frac{bq^{N - 1} + \delta L(q) - \mu_{P}}{\sigma_{P}})\frac{\left\lbrack b(N - 1)q^{N - 2} + \delta L'(q) \right\rbrack}{\sigma_{P}}.
\end{equation}See main Methods (``Additional response, agenda and private-type
variants'') for the model and procedures.

\textbf{Scope of the fixed-point claim}

The displayed constant, proportional, convex, and threshold continuation functions produce one or three fixed points over the reported parameter ranges. This count is not a theorem for every differentiable continuation function. The equilibrium correspondence leaves selection open.

\begin{figure}[htbp]
\centering
\includegraphics[width=1\linewidth,height=0.65\textheight,keepaspectratio]{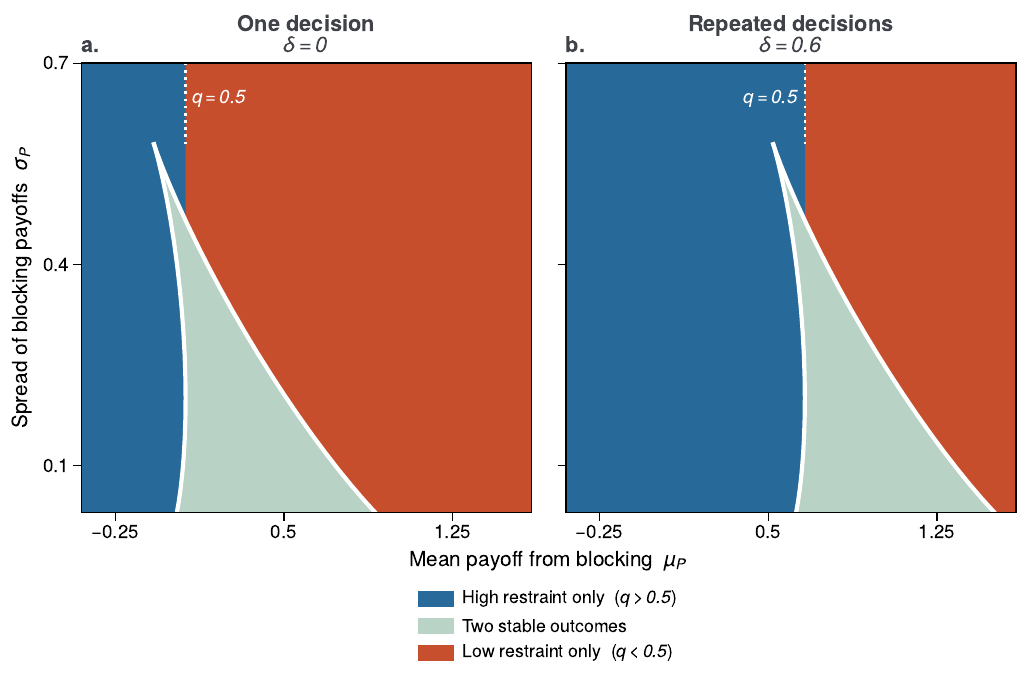}
\caption{Normal private-type phase map. Colour separates a unique low-restraint equilibrium, coexistence of low- and high-restraint equilibria, and a unique high-restraint equilibrium. The dotted line marks where the high-restraint equilibrium crosses $q = 0.5$. The comparison fixes $N = 5$, $b = 1$, and $L(q) = 1$, with $\delta = 0$ and $0.6$. The imposed continuation term shifts the boundaries through $\delta L$.}
\label{fig:archive-private-alignment}
\end{figure}

\begin{figure}[htbp]
\centering
\includegraphics[width=1\linewidth,height=0.65\textheight,keepaspectratio]{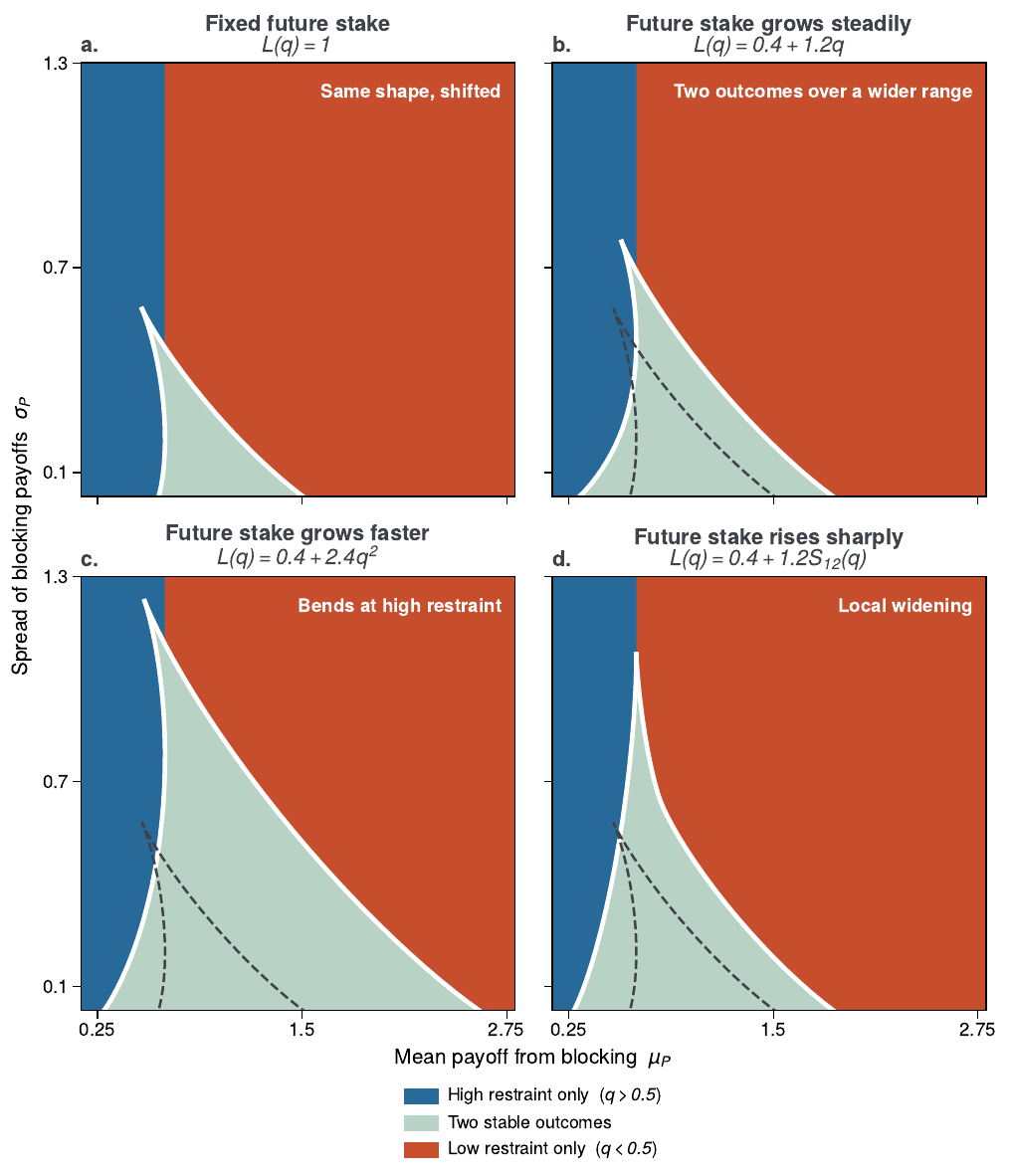}
\caption{Continuation-shape comparison at fixed $\delta = 0.6$ and $L(0.5) = 1$. A constant future value translates the response map. Proportional, convex, and threshold functions also change its slope and reshape the coexistence region.}
\label{fig:archive-continuation-shapes}
\end{figure}

\begin{figure}[htbp]
\centering
\includegraphics[width=1\linewidth,height=0.65\textheight,keepaspectratio]{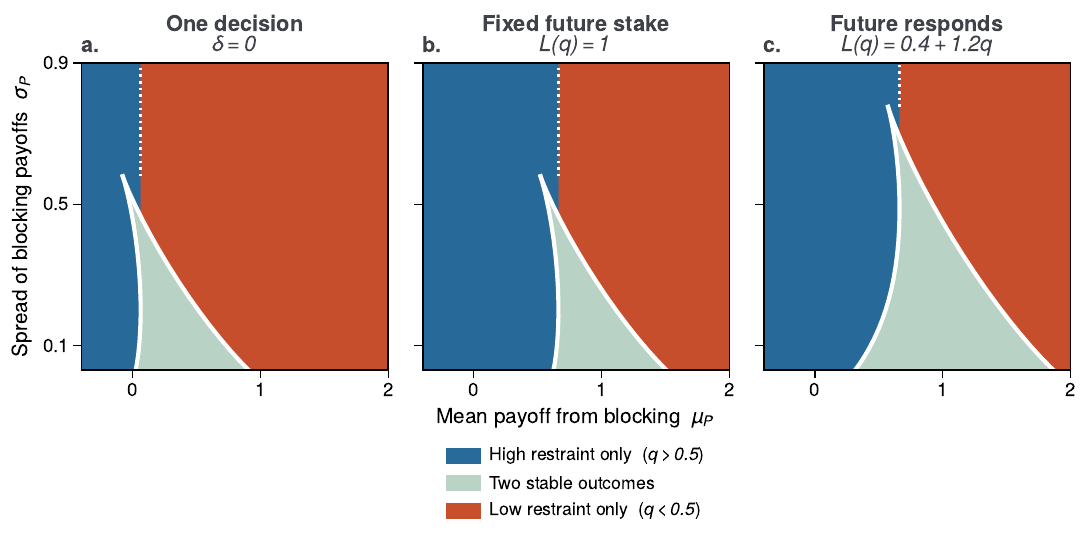}
\caption{Fold geometry under three continuation specifications. Moving from the one-shot game to a constant future value shifts the coordination region. Belief-dependent continuation also changes its shape. The comparison fixes $\delta = 0.6$ and $L(0.5) = 1$ where indicated.}
\label{fig:archive-continuation-cusp}
\end{figure}

\section{Basin weights make selection explicit}

Let the regular fixed points alternate between stable and unstable values. Each stable fixed point receives the probability mass of initial common beliefs in its basin under the stated iteration. For the common three-root case $q_{L} < q_{u} < q_{H}$, use

\begin{equation}
q_{0} \sim \text{ Beta}\left( m\kappa,(1 - m)\kappa \right).
\end{equation}Here $m \in (0,1)$ is the prior mean and $\kappa > 0$ is its
concentration. Writing $I_{z}$ for the regularised incomplete beta function, or beta cumulative distribution function at $z$, the basin weights are

\begin{equation}
w_{L} = I_{q_{u}}\left( m\kappa,(1 - m)\kappa \right),\quad w_{H} = 1 - w_{L}.
\end{equation}A unique stable fixed point receives weight one. Fold equalities remain
boundaries. The Beta law represents analyst uncertainty about an unobserved common starting belief across decision episodes. It is not a distribution of private beliefs, a Bayesian-learning model, or an equilibrium refinement.

For common costs $\Delta_{i} = \Delta$ and $c_{V,i} = c_{V}$, the net veto payoff $s_{i} = - \left( \Delta + h_{i} + c_{V} \right)$ has continuous cumulative distribution $F_{P}$. Since active support requires $h_{i} > \Delta$, equivalently $s_{i} < - 2\Delta - c_{V}$, the same type distribution gives both active-support and current-payoff restraint probabilities at a selected stable equilibrium $q_{k}$:

\begin{equation}
p_{A} = F_{P}\left( - 2\Delta - c_{V} \right),\quad r_{k} = F_{P}\left( bq_{k}^{N - 1} \right).
\end{equation}These probabilities are ordered by set inclusion. Because $b \geq 0$,
active support, $h_{i} > \Delta$, implies $s_{i} < 0 \leq bq_{k}^{N - 1}$ and so inclusion among current-payoff refrainers. Because the maintained continuation advantage satisfies $\delta L\left( q_{k} \right) \geq 0$, the current threshold does not exceed the equilibrium threshold. So $p_{A} \leq r_{k} \leq q_{k}$. Conditional independence and basin weights $w_{k}$ give the partition:

\begin{longtable}[]{@{}
  >{\raggedright\arraybackslash}p{(\linewidth - 4\tabcolsep) * \real{0.1000}}
  >{\raggedright\arraybackslash}p{(\linewidth - 4\tabcolsep) * \real{0.5302}}
  >{\raggedright\arraybackslash}p{(\linewidth - 4\tabcolsep) * \real{0.3701}}@{}}
\caption{Partition of passage probabilities by active support,
current-payoff acquiescence and continuation-supported
restraint.}\label{tab:si-composition-probabilities}\tabularnewline
\toprule\noalign{}
\begin{minipage}[b]{\linewidth}\raggedright
\textbf{Term}
\end{minipage} & \begin{minipage}[b]{\linewidth}\raggedright
\textbf{Passage composition}
\end{minipage} & \begin{minipage}[b]{\linewidth}\raggedright
\textbf{Probability}
\end{minipage} \\
\midrule\noalign{}
\endfirsthead
\toprule\noalign{}
\begin{minipage}[b]{\linewidth}\raggedright
\textbf{Term}
\end{minipage} & \begin{minipage}[b]{\linewidth}\raggedright
\textbf{Passage composition}
\end{minipage} & \begin{minipage}[b]{\linewidth}\raggedright
\textbf{Probability}
\end{minipage} \\
\midrule\noalign{}
\endhead
\bottomrule\noalign{}
\endlastfoot
$P_{A}$ & Every party actively supports & $p_{A}^{N}$ \\
$P_{Q}$ & At least one current-payoff acquiescer and no future-induced
restrainer & $\sum_{k}w_{k}\left( r_{k}^{N} - p_{A}^{N} \right)$ \\
$P_{C}$ & At least one future-induced restrainer &
$\sum_{k}w_{k}\left( q_{k}^{N} - r_{k}^{N} \right)$ \\
\end{longtable}

These components satisfy

\begin{equation}
P_{\text{pass}} = P_{A} + P_{Q} + P_{C} = \sum_{k}w_{k}q_{k}^{N},\quad P_{\text{app}} = P_{Q} + P_{C} = P_{\text{pass}} - P_{A}.
\end{equation}$P_{A}$ is passage with active support from every party. $P_{Q}$ is
passage that needs at least one current-payoff acquiescer and no future-induced restrainer. $P_{C}$ is passage in which continuation changes at least one interim action from veto to restraint. $P_{C}$ establishes a model mechanism but does not establish expressed opposition, substantive dislike, or an ex-post preference for veto after every other party restrains.

The illustrative private-type comparison gives a largest selected $P_{C}$ of about $0.46$. This magnitude depends on the imposed type distribution, continuation function and basin protocol specified in main Methods. A conditional share $\frac{P_{C}}{P_{\text{pass}}}$ can be large when passage is rare, so it must not be confused with the unconditional mechanism probability.

The numerical tables preserve the response-map and basin-selection results. Three legacy graphics remain in \nolinkurl{figures/} for provenance, but their axis label \nolinkurl{Pr(AC)} predates the current terminology and denotes the narrow continuation-supported quantity $P_{C}$. These graphics are omitted here to avoid conflating $P_{C}$ with apparent consensus $P_{\text{app}}$.

\section{The public outcome discards composition}

Let $Y = 1$ when no party chooses $V$ and $Y = 0$ otherwise.

\begin{proposition}[A binary consensus record cannot recover latent actions]

For $N$ parties with action set $\left\{ A,Q,V \right\}$, the record maps all $2^{N}$ profiles in $\left\{ A,Q \right\}^{N}$ to passage. One profile is unanimous active support, and the other $2^{N} - 1$ profiles are apparent consensus. The remaining $3^{N} - 2^{N}$ profiles map to breakdown. Under unrestricted heterogeneous strict incentives, every profile in both observed categories can arise as a best-response profile for some payoff vector.

\end{proposition}

\begin{proof}

Each party has three latent actions, which creates $3^{N}$ profiles. Passage permits two actions for each party, which creates $2^{N}$ passage profiles. Only $(A,\ldots,A)$ contains no acquiescence. Every other profile in $\left\{ A,Q \right\}^{N}$ contains at least one $Q$ and is apparent consensus. Every remaining profile outside $\left\{ A,Q \right\}^{N}$ contains at least one veto and maps to breakdown. To rationalize any strict profile, choose each party's alignment and visible-action costs so its prescribed action is strictly optimal conditional on the fixed actions of the others.

\end{proof}

Repeated binary records do not restore the discarded labels without additional restrictions or variation. Useful additions include separately recorded stand-asides, signed statements, proposal-linked amendments, exogenous membership changes, or a justified selection model.

\section{One veto can conceal opponents}

The original payoff rewards taking a veto position even after another party blocks passage. A separate complete-information extension attaches opposition to the policy outcome and makes public veto costly. Hold non-opposed parties silent. Opposed party $i$ loses $g_{i} > 0$ if everyone remains silent and the proposal passes, receives zero when another party vetoes, and pays $c_{i} > 0$ whenever it vetoes. If $\pi_{i}$ is the probability that every other party remains silent, silence gives $- \pi_{i}g_{i}$ and veto gives $- c_{i}$.

\begin{proposition}[Breakdown can hide additional opposition]

Party $i$ strictly prefers veto exactly when $\pi_{i} > \frac{c_{i}}{g_{i}}$. Suppose types and costs are common knowledge among the parties, at least one party opposes the proposal, and every opposed party satisfies $g_{i} > c_{i}$. A pure-strategy profile is a Nash equilibrium exactly when one opposed party vetoes. If at least two parties oppose, the public veto record reveals one maintained objection and conceals the others.

\end{proposition}

\begin{proof}

The first comparison follows from $- c_{i} > - \pi_{i}g_{i}$. In complete information, a sole vetoer pays $c_{i}$ and would lose $g_{i}$ after withdrawing the pivotal veto, so it maintains the veto. Every other objector receives zero through silence and would pay $c_{j}$ after adding a redundant veto. A no-veto profile invites an opposed party to block passage. A profile with several vetoes lets any one vetoer withdraw without changing breakdown.

\end{proof}

This result belongs to the outcome-opposition payoff extension. It does not apply to every parameterisation of the original expressive-veto payoff.

\section{Antarctic records used in the main text}

\protect\phantomsection\label{sec:si-record}{}

See main Methods (``Documentary sources and reconstruction'') for the model and procedures.

\subsection{Revisions and new proposals}

\protect\phantomsection\label{sec:si-revisions}{}

\subsubsection{Title classifications and predecessor consistency}

See main Methods (``Working-paper titles, outcomes and follow-up'') for the model and procedures.

These rules classify 289 papers as revisions and 109 as new proposals, 183 as reports, 46 as procedural papers and 968 as other papers. The labels describe title language and need not identify the legal novelty of every requested action. As a check against the release's operative instrument links, 186 of 218 originally adopted revisions act on a predecessor, while 22 of 28 originally adopted new proposals have no predecessor. This is agreement between documentary classifications, not independent validation of the titles or outcomes.

\subsubsection{Outcomes and follow-up results}

The source assessment retains all 13 positive paper-level corrections from the earlier audit (six revisions and seven new proposals). Among the 124 papers in the bounded review, it supports adoption of an identified requested instrument object in 119, rejects four same-meeting revision adoptions, and leaves one new proposal unresolved. The remaining classified papers retain inherited outcomes. This is not a completed independent validation of the full frame. Main Methods (``Working-paper titles, outcomes and follow-up'') specifies the assessment and denominator rules.

The reversals concern different requested actions and do not reflect disagreement between CEP advice and a later ATCM adoption. ATCM 21 WP 24 proposed document-circulation and translation amendments. Paragraph 80 deferred these amendments, while paragraph 81 attributed Decision 1 (1997) to other papers. ATCM 24 WP 22 proposed a revised Byers Peninsula plan, whereas Measure 3 (2001) extended the existing plan. ATCM 28 WPs 15 and 37 proposed revised Haswell and Ardley Island plans, and Measure 4 (2005) extended their existing plans. The report explicitly separated those extensions from the revisions awaiting intersessional review. ATCM 16 WP 41 remains unresolved because the adopted subgroup tourism draft cannot be mapped to that paper's individual requested provisions. The 276 revisions and 90 new proposals retained in the primary comparison have the following outcomes.

\begin{longtable}[]{@{}
  >{\raggedright\arraybackslash}p{(\linewidth - 4\tabcolsep) * \real{0.5001}}
  >{\raggedright\arraybackslash}p{(\linewidth - 4\tabcolsep) * \real{0.2501}}
  >{\raggedright\arraybackslash}p{(\linewidth - 4\tabcolsep) * \real{0.2501}}@{}}
\caption{Reviewed working-paper outcomes underlying main-text Fig. 5a.
The comparison retains 13 positive corrections, reverses four mismatched
instrument adoptions and excludes one unresolved proposal. Missing
outcomes and withdrawals are excluded. Recorded non-agreement is not a
complete census of vetoes.}\label{tab:si-record-outcomes}\tabularnewline
\toprule\noalign{}
\begin{minipage}[b]{\linewidth}\raggedright
\textbf{Submission-meeting outcome}
\end{minipage} & \begin{minipage}[b]{\linewidth}\raggedright
\textbf{Revision}
\end{minipage} & \begin{minipage}[b]{\linewidth}\raggedright
\textbf{New proposal}
\end{minipage} \\
\midrule\noalign{}
\endfirsthead
\toprule\noalign{}
\begin{minipage}[b]{\linewidth}\raggedright
\textbf{Submission-meeting outcome}
\end{minipage} & \begin{minipage}[b]{\linewidth}\raggedright
\textbf{Revision}
\end{minipage} & \begin{minipage}[b]{\linewidth}\raggedright
\textbf{New proposal}
\end{minipage} \\
\midrule\noalign{}
\endhead
\bottomrule\noalign{}
\endlastfoot
Instrument & 220 & 34 \\
Agreed, no instrument & 11 & 22 \\
Noted & 2 & 1 \\
Deferred or referred & 37 & 27 \\
Not agreed & 6 & 6 \\
Total & 276 & 90 \\
\end{longtable}

The reviewed follow-up chain connects the Byers revision to Measure 1 (2002), one meeting later, and the Haswell revision to Measure 1 (2006), also one meeting later. The Ardley revision proceeds through further reviews and objections before Measure 9 (2009), four meetings later. These chains identify adoption of the revised-plan object after amendments, not preservation of every original clause. No later instrument can yet be assigned to ATCM 21 WP 24 from the checked evidence. It remains a known non-adoption at submission and is excluded from the positive-window follow-up results. The source ledger retains exact report paragraphs, page locators and instrument identifiers for these judgments.

\subsubsection{Standardised comparisons and sensitivities}

See main Methods (``Working-paper statistical comparison'') for standardisation and uncertainty procedures.

Corrected instrument adoption at submission is 220/276 (79.7\%) for revisions and 34/90 (37.8\%) for new proposals. The standardised difference is 28.0 percentage points (95\% interval 17.1 to 41.2), compared with the raw difference of 41.9 points. Within site subjects the standardised difference is 29.1 points (15.8 to 43.0), and outside them it is 25.4 points (2.5 to 50.9). Including unclear outcomes in the denominator, under the original sensitivity convention, gives 29.1 points (18.5 to 41.9).

Among papers coded as seeking an instrument, submission-meeting adoption is 220/260 (84.6\%) versus 34/58 (58.6\%), with a standardised difference of 20.3 points (8.3 to 34.3). With two meetings of resolved follow-up, 233/240 revisions (97.1\%) and 45/54 new proposals (83.3\%) have an instrument, with a standardised difference of 10.3 points (1.3 to 20.2). The corresponding Kaplan--Meier estimates are 96.9\% and 81.7\%. Those follow-up denominators exclude the unresolved tourism object and the unresolved later disposition of the 1997 procedural amendment.

The submission-meeting result is unchanged when only high-confidence matches are used. The two-meeting result among instrument-seeking papers is less secure. High-confidence matching gives a difference of 7.8 points (−1.4 to 18.5). Alternative model-assisted sought/match coding, with the same direct source corrections and exclusions, gives 4.8 points (−5.1 to 15.0) at two meetings. Both intervals include zero.

Counting the ancillary extension requested for Ardley Island as paper-level success, while still recognising that its revised plan did not pass at submission, gives 221/276 revisions (80.1\%) and a standardised difference of 28.3 points (17.4 to 41.6). Assigning the unresolved tourism paper either submission outcome leaves the adjusted difference unchanged because its stratum contains no revisions. Excluding all 111 originally disputed or weak-link-only assignments, while retaining the 13 positive corrections, gives 115/167 revisions (68.9\%) versus 33/89 new proposals (37.1\%), with a standardised difference of 23.1 points (12.3 to 36.3). This restricted comparison changes the population and is not an accuracy-validated sample. Scheduled review, prior selection into adoption, scope and agenda composition remain alternative explanations for the observed differences.

\subsection{Recorded disagreement around ATCM 44}

\protect\phantomsection\label{sec:si-disagreement}{}

This descriptive comparison concerns recorded disagreement around ATCM 44, the first Consultative Meeting after Russia's invasion of Ukraine. Reports also record disputes over participation and Consultative status, including the stalled applications of Belarus and Canada \citep{ATCM2023FinalReport} \citep{ATCM2024FinalReport} \citep{ATCM2025FinalReport}. The model does not represent these geopolitical changes. The comparison leaves changes in future cooperation, current interests, agenda and reporting unresolved.

\begin{figure}[htbp]
\centering
\includegraphics[width=0.55\linewidth,height=0.48\textheight,keepaspectratio]{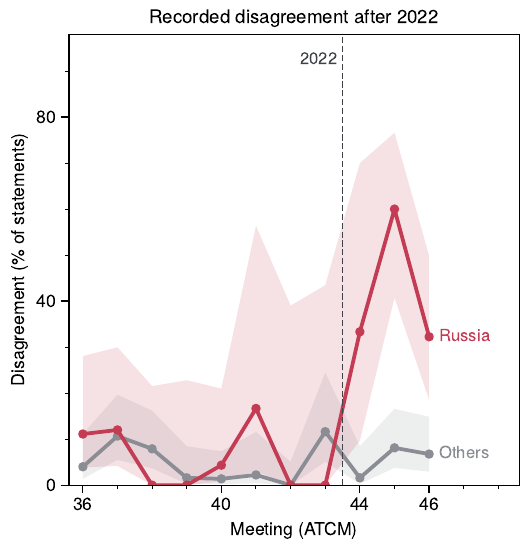}
\caption{Recorded disagreement in sampled statements before and after ATCM 44. Lines show the share of sampled statements coded as disagreement for Russia and the other Consultative Parties, with 95\% Wilson intervals shaded. Disagreement includes reservations that did not block adoption. The dashed line marks the first meeting after Russia's invasion of Ukraine. Russia's increase exceeds the pooled increase for other Parties by 35.9 percentage points (95\% interval 17.0--54.5, resampling agenda segments). These meeting-specific bands and the contrast interval describe different quantities. The comparison does not identify the cause of the change.}
\label{fig:si-recorded-disagreement}
\end{figure}

In the sampled statements, Russia's disagreement rate rises from 7\% (8/119, ATCM 36--43) to 44\% (27/62, ATCM 44--46), while the pooled rate for other Parties changes from 5\% to 6\% (Figure~\ref{fig:si-recorded-disagreement}). The difference in differences is 36 percentage points (95\% interval 17 to 54, resampling agenda segments). These statements include reservations that did not block adoption, so the comparison does not measure veto rates.

\subsubsection{Statement extraction, masking and coding}

\protect\phantomsection\label{sec:methods-statements}{}

The statement frame uses \nolinkurl{agenda_item} text from the repaired final-report segments for ATCM 28--46. A candidate is a Party--sentence pair in which a Consultative-Party alias is followed within 40 characters by a reporting verb. The extraction vocabulary covers supportive, critical and neutral speech. Table rows are skipped, and a Party is counted at most once per sentence. One sentence can generate items for more than one Party. The previous sentence within the paragraph is retained as context. This extraction does not establish that every statement was recovered or correctly attributed.

The fixed sample retains every extracted Russian item and at most four randomly selected items per other Party per meeting (seed 20261009). This gives 403 Russian and 1,539 other items, totalling 1,942. Conditional on extraction, the inclusion probability is one for Russia and $\min(1,\frac{4}{N_{p,m}})$ for an item from another Party, where $N_{p,m}$ is that Party's extracted-item count at meeting $m$. Selection within Party--meeting cells is without replacement and is not stratified by agenda. Other Parties are pooled at the statement level. The rate therefore describes sampled, attributable statements, not all speech, an equal-weight average of Parties or the probability that a proposal is vetoed.

The active analysis uses the recoding frozen on 1 November 2026, retaining the original sample and comparison windows. Direct country names, aliases and demonyms are replaced by anonymous tokens that change between rows. The focal participant is marked as the target. Dates and meeting/document identifiers are masked. Two separately prompted passes of the same inherited model each code every item, in eight shuffled batches with a fresh context for each batch (shuffle seed 20261101). Each context receives only its masked excerpts and the frozen rubric, without the hypothesis, identity key, old labels or the other pass's decisions. The retained execution records do not identify the inherited model version or decoding settings. The masking/shuffle seed is not a model sampling seed. Separate execution does not imply independent model errors. Subject matter can still reveal identity or period. This post-audit robustness recoding assesses coding consistency and is not an independently preregistered replication. The later panel separately examined extraction coverage in sampled repaired-report passages (Supplementary Section S22.5).

Each pass first judges whether the excerpt reports the target's own speech or position, including explicit joint speech, and assigns YES, NO or UNCLEAR. Mentioning a Party in another speaker's statement or a paper title is insufficient. A NO or UNCLEAR attribution requires an UNCLEAR stance. Stance labels are DISAGREE (opposition, rejection, reservation or inability to support a proposal, text, interpretation or statement, including procedural objections), SUPPORT (support, agreement, endorsement, welcome or co-sponsorship), NEUTRAL (reporting, presenting, information requests or thanks without a position), and UNCLEAR. Presenting a paper alone is neutral. An explicit objection takes precedence over accompanying thanks.

The primary analysis retains a stance only when both passes attribute the statement to the target and return the same stance. Matching UNCLEAR labels remain in its denominator as non-DISAGREE. Dropping them is a sensitivity. Attribution exclusions, stance differences and matching UNCLEAR labels form the review list, and none was adjudicated into the primary result. We report raw agreement and Cohen's $\kappa$ for attribution, four-category stance and DISAGREE versus every other label, distinguishing all items from jointly attributed items. The binary agreement measure includes UNCLEAR in the other category. It is not the estimated disagreement rate. High agreement cannot rule out shared errors. The subsequent panel supplied a separately coded masked sample and a passage-based coverage check, with their results and scope reported below and in Supplementary Section S22.5.

\subsubsection{Disagreement contrast and uncertainty}

\protect\phantomsection\label{sec:methods-statement-statistics}{}

The pre-period is ATCM 36--43 and the post-period ATCM 44--46. Let $s_{g,t}$ denote the fraction of retained statements labelled DISAGREE in group $g$ and period $t$. We report the change for Russia minus the pooled change for other Parties as

\begin{equation}
\begin{aligned}
D_{\text{shock}} & = \left( s_{\text{Russia, post}} - s_{\text{Russia, pre}} \right) \\
 & \quad - \left( s_{\text{others, post}} - s_{\text{others, pre}} \right).
\end{aligned}
\end{equation}Meeting-specific plotted shares use 95\% Wilson binomial intervals with
$z = 1.96$. These bands differ from the before/after contrast interval. The original Protocol v2 calculation independently resamples statements within each of the four group-by-period cells, using 4,000 draws and seed 20261009. The primary descriptive comparison uses the post hoc agenda-segment bootstrap. We draw segment identifiers with replacement from retained statements at ATCM 36--46, retain every statement in each drawn segment, recalculate the four means, and omit draws missing a cell. A separate meeting bootstrap draws pre-period and post-period meetings within their own periods. Both use 4,000 draws, the same seed and percentile endpoints. These intervals condition on the extracted sample and coding. None incorporates unknown extraction or shared model error.

We add post hoc sampling-weight and agenda checks using the active 1 November labels. The design weight is the inverse of the original Party--meeting inclusion probability, conditional on extraction. Weighted disagreement rates are ratios of weighted DISAGREE counts to weighted retained-label counts. They are not full-frame prevalence estimates without assumptions about label retention. We do not estimate extraction or label-retention propensities. Both weighted and unweighted checks retain matching UNCLEAR labels in the denominator.

For the agenda check, a stratum combines forum with exact agenda title after Unicode compatibility normalisation, case folding and whitespace normalisation. Differently named or punctuated headings are not merged, and changing item numbers are not treated as stable identities. Seven strata contain retained statements in all four group--period cells. One comparison restricts to these strata while retaining each cell's observed agenda mix. A separate fixed-mix diagnostic uses the equal-weight average of the four extracted-frame agenda distributions after each is restricted to common support. The target shares are fixed from metadata and sample support, without using stance outcomes. The shared-support subset retains 36--50\% of each cell's statements. Its broad headings do not establish exchangeability of topics, speakers or reporting within each heading.

These checks use the same 4,000 agenda-segment bootstrap draws as the primary contrast, with original weights, support and target shares fixed. The intervals describe conditional segment-resampling variability, not finite-population sampling variance or uncertainty in extraction, labels or the selected agenda support. Only 32 of 4,000 draws retain all cells needed for the fixed-mix estimate, so we report its point estimate without a percentile interval rather than condition an interval on those surviving draws. The code, exact sample reconstruction, input hashes and checks are archived in \nolinkurl{reviews/results_revision_2026-11-06_130225/statement_sensitivity/}.

Direct war flags search the statement and preceding context, case-insensitively, for the stem Ukrain, the word war, aggression or invasion. Sensitivities remove flagged items from all Parties and then also remove the separately specified fallout list of six Russian disagreement items concerning visas or attendance at ATCM 45, Belarus, politicisation or the rules-based order at ATCM 46. This interpretive list is not a validated exhaustive classifier of indirect war effects. Other sensitivities use each pass alone conditional on that pass's YES attribution, drop UNCLEAR, or omit each post-period meeting in turn. Adversarial assignment of unresolved or UNCLEAR stances gives a range conditional on both passes' YES attribution. It does not restore attribution exclusions and is not a confidence interval.

The historical Protocol v2, dated 9 October 2026, required Russia to have the largest change among Parties with at least eight retained statements in each period and cutoff 44 to have the largest change across its placebo series. The placebo calculation uses four meetings before a cutoff and three from the cutoff, with at least five statements in each window, over cutoffs 32--44. It therefore differs from the primary eight-meeting baseline. Cutoffs 42 and 43 include post-shock meetings. The descriptive pre-shock comparison restricts cutoffs to at most 41. The post hoc pooled Party--cutoff reference distribution also uses cutoffs at most 41. Overlapping windows and repeated Parties are not independent randomisation units. Archived per-Party tests use one-sided Fisher exact comparisons with Holm correction across eligible Parties. These statement-level tests do not account for agenda clustering. The literal historical joint success criterion is not met by the active recoding, because cutoff 44 is not the largest placebo change.

The original identity-visible coding and earlier lexicon analysis are historical. The current analysis uses the 1 November attribution screen. Protocol v3 addresses whether disagreement episodes passed or stalled. Its frozen outcome tables have not been recoded using the active disagreement labels. Recorded disagreement can accompany passage. The present contrast cannot distinguish changes in future cooperation from agenda, delegation, reporting or selection changes.

\subsubsection{Statement frame and recoding sample}

See the statement extraction and coding procedures above.

The fixed recoding sample contains 403 Russian and 1,539 other items, totalling 1,942. Every extracted Russian item is retained, and up to four items are selected randomly without replacement per other Party--meeting cell. Conditional inclusion probabilities are one for Russia and $\min(1,\frac{4}{N_{p,m}})$ otherwise. These probabilities concern the extracted frame, not all statements in the original reports. The primary rates are unweighted over retained statements, and there is no agenda stratification.

\subsubsection{Attribution and coding agreement}

See the statement extraction and coding procedures above.

Of 1,942 items, 1,763 pass both attribution judgments and 179 are excluded. Among the attributed items, 108 have different stance labels. This leaves 1,655 matching labels, comprising 112 DISAGREE, 444 SUPPORT, 1,096 NEUTRAL and three UNCLEAR. The 290 review items comprise attribution exclusions, stance differences and the three matching UNCLEAR items. None was adjudicated into the primary result. The inclusion rule and treatment of matching UNCLEAR are specified above.

Across all items, attribution agreement is 99.79\% (Cohen's $\kappa = 0.988$), and four-category stance agreement is 94.34\% ($\kappa = 0.902$). Among the 1,763 jointly attributed items, four-category agreement is 93.87\% ($\kappa = 0.877$), and agreement on DISAGREE versus every other label is 98.47\% ($\kappa = 0.884$). The latter includes UNCLEAR in the other category and is an agreement statistic, not the estimated disagreement rate. High agreement cannot rule out shared model errors. The original prompted passes provide a consistency check within one model. Section S22.5 reports the subsequent masked recoding and sampled passage checks.

\subsubsection{Disagreement contrasts and intervals}

See the disagreement contrast and uncertainty procedures above.

Russia's rate rises from 8/119 (6.7\%) to 27/62 (43.5\%), while the pooled rate for other Parties changes from 25/522 (4.8\%) to 12/211 (5.7\%). The contrast is 35.9 percentage points. The meeting-specific bands in Figure~\ref{fig:si-recorded-disagreement} are distinct from the contrast intervals reported below.

The statement-resampling interval is 22.8 to 49.3 percentage points. The post hoc agenda-segment interval for the primary comparison is 17.0 to 54.5 points, and resampling meetings gives 22.6 to 51.7 points. These intervals describe uncertainty conditional on the extracted sample and coding. None incorporates unknown extraction or shared coding error. Resampling designs are specified above.

\subsubsection{Robustness and historical protocols}

\begin{longtable}[]{@{}
  >{\raggedright\arraybackslash}p{(\linewidth - 6\tabcolsep) * \real{0.4386}}
  >{\raggedright\arraybackslash}p{(\linewidth - 6\tabcolsep) * \real{0.1404}}
  >{\raggedright\arraybackslash}p{(\linewidth - 6\tabcolsep) * \real{0.2105}}
  >{\raggedright\arraybackslash}p{(\linewidth - 6\tabcolsep) * \real{0.2105}}@{}}
\caption{Frozen shock sensitivities. Differences and intervals are in
percentage points. Individual-pass results require that pass's YES
attribution, so the samples and interval designs differ.}\label{tab:si-record-shock}\tabularnewline
\toprule\noalign{}
\begin{minipage}[b]{\linewidth}\raggedright
\textbf{Analysis}
\end{minipage} & \begin{minipage}[b]{\linewidth}\raggedright
\textbf{Difference}
\end{minipage} & \begin{minipage}[b]{\linewidth}\raggedright
\textbf{95\% interval}
\end{minipage} & \begin{minipage}[b]{\linewidth}\raggedright
\textbf{Resampling unit}
\end{minipage} \\
\midrule\noalign{}
\endfirsthead
\toprule\noalign{}
\begin{minipage}[b]{\linewidth}\raggedright
\textbf{Analysis}
\end{minipage} & \begin{minipage}[b]{\linewidth}\raggedright
\textbf{Difference}
\end{minipage} & \begin{minipage}[b]{\linewidth}\raggedright
\textbf{95\% interval}
\end{minipage} & \begin{minipage}[b]{\linewidth}\raggedright
\textbf{Resampling unit}
\end{minipage} \\
\midrule\noalign{}
\endhead
\bottomrule\noalign{}
\endlastfoot
Matching stance, both attributed & 35.9 & 17.0 to 54.5 & Agenda
segment \\
Exclude direct war flags & 30.1 & 12.5 to 47.9 & Agenda segment \\
Also exclude coded fallout & 22.6 & 6.2 to 40.7 & Agenda segment \\
Pass A alone & 34.4 & 21.4 to 47.7 & Statement \\
Pass B alone & 30.6 & 17.2 to 44.1 & Statement \\
Drop UNCLEAR & 36.6 & 22.8 to 50.4 & Statement \\
\end{longtable}

Excluding direct war flags removes six of Russia's 27 post-period disagreements, leaving 21/56 statements. Excluding the six separately specified fallout items too leaves 15/50. This list is not a validated exhaustive classifier of indirect war effects. Omitting each post-period meeting one at a time gives contrasts of 35.3, 26.1 and 47.8 points. The adversarial unresolved-stance range is 21.5 to 47.6 points, conditional on both passes' YES attribution. It does not restore attribution exclusions and is not a confidence interval.

Under the historical Protocol v2 criteria, Russia ranks first among 17 eligible Parties in the active recoding, while cutoff 44 ranks second of 13 cutoffs (32--44), behind cutoff 43. The literal joint success criterion is not met. Cutoffs 42 and 43 include post-shock meetings. Restricting the descriptive placebo comparison to cutoffs at or before 41 gives a maximum prior Russian change of 5.1 points, against 38.5 points at cutoff 44 on the four-meeting baseline. The primary comparison uses a different, eight-meeting baseline.

The post hoc pooled comparison contains 218 Party--cutoff cells with cutoffs at or before 41. Its maximum is 33.3 points and its 95th percentile 12.8 points. Russia and China exceed the maximum at cutoff 44, and Japan also exceeds the 95th percentile. China's four-meeting change is 34.6 points (2/13 to 5/10), whereas its eight-meeting change is 27.8 points (6/27 to 5/10), below that pooled maximum. The overlapping windows and repeated Parties mean these are descriptive reference distributions, not independent randomisation units. In the archived per-Party Fisher comparisons with Holm adjustment, Russia has adjusted $p = 1.36 \times 10^{- 7}$, while China's adjusted $p = 1$. These statement-level tests do not account for agenda clustering.

The historical outcome-split analysis has not been recoded with the active labels and cannot establish that the rise in recorded disagreement is a rise in vetoes.

\subsubsection{Sampling-weight and shared-agenda sensitivities}

The active recoding reproduces the unweighted contrast of 35.9278 percentage points before applying weights. Conditional inclusion probabilities are recovered from the original Party--meeting sample, with probability one for Russia and $\min(1,\frac{4}{N_{p,m}})$ for other Parties. The weighted rate is the ratio of weighted disagreement counts to weighted retained-label counts. It does not correct unknown extraction or label-retention mechanisms. Russia's rates remain 6.7227\% before and 43.5484\% after. Other Parties' rates change from 4.7893\% and 5.6872\% unweighted to 6.4818\% and 6.8182\% weighted.

\begin{longtable}[]{@{}
  >{\raggedright\arraybackslash}p{(\linewidth - 4\tabcolsep) * \real{0.5111}}
  >{\raggedright\arraybackslash}p{(\linewidth - 4\tabcolsep) * \real{0.1778}}
  >{\raggedright\arraybackslash}p{(\linewidth - 4\tabcolsep) * \real{0.3111}}@{}}
\caption{Post hoc sensitivities using the active statement recoding.
Weights refer to selection from the extracted frame. The shared-heading
restriction and fixed-mix diagnostic change the comparison population
and do not identify a causal effect.}\label{tab:si-active-statement-weights}\tabularnewline
\toprule\noalign{}
\begin{minipage}[b]{\linewidth}\raggedright
\textbf{Active-label comparison}
\end{minipage} & \begin{minipage}[b]{\linewidth}\raggedright
\textbf{Contrast (pp)}
\end{minipage} & \begin{minipage}[b]{\linewidth}\raggedright
\textbf{95\% segment interval}
\end{minipage} \\
\midrule\noalign{}
\endfirsthead
\toprule\noalign{}
\begin{minipage}[b]{\linewidth}\raggedright
\textbf{Active-label comparison}
\end{minipage} & \begin{minipage}[b]{\linewidth}\raggedright
\textbf{Contrast (pp)}
\end{minipage} & \begin{minipage}[b]{\linewidth}\raggedright
\textbf{95\% segment interval}
\end{minipage} \\
\midrule\noalign{}
\endhead
\bottomrule\noalign{}
\endlastfoot
Unweighted primary & 35.93 & 17.04 to 54.46 \\
Inverse-inclusion weighted & 36.49 & 16.53 to 55.68 \\
Weighted, shared agenda headings & 24.81 & -6.22 to 50.72 \\
Weighted, fixed agenda mix & 22.35 & Not reported because support is sparse \\
\end{longtable}

Shared support contains seven forum--normalised-title strata represented in all four group--period cells. Exact-title normalisation changes Unicode compatibility, case and whitespace only. The restriction retains 188/522 other-Party pre-period items, 82/211 post-period items, 50/119 Russian pre-period items and 31/62 post-period items, respectively 36.0\%, 38.9\%, 42.0\% and 50.0\%. It retains the observed within-cell agenda mix. The fixed-mix diagnostic instead gives every cell the same target distribution, namely the equal-weight average of the four extracted-frame distributions within shared support, fixed without using stance labels. Several Russian agenda cells contain only one to three statements. Only 32 of 4,000 segment-resampling draws retain every required cell, so a percentile interval for the fixed-mix point estimate is not reported. The point and missing-draw count remain visible.

The 4,000 segment draws use seed 20261009 and reproduce the primary unweighted segment interval. Weights, labels, support and target shares are held fixed. No interval covers extraction omissions, coding errors, unknown retention, uncertainty in selected support, dependence between segments within meetings or post hoc analysis selection. The original sample reconstruction, coding hashes, inclusion counts, weight identities and exact point-estimate checks are archived with the runnable sensitivity. These numerical checks address weighting and uncertainty under the retained labels. The subsequent panel examined sampled attribution and stance labels and assessed missed statements in sampled report passages (Section S22.5).

\subsubsection{Consultative-status disputes in the sampled period}

ATCM 44 in Berlin was the first Consultative Meeting after Russia's invasion of Ukraine. The final reports document disputes over participation and admission, without measuring the expected value of future Antarctic cooperation. Belarus, which supports the invasion, has been refused Consultative status since 2023 \citep[p. 34]{ATCM2023FinalReport}. At ATCM 46 many Parties held that under ``current political circumstances'' no agreement could be reached \citep[p. 36]{ATCM2024FinalReport}, and at ATCM 47 Ukraine opposed admission because of Belarus's support for the war \citep[p. 34]{ATCM2025FinalReport}. Canada's application stalled over the same years \citep[p. 152]{ATCM2022FinalReport} \citep[p. 34]{ATCM2025FinalReport}. Two Parties judged its research insufficient and described Consultative status as a legal and scientific matter. These records establish reported positions and outcomes, rather than identifying whether a particular Party valued future cooperation less.

\subsection{Four proposal histories}

\protect\phantomsection\label{sec:si-histories}{}

See main Methods (``Selection and interpretation of proposal histories'') for the model and procedures.

\subsubsection{Emperor penguin and Amanda Bay}

Designation remained unagreed in 2022--2024. The 2022 CEP account records no consensus and one exception to support. The 2023 passages recall that earlier result and record continuing disagreement, without establishing a separate new listing veto. The 2024 report explicitly records no consensus on the tabled proposal \citep{ATCM2022FinalReport} \citep{ATCM2023FinalReport} \citep{ATCM2024FinalReport}. The Amanda Bay colony has a separate site-plan history, with designation in 2008, revision in 2014 and a revised plan jointly proposed by China and Australia adopted as Measure 2 (2025). In 2025 China referred to that site's reported population in disputing the evidence for species designation, while Russia raised research restrictions. No new species-designation proposal was tabled that year \citep[CEP paras. 251--255; Measure 2]{ATCM2025FinalReport}. In 2025 the record shows continued disagreement without a new listing decision, and no observed veto on a submitted listing.

The site plan is a narrower, separate proposal. Its adoption is not a demonstrated amendment of the species-listing draft or proof that its proponents substituted it to obtain agreement. These histories concern different legal objects and fallbacks, so their juxtaposition does not identify accommodation of the species-listing proposal. For the earlier blocked listing versions, the objecting Party did not stand aside. The record does not classify how each Party accepted the later site plan.

The available 2026 report records continuing disagreement. China considered species designation premature and Russia supported that intervention. The ATCM agreed to continue considering appropriate management measures at ATCM 49 \citep[paras. 93, 97; pp. 28--29]{ATCM2026FinalReport}. This is a later narrative update, outside the working-paper and statement samples and the plotted timeline.

\subsubsection{ASPA 152 and 153 to ASPA 182}

Both existing marine areas had their plans revised in 2009 and 2015. The 2022 proposal combined them and expanded their coverage. China and Russia objected at CCAMLR-41 that the expansion was substantial and insufficiently justified, and consensus was not reached \citep[p. 31]{CCAMLR2022Report}. The subsequent plan retained the original horizontal extent and placed the upper boundary 20 m below the sea surface. CCAMLR approved it in 2023, followed by ATCM adoption as ASPA 182 in Measure 17 (2024) \citep[p. 43]{CCAMLR2023Report} \citep[Measure 17, pp. 207--208]{ATCM2024FinalReport}.

This objection--revision--adoption sequence supports accommodation. The 2022 objections exclude restraint or reluctant acceptance for those objectors on the blocked version. They do not exclude either pathway for the revised plan in 2023--2024, and the records do not supply a complete final response profile. Main Figure 5b gives the blocked plan and the later revised plan separate matrix rows. The crosses concern the objectors on the 2022 version, while the accommodation tick concerns the revision followed by approval and adoption in 2023--2024. The final response pathways remain open.

\subsubsection{Ross Sea}

The Ross Sea proposals were negotiated from 2012 to adoption of CM 91-05 in 2016. Russia proposed boundary and research-zone changes in 2013, and Russia and China raised fishing concerns in 2014 \citep[p. 26]{CCAMLR2013Report} \citep[pp. 44--45]{CCAMLR2014Report}. In 2015 the proponents linked revisions to China's concerns, and China supported the revised text as a basis for further consideration while Russia sought further consultation. This was not adoption \citep[paras. 8.107--8.110]{CCAMLR2015Report}. In 2016 the proponents expressly addressed Russia's remaining concerns. Russia acknowledged the changes, and the Commission adopted the measure \citep[paras. 8.38--8.48]{CCAMLR2016Report}. The sequence supports accommodation, but neither establishes unchanged private preferences nor identifies the final stance of every Member. Fishing access, boundaries, research arrangements and duration changed together, so the history supplies no measured scalar bite or uniquely identified limiting responder.

\subsubsection{East Antarctica}

The East Antarctic proposal began with seven areas in 2012, moved to an initial implementation stage of four areas in 2013 \citep[para. 7.2]{CCAMLR2013Report} and was reduced to three in 2015. By 2016 the proponents also described changes to activity defaults and management provisions, while Russia, China and Japan continued to request work on objectives, plans or duration. Adoption did not follow \citep[paras. 7.80--7.82]{CCAMLR2012Report} \citep{CCAMLR2015Report} \citep[paras. 8.74--8.84]{CCAMLR2016Report}. The 2025 report still records a draft, with Russia seeking three separate proposals and China questioning baseline information \citep[pp. 38--39]{CCAMLR2025Report}. Accommodation was attempted but did not suffice over this observation period. Where an objection prevents agreement, stand-aside pathways are excluded only for that objector and version. Non-adoption does not identify every Member as an objector, nor establish that accommodation can never succeed. Differences in geography, science, objectives and terms also prevent treating Ross Sea and East Antarctica as a controlled comparison.

\subsection{Proposal thresholds and recorded outcomes}
\label{sec:si-proposal-thresholds}
The accommodation threshold in Section S6 permits passage to persist as membership grows if the proposer can reduce its demand to satisfy the most restrictive Party. With existing thresholds fixed, adding a Party can lower the largest acceptable demand. This result does not establish a stability advantage for gradual membership growth, since the formation and transmission of learned conventions require additional assumptions.

We examined whether a statistical model motivated by this threshold could approximate the reviewed working-paper outcomes. Let $R_j=\mu_j+\epsilon_j$ be an unobserved residual demand after prior adjustment, with standard logistic $\epsilon_j$. The model assigns agreement when $R_j\leq0$, deferral or referral when $0<R_j\leq a$, and recorded non-agreement when $R_j>a$, where $a>0$. Writing $F$ for the logistic distribution function gives
\[
P_A=F(-\mu_j),\qquad P_D=F(a-\mu_j)-F(-\mu_j),\qquad P_B=1-F(a-\mu_j).
\]
This ordering is an observation assumption, since procedural referrals need not indicate intermediate opposition. A separate logistic regression distinguishes instrument adoption from agreement without an instrument, and a constant probability accounts for noting. Both regressions use proposal type, agenda designation, site subject and the era categories in Section S22.1. Slopes receive a quadratic penalty of one, with intercepts unpenalised. Half-count smoothing is applied to the noted probability and categorical comparison models. The index does not identify individual veto incentives, the amount of accommodation or negotiation duration.

The sample contains the same 276 revisions and 90 new proposals as Section S22.1. Holding out each whole meeting in turn gives predicted instrument-adoption shares of 78.7\% for revisions and 40.9\% for new proposals, against recorded shares of 79.7\% and 37.8\%. Predicted non-agreement shares are 2.1\% and 6.9\%, against 2.2\% and 6.7\%. These are recorded dispositions rather than measured veto rates. Mean negative log predictive probability is 0.816, compared with 0.890 for outcome frequencies by proposal type and 0.792 for an unrestricted multinomial model using the same covariates. Lower values indicate better prediction.

Training on ATCM 16--43 and predicting ATCM 44--47 shows larger discrepancies among the 13 new proposals. The model predicts 43.6\% instrument adoption against 7.7\% observed, and 2.7\% non-agreement against 30.8\% observed. The corresponding counts are one adoption and four non-agreements. The 55 revisions have 89.1\% recorded adoption against 79.6\% predicted. These discrepancies remain after removing training papers with matching normalised titles. The chronological log loss changes from 0.789 to 0.810, while the unrestricted model scores 0.698 and 0.711.

These checks hold out model-fitting observations, while the hypotheses and retrospective coding were developed with the records available. Repeated proposal histories may still cross the splits under different titles. The fit concerns the selected known-outcome cohort and does not establish a mechanism for membership effects. Code and paper-level predictions are retained in \texttt{scripts/fit\_proposal\_thresholds.py} and \texttt{data/proposal\_thresholds\_2026-11-06/}.

\subsection{Additional documentary checks}
\label{sec:si-documentary-panel}
For statement coding, we sampled 50 items without replacement from each combination of Russia or other Parties and the pre- or post-period, giving 200 items from the 1,090 sampled statements at ATCM 36--46. Selection used seed 20261106 and did not depend on existing labels. One human coder and one AI agent received the same identity-masked excerpts and coding rubric in different orders, without the identity key, study hypothesis or previous labels. A further AI agent separately coded the 21 items where the new readings differed from each other or from the original inclusion or label. We evaluated the contrast after substituting each new reading for the selected items, after retaining only matching readings, and after majority classification of the selected discrepancies. The remaining items kept their original labels. These substitutions are sensitivity analyses, and the primary analysis retains its frozen coding.

Extraction coverage was assessed separately in 66 repaired-report paragraphs, sampled uniformly within each meeting at six paragraphs per meeting from ATCM 36--46, using seed 2026110602. Eligible paragraphs contained 80--3,500 characters and did not begin with a table separator. Two further agents enumerated attributable Party--sentence units without seeing the extractor or its output. We then matched those units to the original extractor's output, distinguishing direct named speech from contextual continuations. The sampling frame contained 10,204 eligible paragraphs.

The AI agents worked in separate contexts and ran on the same language model as the original statement coding. Agreement among coders does not exclude shared errors. The coverage check concerns sampled repaired text and does not establish completeness against the original PDFs or across the full corpus. Results and source records are reported in Supplementary Section S22.5.

The completed documentary review examined source support for working-paper outcomes, the consistency of statement coding and extraction coverage in sampled report passages. Main Methods describes the working-paper source review. Statement sampling and review procedures are given above. The original outcomes and statement labels remain frozen, with the new readings reported as checks and sensitivity analyses.

The source reviewers supported retaining all 124 existing object/stage judgments, including the unresolved tourism paper's exclusion. They preserved the distinction between extending an existing management plan and adopting its submitted revision, and between CEP advice and final ATCM action. This result concerns the previously reviewed cases and the available cited passages. It is not an estimate of accuracy across all 1,595 working papers or all extracted relationships.

In the 200-item masked statement sample, the new coders agreed on attribution for 199 items and on stance for 191 items, giving 99.5\% attribution agreement and 95.5\% stance agreement. Four-category stance agreement had Cohen's $\kappa=0.927$. The new coders agreed with each other and with the original inclusion and label on 179 items, and agreed with each other but differed from the original classification on 12 items. These agreement measures do not identify which reading is correct when classifications differ. The third reading supplied a further sensitivity for the 21 selected discrepancies rather than an independent ground truth.

Replacing only the sampled classifications with the first or second new reading gives disagreement contrasts of 35.85 and 35.12 percentage points, compared with 35.93 points in the original analysis. Retaining only the new coders' matching attributable labels gives 36.19 points, while majority classification using the third reading for the selected discrepancies gives 35.13 points. Unsampled classifications are unchanged in each comparison. The narrow variation supports the descriptive contrast's robustness to these sampled coding alternatives, without resolving shared errors or extraction omissions.

The coverage reviewers identified the same 27 Party--sentence units in nine of the 66 sampled paragraphs after normalising Party aliases. The remaining 57 paragraphs had no individually attributable unit under their reading rules. The original extractor recovered 14 of those 27 units. Of the 16 units classified as direct named speech, 14 were recovered. The other units comprised four pronoun continuations, five statements inferred from the content of a presented paper and two ambiguous continuations. None of these contextual units was recovered. The extractor also returned two additional units that both reviewers attributed to another speaker. These concerned a recipient of thanks and a country mentioned before the actual speaker in a longer sentence.

The coverage check therefore documents both recovered statements and omissions, with the result depending on whether contextual continuations are included in the reference definition. The counts concern the sampled repaired-report paragraphs and do not estimate corpus-wide recall. They support interpreting the main comparison as a contrast among extracted and retained statements. The check does not establish whether omissions differ systematically between Parties or periods.

The panel inputs, source quotations, coding records and comparison outputs are archived under \nolinkurl{reviews/latex_documentary_panel_2026-11-06/}. The summary is reproduced by \nolinkurl{scripts/summarize_documentary_panel.py}. Input hashes and sampling records identify the material supplied to each reviewer.

\section{ATS documents supply unsigned magnitudes}

\protect\phantomsection\label{sec:si-unsigned}{}

See main Methods (``Documentary prominence and sign scenarios'') for the model and procedures.

Twenty-eight of the 29 parties authored at least one working paper in the window. The all-zero party remains in the full institutional response system.

The ratio measures relative documentary specialisation. A party with few total submissions can attain a high value from one topic paper. On tourism, the parties ranked second and third by this measure differ tenfold in their number of working papers. The ranking tracks specialisation, which is distinct from total activity. The measure does not identify preference direction, influence, expertise, policy stakes, bargaining power, or a position on an adopted text. A zero cell leaves political direction open.

\begin{figure}[htbp]
\centering
\includegraphics[width=1\linewidth,height=0.65\textheight,keepaspectratio]{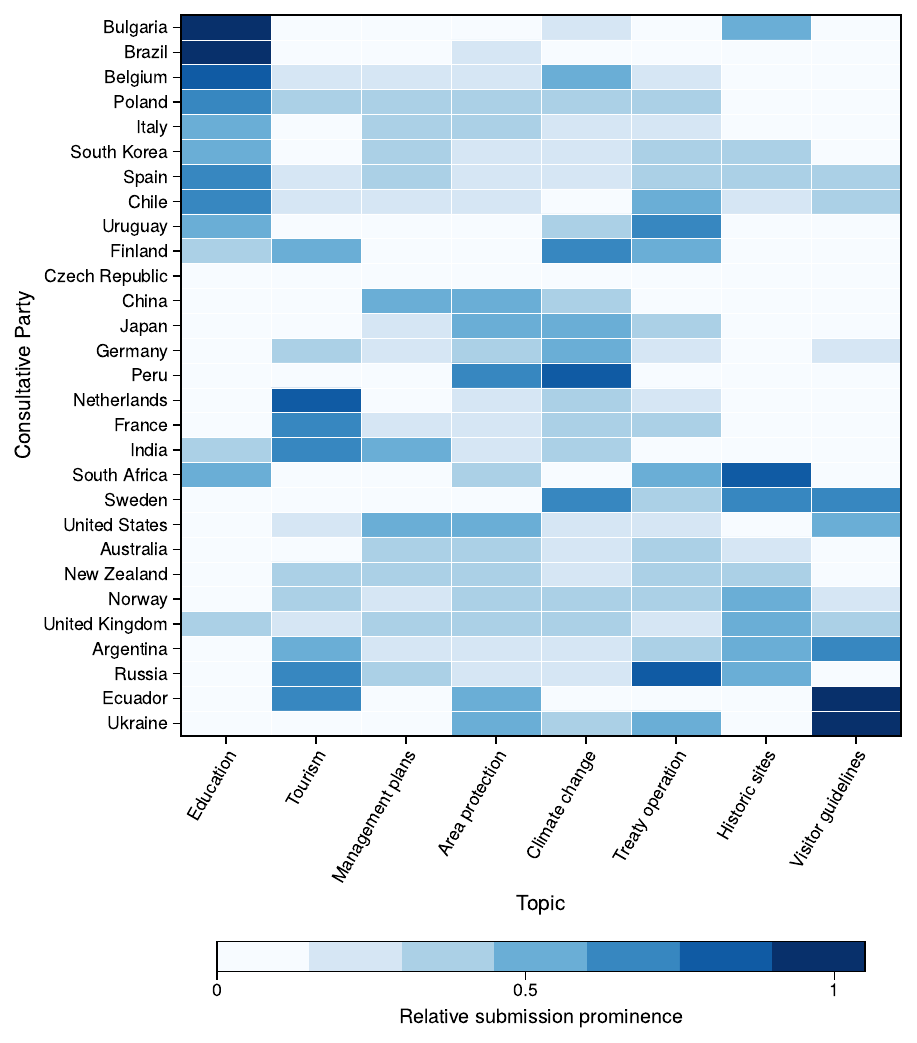}
\caption{Unsigned relative topic specialisation for 29 Consultative Parties and eight retained topics. Each cell reports the transformed and scaled working-paper measure. White cells mark no observed authorship on that topic during 2015 through 2024. Clustering makes similar documentary profiles adjacent and does not identify political blocs.}
\label{fig:si-prominence-values}
\end{figure}

\begin{longtable}[]{@{}
  >{\raggedright\arraybackslash}p{(\linewidth - 6\tabcolsep) * \real{0.2500}}
  >{\raggedright\arraybackslash}p{(\linewidth - 6\tabcolsep) * \real{0.2500}}
  >{\raggedright\arraybackslash}p{(\linewidth - 6\tabcolsep) * \real{0.2500}}
  >{\raggedright\arraybackslash}p{(\linewidth - 6\tabcolsep) * \real{0.2500}}@{}}
\caption{Observed documentary quantities, imposed assumptions and
unidentified political quantities.}\label{tab:si-observables}\tabularnewline
\toprule\noalign{}
\begin{minipage}[b]{\linewidth}\raggedright
\textbf{Observed}
\end{minipage} & \begin{minipage}[b]{\linewidth}\raggedright
\textbf{Assumed}
\end{minipage} & \begin{minipage}[b]{\linewidth}\raggedright
\textbf{Model output}
\end{minipage} & \begin{minipage}[b]{\linewidth}\raggedright
\textbf{Unidentified}
\end{minipage} \\
\midrule\noalign{}
\endfirsthead
\toprule\noalign{}
\begin{minipage}[b]{\linewidth}\raggedright
\textbf{Observed}
\end{minipage} & \begin{minipage}[b]{\linewidth}\raggedright
\textbf{Assumed}
\end{minipage} & \begin{minipage}[b]{\linewidth}\raggedright
\textbf{Model output}
\end{minipage} & \begin{minipage}[b]{\linewidth}\raggedright
\textbf{Unidentified}
\end{minipage} \\
\midrule\noalign{}
\endhead
\bottomrule\noalign{}
\endlastfoot
Working-paper counts by paper, party, and topic & Baseline
$\mu_{0}(P)$ and scale $\eta$ & Conditional passage probability &
Direction of each party's stake \\
Derived prominence $a_{iP}$ & Maximal transfer $\rho_{P} = 1$ &
Continuation-supported passage $P_{C}$ & Topic relevance
$\rho_{P}$ \\
Party set and $N = 29$ & Belief prior $m,\kappa$ and residual scale
$\tau$ & Transition $\delta_{50}$ & Institutional discount factor \\
Working-paper co-sponsorship & Common continuation function $L(q) = 1$
& Basin weights under the stated iteration & Party-specific continuation
stakes \\
\end{longtable}

\protect\phantomsection\label{tbl:si-observability}{}

The table distinguishes documentary measurement from analyst choices. Every ATS quantity below is conditional on the assumptions in its second column.

The illustrative ATS baseline is fixed across topics:

\begin{longtable}[]{@{}
  >{\raggedright\arraybackslash}p{(\linewidth - 2\tabcolsep) * \real{0.3000}}
  >{\raggedright\arraybackslash}p{(\linewidth - 2\tabcolsep) * \real{0.7000}}@{}}
\caption{Baseline specification for the unsigned documentary sensitivity
exercise.}\label{tab:si-ats-parameters}\tabularnewline
\toprule\noalign{}
\begin{minipage}[b]{\linewidth}\raggedright
\textbf{Model component}
\end{minipage} & \begin{minipage}[b]{\linewidth}\raggedright
\textbf{Baseline specification}
\end{minipage} \\
\midrule\noalign{}
\endfirsthead
\toprule\noalign{}
\begin{minipage}[b]{\linewidth}\raggedright
\textbf{Model component}
\end{minipage} & \begin{minipage}[b]{\linewidth}\raggedright
\textbf{Baseline specification}
\end{minipage} \\
\midrule\noalign{}
\endhead
\bottomrule\noalign{}
\endlastfoot
Response system & $N = 29$, $b = 1$ \\
Prominence-to-type mapping & $\mu_{0}(P) = 0.6$, $\eta = 0.4$,
$\rho_{P} = 1$ \\
Independent normal residuals & $\tau = 0.15$ \\
Common-belief prior & $m = 0.5$, $\kappa = 8$ \\
Expected continuation advantage & $L(q) = 1$ \\
\end{longtable}

\protect\phantomsection\label{tbl:si-ats-baseline}{}

These values are illustrative assumptions rather than estimates of historical preferences or continuation payoffs.

\section{Hidden signs shift restraint}

Let $z_{iP} \in \left\{ - 1, + 1 \right\}$ be an analyst-assigned direction and let $\rho_{P} \in \lbrack 0,1\rbrack$ describe how much topic prominence maps to the magnitude of a political stake. The stress-test location model is

\begin{equation}
s_{iP} = \mu_{0}(P) - \eta\rho_{P}a_{iP}z_{iP} + \varepsilon_{iP},\quad\varepsilon_{iP} \sim \mathcal{N}(0,\tau^{2}).
\end{equation}Conditional on topic $P$ and the assigned sign vector $\mathbf{z}$,
the residuals $\varepsilon_{iP}$ are independent across parties. This maintained conditional-independence assumption licenses the products of party-specific restraint probabilities below.

A positive sign lowers the veto-payoff location and moves the party toward restraint. A negative sign raises the location and moves it toward veto. The documents identify neither sign nor $\rho_{P}$. The reported maximal-transfer calculation sets $\rho_{P} = 1$. Lower values move all party locations toward $\mu_{0}(P)$ and compress the sign sensitivity within this mapping.

For a fixed sign vector $\mathbf{z}$,

\begin{equation}
F_{iP,\mathbf{z}}(s) = \Phi(\frac{s - \left\lbrack \mu_{0}(P) - \eta\rho_{P}a_{iP}z_{iP} \right\rbrack}{\tau}).
\end{equation}See main Methods (``Documentary prominence and sign scenarios'') for the
model and procedures.

The response map is increasing in every other party's restraint probability. This order property alone does not establish convergence of simultaneous vector iteration, which can cycle. Selected-limit statements below are conditional on convergence under the numerical protocol in main Methods.

At a selected vector fixed point $\mathbf{q}$,

\begin{equation}
P_{\text{pass}}\left( \mathbf{z}|\mathbf{q} \right) = \prod_{i}q_{i}.
\end{equation}Holding that fixed point constant and removing continuation gives

\begin{equation}
P_{\text{current}}\left( \mathbf{z}|\mathbf{q} \right) = \prod_{i}F_{iP,\mathbf{z}}\left( b\prod_{j \neq i}q_{j} \right).
\end{equation}Their difference is continuation-supported passage conditional on the
selected fixed point. Re-solving the system at $\delta = 0$ would answer a different counterfactual.

\section{Composition matters at fixed positive-prominence count}

Write $\ell_{i} = \mu_{0}(P) - \eta\rho_{P}a_{iP}z_{iP}$ for party $i$`s veto-payoff location.

Let $\mathcal{N}_{+} = \left\{ i:a_{iP} > 0 \right\}$ and let $k$ count negative signs only within $\mathcal{N}_{+}$. Parties with $a_{iP} = 0$ remain outside this positive-prominence opposition count because their assigned signs do not change outcomes under the maximal-transfer mapping. Every count-controlled statement below uses this definition of $k$.

\begin{proposition}[Ordered placements give sharp transition endpoints]

Let $\mathbf{\ell} \leq \mathbf{\ell'}$ componentwise, and start the two adaptive systems from the same $\mathbf{q}^{0}$. Then $\mathbf{q}^{t}\left( \mathbf{\ell'} \right) \leq \mathbf{q}^{t}\left( \mathbf{\ell} \right)$ at every iteration and at the selected limits whenever both exist. Under convergence, selected passage is weakly lower under $\mathbf{\ell'}$ at every $\delta$. For the endpoint statement, impose the maintained permutation-invariant selection protocol, in which parties share the same response parameters apart from $\ell_{i}$, every sign assignment uses common-scalar starts $\mathbf{q}^{0} = q_{0}\mathbf{1}$, and the same weighted draws of $q_{0}$ are used across assignments. Require convergence for the compared assignments and continuation values. Conditional on positive-prominence opposition count $k$, assigning negative signs to the $k$ most prominent parties then yields the latest passage transition, while assigning them to the $k$ least prominent parties in $\mathcal{N}_{+}$ yields the earliest. These attainable assignments are sharp endpoints for $\delta_{50}$ under that mapping and selection protocol, with unresolved crossings retained as boundary cases.

\end{proposition}

\begin{proof}

The normal response $T_{i}\left( \mathbf{q};\ell_{i} \right) = \Phi(\frac{b\prod_{j \neq i}q_{j} + \delta L - \ell_{i}}{\tau})$ is nondecreasing in every $q_{j}$ and decreasing in $\ell_{i}$. Monotonicity and induction preserve the componentwise order from a common start. The product of restraint probabilities preserves that order, which passes to convergent selected limits and their common weighted average. Every fixed passage threshold occurs weakly later under the higher location vector.

For the endpoint claim, consider prominences $a \geq b \geq 0$ and absorb $\rho_{P}$ into the common nonnegative scale $\eta$. Moving the negative sign from $b$ to $a$ changes the descending location pair from $\left( \mu_{0} + \eta b,\mu_{0} - \eta a \right)$ to $\left( \mu_{0} + \eta a,\mu_{0} - \eta b \right)$, increasing both coordinates. Repeated swaps give the largest sorted location vector by opposing the largest $k$ prominences, and the smallest by opposing the smallest positive $k$. Permutation invariance lets these sorted vectors be compared from the same scalar starts. Both extreme assignments belong to the admissible sign set, so the endpoints are attained.

\end{proof}

The ordering does not give sharp bounds for $P_{C}$. That quantity subtracts current-payoff passage, and raising a veto-payoff location can lower total passage while increasing the difference between total and current-payoff passage.

See main Methods (``Documentary prominence and sign scenarios'') for the model and procedures.

At tourism positive-prominence count $k = 3$, opposition among the three most specialised parties gives $\delta_{50} = 0.794$. Opposition among the eleven least specialised parties in the positive-prominence ranking gives $\delta_{50} = 0.797$. The assignments impose different positive-prominence objector counts but produce nearly the same passage transition. Every party retains an equal formal veto right, while the maximal-transfer mapping gives documentary specialisation a larger possible stake.

At fixed positive-prominence count, the ordered tourism gap averages $0.078$ in $\delta_{50}$, or $38.2\%$ of the full $0.203$ transition bound. The largest gap occurs at $k = 6$ and equals $0.115$, or $56.5\%$ of the full bound. Random placements usually lie between the endpoints. Their bands describe a constructed assignment distribution and are not confidence intervals.

\begin{figure}[htbp]
\centering
\includegraphics[width=1\linewidth,height=0.65\textheight,keepaspectratio]{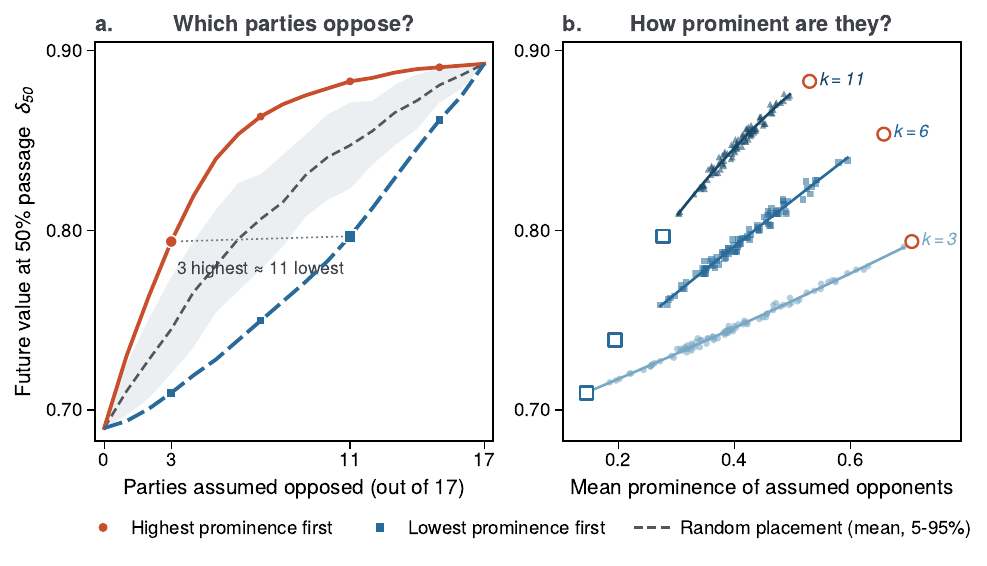}
\caption{Tourism stress test conditional on positive-prominence opposition count. The analyst imposes signs, and relative topic specialisation has no political sign. The ordered curves place opposition on the most or least specialised parties among the 17 with positive tourism prominence. Random placements show a constructed assignment distribution. The figure measures sensitivity and does not recover historical opposition.}
\label{fig:si-objector-composition}
\end{figure}

Prominence concentration adds information beyond count. For a nonempty opposed set $\mathcal{O} \subseteq \mathcal{N}_{+}$ of size $k$, define $w_{i} = \frac{a_{iP}}{\sum_{j \in \mathcal{O}}a_{jP}}$. Its Herfindahl concentration is

\begin{equation}
\sum_{i \in \mathcal{O}}w_{i}^{2} = \frac{1 + c^{2}}{k},
\end{equation}where $c$ is the coefficient of variation of prominence within the
opposed set, with variance computed using divisor $k$. The count fixes $\frac{1}{k}$, while within-set dispersion determines the remaining concentration.

\section{Every topic leaves a composition gap}

\begin{longtable}[]{@{}
  >{\raggedright\arraybackslash}p{(\linewidth - 8\tabcolsep) * \real{0.4393}}
  >{\centering\arraybackslash}p{(\linewidth - 8\tabcolsep) * \real{0.1028}}
  >{\centering\arraybackslash}p{(\linewidth - 8\tabcolsep) * \real{0.1495}}
  >{\centering\arraybackslash}p{(\linewidth - 8\tabcolsep) * \real{0.1495}}
  >{\centering\arraybackslash}p{(\linewidth - 8\tabcolsep) * \real{0.1589}}@{}}
\caption{Composition sensitivity at fixed counts of positive-prominence
parties across topics.}\label{tab:si-topic-bounds}\tabularnewline
\toprule\noalign{}
\begin{minipage}[b]{\linewidth}\raggedright
\textbf{Topic}
\end{minipage} & \begin{minipage}[b]{\linewidth}\centering
$n_{+}$
\end{minipage} & \begin{minipage}[b]{\linewidth}\centering
\textbf{Bound width}
\end{minipage} & \begin{minipage}[b]{\linewidth}\centering
\textbf{Mean share}
\end{minipage} & \begin{minipage}[b]{\linewidth}\centering
\textbf{Widest share}
\end{minipage} \\
\midrule\noalign{}
\endfirsthead
\toprule\noalign{}
\begin{minipage}[b]{\linewidth}\raggedright
\textbf{Topic}
\end{minipage} & \begin{minipage}[b]{\linewidth}\centering
$n_{+}$
\end{minipage} & \begin{minipage}[b]{\linewidth}\centering
\textbf{Bound width}
\end{minipage} & \begin{minipage}[b]{\linewidth}\centering
\textbf{Mean share}
\end{minipage} & \begin{minipage}[b]{\linewidth}\centering
\textbf{Widest share}
\end{minipage} \\
\midrule\noalign{}
\endhead
\bottomrule\noalign{}
\endlastfoot
Site Guidelines for Visitors & 12 & $0.187$ & $50\%$ & $67\%$ \\
Historic Sites and Monuments & 13 & $0.174$ & $36\%$ & $48\%$ \\
Educational issues & 17 & $0.261$ & $50\%$ & $70\%$ \\
Tourism and non-governmental activities & 17 & $0.203$ & $38\%$ &
$57\%$ \\
Management Plans & 19 & $0.178$ & $26\%$ & $38\%$ \\
Operation of the ATS & 22 & $0.238$ & $30\%$ & $43\%$ \\
Area Protection and Management Plans & 24 & $0.230$ & $27\%$ &
$40\%$ \\
Climate Change & 25 & $0.262$ & $33\%$ & $49\%$ \\
\end{longtable}

\protect\phantomsection\label{tbl:si-topic-bounds}{}

$n_{+}$ counts parties with positive prominence. Across eight topics, fixing the positive-prominence opposition count leaves an average of $26\%$ to $50\%$ of the full transition range open, and the widest count leaves $38\%$ to $70\%$ open. The table's mean and widest shares use the definitions in main Methods.

Under the four named sign scenarios, assigned signs account for $90.1\%$ of variation in $\delta_{50}$ and $81.5\%$ of variation in maximum $P_{C}$. Across eight alternative fair-sign menus, the corresponding shares range from $29.4\%$ to $82.4\%$ and from $12.0\%$ to $70.2\%$. These shares diagnose the selected scenario menu. They do not measure how much information the documents contribute.

The topic traces move largely in parallel across the named scenarios. Holding one scenario fixed, topics span $0.076$ in $\delta_{50}$ on average. Reassigning signs on tourism spans the sharp $0.203$ transition range. For maximum $P_{C}$, the corresponding descriptive ranges are $0.368$ across topics and $0.756$ across the ordered tourism paths. These scale comparisons depend on the chosen topic and sign menus and do not bound measurement error.

\section{Directional anchors narrow the range}

See main Methods (``Documentary prominence and sign scenarios'') for the model and procedures.

\begin{longtable}[]{@{}
  >{\raggedright\arraybackslash}p{(\linewidth - 8\tabcolsep) * \real{0.4145}}
  >{\centering\arraybackslash}p{(\linewidth - 8\tabcolsep) * \real{0.1093}}
  >{\centering\arraybackslash}p{(\linewidth - 8\tabcolsep) * \real{0.1587}}
  >{\centering\arraybackslash}p{(\linewidth - 8\tabcolsep) * \real{0.1675}}
  >{\centering\arraybackslash}p{(\linewidth - 8\tabcolsep) * \real{0.1499}}@{}}
\caption{Directional information supplied by co-sponsorship components
under the imposed equality restriction.}\label{tab:si-direction-anchors}\tabularnewline
\toprule\noalign{}
\begin{minipage}[b]{\linewidth}\raggedright
\textbf{Component}
\end{minipage} & \begin{minipage}[b]{\linewidth}\centering
\textbf{Members}
\end{minipage} & \begin{minipage}[b]{\linewidth}\centering
\textbf{Largest prominence}
\end{minipage} & \begin{minipage}[b]{\linewidth}\centering
\textbf{Worst remaining range}
\end{minipage} & \begin{minipage}[b]{\linewidth}\centering
\textbf{Share closed}
\end{minipage} \\
\midrule\noalign{}
\endfirsthead
\toprule\noalign{}
\begin{minipage}[b]{\linewidth}\raggedright
\textbf{Component}
\end{minipage} & \begin{minipage}[b]{\linewidth}\centering
\textbf{Members}
\end{minipage} & \begin{minipage}[b]{\linewidth}\centering
\textbf{Largest prominence}
\end{minipage} & \begin{minipage}[b]{\linewidth}\centering
\textbf{Worst remaining range}
\end{minipage} & \begin{minipage}[b]{\linewidth}\centering
\textbf{Share closed}
\end{minipage} \\
\midrule\noalign{}
\endhead
\bottomrule\noalign{}
\endlastfoot
Main component & 14 & $0.777$ & $0.067$ & $67.1\%$ \\
Ecuador & 1 & $0.670$ & $0.190$ & $6.3\%$ \\
Russia & 1 & $0.628$ & $0.192$ & $5.3\%$ \\
Australia & 1 & $0.075$ & $0.202$ & $0.5\%$ \\
\end{longtable}

\protect\phantomsection\label{tbl:si-direction-value}{}

The reference is the $0.203$ co-sponsorship-equality transition range for tourism. Any observation in the fourteen-party component closes at least $67.1\%$ of that range. Observing a prominent isolated party closes much less. Members of the same component share an information value because the restriction propagates their observed sign.

\section{Robustness preserves the placement result}

The alternative topic-share measure narrows the full tourism transition bound from $0.203$ to $0.116$, while the mean fixed-positive-prominence-count share rises from $38.2\%$ to $45.4\%$ and the widest share rises from $56.5\%$ to $64.4\%$. For maximum $P_{C}$, the corresponding mean and widest shares rise from $54\%$ to $59\%$ and from $77\%$ to $79\%$. Placement remains consequential under an activity-share measure.

The levelled-profile comparison assigns $91.9\%$ of $\delta_{50}$ variation and $93.8\%$ of maximum-$P_{C}$ variation to signs. The five synthetic-profile comparisons give $85.0\%$ and $70.9\%$. Two of those five profiles never cross the passage threshold, so the first figure averages the three resolved draws and the second uses all five. These checks preserve the qualitative named-menu result and do not validate the documentary mapping.

One-at-a-time changes in $\mu_{0}$, $\eta$, $\tau$, $m$, and $\kappa$ leave the mean transition share between $32.7\%$ and $44.1\%$ and the widest share between $48.5\%$ and $66.8\%$. The prior mean and concentration do not move the fold-based $\delta_{50}$ shares to three decimals in this design. Raising $\eta$ widens the sign-placement effect, while raising $\tau$ smooths and narrows the effect.

Uniform random placement at fixed positive-prominence count accounts for $8.9\%$ of $\delta_{50}$ variance and $18.2\%$ of maximum-$P_{C}$ variance within the stratified ensemble. These variance shares depend on sweeping the full positive-prominence count range and do not estimate population contributions.

The ordered endpoints are attainable but atypical under uniform random placement. The concentrated-opposition scenario sits on the most-prominent-first frontier at positive-prominence count six and has $\delta_{50} = 0.854$. The split-salient-interests scenario lies inside the random band at positive-prominence count nine and has $\delta_{50} = 0.834$.

Within a fixed positive-prominence count, $\delta_{50}$ rises with mean opposed prominence. The fitted sensitivity grows from $0.052$ at $k = 1$ to about $0.38$ at $k = 16$. The explained-variance decomposition assigns $86.2\%$ of explained $\delta_{50}$ variation to that count, $10.5\%$ to mean opposed prominence, and $3.3\%$ to dispersion at $R^{2} = 0.969$. For maximum $P_{C}$, the shares are $75.4\%$, $20.0\%$, and $4.6\%$ at $R^{2} = 0.960$. The constructed ensemble determines these summaries.

Subset sampling creates a separate theoretical shape artefact. Let $\gamma_{\text{pop}}$ denote the skew of the $n_{+} = 17$ positive prominence values, computed from population central moments. For a simple random sample of $k$ parties without replacement, the exact skew of the sample mean is

\begin{equation}
\gamma_{k} = \gamma_{\text{pop }}\frac{\sqrt{n_{+}\left( n_{+} - 1 \right)}}{n_{+} - 2}\frac{1 - 2f}{\sqrt{k(1 - f)}},\quad f = \frac{k}{n_{+}},\quad 1 \leq k < n_{+}.
\end{equation}The underlying prominence distribution has right skew, so $\gamma_{k}$
is positive for $1 \leq k \leq 8$ and negative for $9 \leq k \leq 16$. At $k = 17$, the sample mean is constant and its standardised skew is undefined. The empirical skew of the sampled subset clouds remains noisy. This sign change belongs to the assignment design and does not measure a political effect of opposition.

The fair-sign integration has largest across-batch $t$ half-width $0.030$, and largest difference between the 32-profile and 64-profile estimates $0.029$. These are numerical integration diagnostics for the constructed design, and its quadrature and scrambled-batch procedures are in main Methods.

\section{Numerical checks}

\protect\phantomsection\label{sec:si-numerical}{}

See main Methods (``Numerical solution and checks'') for the model and procedures.

The heterogeneous system raises $\delta_{50}$ by $0.011$ to $0.014$ across the three opposition scenarios and by $0.002$ under broad alignment. Passage thresholds $0.1$, $0.5$ and $0.9$ give the same transitions at reported precision because selected passage jumps across that interval at the fold.

Every sign profile in the batched placement calculation crosses within $\lbrack 0,0.999\rbrack$, so none is censored. The 36 ordered maximum-$P_{C}$ paths differ by less than $8 \times 10^{- 10}$ between the tested grid resolutions and the 241-point reference. Numerical grids and tolerances are specified in main Methods.

Beta basin weights remain nonnegative and sum to one. Outcome measures satisfy

\begin{equation}
0 \leq P_{C} \leq P_{\text{pass}} \leq 1.
\end{equation}$P_{C}$ equals zero when $\delta = 0$ or $L(q) = 0$. The Monte
Carlo comparison agrees with analytic basin weights and outcome probabilities within simulation error. All $2^{17} = 131,072$ tourism sign vectors satisfy the ordered location bounds in the exhaustive check. The proposition extends these location bounds to transitions, although this check does not solve every sign vector's transition.

As $\kappa$ grows, the Beta selection protocol concentrates on the stable basin containing its mean $m$, except when $m$ lies on a basin boundary. The general solver agrees with the constant and linear continuation wrappers in the numerical comparison.

\textbf{Empirical interpretation}

The unsigned ATS stress test identifies sensitivity to imposed direction under a maximal-transfer mapping. It does not estimate historical preferences, opposition, support, patience, or causal effects of documentary prominence. Ordered placements are attainable extrema within the stated sign set. The analyst constructs the random bands and named scenarios. Co-sponsorship narrows a range only after the analyst imposes component equality and obtains proposal-level direction.

\section{What the results establish}

The complete analysis has two hidden adjustment margins. Proposal-side accommodation changes the proposal before the response stage. Response-side restraint changes how parties act toward the proposal that reaches the table. A no-veto record can pool either margin with active support.

The one-shot game establishes observational equivalence and equilibrium multiplicity. The repeated games then impose a stronger requirement. A current veto must change future value, and every response and proposal prescribed after that veto must remain optimal. Strategic proposal choice creates the central interaction. The responder may accept a hard cooperative proposal because future cooperation matters, while the punishment proposer may escape the supporting threat by choosing a weaker accepted proposal. Menu restrictions, a bite floor, or a proposal-deviation penalty can close that escape under stated conditions.

Forgiveness creates a second non-monotonicity. Some continuation makes restraint worthwhile. More continuation can make an off-path veto unattractive when passage repairs the relationship. The maintained learning process adds selection. Parties can settle on harder proposals and fewer vetoes even when the modal rule fails the full off-path audit. Vetoes during the transition can filter the next menu or move the relationship into punishment. Apparent consensus can rise as accommodation falls under this learning process.

The ATS exercise addresses identification after the theory establishes these mechanisms. Unsigned documentary specialisation supplies heterogeneous magnitudes and leaves direction open. Counting assumed objectors narrows the set, but the prominence composition of that set still changes the passage transition. A proposal-level directional record can narrow the range, especially when an imposed component-equality restriction lets that observation anchor a large co-sponsorship component.

The analysis separates what a public record identifies, which complete strategies remain credible, and which convention the stated adaptive process selects. The strategic-proposer trigger theorem permits multiple responders, while the three-mode learning construction has two alternating parties. Extending the latter to multiple responders, private information, mixed or asymmetric strategies, endogenous legitimacy, imperfect public monitoring, or proposal-dependent agenda transitions requires a separate analysis.

\section{The model connects three literatures}

Non-opposition and unanimous private assent are already distinct concepts \citep{Urfalino2014} \citep{Kuran1995}. Costly dissent also has a formal treatment in which judges join a majority despite preferring the opposite disposition \citep{Fischman2011}. Judicial dissent does not block that majority's outcome, whereas veto blocks adoption here. Formal voting models already distinguish support, nonblocking opposition, and veto, but require an affirmative quorum as well as the absence of veto \citep{Bouton2018}. Our non-opposition rule permits an all-$Q$ response profile to pass. These institutional differences matter, but neither concealed disagreement nor three response labels is new.

The continuous-bite threshold applies the acceptance-constraint logic of agenda setting and legislative bargaining \citep{RomerRosenthal1978} \citep{BaronFerejohn1989} \citep{BondEraslan2010}. Under non-opposition, the acceptable set is the intersection of responder thresholds. Separating $A$ from $Q$ records whether passage contains active endorsement, but does not automatically add a strategic mechanism. When implementation and continuation coincide for $A$ and $Q$ after every relevant history and profile of others' responses, their fixed payoff ranking lets one non-veto action be eliminated outside ties when predicting passage. This reduction does not cover the discrete-instrument game, where support and acquiescence implement different outcomes under the binding instrument.

Future-supported participation and the interaction between bargaining and enforcement also have established foundations \citep{MaggiMorelli2006} \citep{Fearon1998}. The former studies self-enforcing voting rules with a given collective action and private participation costs. We instead fix non-opposition and examine strategic proposal demands and responses, without extending its private-information analysis. Dynamic legislative bargaining already makes acceptable revisions and future bargaining positions jointly determine current choices. Reconsideration replaces an approved default policy \citep{DiermeierFong2011}, while dynamic veto bargaining can retain the previous allocation after rejection \citep{Nunnari2021}. Our strategic-proposer construction instead tests a current offer within a prescribed public punishment state. Its institutional comparison concerns which proposal menus and deviation costs preserve that specified trigger, not the introduction of forward-looking proposal choice.

Proposal redesign must also be distinguished from renegotiation-proofness. A punishment can be subgame perfect yet vulnerable to mutually preferred replacement continuations \citep{FarrellMaskin1989}. Our redesign condition tests a profitable unilateral proposal deviation while holding the prescribed responder strategies and public transition rule fixed. Such an alternative offer can invalidate the candidate at the level of ordinary subgame perfection, without an agreement to replace the continuation strategy. We do not establish renegotiation-proofness. The forgiveness window also rejects one prescribed rule at high continuation weights. It does not reject every cooperative equilibrium or every punishment.

The analytical result is a conditional comparison of proposal institutions. In the normalised alternating construction, free redesign, a restricted instrument menu, a minimum substantive bite, and a proposal-deviation penalty produce different bounds for sustaining the same target rule. The learning audit then separates observed play from full incentive compatibility within its tested class. The unsigned ATS exercise addresses a different question, how much the modelled transition can vary when documentary magnitudes are known but opposition is assigned rather than observed. Neither the combination of established mechanisms nor that exercise establishes that the mechanisms caused particular historical decisions.

\section{Process records can narrow ambiguity}

For Antarctic decisions, researchers should pair a final instrument with proposal text, amendments, reservations, delays, changes in instrument strength, negotiating interventions, and later agenda recurrence. These records can narrow the candidate mechanism by locating when a proposal changed and which party expressed a direction. Recurrence is a cue to unresolved conflict or adaptive revision. It does not prove either interpretation. Proposal-level coding and triangulation remain necessary before attaching a direction to a party and matter.

The observation result applies to institutions where silence permits closure and maintained opposition can block closure. Related pressures appear in the Council of the European Union, World Trade Organization rounds, and climate negotiations, although their formal rules differ \citep{Heisenberg2005} \citep{Steinberg2002} \citep{Rietig2023}. Continuation-supported restraint requires more structure. Later cooperation must depend on current conduct, and a veto must credibly put enough future value at risk. Proposal-side accommodation requires a proposer with discretion over instrument form or bite.

The model does not rank consensus against majority rule. Consensus protects parties from open defeat and can preserve participation. The analysis adds legibility to the institutional trade-off, in which the same protection compresses support, restraint, and accommodation into one public endpoint.

\section{Provenance fixes every numerical claim}

\protect\phantomsection\label{sec:si-provenance}{}

\begin{longtable}[]{@{}
  >{\raggedright\arraybackslash}p{(\linewidth - 4\tabcolsep) * \real{0.2640}}
  >{\raggedright\arraybackslash}p{(\linewidth - 4\tabcolsep) * \real{0.3441}}
  >{\raggedright\arraybackslash}p{(\linewidth - 4\tabcolsep) * \real{0.3921}}@{}}
\caption{Executable sources and recorded outputs for the result
families.}\label{tab:si-provenance}\tabularnewline
\toprule\noalign{}
\begin{minipage}[b]{\linewidth}\raggedright
\textbf{Result family}
\end{minipage} & \begin{minipage}[b]{\linewidth}\raggedright
\textbf{Executable source}
\end{minipage} & \begin{minipage}[b]{\linewidth}\raggedright
\textbf{Recorded output}
\end{minipage} \\
\midrule\noalign{}
\endfirsthead
\toprule\noalign{}
\begin{minipage}[b]{\linewidth}\raggedright
\textbf{Result family}
\end{minipage} & \begin{minipage}[b]{\linewidth}\raggedright
\textbf{Executable source}
\end{minipage} & \begin{minipage}[b]{\linewidth}\raggedright
\textbf{Recorded output}
\end{minipage} \\
\midrule\noalign{}
\endhead
\bottomrule\noalign{}
\endlastfoot
Fixed response and private types & \nolinkurl{consensus_game.py},
\nolinkurl{identified_set.py} & \nolinkurl{test_consensus_game.py},
\nolinkurl{data/identified_set_*.csv} \\
Repeated and rotating agendas & \nolinkurl{repeated_game.py},
\nolinkurl{repeated_game_examples.py} &
\nolinkurl{test_repeated_game.py} \\
Procedural pruning & \nolinkurl{adaptive_agenda.py},
\nolinkurl{proposal_sequence_analysis.py} &
\nolinkurl{test_adaptive_agenda.py},
\nolinkurl{test_proposal_sequence_analysis.py} \\
Strategic proposer audit & \nolinkurl{strategic_proposer_game.py} &
\nolinkurl{test_strategic_proposer_game.py} \\
Forgiveness and learning & \nolinkurl{emergent_consensus.py},
\nolinkurl{run_emergent_consensus_analysis.py} &
\nolinkurl{data/emergent_*.csv},
\nolinkurl{data/emergent_analysis_summary.json} \\
Main-text learning audit & \nolinkurl{fig02_intervention_data.py},
\nolinkurl{fig02_sankey_data.py}, \nolinkurl{fig02_untested_agreement.py},
\nolinkurl{fig02d_policy_landscape_alt.py} &
\nolinkurl{data/fig02_intervention_deviations.csv},
\nolinkurl{data/fig02_sankey_flows_n1000.csv},
\nolinkurl{data/intervention_experiment/intervention_runs_n1000.csv} \\
Learning sensitivity and horizon &
\nolinkurl{parameter_sweep_learning.py},
\nolinkurl{explore_learning_horizon.py} &
\nolinkurl{data/parameter_sweep_learning/},
\nolinkurl{data/learning_horizon/} \\
Choice noise across continuation values &
\nolinkurl{temperature_continuation_check.py},
\nolinkurl{fig_temperature_continuation.py},
\nolinkurl{test_temperature_continuation_check.py} &
\nolinkurl{data/temperature_continuation_check/},
\nolinkurl{figures/fig_temperature_continuation.pdf} \\
Learning inside the credible window &
\nolinkurl{seeded_credible_check.py}, \nolinkurl{window_horizon_check.py}
& \nolinkurl{data/seeded_credible_check/},
\nolinkurl{data/window_horizon_check/} \\
Resumption after a dispute & \nolinkurl{intervention_experiment.py} &
\nolinkurl{data/intervention_experiment/} \\
Time course after a dispute & \nolinkurl{intervention_trajectories.py} &
\nolinkurl{data/intervention_trajectories/} \\
Disputes across settings & \nolinkurl{robustness_experiment.py},
\nolinkurl{band_edges.py}, \nolinkurl{split_edges.py},
\nolinkurl{fig_robustness.py}, \nolinkurl{scripts/plot_robustness_veto.py},
\nolinkurl{scripts/fig3_per_party_audit.py},
\nolinkurl{scripts/fig3d_repeated_per_party.py} &
\nolinkurl{data/{alignment,asymmetric,recovery,repeated,settings}_pairs.csv.gz},
\nolinkurl{data/band_edges.csv}, \nolinkurl{data/split_edges.csv},
\nolinkurl{data/robustness_2m_2026-11-05/{per_party_audit,repeated_per_party_23of24}.csv.gz} \\
Documented-response bound & \nolinkurl{observability_bound.py} &
\nolinkurl{test_observability_bound.py} \\
Revisions and new proposals & \nolinkurl{lineage_passage.py},
\nolinkurl{lineage_audit.py}, \nolinkurl{lineage_object_review.py} &
\nolinkurl{data/lineage_passage/} \\
Recorded disagreement & \nolinkurl{ats_shock_analysis.py},
\nolinkurl{ats_shock_robustness.py} & \nolinkurl{data/ats_shock_event/},
\nolinkurl{separate_coding/2026-11-01_identity_masked/} \\
Illustrative proposal histories &
\nolinkurl{fig03_record_through_model.py} &
\nolinkurl{data/fig03_case_lineages.csv}, cited meeting reports \\
ATS sign placement & \nolinkurl{empirical_interest_scenarios.py},
\nolinkurl{sign_mechanism.py}, \nolinkurl{topic_bounds.py} &
\nolinkurl{data/sign_mechanism_*.csv}, \nolinkurl{data/topic_bounds.csv} \\
Directional anchors & \nolinkurl{cosponsorship_bound.py},
\nolinkurl{party_linked_direction.py} &
\nolinkurl{data/cosponsorship_bound.csv},
\nolinkurl{data/party_linked_direction_value.csv} \\
Sensitivity and checks & \nolinkurl{measure_robustness.py},
\nolinkurl{parameter_sensitivity.py}, \nolinkurl{scenario_sensitivity.py} &
\nolinkurl{data/measure_robustness.csv},
\nolinkurl{data/parameter_sensitivity.csv},
\nolinkurl{data/scenario_menu_sensitivity.csv} \\
\end{longtable}

\protect\phantomsection\label{tbl:si-provenance}{}

\nolinkurl{BUILD.md} records the clean-build commands. The packaged actor-topic file is \nolinkurl{data/ats_actor_topic_records_2015_2024.csv}. Its unsigned profiles supply no inferred political directions. The separate statement analysis classifies reported speech, not latent preferences. The scripts, cached results and coding records named above distinguish the main-text analyses from the additional model illustrations.